\documentclass{aa}
\usepackage{txfonts}
\usepackage{graphicx}
\usepackage{caption}
\usepackage{subcaption}
\usepackage{color}

\usepackage{tikz}
\usetikzlibrary{arrows,shapes,backgrounds}

\begin{document} 

\titlerunning{}
\authorrunning{Masoura et al.}

\title{The $\it{XXL}$ survey: XL. Obscuration properties of red AGNs in $\it{XXL-N}$}

\author{V. A. Masoura \inst{1,2}, I. Georgantopoulos\inst{1}, G. Mountrichas\inst{3,1}, C. Vignali\inst{4,5}, E. Koulouridis\inst{1,6}, L. Chiappetti\inst{7},  S. Fotopoulou\inst{8}, S. Paltani\inst{9}, M. Pierre\inst{6}}
          
    \institute {National Observatory of Athens, V.  Paulou  \& I.  Metaxa, 15 236 Penteli, Greece
              \email{vmasoura@noa.gr}
           \and
             Section of Astrophysics, Astronomy and Mechanics, Department of Physics, Aristotle University of Thessaloniki, 54 124, Thessaloniki, Greece
             \and
             Instituto de Fisica de Cantabria (CSIC-Universidad de Cantabria), Avenida de los Castros, 39005 Santander, Spain
             \and
             Dipartimento di Fisica e Astronomia, Alma Mater Studiorum, Università degli Studi di Bologna, Via Gobetti 93/2, I-40129 Bologna, Italy
             \and
             INAF - Osservatorio di Astrofisica e Scienza dello Spazio di Bologna, Via Gobetti 93/3, I-40129 Bologna, Italy
             \and
              AIM, CEA, CNRS, Université Paris-Saclay, Université Paris Diderot, Sorbonne Paris Cité, F-91191 Gif-sur-Yvette, France
              \and
             INAF, IASF Milano, via Corti 12, 20133 Milano, Italy            
             \and
             Centre for Extragalactic Astronomy, Department of Physics, Durham University, Durham DH1 3LE, UK
              \and
             Department of Astronomy, University of Geneva, ch. d’ \'Ecogia 16, CH-1290 Versoix, Switzerland}

\abstract {The combination of optical and mid-infrared (MIR) photometry has been extensively used to select red active galactic nuclei (AGNs). Our aim is to explore the obscuration properties of these red AGNs with both X-ray spectroscopy and spectral energy distributions (SEDs). In this study, we re-visit the relation between optical/MIR extinction and X-ray absorption. We use IR selection criteria, specifically the $W1$ and $W2$ WISE bands, to identify 4798 AGNs in the $\it{XMM-XXL}$ area ($\sim 25$\,deg$^2$). Application of optical/MIR colours ($r- W2 > 6$) reveals 561 red AGNs (14$\%$). Of these, 47 have available X-ray spectra with at least 50 net (background-subtracted) counts per detector. For these sources, we construct SEDs from the optical to the MIR using the CIGALE code. The SED fitting shows that 44 of these latter 47 sources present clear signs of obscuration based on the AGN emission and the estimated inclination angle. Fitting the SED also reveals ten systems ($\sim20\%$) which are dominated by the galaxy. In these cases, the red colours are attributed to the host galaxy rather than AGN absorption. Excluding these ten systems from our sample and applying X-ray spectral fitting analysis shows that up to $76\%$ (28/37) of the IR red AGNs present signs of X-ray absorption. Thus, there are nine sources ($\sim20\%$ of the sample) that although optically red, are not substantially X-ray absorbed. Approximately $50\%$ of these sources present broad emission lines in their optical spectra. We suggest that the reason for this apparent discrepancy is that the r-W2 criterion is sensitive to  smaller amounts of obscuration relative to the X-ray spectroscopy. In conclusion, it appears that the majority of red AGNs present considerable obscuration levels as shown by their SEDs. Their X-ray absorption is moderate with a mean of $\rm N_H \sim 10^{22}\, \rm{cm^{-2}}$.}



\maketitle   
\section{Introduction}
The vast majority of galaxies host a supermassive black hole (SMBH) in their centre \citep[e.g.][]{Kormendy1996, johnkormendy1996}. The masses of these SMBHs range from hundreds of thousands to billions of solar masses ($\approx 10^5 - 10^{10}M_{\bigodot}$). When a SMBH grows, it releases huge amounts of energy across the electromagnetic spectrum and becomes visible as an AGN. These latter objects have characteristic properties such as high luminosities (up to $L_{\rm bol}\simeq {10}^{48}~$ $\rm{erg s^{-1}}$) and rapid variability \citep[e.g.][]{Ulrich1997}.

According to the unification scheme \citep{Antonucci1993} the central source of the AGN structure consists of the SMBH, the corona, and the accretion disc. These regions are surrounded by an axisymmetric dusty structure that is toroidal in shape \citep[e.g.][]{Netzer2015}. Starting from the inner part of the source, the X-ray and the optical/ultraviolet (UV) wavelengths are related to the corona and the accretion disc respectively \citep[e.g.][]{Wilkins2015, Haardt1991}. The torus is responsible for the infrared (IR) emission  \citep[e.g.][]{Edelson1986}, because the hot dust is related to the radiation at these wavelengths. There are different torus models based on the distribution of dust, such as smooth \citep[e.g.][]{Fritz2006}, clumpy \citep{Nenkova2008, Nenkova2008a, Hoenig2010, Hoenig2010a, Hoenig2017}, and a combination of the two \citep{Siebenmorgen2015, Stalevski2016}. Unified models propose that the different observational classes of AGNs are nevertheless a single type of physical object observed at different viewing angles (e.g. different orientations of the observer with respect to the torus). Thus, Type 1 and Type 2 AGNs refer to the face-on (unobscured) and edge-on (obscured) AGNs, respectively. However, in the case of a clumpy torus, the classification of AGN type is not necessarily a consequence of a difference in viewing angle, because the dust is not distributed homogeneously. Based on their optical/UV spectrum, Type 1 AGNs present broad permitted emission lines, while Type 2 objects show only narrow lines. There is a continuum of objects in between (intermediate types), where the broad line components are increasingly difficult to observe.

Despite the huge radiative power of AGNs, obscuration presents an important challenge for uncovering their complete population and explaining the complicated mechanisms that regulate these systems \citep{Hickox2018}. Regarding the different obscuring material, X-ray energies are obscured by gas, whereas UV, optical, and IR wavelengths are extincted by dust. A reliable method to identify and quantify obscuration is through X-ray observations. Modelling of the X-ray spectrum provides a robust estimation of the $\rm N_H$ column density.

The combination of optical and IR data also provides us with a powerful tool to search for obscured AGNs. The optical emission of a SMBH is attenuated by dust and is re-emitted in the NIR to MIR wavelengths. In this case the galaxy appears faint in the optical but bright in the IR \citep{LaMassa2016a}. Various optical and IR colour criteria have been used in the literature to identify red AGNs, using either $\it{Spitzer}$ \citep[e.g.][]{Fiore2009, Hickox2007, LaMassa2016a, Donoso2014} or WISE \citep{Yan2013} IR bands.

Previous studies have found a correlation between the optical/IR colours and X-ray absorption  \citep[e.g.][]{Civano2012}. On the other hand, \cite{koutoulidis2018} found that reddened AGNs are equally divided into X-ray absorbed and unabsorbed. Specifically, these latter authors used about 1500 X-ray AGNs from five deep {\it{Chandra}} fields (CDF-N, CDF-S, ECDF-S, COSMOS, AEGIS) and divided them into obscured and unobscured sources using X-ray (Hardness Ratio) and optical/mid-infrared(MIR) criteria (R-[4.5]= 6.1). Nevertheless their sources span low luminosities ($10^{42} - 10^{43}$  $\rm{erg s^{-1}}$). Based on their findings Koutoulidis et al. suggested that host galaxy contamination in the MIR bands affects the optical/MIR AGN classification.

We examine whether or not red AGNs are also absorbed in X-rays. To this aim, we use X-ray AGNs in the $\it{XXL}$ survey \citep[][hereafter XXL paper I]{Pierre2016} and select IR AGN candidates by applying the criteria of \cite{Assef2018}. We use optical and MIR colours \citep{Yan2013} to select optically red sources. The final sample is restricted to those AGNs for which X-ray spectroscopy is available (47 sources). In section 3.1 we fit the X-ray spectra to quantify the X-ray absorption of the sources. In addition we construct spectral energy distributions (SEDs) using optical, near-infrared(NIR), and MIR photometry. We use the CIGALE code \citep{Ciesla2015} to fit the SEDs and get an estimation of various absorption indices (AGN emission, torus inclination). We also complement our analysis with available optical spectra (see section 3.3). Finally, we compare the results of different wavelengths and techniques for each red AGN.

\section{Data}
The data of this study come from the $\it{XXL}$ survey. In this section we describe the survey as well as 
the sample selection.

\subsection{$\it{XXL}$}
The $\it{XXL}$ survey is a medium-depth X-ray survey that covers a total area of 50 deg$^2$. It covers two fields of nearly equal size, the $\it{XXL}$ North ($\it{XXL-N}$) and the $\it{XXL}$ South ($\it{XXL-S}$). Furthermore, it is the largest XMM-Newton project approved to date (>6 Msec) with median exposure at 10.4 ks and a depth of $\sim$ 6 $\times$ 10$^{-15}$ erg~sec$^{-1}$ cm$^{-2}$ for point sources at the 90$\%$ completeness limit in the [0.5-2] keV band (XXL paper I). In this study, we use the $\it{XXL-N}$ sample that consists of 14\,168 sources, including extended objects. To identify the X-ray detections at other wavelengths the X-ray counterparts have been crossmatched with optical, NIR, and MIR surveys \citep[for more details see][hereafter XXL paper XXVII]{Chiappetti2018}.

\subsection{Infrared AGN candidates} 
WISE \citep{Wright2010} completed an all-sky coverage in four MIR bands: 3.4, 4.6, 12, and 22 $\mu$m ($W1$, $W2$, $W3$ and $W4$ bands, respectively). Various colour criteria used these IR bands to efficiently identify AGN candidates. For instance, \cite{Mateos2012} suggested a selection method using three WISE colours. \cite{Stern2012} used the $W1$ and $W2$ bands and applied the criterion $W1 - W2 \geq 0.8$ to select AGNs with $W2 < 15.05$ in the COSMOS field. \cite{Assef2013} extended the latter criterion and provided a selection of AGNs for fainter WISE sources using the WISE All-Sky data release catalogue. \cite{Assef2018} modified these criteria to incorporate data obtained during the post-cryogenic main mission extension (AllWISE catalogue).

In this study, we use approximately 500\,000 WISE detections included in the latest WISE catalogue (AllWISE) that lie within the XMM-XXL area to select IR AGN candidates, applying the criteria of \cite{Assef2018}. Specifically, we use\\[0.3cm]
\begin{equation}
W1-W2 \geq \alpha_R,  \,\,\,\,\,\,W2 \leq \gamma_{R}
,\end{equation}
\begin{equation}
W1-W2 > \alpha_R~ \exp~[~{\beta_{R}~{(W2 - \gamma_{R})}^2}~],\,\,W2 > \gamma_{R} 
,\end{equation} \\
where ($\alpha_R$ , $\beta_{R}$ ,  $\gamma_{R}$) = (0.650, 0.153, 13.86), to select AGNs with 90$\%$ reliability. 

\subsection{Red AGNs}
\label{sec_red}

To select optically red sources from the IR AGN candidates, we apply  the \cite{Yan2013} criterion ($r- W2 > 6$). This analysis reveals 561 red AGNs. We cross-match this subsample with the $\it{XXL}$ catalogue to identify the X-ray counterparts of these sources. For the cross-match, we use a radius of 3 arcsecs. Within this radius and given the X-ray source sky density, we find that a fraction of $\sim$ 0.2$\%$ of spurious matches is expected \citep{Mountrichas2017}. To study the X-ray absorption of the optically red AGN population, we apply an X-ray spectral fitting analysis. As one of our main goals is to quantify the obscuration in the red AGNs, we chose to analyse only the X-ray spectra of the sources with reliable photon statistics. For this reason, we keep only the sources with 50  or more net counts per detector \citep[see][]{Corral2015}.
There are 47 red AGNs that meet the aforementioned X-ray criteria. Table \ref{sample} presents the number of sources in the various subsamples. Table \ref{general_properties} presents the identity (ID), the redshift, and the r and W2 magnitudes in the Vega system for the 47 optically red AGNs. In addition to optical and MIR photometry, 45 out of the 47 optically red AGNs have also been detected in the NIR \citep[VISTA;][]{Emerson2006, Dalton2006}. 
The XXL field has been observed by the SPIRE instrument onboard {\it Herschel} mission at far-infrared (FIR) wavelengths \citep{Oliver2012}. To identify the FIR counterparts, we used the ARCHES cross-correlation tool {\sc XMATCH} \citep{Pineau2017}. The details of the cross-match procedure are given in Section 2.2 of \cite{Masoura2018}.
We find nine sources with {\it Herschel} photometry. Finally, optical spectra are available for 33 out of the 47 red AGNs. The vast majority of them (30 out of 33) come from the SDSS (DR15) survey. The remaining optical spectra are from the Galaxy And Mass Assembly \citep[GAMA;][]{Driver2011, Baldry2010} and the VIPERS \citep{Guzzo2014,Scodeggio2018} surveys. In our analysis, we use spectroscopic redshifts for 33 sources. For the remaining AGNs we use their photometric redshifts, estimated in \citet[hereafter XXL paper VI]{Fotopoulou2016}. The photometric redshift accuracy is 0.095 (for the full XMM-XXL catalogue). In our analysis, we incorporate the full probability density function (PDF) of the photometric redshifts when we calculate the uncertainties of the various parameters.

\begin{table*}
\caption{Number of IR-selected AGNs with X-ray and optical observations and spectra in the XMM-XXL field.}

\centering
\setlength{\tabcolsep}{2.3mm}
\begin{tabular}{cccc}
      \hline
&Total number& X-ray Detections &X-ray Spectra\\
       \hline
IR AGNs&4798&1503&312\\
\hline
IR AGNs with SDSS&2652&1268&262\\
\hline
Red AGNs&561&135&47\\
\hline
\label{sample}
\end{tabular}
\end{table*}

\begin{table*}
\caption{General properties of the red AGN sample.}
\centering
\setlength{\tabcolsep}{2.3mm}
\begin{tabular}{ccccc}
      \hline
Object&3XLSS ID&redshift&r (Vega)&W$_2 (\rm{Vega})$\\
       \hline
1&J021835.7-053758&0.387&19.84&13.56\\
2&J022848.4-044426&1.046&21.14&15.03\\
3&J022928.4-051124&0.307&17.76&11.65\\
4&J022809.0-041235&0.879&19.46&13.01\\
5&J020654.9-064552&1.412&21.12&14.82\\
6&J020410.4-063924&0.414&21.83&14.65\\
7&J023315.5-054747&0.598&20.91&13.70\\
8&J021337.9-042814&0.419&18.54&12.17\\
9&J020436.4-042833&0.827&19.77&13.46\\
10&J020543.0-051656&0.653&20.78&13.93\\
11&J020517.3-051024&0.792&21.05&14.92\\
12&J022244.3-030525&0.637&21.74&14.39\\
13&J022323.4-031157&0.691&20.42&13.90\\
14&J022209.6-025023&0.400&21.36&14.51\\
15&J022750.7-052232&0.804&22.26&13.98\\
16&J022758.4-053306&0.956&21.49&13.74\\
17&J022453.2-054050&0.488&20.76&13.32\\
18&J022258.8-055757&0.732&21.60&13.70\\
19&J021844.6-054054&0.671&21.19&13.88\\
20&J021523.2-044337&0.860&19.96&13.21\\
21&J022321.9-045739&0.779&21.34&13.83\\
22&J022443.6-050905&0.943&21.55&14.93\\
23&J023418.0-041833&0.582&22.57&15.07\\
24&J020543.7-063807&0.772&22.72&14.79\\
25&J022932.6-055438&1.263&19.70&13.53\\
26&J021239.2-054816&0.711&22.32&15.10\\
27&J021511.4-060805&0.772&22.34&14.19\\
28&J021808.8-055630&0.572&20.05&14.06\\
29&J022538.9-040821&0.733&21.28&14.68\\
30&J020845.1-051354&$\it{1.46}$&23.39&15.16\\
31&J020953.9-055102&0.642&20.91&14.11\\
32&J020806.6-055739&0.539&21.34&13.83\\
33&J021509.0-054305&$\it{0.70}$&22.48&14.53\\
34&J021512.9-060558&$\it{1.14}$&23.36&15.14\\
35&J020529.5-051100&$\it{1.20}$&23.40&14.94\\
36&J022404.0-035730&$\it{0.17}$&22.79&14.88\\
37&J023501.0-055234&$\it{0.85}$&22.49&15.06\\
38&J020210.6-041129&$\it{2.71}$&21.09&14.08\\
39&J022650.3-025752&$\it{1.21}$&22.79&14.63\\
40&J021303.7-040704&$\it{1.35}$&22.20&15.32\\
41&J023357.7-054819&$\it{1.60}$&24.20&14.69\\
42&J020335.0-064450&$\it{0.67}$&23.28&15.05\\
43&J020823.4-040652&$\it{2.23}$&24.68&14.92\\
44&J020135.4-050847&$\it{0.73}$&23.14&15.16\\
45&J021837.2-060654&0.943&21.47&15.11\\
46&J022149.9-045920&1.461&23.05&14.79\\
47&J020311.3-063534&$\it{1.05}$&22.96&15.19\\
\hline
\label{general_properties}
\end{tabular}
\tablefoot{r is the SDSS photometric band \citep[SDSS DR15;][]{Aguado2019}. r and W$_2$ magnitudes are on the Vega system. We use three decimal points for spectroscopic redshifts and photometric redshifts are in italics with two decimal points.}
\end{table*}

\section{Analysis}
We classify the 47 optically red AGNs of our sample as obscured or unobscured using different criteria. Specifically, we examine their X-ray spectra, SEDs, and optical spectra.

\begin{table*}
\caption{X-ray properties of the red AGN sample.}
\centering
\setlength{\tabcolsep}{2.8mm}
\begin{tabular}{ccccccc}
      \hline
Object&3XLSS ID&$\rm N_H\times10^{22}cm^{-2}$& $\rm N_H\times10^{22}cm^{-2}$ &$\Gamma$&$\rm log~L_X (\rm 2-10~keV)$&C-stat/d.f.\\
&&& $\Gamma=1.8$&&$\rm{erg~s^{-1}}$&\\
       \hline
1&J021835.7-053758&0.15$^{+0.06}_{-0.06}$&-&1.85$^{+0.15}_{-0.15}$&8.7$\times10^{43}$&802/1027\\
2&{\it{J022848.4-044426}}&$<1.46$&-&1.77$^{+1.40}_{-0.60}$&1.1$\times10^{44}$&33/38\\
3&{\bf{J022928.4-051124}}&1.60$^{+0.10}_{-0.08}$&-&2.24$^{+1.20}_{-0.30}$&2.9$\times10^{43}$&512/578\\
4&J022809.0-041235&0.14$^{+0.12}_{-0.11}$&-&1.87$^{+0.21}_{-0.20}$&1.8$\times10^{45}$&488/577\\
5&J020654.9-064552&$<0.23$&-&1.75$^{+0.26}_{-0.25}$&3.8$\times10^{43}$&103/155\\
6&{\it{J020410.4-063924}}&0.81$^{+0.89}_{-0.68}$&-&1.86$^{+0.70}_{-0.58}$&7.8$\times10^{43}$&106/138\\
7&J023315.5-054747&$<0.62$&-&1.94$^{+0.70}_{-0.40}$&3.2$\times10^{43}$&89/107\\
8&{\bf{J021337.9-042814}}&12.1$^{+7.20}_{-5.37}$&-&1.60$^{+0.60}_{-0.30}$ &4.8$\times10^{43}$&63/70\\
9&J020436.4-042833&$<0.45$&-&1.86$^{+0.25}_{-0.24}$&3.7$\times10^{44}$&510/550\\
10&{\bf{J020543.0-051656}}&$<48.0$&-&1.80$^{+8.10}_{-1.50}$&4.4$\times10^{43}$&17/18\\
11&{\it{J020517.3-051024}}&0.78$^{+1.10}_{-0.70}$&-&1.24$^{+3.10}_{-0.90}$&2.5$\times10^{44}$&22/46\\
12&{\bf{J022244.3-030525}}&23.8$^{+18.30}_{-12.70}$&-&1.51$^{+1.20}_{-1.0}$&1.4$\times10^{44}$&112/117\\
13&{\bf{J022323.4-031157}}&17.4$^{+23.60}_{-13.00}$&-&2.70$^{+0.40}_{-0.20}$&7.4$\times10^{43}$&27/39\\
14&{\it{J022209.6-025023}}&0.80$^{+0.40}_{-0.30}$&-&1.90$^{+0.40}_{-0.30}$&3.5$\times10^{43}$&211/270\\
15&{\bf{J022750.7-052232}}&5.00$^{+3.10}_{-2.00}$&-&2.40$^{+0.80}_{-0.70}$&1.2$\times10^{44}$&154/179\\
16&J022758.4-053306&$<0.10$&-&1.40$^{+0.90}_{-0.50}$&1.6$\times10^{44}$&33/44\\
17&{\bf{J022453.2-054050}}&5.80$^{+4.20}_{-3.00}$&-&1.40$^{+0.70}_{-0.20}$&3.4$\times10^{43}$&56/66\\
18&J022258.8-055757&$<0.20$&-&1.60$^{+2.60}_{-0.90}$&5.0$\times10^{43}$&65/69\\
19&J021844.6-054054&$<0.74$&-&1.80$^{+0.70}_{-0.40}$&1.1$\times10^{44}$&146/157\\
20&J021523.2-044337&$<0.28$&-&2.10$^{+0.80}_{-0.20}$&3.6$\times10^{43}$&130/141\\
21&{\bf{J022321.9-045739}}&15.10$^{+3.1}_{-1.40}$&-&1.67$^{+2.80}_{-1.80}$&1.5$\times10^{44}$&93/92\\
22&{\it{J022443.6-050905}}&$<1.10$&-&1.33$^{+0.48}_{-0.35}$&8.1$\times10^{43}$&142/172\\
23&J023418.0-041833&$<0.40$&-&1.84$^{+0.36}_{-0.45}$&2.3$\times10^{43}$&15/16\\
24&{\bf{J020543.7-063807}}&$<3.90$&8.1$^{+2.2}_{-5.0}$&1.18$^{+0.54}_{-0.36}$&6.5$\times10^{44}$&135/145\\
25&J022932.6-055438&$<0.43$&-&1.50$^{+0.50}_{-0.40}$&2.3$\times10^{44}$&62/82\\
26&{\bf{J021239.2-054816}}&11.40$^{+0.30}_{-0.20}$&-&1.80$^{+0.60}_{-0.20}$&6.0$\times10^{43}$&10/11\\
27&{\bf{J021511.4-060805}}&$<22$&16.0$^{+3.8}_{-9.2}$&1.00$^{+1.80}_{-1.50}$&2.4$\times10^{44}$&48/50\\
28&J021808.8-055630&0.16$^{+0.07}_{-0.06}$&-&1.98$^{+0.22}_{-0.24}$&9.2$\times10^{43}$&92/104\\
29&{\bf{J022538.9-040821}}&4.95$^{+4.80}_{-4.30}$&-&1.60$^{+0.80}_{-0.30}$&2.3$\times10^{43}$&31/34\\
30&{\it{J020845.1-051354}}&$<1.00$&-&1.90$^{+0.70}_{-0.40}$&2.5$\times10^{44}$&73/89\\
31&{\bf{J020953.9-055102}}&$<28$&-&1.60$^{+0.6}_{-0.8}$&8.7$\times10^{43}$&37/30\\
32&{\bf{J020806.6-055739}}&1.80$^{+0.30}_{-0.18}$&-&2.30$^{+1.20}_{-0.40}$&2.0$\times10^{43}$&80/56\\
33&{\bf{J021509.0-054305}}&$<33.00$&-&1.33$^{+3.10}_{-1.90}$&3.3$\times10^{43}$&30/28\\
34&{\it{J021512.9-060558}}&$<1.70$&-&1.71$^{+0.70}_{-0.50}$&1.7$\times10^{44}$&103/123\\
35&{\it{J020529.5-051100}}&$<1.47$&-&1.35$^{+0.80}_{-0.40}$&1.6$\times10^{44}$&86/90\\
36&{\bf{J022404.0-035730}}&5.30$^{+7.15}_{-3.09}$&-&1.82$^{+0.40}_{-0.70}$&9.3$\times10^{43}$&52/59\\
37&{\it{J023501.0-055234}}&<5.40&-&1.42$^{+0.30}_{-0.60}$&3.5$\times10^{43}$&33/23\\
38&{\it{J020210.6-041129}}&<4.20&-&1.22$^{+0.90}_{-0.40}$&1.8$\times10^{45}$&8/12\\
39&{\bf{J022650.3-025752}}&9.00$^{+1.37}_{-2.94}$&-&1.83$^{+0.8}_{-0.4}$&3.5$\times10^{44}$&56/70\\
40&{\bf{J021303.7-040704}}&$<14.0$&19.2$^{+3.9}_{-3.0}$&0.70 $^{+1.10}_{-0.80}$&1.8$\times10^{44}$&32/46\\
41&{\it{J023357.7-054819}}&$<2.30$&-&1.60$^{+0.40}_{-0.40}$&1.1$\times10^{45}$&192/259\\
42&{\it{J020335.0-064450}}&$<2.50$&-&1.50$^{+1.38}_{-0.42}$&5.4$\times10^{43}$&44/57\\
43&{\bf{J020823.4-040652}}&$<0.50$&1.9$^{+0.6}_{-1.6}$&1.20$^{+1.00}_{-0.70}$&3.2$\times10^{44}$&55/58\\
44&J020135.4-050847&$<0.81$&-&1.35$^{+0.80}_{-0.30}$&5.6$\times10^{43}$&86/90\\
45&{\it{J021837.2-060654}}&0.93$^{+0.05}_{-0.28}$&-&2.00$^{+0.40}_{-0.30}$&3.8$\times10^{43}$&158/172\\
46&{\bf{J022149.9-045920}}&$<1.70$&2.6$^{+0.5}_{-1.2}$&1.10$^{+1.20}_{-0.80}$&3.1$\times10^{42}$&40/49\\
47&{\bf{J020311.3-063534}}&3.20$^{+14.42}_{-2.85}$&-&1.60$^{+1.50}_{-0.90}$&9.3$\times10^{43}$&20/30\\
\hline
\label{x_ray}
\end{tabular}
\tablefoot {When $\Gamma \leq 1.2$, we quote the $\rm N_H$ estimations with $\Gamma$ fixed to $\Gamma=1.8$  and present the $\rm N_H$ estimations in column 4. Sources classified as X-ray obscured based on our strict criteria (see text for more details) are presented in bold. Sources that satisfy the loosened X-ray criteria (see text) are shown in italics. The energy band of $L_X$ is [2-10keV].}
\end{table*}

\begin{table*}
\caption{The models and the values for their free parameters used by CIGALE for the SED fitting of our X-ray samples.}
\centering
\setlength{\tabcolsep}{0.1mm}
\begin{tabular}{lc}
       \hline
Parameter &  Model/values \\
        \hline
\multicolumn{2}{c}{stellar population synthesis model} \\
\\
initial mass function & Salpeter \\
metallicity & 0.02 (Solar) \\
single stellar population library & Bruzual \& Charlot (2003) \\
\hline
\multicolumn{2}{c}{double exponentially decreasing (2$\tau$ dec) model} \\
\\
$\tau$ & 100, 1000, 5000, 10000 \\ 
age & 500, 2000, 5000, 10000, 12000 \\
burst age & 100, 200, 400 \\
\hline
\multicolumn{2}{c}{Dust extinction} \\
\\
dust attenuation law & \cite{Calzetti2000} \\
reddening E(B-V) & 0.01, 0.1, 0.3, 0.5, 0.8, 1.2 \\ 
E(B-V) reduction factor between old and young stellar population & 0.44 \\
\hline
\multicolumn{2}{c}{\cite{Fritz2006} model for AGN emission} \\ 
\\
ratio between outer and inner dust torus radii & 10, 60, 150 \\
9.7 $\mu m$ equatorial optical depth & 0.1, 0.3, 1.0, 2.0, 6.0, 10.0 \\
$\beta$ & -0.5 \\
$\gamma$ & 0.0, 2.0, 6.0 \\
$\Theta$ & $100^{\rm o}$ \\
$\Psi$ & $0^{\rm o}$, $10^{\rm o}$, $20^{\rm o}$, $30^{\rm o}$, $40^{\rm o}$, $50^{\rm o}$, $60^{\rm o}$, $70^{\rm o}$, $80^{\rm o}$, $90^{\rm o}$\\
AGN fraction & 0.1, 0.2, 0.3, 0.4, 0.5, 0.6, 0.7, 0.8, 0.9 \\
\hline
\label{table_cigale}
\end{tabular}
\tablefoot{$\tau$ is the e-folding time of the main stellar population model in Myr, age is defined as the age of the main stellar population in the galaxy in Myr, and burst age is the age of the late burst in Myr. Here, $\beta$ and $\gamma$ are the parameters used to define the law for the spatial behaviour of density of the torus density. The functional form of the latter is $\rho (r, \theta) \propto  r^\beta e^{-\gamma | cos \theta |}$, where r and $\theta$ are the radial distance and the polar distance, respectively. $\Theta$ is the opening angle and $\Psi$ the viewing angle of the torus. Type 2 AGNs have $\Psi \leq 30^{\rm o}$ (edge-on) and Type 1 AGN have $\Psi \geq 70^{\rm o}$ (face-on). $40^{\rm o} \leq \Psi \leq 60^{\rm o}$ corresponds to intermediate-type AGNs. The AGN fraction is measured as the AGN emission relative to IR luminosity ($1-1000\,\rm\mu m$).}
\end{table*}

\begin{table*}
\caption{SED properties of the red AGN sample.}
\centering
\setlength{\tabcolsep}{2.3mm}
\begin{tabular}{ccccc}
      \hline
Object&3XLSS ID&$\Psi (\rm{degrees)}$ & $\rm{frac_{AGN}}$ & $\chi^2_{red}$\\
       \hline
1&J021835.7-053758&$20$&  $0.30^{+0.10}_{-0.10}$&  0.43\\
2&J022848.4-044426&$40$ & $0.30^{+0.10}_{-0.10}$ &6.35\\
3&J022928.4-051124&     $20$ & $0.40^{+0.10}_{-0.10}$&4.42\\
4&J022809.0-041235& $80$ & $0.30^{+0.15}_{-0.15}$&1.52\\
5&J020654.9-064552 &$20$& $0.30^{+0.15}_{-0.15}$&4.23\\
6&J020410.4-063924&$40$& $0.50^{+0.15}_{-0.15}$&2.99\\
7&J023315.5-054747&$20$& $0.30^{+0.10}_{-0.10}$&0.48\\
8&J021337.9-042814&$10$& $0.60^{+0.05}_{-0.05}$&2.50\\
9&J020436.4-042833&$20$& $0.60^{+0.15}_{-0.15}$&3.17\\
10&J020543.0-051656&$20$ &$0.50^{+0.05}_{-0.05}$&0.47\\
11&J020517.3-051024&$70$&$0.20^{+0.10}_{-0.10}$&5.09\\
12&J022244.3-030525 &$40$&$0.40^{+0.15}_{-0.15}$&0.43\\
13&J022323.4-031157&$20$&$0.40^{+0.20}_{-0.20}$&3.00\\
14&J022209.6-025023&$30$&$0.30^{+0.15}_{-0.15}$&2.71\\
15&J022750.7-052232&$10$&$0.30^{+0.10}_{-0.10}$&3.28\\
16&J022758.4-053306&$0$&$0.60^{+0.05}_{-0.05}$ &3.13\\
17&J022453.2-054050&$10$&$0.40^{+0.15}_{-0.15}$&0.70\\
18&J022258.8-055757&$20$&$0.50^{+0.05}_{-0.05}$&2.72\\
19&J021844.6-054054&$30$&$0.70^{+0.10}_{-0.10}$&1.82\\
20&J021523.2-044337&$30$&$0.50^{+0.15}_{-0.15}$&3.04\\
21&J022321.9-045739&$30$&$0.60^{+0.10}_{-0.10}$&2.65\\
22&J022443.6-050905&$10$&$0.30^{+0.10}_{-0.10}$&0.56\\
23&J023418.0-041833&$10$&$0.70^{+0.20}_{-0.20}$&2.42\\
24&J020543.7-063807&$40$&$0.60^{+0.10}_{-0.10}$&0.13\\
25&J022932.6-055438&$10$&$0.60^{+0.10}_{-0.10}$&1.25\\
26&J021239.2-054816&$50$&$0.80^{+0.20}_{-0.20}$&3.34\\
27&J021511.4-060805&$10$&$0.70^{+0.20}_{-0.20}$&1.02\\
28&J021808.8-055630&$80$&$0.30^{+0.10}_{-0.10}$&2.30\\
29&J022538.9-040821&$10$&$0.40^{+0.10}_{-0.10}$&1.30\\
30&J020845.1-051354&$50$&$0.30^{+0.10}_{-0.20}$&1.28\\
31&J020953.9-055102&$30$&$0.60^{+0.10}_{-0.10}$&2.68\\
32&J020806.6-055739&$40$&$0.60^{+0.10}_{-0.10}$&0.92\\
33&J021509.0-054305&$30$&$0.70^{+0.20}_{-0.20}$&4.89\\
34&J021512.9-060558&$0$&$0.80^{+0.10}_{-0.10}$&0.18\\
35&J020529.5-051100&$10$&$0.50^{+0.20}_{-0.20}$&4.19\\
36&J022404.0-035730&$20$&$0.30^{+0.20}_{-0.20}$&0.60\\
37&J023501.0-055234&$10$&$0.30^{+0.10}_{-0.20}$&0.37\\
38&J020210.6-041129&$10$&$0.30^{+0.05}_{-0.05}$&1.46\\
39&J022650.3-025752&$40$&$0.50^{+0.20}_{-0.20}$&1.67\\
40&J021303.7-040704&$30$&$0.80^{+0.10}_{-0.10}$&0.19\\
41&J023357.7-054819&$10$&$0.70^{+0.10}_{-0.20}$&0.62\\
42&J020335.0-064450&$10$&$0.70^{+0.20}_{-0.20}$&0.03\\
43&J020823.4-040652&$10$&$0.80^{+0.10}_{-0.20}$&0.53\\
44&J020135.4-050847&$20$&$0.20^{+0.20}_{-0.10}$&0.62\\
45&J021837.2-060654&$20$&$0.40^{+0.20}_{-0.20}$&1.00\\
46&J022149.9-045920&$30$&$0.80^{+0.10}_{-0.10}$&0.36 \\
47&J020311.3-063534&$10$&$0.40^{+0.15}_{-0.15}$&0.16\\
\hline
\label{sed_properties}
\end{tabular}
\tablefoot{We consider sources with $\Psi \leq 30^{\rm o}$ as Type 2 (edge-on), $40^{\rm o} \leq \Psi \leq 60^{\rm o}$ as intermediate and $\Psi \geq 70^{\rm o}$ as Type 1 AGNs, based on their best-fit  values.}
\end{table*}

\begin{table*}
\caption{Optical properties of the red AGN sample.}
\centering
\setlength{\tabcolsep}{2.3mm}
\begin{tabular}{ccc}
      \hline
Object&3XLSS ID&Optical\\
       \hline
1&J021835.7-053758&BL\\
2&J022848.4-044426&BL, flat\\
3&J022928.4-051124&BL\\
4&J022809.0-041235&BL, flat\\
5&J020654.9-064552&NL\\
6&J020410.4-063924&NL\\
7&J023315.5-054747&NL\\
8&J021337.9-042814&BL, blue\\
9&J020436.4-042833&BL\\
10&J020543.0-051656&BL\\
11&J020517.3-051024&BL\\
12&J022244.3-030525&NL, red\\
13&J022323.4-031157&IMD\\
14&J022209.6-025023&NL\\
15&J022750.7-052232&IMD\\
16&J022758.4-053306&NL\\
17&J022453.2-054050&IMD\\
18&J022258.8-055757&IMD\\
19&J021844.6-054054&IMD\\
20&J021523.2-044337&NL\\
21&J022321.9-045739&NL\\
22&J022443.6-050905&IMD\\
23&J023418.0-041833&noisy\\
24&J020543.7-063807&noisy\\
25&J022932.6-055438&BL\\
26&J021239.2-054816&noisy\\
27&J021511.4-060805&NL\\
28&J021808.8-055630&BL\\
29&J022538.9-040821&BL\\
30&J020845.1-051354&-\\
31&J020953.9-055102&BL\\
32&J020806.6-055739&NL, red\\
33&J021509.0-054305&-\\
34&J021512.9-060558&-\\
35&J020529.5-051100&-\\
36&J022404.0-035730&-\\
37&J023501.0-055234&-\\
38&J020210.6-041129&-\\
39&J022650.3-025752&-\\
40&J021303.7-040704&-\\
41&J023357.7-054819&-\\
42&J020335.0-064450&-\\
43&J020823.4-040652&-\\
44&J020135.4-050847&-\\
45&J021837.2-060654&noisy\\
46&J022149.9-045920&BL\\
47&J020311.3-063534&-\\
\hline
\label{optical_properties}
\end{tabular}
\tablefoot{Optically unobscured sources (type 1) are considered to be those with broad emission lines (BL). Whereas obscured sources (type 2) exhibit only narrow emission lines (NL) or a red continuum. The AGNs are characterised as obscured, unobscured, or intermediate type (IMD) based on visual inspection of their optical spectra. We note that in the optical spectra of sources  13 and 22, while the MgII lines are broad, all other lines are narrow. This is attributed to the wider area from which the MgII originates with respect to the Balmer lines. These sources are characterized as IMD.}
\end{table*}

\begin{figure}
\includegraphics[height=0.97\columnwidth]{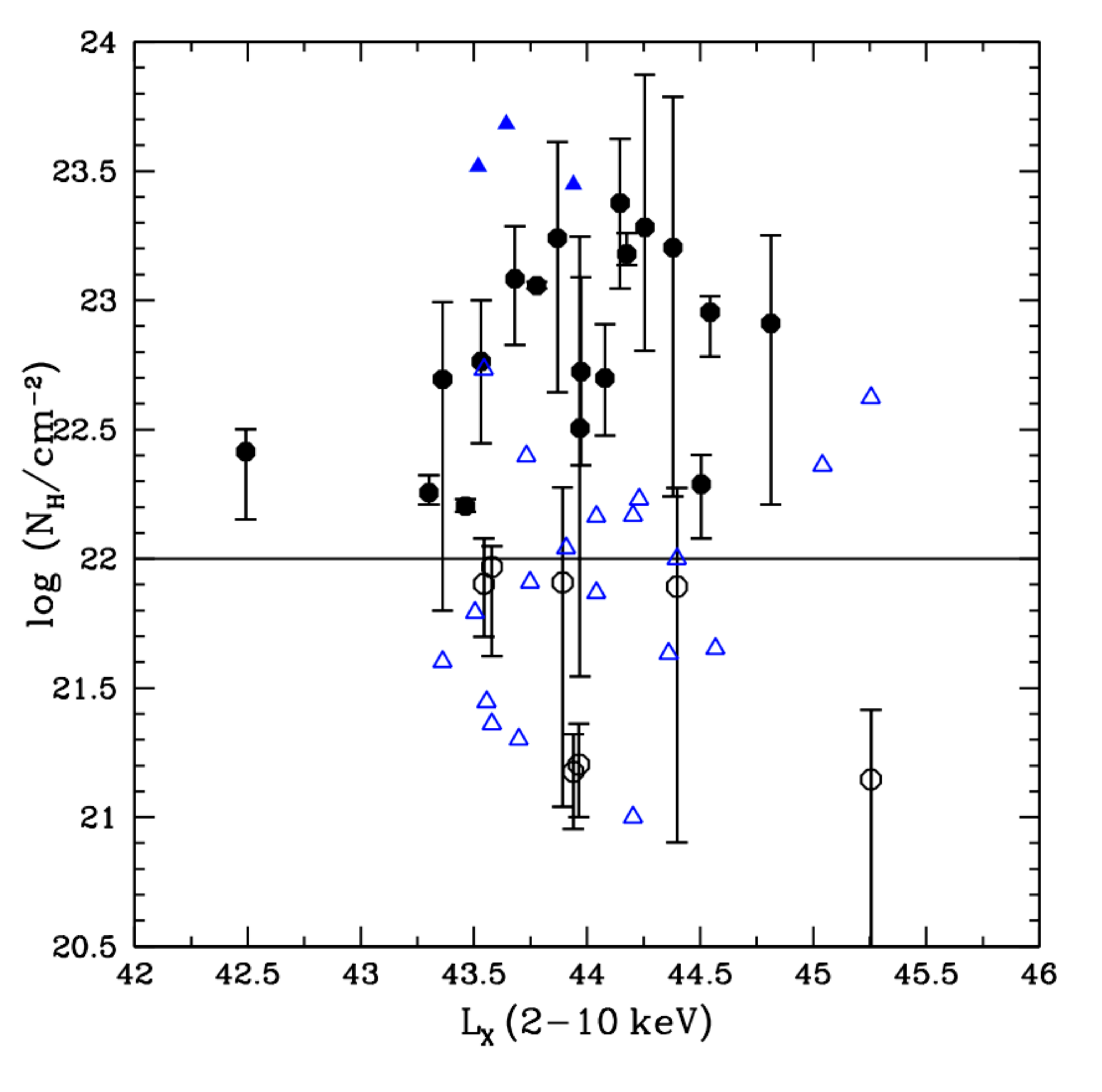}
\caption{ $\rm N_H$ as a function of X-ray luminosity for the 47 optical red AGNs. Those sources for which the X-ray spectral fitting could not constrain their $\rm N_H$ are presented with a triangle (upper limit estimations). The 21 AGNs that satisfy our X-ray absorption criteria are presented with filled symbols (see text for more details).} 
\label{nh_vs_lx}
\end{figure}

\subsection{X-ray spectra}
\label{sec_xray_spectra}
We examine the X-ray absorption of the 47 red AGNs in our sample. To achieve that, we apply X-ray spectral-fitting using the XSPEC v12.10 software \citep{Arnaud1996}. We use the Cash statistical analysis \citep[C-stat;][]{Cash1979} on the spectra binned to 1 count/bin, which has been shown to recover the actual spectral parameters in the most accurate way even for very low-count spectra \citep[e.g.][]{Krumpe2008}. The model applied to the spectral data is a simple power law, $A(E) = KE^{-\Gamma}$, absorbed by both the Galactic photoelectric absorption and the intrinsic absorption.
$\Gamma$ is the photon index of the power law and K is the flux at 1\,keV in units of photons/keV/cm$^2$/s. The Galactic absorption is taken from  the Leiden/Argentine/Bonn (LAB) Survey of Galactic HI \cite{Kalberla2005}. The X-ray spectra for the 47 AGNs are presented  in Figures  \ref{4}- \ref{50}. The best-fit values are presented in Table \ref{x_ray}. In Fig. \ref{nh_vs_lx} we present $\rm N_H$ versus X-ray luminosity.

\begin{figure}
\includegraphics[height=0.97\columnwidth]{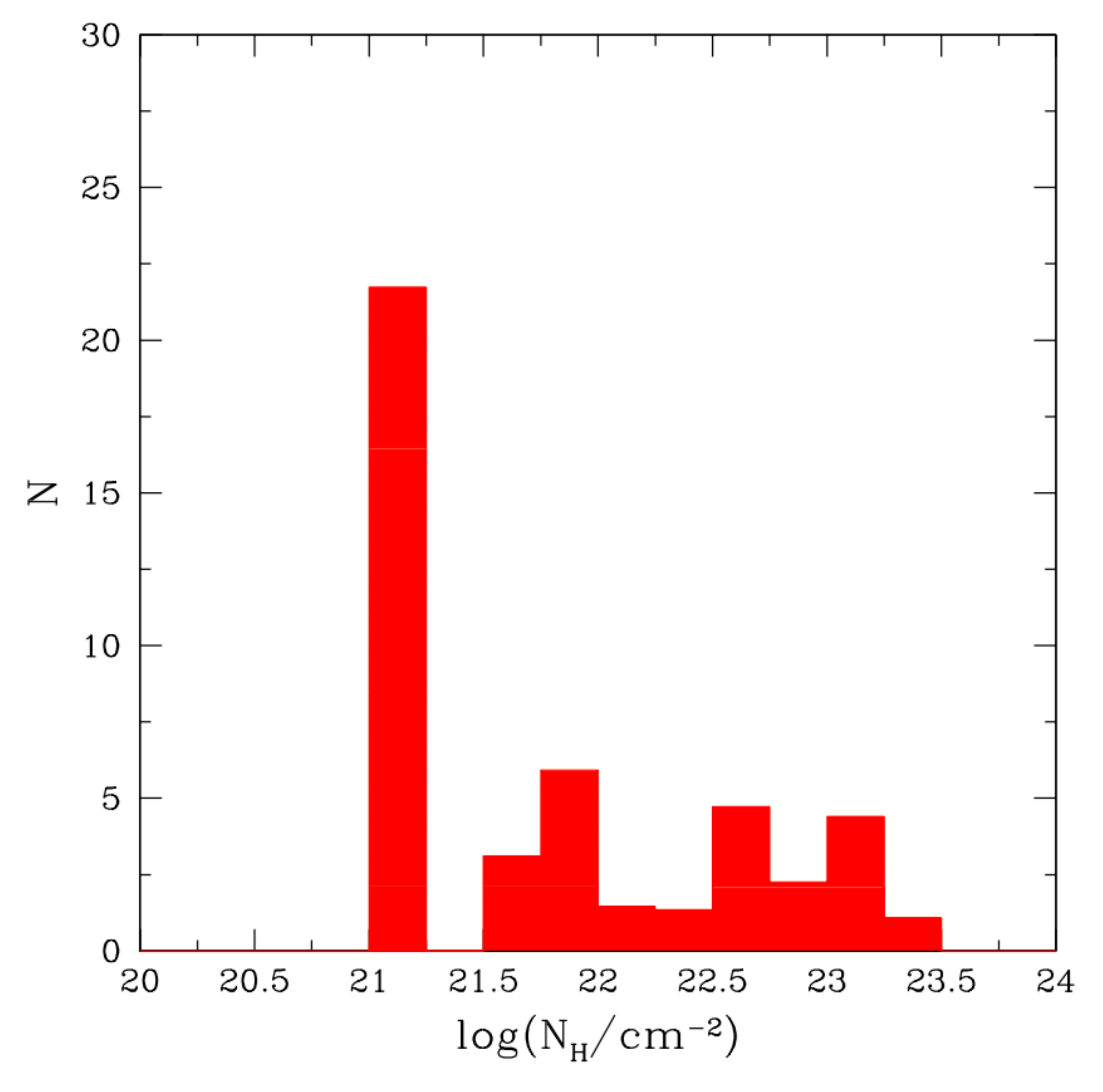}
\caption{$\rm N_H$ distribution taking into account $\rm N_H$ upper limits; see text for more details.}
\label{C}
\end{figure}

Figure \ref{C} presents the $\rm N_H$ distribution of the sources, taking into account the $\rm N_H$ upper limits \citep{Isobe1986}. For that purpose, we use the Astronomical SURVival Statistics \citep[ASURV;][]{Lavalley1992} software package, which adopts the maximum-likelihood Kaplan-Meier estimator to take into account  censored data \citep{Feigelson1985}. The derived mean value is  $\rm log\,N_H/cm^{-2}$ = $21.80\, \pm0.13$. 
The distribution shows that the red sources are almost equally divided between obscured ($\rm N_H>10^{22} cm^{-2}$) and unobscured AGNs. The limit of $\rm N_H=10^{22}cm^{-2}$ is often used in X-rays because lower column densities can be produced by dust lanes in the galaxy instead of the torus \citep{Malkan1998}.  
However, we note that \cite{Merloni2013} propose an alternative dividing line of $3\times10^{21}\rm{cm^{-2}}$ in the sense that this provides a much better agreement when compared with the optical classification between type-1 and type-2 AGNs. From Figures \ref{nh_vs_lx} and \ref{C} it appears that the red r-W2 colour alone cannot guarantee that the source will be classified as obscured in X-rays.  

\subsection{SED fitting}
We calculate the contribution of the AGNs to the power output of the host galaxy using the Code Investigating GALaxy Emission \cite[CIGALE;][]{Burgarella2005, Noll2009}. We provide the CIGALE code version 2018.0 with multi-wavelength flux densities, using optical, NIR, and MIR photometry. 
Table \ref{table_cigale} presents the models and the parameters used by CIGALE for the SED fitting of our X-ray sample. The \cite{Fritz2006} library of templates was used to model the AGN emission. The AGN fraction is measured as the AGN emission relative to IR luminosity ($1-1000\,\rm\mu m$). Here, $\tau$ is the e-folding time of the main stellar population model in millions of years. Age is defined as the age of the main stellar population, also in millions
of years. The extinction of a source is derived from the viewing angle, $\Psi$, of the torus: $\Psi \leq 30^{\rm o}$ for Type 2 (edge-on), $40^{\rm o}\leq \Psi=60^{\rm o}$ for intermediate, and $\Psi \geq 70^{\rm o}$ for Type 1 AGNs \citep[face-on;][]{Ciesla2015}, respectively. No intermediate values are used. The results from these fits appear in Table \ref{2_table} and the SEDs are shown in Figures \ref{4}- \ref{50} \citep[for more details on the SED fitting, see also][]{Masoura2018}.

As mentioned in the previous section, Herschel photometry is available for nine of the sources in our sample. We fit the SEDs of these sources with and without FIR data to check whether the lack of FIR photometric bands affects the SED fitting measurements. Specifically, we are interested in the inclination angle $\Psi$, since this parameter is used as a proxy to determine the obscuration of an AGN. We find that the inclusion of FIR photometry in the SED fitting does not change the AGN emission or the estimated inclination angle of the sources. Thus, we conclude that a lack of FIR photometry for the remaining  sources does not affect our analysis.

\subsection{Optical spectra}
The optical spectra for 33 out of 47 sources are presented in Figures \ref{4}- \ref{50}. Optically unobscured sources (type 1) are considered to be those with broad emission lines (BL), whereas obscured sources (type 2) exhibit only narrow emission lines (NL) or a red continuum. The AGNs are characterised as obscured, unobscured, or intermediate (IMD) type, based on visual inspection of their optical spectra. Table \ref{optical_properties} contains the classification according to the optical spectrum for each object.

To verify the redshift of our sources and assess their obscuration in the optical band we used the optical spectra provided by the SDSS. We recomputed the spectroscopic redshift and for a small number of cases we provide in Table \ref{general_properties} a different measurement from the one in SDSS.

\begin{figure}
\includegraphics[height=1\columnwidth]{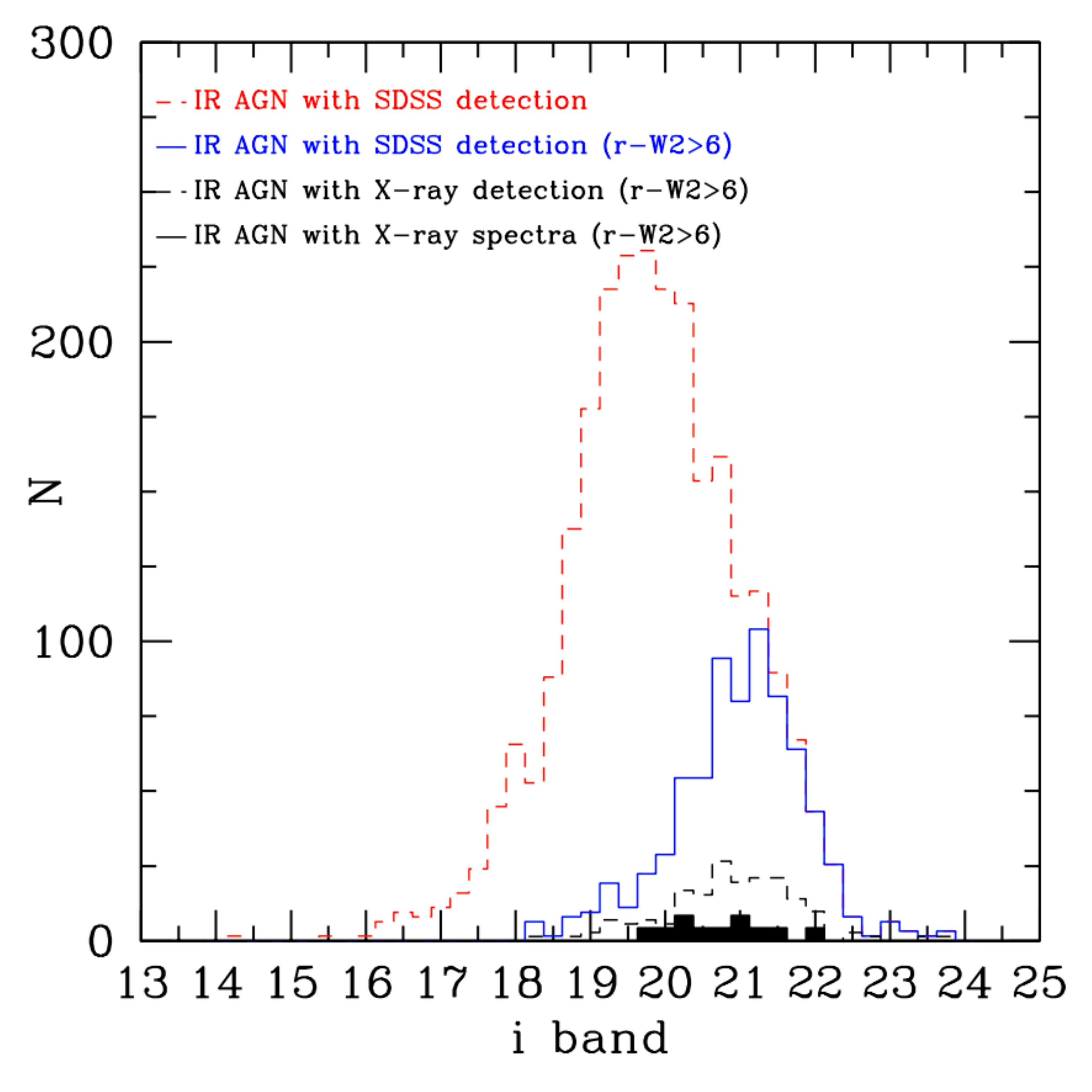}
\caption{i-band distribution of our sample. Red sources occupy the faint part of the i-band distribution}
\label{i_distrib}
\end{figure}

\begin{figure}
\includegraphics[height=1.\columnwidth]{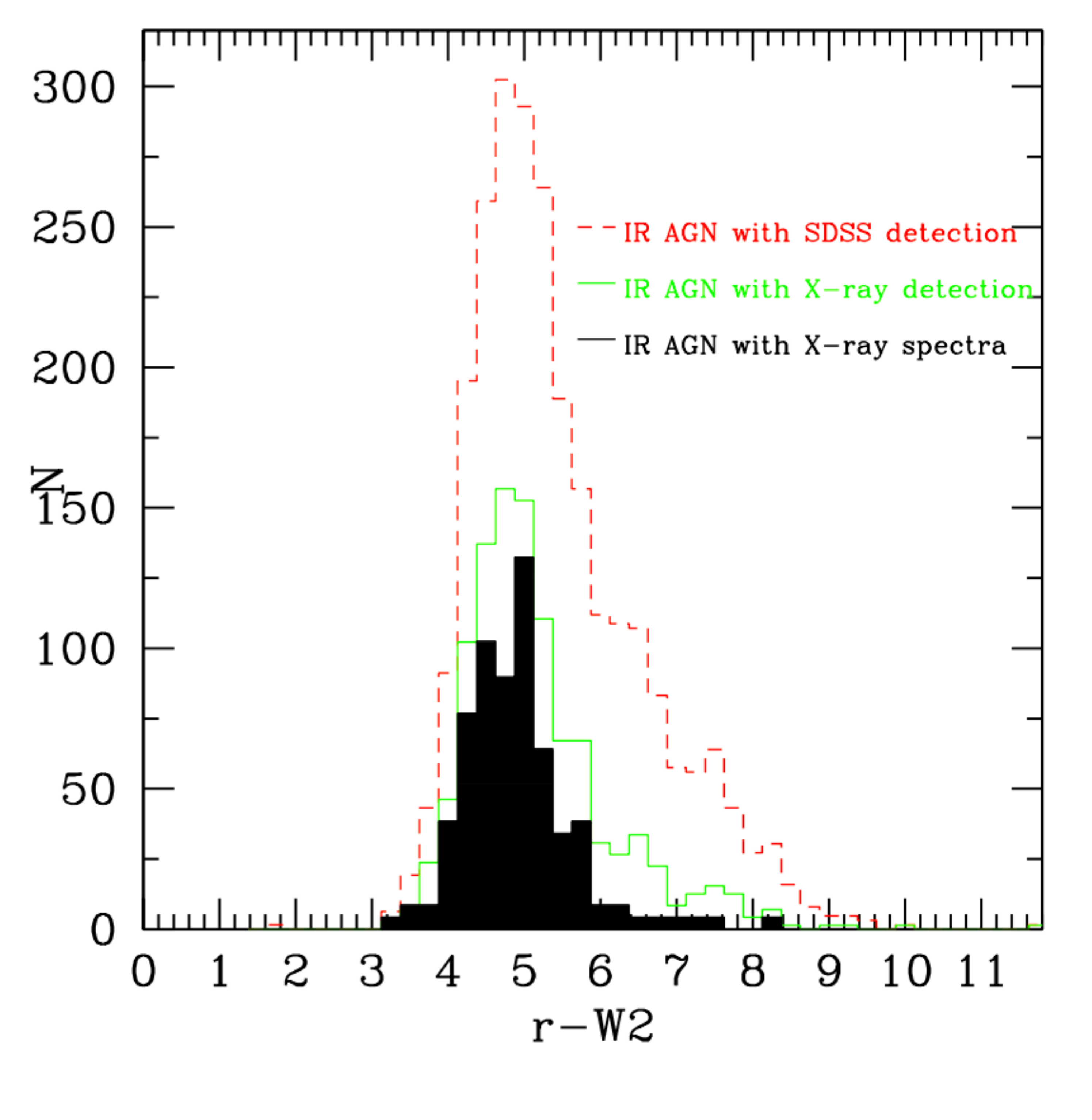}
\caption{r-W2 distribution of our sample. The AGN sample with available X-ray spectra lacks sources with the highest $r-W2$ values.}
\label{r_w2_distrib}
\end{figure}

 



\begin{table*}
\caption{Comparison of the X-ray, SED, and optical properties.}
\centering
\setlength{\tabcolsep}{2.8mm}
\begin{tabular}{cccccc}
      \hline
Object&ID&Classification\\
&3XLSS&\\
&  & Xray~~~~SED~~~~optical \\
    \hline
1&J021835.7-053758&1~~~~~~~~~~~~~2~~~~~~~~~~~~~1\\
2&{\it{J022848.4-044426}}&2~~~~~~~~~~~~~2~~~~~~~~~~~~~1\\
3&{\bf{J022928.4-051124}}&2~~~~~~~~~~~~~2~~~~~~~~~~~~~1\\
4&J022809.0-041235&1~~~~~~~~~~~~~1~~~~~~~~~~~~~1\\
5&J020654.9-064552&1~~~~~~~~~~~~~2~~~~~~~~~~~~~2\\
6&{\it{J020410.4-063924}}&2~~~~~~~~~~~~~2~~~~~~~~~~~~~2\\
7&J023315.5-054747&1~~~~~~~~~~~~~2~~~~~~~~~~~~~2\\
8&{\bf{J021337.9-042814}}&2~~~~~~~~~~~~~2~~~~~~~~~~~~~1\\
9&J020436.4-042833&1~~~~~~~~~~~~~2~~~~~~~~~~~~~1\\
10&{\bf{J020543.0-051656}}&2~~~~~~~~~~~~~2~~~~~~~~~~~~~1\\
11&{\it{J020517.3-051024}}&2~~~~~~~~~~~~~1~~~~~~~~~~~~~1\\
12&{\bf{J022244.3-030525}}&2~~~~~~~~~~~~~2~~~~~~~~~~~~~2\\
13&{\bf{J022323.4-031157}}&2~~~~~~~~~~~~~2~~~~~~~~~~~~~{\bf{3}}\\
14&{\it{J022209.6-025023}}&2~~~~~~~~~~~~~2~~~~~~~~~~~~~2\\
15&{\bf{J022750.7-052232}}&2~~~~~~~~~~~~~2~~~~~~~~~~~~~{\bf{3}}\\
16&J022758.4-053306&1~~~~~~~~~~~~~2~~~~~~~~~~~~~2\\
17&{\bf{J022453.2-054050}}&2~~~~~~~~~~~~~2~~~~~~~~~~~~~{\bf{3}}\\
18&J022258.8-055757&1~~~~~~~~~~~~~2~~~~~~~~~~~~~{\bf{3}}\\
19&J021844.6-054054&1~~~~~~~~~~~~~2~~~~~~~~~~~~~{\bf{3}}\\
20&J021523.2-044337&1~~~~~~~~~~~~~2~~~~~~~~~~~~~2\\
21&{\bf{J022321.9-045739}}&2~~~~~~~~~~~~~2~~~~~~~~~~~~~2\\
22&{\it{J022443.6-050905}}&2~~~~~~~~~~~~~2~~~~~~~~~~~~~{\bf{3}}\\
23&J023418.0-041833&1~~~~~~~~~~~~~2~~~~~~~~~~~~~0\\
24&{\bf{J020543.7-063807}}&2~~~~~~~~~~~~~2~~~~~~~~~~~~~0\\
25&J022932.6-055438&1~~~~~~~~~~~~~2~~~~~~~~~~~~~1\\
26&{\bf{J021239.2-054816}}&2~~~~~~~~~~~~~2~~~~~~~~~~~~~0\\
27&{\bf{J021511.4-060805}}&2~~~~~~~~~~~~~2~~~~~~~~~~~~~1\\
28&J021808.8-055630&1~~~~~~~~~~~~~1~~~~~~~~~~~~~1\\
29&{\bf{J022538.9-040821}}&2~~~~~~~~~~~~~2~~~~~~~~~~~~~1\\
30&{\it{J020845.1-051354}}&2~~~~~~~~~~~~~2~~~~~~~~~~~~~-\\
31&{\bf{J020953.9-055102}}&2~~~~~~~~~~~~~2~~~~~~~~~~~~~1\\
32&{\bf{J020806.6-055739}}&2~~~~~~~~~~~~~2~~~~~~~~~~~~~2\\
33&{\bf{J021509.0-054305}}&2~~~~~~~~~~~~~2~~~~~~~~~~~~~-\\
34&{\it{J021512.9-060558}}&2~~~~~~~~~~~~~2~~~~~~~~~~~~~-\\
35&{\it{J020529.5-051100}}&2~~~~~~~~~~~~~2~~~~~~~~~~~~~-\\
36&{\bf{J022404.0-035730}}&2~~~~~~~~~~~~~2~~~~~~~~~~~~~-\\
37&{\it{J023501.0-055234}}&2~~~~~~~~~~~~~2~~~~~~~~~~~~~-\\
38&{\it{J020210.6-041129}}&2~~~~~~~~~~~~~2~~~~~~~~~~~~~-\\
39&{\bf{J022650.3-025752}}&2~~~~~~~~~~~~~2~~~~~~~~~~~~~-\\
40&{\bf{J021303.7-040704}}&2~~~~~~~~~~~~~2~~~~~~~~~~~~~-\\
41&{\it{J023357.7-054819}}&2~~~~~~~~~~~~~2~~~~~~~~~~~~~-\\
42&{\it{J020335.0-064450}}&2~~~~~~~~~~~~~2~~~~~~~~~~~~~-\\
43&{\bf{J020823.4-040652}}&2~~~~~~~~~~~~~2~~~~~~~~~~~~~-\\
44&J020135.4-050847&1~~~~~~~~~~~~~2~~~~~~~~~~~~~-\\
45&{\it{J021837.2-060654}}&2~~~~~~~~~~~~~2~~~~~~~~~~~~~0\\
46&{\bf{J022149.9-045920}}&2~~~~~~~~~~~~~2~~~~~~~~~~~~~2\\
47&{\bf{J020311.3-063534}}&2~~~~~~~~~~~~~2~~~~~~~~~~~~~-\\

\hline
\label{2_table}
\end{tabular}
\tablefoot{Sources classified as X-ray obscured based on our strict criteria (see text for more details) are presented in bold. Sources that satisfy the loosened X-ray criteria (see text) are shown in italics. Numbers 1 (unobscured) and 2 (obscured) are used to denote the classification of the sources based on the various criteria. Numbers 0 and 3 refer to optical spectra that are too noisy to allow us to classify them and are IMD type, respectively. The first number refers to the X-ray classification, while the second and the third numbers correspond to the SED- and optical-based characterisation.}
\end{table*}

\section{Discussion}

Our IR selection criteria identify 4798 AGN candidates (Table \ref{sample}). Of these, 2652  have been observed by SDSS. Approximately $20\%$ of these sources are optically red using the criteria of \cite{Yan2013}; see Table \ref{sample}. Figures \ref{i_distrib} and \ref{r_w2_distrib} show the i-band and the $r- W2$ distributions of the various samples. As expected, red sources occupy the faint part of the i-band distribution while the AGN sample with available X-ray spectra lacks sources with the highest $r-W2$ values.

In our analysis, we explore the X-ray spectral properties of the 47 optically red AGNs and examine whether these sources are also X-ray absorbed. Furthermore, we search for indications of extinction either on their SEDs or in their optical spectra. The results of the X-ray spectral fitting are presented in Table {\ref{x_ray}}. For those cases where the photon index ($\Gamma$) is unphysically low ($\Gamma<1.2$) we quote the $\rm{N_H}$ estimations with $\Gamma$ fixed to $\Gamma=1.8$ \citep{Nandra1994}. The latter $\rm N_H$  estimation is used to characterise the X-ray absorption of the source. Our analysis reveals 18 sources with best-fit intrinsic column densities $\rm N_H>10^{22}\, \rm{cm^{-2}}$. We also consider as possibly absorbed a source whose upper limit lies above the arbitrary chosen limit of $\rm N_H=10^{23}\, \rm{cm^{-2}}$ (i.e. J020543.0-051656, J020953.9-055102, J021509.0-054305). The latter increases the number of candidate X-ray-absorbed sources to 21. We characterise these systems as  X-ray-absorbed sources and mark them in bold in Table \ref{2_table}. However, there are AGNs that present mild absorption and could be classified as X-ray absorbed if we eased our X-ray criteria. Thus, lowering our X-ray absorption criteria, that is,  $\rm N_H>10^{21.5}\, \rm{cm^{-2}}$ \citep{Merloni2013} or upper limit that lies above the $\rm N_H$ value $\rm N_H=10^{22}\, \rm{cm^{-2}}$, we find 13 additional sources that present signs of X-ray absorption. These additional sources are shown in italics in Table \ref{2_table}. Therefore, up to 34 AGNs ($72\%$) in the sample present some signs of X-ray absorption.

The vast majority of our optically red AGNs have SEDs that show clear signs of obscuration in their AGN emission (green lines is Figures \ref{4}-{50}). However, the AGN emission of the source J022809.0-041235 (no. 4 in Table \ref{sed_properties}) extends to the optical part of the spectrum without presenting signs of even mild obscuration. This source has also large $\Psi$ ($\Psi=80^{\rm o}$) that also indicates a type 1 AGN (X-ray and optical spectra also confirm this is an unobscured AGN). There are two more sources with a large $\Psi$ value (i.e. observed face-on), namely sources J020517.3-051024 and J021808.8-055630 (no.11 and 28 in Table \ref{sed_properties}). Their SEDs reveal that even though the AGN emission extends to the optical part of the SED, the emission of the galaxy is a dominant component of the SED at the optical wavelengths. We refer to the latter two systems as galaxy dominated. We attribute this term to sources that satisfy the following criteria: (i) the IR (W2) emission of the system is due to the AGN emission, and (ii) there is a strong galaxy emission in the optical part of the SED even though the AGN emission is only mildly obscured (or not obscured at all). Specifically, the ratio of the galaxy emission to the AGN emission in the r band is $\geq 1$. Although source no. 28 is not X-ray absorbed, source no. 11 presents signs of mild X-ray absorption ($\rm N_H=0.78^{+1.10}_{-0.70}\times10^{22}\,{cm^{-2}}$). The optical spectra of these two systems have broad emission lines, corroborating the assertion that these are unobscured sources. We conclude that the SEDs for 44 out of the 47 sources in our sample present clear signs of obscuration based on their AGN emission and their $\Psi$ values. 

In addition to these  two AGNs (i.e. sources 11 and 28 in Table \ref{sed_properties}), there are eight more sources that satisfy our aforementioned criteria and are therefore considered galaxy-dominated systems (nos. 5, 7, 10, 26, 35, 39, 44, and 47). In these systems, the galaxy and/or AGN emission in the r band ranges from $\approx 1$ (e.g. no. 28) to $\geq10$ (e.g. nos. 10, 39, and 47). Five of these sources are among the AGNs that are also X-ray absorbed (nos. 10, 26, 35, 39, and 47 in Table \ref{2_table}). Their AGN fraction, $\rm{frac_{AGN}}=40-80\%$ (Table \ref{sed_properties}). These suggest that these systems are galaxy dominated because the AGN is obscured and not because the active SMBH is intrinsically weak. However, for the remaining three galaxy-dominated systems (nos. 5, 7, and 44) that are not X-ray absorbed, it could be that they are characterised as optically red (r-W2>6) due to an intrinsically weak AGN (AGN fraction, $\rm{frac_{AGN}}=20-30\%$, Table \ref{sed_properties}) rather than an obscured AGN. Galaxy-dominated systems are known to confuse optical and MIR criteria \citep{koutoulidis2018}. 

Excluding the ten galaxy-dominated systems from our sample of red IR AGNs reduces the total number of red sources to 37. Of these, 17 (46$\%$ of the total sample) and 28 (76$\%$ of the total sample) are X-ray absorbed using the strict and loose criteria, respectively. Therefore, there are nine ($\sim 20\%$) IR red AGNs in our sample that present no signs of X-ray absorption. The optical spectra for two of these sources present narrow emission lines (nos. 16 and 20 in Table  \ref{2_table}), four present broad emission lines (nos. 1, 3, 9, and 25 in Table  \ref{2_table}) and two are of intermediate type (nos. 17 and 18 in Table  \ref{2_table}) based on our visual inspection. The optical spectrum of source no. 23 is too noisy to extract any information. Previous studies have found that selection of obscured AGNs based on optical and/or MIR colours are 80$\%$ reliable \citep{Hickox2007}. This could explain why at least some of these nine AGNs, although optically red, present broad emission lines in their spectra and no substantial X-ray absorption. 

Recently, \cite{Jaffarian2020} studied the optical E(B-V) reddening versus the X-ray absorbing column density for a large sample of Seyfert galaxies with both E(B-V) and X-ray absorption column density measurements. These latter authors find a significant correlation between E(B-V) and $N_H,$ albeit with a large scatter of $\pm$dex, which they claim could be attributed to X-ray column density variability. Regardless of the origin of the physical interpretation, the presence of such a large scatter can readily explain why some of our red sources show no sign of significant absorption in X-rays. The fact that the average absorbing column density derived for the red $\rm r-W2>6$ sources is only $\rm N_H=21.8\pm0.13$ suggests that the above obscuration criterion selects sources with lower absorption compared to those selected in X-rays using $\rm N_H>10^{22} cm^{-2}$ \citep[e.g.][]{Akylas2006, Ueda2014}. In other words, the optically red source selection does not correspond to the same level of obscuring column densities as those seen for X-ray wavelengths. \cite{Hickox2017} proposed an additional criterion for the selection of obscured AGNs based on the combination of the {$\it u$} and WISE  W3 band. Since there is not a unique definition of how obscured an AGN must be to be classified as absorbed, the various selection criteria target different ranges of obscuration. Specifically, the application of the `usual' X-ray criterion $\rm N_H>10^{22} cm^{-2}$ selects the most obscured part of the obscured AGN distribution, while the r-W2 criterion extends to lower levels of obscuration. Finally, the Hickox et al. {$\it u$}-$W3$ criterion is sensitive to the lowest levels of obscuration. This is due to the use of the {$\it u$} band that is easily affected by even small amounts of reddening. The different selection criteria indeed result in different obscured AGN samples. This is most probably the reason why only 76$\%$ (at most) of the red AGN sample is absorbed in X-rays with column densities $\rm log\,N_H/cm^{-2}>21.5$.

\section{Summary}

Here, we applied the \cite{Assef2018} criteria based on the W1 and W2 WISE bands to select IR AGNs in the XXL-North 25$\rm deg^2$ survey area. The above criteria yield 4798 AGNs. Moreover, we applied the optical MIR colour criterion r-W2 \citep{Yan2013} in order to select obscured AGNs. We identify 561 red AGNs, of which 135 are detected in X-rays while 47 have good photon statistics (at least 50 net counts per detector), and derive a reliable estimate of the absorbing column density. We used XSPEC to fit the X-ray spectra of the 47 sources and quantify their X-ray absorption. Of these 47 sources, 76\%  may show signs of absorption higher than  $\rm logN_H/cm^{-2}=21.5$. Furthermore, we applied the CIGALE code in order to study the SEDs of our 47 red sources. The SED analysis revealed that 10/47 sources are galaxy-dominated systems as the red colours are attributed to the host galaxy rather than absorption. In the vast majority of the remaining 37 sources, the SEDs confirm the presence of significant obscuration. The exact nature of the red sources that show low levels of X-ray absorption (25\%) in apparent contradiction with their SEDs is unclear. The large scatter in the r-W2 colour versus column density, possibly caused by variability of the column density, is likely to offer an  explanation.
Our work shows that the r-W2 selection provides a robust method for selecting obscured AGNs. However, it may  also identify sources that present lower levels of obscuration when compared with the widely used X-ray spectroscopic criteria which usually define obscured sources as those with $\rm logN_H/cm^{-2}>21.5$.

\begin{acknowledgements}
The authors are grateful to the anonymous referee for helpful comments.
VAM and GM acknowledge support of this work by the PROTEAS II project (MIS 5002515), which is implemented under the "Reinforcement of the Research and Innovation Infrastructure" action, funded by the "Competitiveness, Entrepreneurship and Innovation" operational programme (NSRF 2014-2020) and co-financed by Greece and the European Union (European Regional Development Fund). GM also acknowledges support by the Agencia Estatal de Investigación, Unidad de Excelencia María de Maeztu, ref. MDM-2017-0765. The Saclay group acknowledges long-term support from the Centre National d'Etudes Spatiales (CNES).
\\

\\
$\it{XXL}$  is  an  international  project  based  around  an  $\it{XXM}$  Very Large Programme surveying two 25 deg$^2$ extragalactic fields at a depth of $\sim$  6 $\times$ $10^{-15}$ erg cm $^{-2}$ s$^{-1}$ in the [0.5-2] keV band forpoint-like sources. The XXL website ishttp://irfu.cea.fr/xxl/.  Multi-band  information  and  spectroscopic  follow-up  ofthe X-ray sources are obtained through a number of survey programmes,  summarised  at http://xxlmultiwave.pbworks.com/.
\\
This research has made use of data obtained from the 3XMM XMM-\textit{Newton} 
serendipitous source catalogue compiled by the 10 institutes of the XMM-\textit{Newton} 
Survey Science Centre selected by ESA.
\\
The Saclay group acknowledges long-term support from the Centre National d'Etudes Spatiales (CNES).
\\
This work is based on observations made with XMM-\textit{Newton}, an ESA science 
mission with instruments and contributions directly funded by ESA Member States 
and NASA. 
\\
Funding for the Sloan Digital Sky Survey IV has been provided by the Alfred P. Sloan Foundation, the U.S. Department of Energy Office of Science, and the Participating Institutions. SDSS-IV acknowledges
support and resources from the Center for High-Performance Computing at
the University of Utah. The SDSS web site is \url{www.sdss.org}.
\\
SDSS-IV is managed by the Astrophysical Research Consortium for the 
Participating Institutions of the SDSS Collaboration including the 
Brazilian Participation Group, the Carnegie Institution for Science, 
Carnegie Mellon University, the Chilean Participation Group, the French Participation Group, Harvard-Smithsonian Center for Astrophysics, 
Instituto de Astrof\'isica de Canarias, The Johns Hopkins University, 
Kavli Institute for the Physics and Mathematics of the Universe (IPMU) / 
University of Tokyo, Lawrence Berkeley National Laboratory, 
Leibniz Institut f\"ur Astrophysik Potsdam (AIP),  
Max-Planck-Institut f\"ur Astronomie (MPIA Heidelberg), 
Max-Planck-Institut f\"ur Astrophysik (MPA Garching), 
Max-Planck-Institut f\"ur Extraterrestrische Physik (MPE), 
National Astronomical Observatories of China, New Mexico State University, 
New York University, University of Notre Dame, 
Observat\'ario Nacional / MCTI, The Ohio State University, 
Pennsylvania State University, Shanghai Astronomical Observatory, 
United Kingdom Participation Group,
Universidad Nacional Aut\'onoma de M\'exico, University of Arizona, 
University of Colorado Boulder, University of Oxford, University of Portsmouth, 
University of Utah, University of Virginia, University of Washington, University of Wisconsin, 
Vanderbilt University, and Yale University.
\\
This publication makes use of data products from the Wide-field Infrared Survey 
Explorer, which is a joint project of the University of California, Los Angeles, 
and the Jet Propulsion Laboratory/California Institute of Technology, funded by 
the National Aeronautics and Space Administration. 
\\
The VISTA Data Flow System pipeline processing and science archive are described 
in \cite{Irwin2004}, \cite{Hambly2008} and \cite{Cross2012a}. Based on 
observations obtained as part of the VISTA Hemisphere Survey, ESO Program, 
179.A-2010 (PI: McMahon). We have used data from the 3rd data release.
\\
\end{acknowledgements}

\bibliography{mybib}{}

\begin{thebibliography}{65}
\expandafter\ifx\csname natexlab\endcsname\relax\def\natexlab#1{#1}\fi

\bibitem[{Aguado {et~al.}(2019)Aguado, Hern{\'{a}}ndez, Prieto, \&
  Rebolo}]{Aguado2019}
Aguado, D.~S., Hern{\'{a}}ndez, J. I.~G., Prieto, C.~A., \& Rebolo, R. 2019,
  ApJ, 874, L21

\bibitem[{{Akylas} {et~al.}(2006){Akylas}, {Georgantopoulos}, {Georgakakis},
  {Kitsionas}, \& {Hatziminaoglou}}]{Akylas2006}
{Akylas}, A., {Georgantopoulos}, I., {Georgakakis}, A., {Kitsionas}, S., \&
  {Hatziminaoglou}, E. 2006, A\&A, 459, 693

\bibitem[{Antonucci(1993)}]{Antonucci1993}
Antonucci, R. 1993, ARA\&A, 31, 473

\bibitem[{Arnaud(1996)}]{Arnaud1996}
Arnaud, K.~A. 1996, ASPC, 101, 17

\bibitem[{Assef {et~al.}(2013)Assef, Stern, Kochanek, Blain, Brodwin, Brown,
  Donoso, Eisenhardt, Jannuzi, Jarrett, Stanford, Tsai, Wu, \& Yan}]{Assef2013}
Assef, R.~J., Stern, D., Kochanek, C.~S., {et~al.} 2013, ApJ, 772, 26

\bibitem[{Assef {et~al.}(2018)Assef, Stern, Noirot, Jun, Cutri, \&
  Eisenhardt}]{Assef2018}
Assef, R.~J., Stern, D., Noirot, G., {et~al.} 2018, ApJS, 234, 23

\bibitem[{Baldry {et~al.}(2010)Baldry, Robotham, Hill, Driver, Liske, Norberg,
  Bamford, Hopkins, Loveday, Peacock, Cameron, Croom, Cross, Doyle, Dye, Frenk,
  Jones, van Kampen, Kelvin, Nichol, Parkinson, Popescu, Prescott, Sharp,
  Sutherland, Thomas, \& Tuffs}]{Baldry2010}
Baldry, I.~K., Robotham, A. S.~G., Hill, D.~T., {et~al.} 2010, MNRAS, 404, 86

\bibitem[{Burgarella {et~al.}(2005)Burgarella, Buat, \&
  Iglesias-P{\'{a}}ramo}]{Burgarella2005}
Burgarella, D., Buat, V., \& Iglesias-P{\'{a}}ramo, J. 2005, MNRAS, 360, 1413

\bibitem[{{Calzetti} {et~al.}(2000){Calzetti}, {Armus}, {Bohlin}, {Kinney},
  {Koornneef}, \& {Storchi-Bergmann}}]{Calzetti2000}
{Calzetti}, D., {Armus}, L., {Bohlin}, R.~C., {et~al.} 2000, ApJ, 533, 682

\bibitem[{{Cash}(1979)}]{Cash1979}
{Cash}, W. 1979, ApJ, 228, 939

\bibitem[{Chiappetti {et~al.}(2018)Chiappetti, Fotopoulou, Lidman, Faccioli,
  Pacaud, Elyiv, Paltani, Pierre, Plionis, Adami, Alis, Altieri, Baldry,
  Bolzonella, Bongiorno, Brown, Driver, Elmer, Franzetti, Grootes, Guglielmo,
  Iovino, Koulouridis, Lef{\`{e}}vre, Liske, Maurogordato, Melnyk, Owers,
  Poggianti, Polletta, Pompei, Ponman, Robotham, Sadibekova, Tuffs, Valtchanov,
  Vignali, \& Wagner}]{Chiappetti2018}
Chiappetti, L., Fotopoulou, S., Lidman, C., {et~al.} 2018, A\&A, 620, A12

\bibitem[{Ciesla {et~al.}(2015)Ciesla, Charmandaris, Georgakakis, Bernhard,
  Mitchell, Buat, Elbaz, LeFloc'h, Lacey, Magdis, \& Xilouris}]{Ciesla2015}
Ciesla, L., Charmandaris, V., Georgakakis, A., {et~al.} 2015, A\&A, 576, A10

\bibitem[{{Civano} {et~al.}(2012){Civano}, {Elvis}, {Brusa}, {Comastri},
  {Salvato}, {Zamorani}, {Aldcroft}, {Bongiorno}, {Capak}, {Cappelluti},
  {Cisternas}, {Fiore}, {Fruscione}, {Hao}, {Kartaltepe}, {Koekemoer}, {Gilli},
  {Impey}, {Lanzuisi}, {Lusso}, {Mainieri}, {Miyaji}, {Lilly}, {Masters},
  {Puccetti}, {Schawinski}, {Scoville}, {Silverman}, {Trump}, {Urry},
  {Vignali}, \& {Wright}}]{Civano2012}
{Civano}, F., {Elvis}, M., {Brusa}, M., {et~al.} 2012, ApJS, 201, 30

\bibitem[{{Corral} {et~al.}(2015){Corral}, {Georgantopoulos}, {Watson},
  {Rosen}, {Page}, \& {Webb}}]{Corral2015}
{Corral}, A., {Georgantopoulos}, I., {Watson}, M.~G., {et~al.} 2015, A\&A, 576,
  A61

\bibitem[{Cross {et~al.}(2012)Cross, Collins, Mann, Read, Sutorius, Blake,
  Holliman, Hambly, Emerson, Lawrence, \& Noddle}]{Cross2012a}
Cross, N. J.~G., Collins, R.~S., Mann, R.~G., {et~al.} 2012, A\&A, 548, A119

\bibitem[{Dalton {et~al.}(2006)Dalton, Caldwell, Ward, Whalley, Woodhouse,
  Edeson, Clark, Beard, Gallie, Todd, Strachan, Bezawada, Sutherland, \&
  Emerson}]{Dalton2006}
Dalton, G.~B., Caldwell, M., Ward, A.~K., {et~al.} 2006, SPIE, 6269

\bibitem[{Donoso {et~al.}(2014)Donoso, Yan, Stern, \& Assef}]{Donoso2014}
Donoso, E., Yan, L., Stern, D., \& Assef, R.~J. 2014, ApJ, 789, 44

\bibitem[{Driver {et~al.}(2011)Driver, Hill, Kelvin, Robotham, Liske, Norberg,
  Baldry, Bamford, Hopkins, Loveday, Peacock, Andrae, Bland-Hawthorn, Brough,
  Brown, Cameron, Ching, Colless, Conselice, Croom, Cross, de~Propris, Dye,
  Drinkwater, Ellis, Graham, Grootes, Gunawardhana, Jones, van Kampen,
  Maraston, Nichol, Parkinson, Phillipps, Pimbblet, Popescu, Prescott,
  Roseboom, Sadler, Sansom, Sharp, Smith, Taylor, Thomas, Tuffs, Wijesinghe,
  Dunne, Frenk, Jarvis, Madore, Meyer, Seibert, Staveley-Smith, Sutherland, \&
  Warren}]{Driver2011}
Driver, S.~P., Hill, D.~T., Kelvin, L.~S., {et~al.} 2011, MNRAS, 413, 971

\bibitem[{Edelson \& Malkan(1986)}]{Edelson1986}
Edelson, R.~A. \& Malkan, M.~A. 1986, ApJ, 308, 59

\bibitem[{Emerson {et~al.}(2006)Emerson, McPherson, \&
  Sutherland}]{Emerson2006}
Emerson, J., McPherson, A., \& Sutherland, W. 2006, Msngr, 126, 41

\bibitem[{Feigelson \& Nelson(1985)}]{Feigelson1985}
Feigelson, E.~D. \& Nelson, P.~I. 1985, ApJ, 293, 192

\bibitem[{{Fiore} {et~al.}(2009){Fiore}, {Puccetti}, {Brusa}, {Salvato},
  {Zamorani}, {Aldcroft}, {Aussel}, {Brunner}, {Capak}, {Cappelluti}, {Civano},
  {Comastri}, {Elvis}, {Feruglio}, {Finoguenov}, {Fruscione}, {Gilli},
  {Hasinger}, {Koekemoer}, {Kartaltepe}, {Ilbert}, {Impey}, {Le Floc'h},
  {Lilly}, {Mainieri}, {Martinez-Sansigre}, {McCracken}, {Menci}, {Merloni},
  {Miyaji}, {Sanders}, {Sargent}, {Schinnerer}, {Scoville}, {Silverman},
  {Smolcic}, {Steffen}, {Santini}, {Taniguchi}, {Thompson}, {Trump}, {Vignali},
  {Urry}, \& {Yan}}]{Fiore2009}
{Fiore}, F., {Puccetti}, S., {Brusa}, M., {et~al.} 2009, ApJ, 693, 447

\bibitem[{Fotopoulou {et~al.}(2016)Fotopoulou, Pacaud, Paltani, Ranalli,
  Ramos-Ceja, Faccioli, Plionis, Adami, Bongiorno, Brusa, Chiappetti, Desai,
  Elyiv, Lidman, Melnyk, Pierre, Piconcelli, Vignali, Alis, Ardila, Arnouts,
  Baldry, Bremer, Eckert, Guennou, Horellou, Iovino, Koulouridis, Liske,
  Maurogordato, Menanteau, Mohr, Owers, Poggianti, Pompei, Sadibekova,
  Stanford, Tuffs, \& Willis}]{Fotopoulou2016}
Fotopoulou, S., Pacaud, F., Paltani, S., {et~al.} 2016, A\&A, 592, A5

\bibitem[{{Fritz} {et~al.}(2006){Fritz}, {Franceschini}, \&
  {Hatziminaoglou}}]{Fritz2006}
{Fritz}, J., {Franceschini}, A., \& {Hatziminaoglou}, E. 2006, MNRAS, 166, 767

\bibitem[{Guzzo {et~al.}(2014)Guzzo, Scodeggio, Garilli, Granett, Fritz, Abbas,
  Adami, Arnouts, Bel, Bolzonella, Bottini, Branchini, Cappi, Coupon, Cucciati,
  Davidzon, Lucia, de~la Torre, Franzetti, Fumana, Hudelot, Ilbert, Iovino,
  Krywult, Brun, F{\`{e}}vre, Maccagni, Ma{\l}ek, Marulli, McCracken, Paioro,
  Peacock, Polletta, Pollo, Schlagenhaufer, Tasca, Tojeiro, Vergani, Zamorani,
  Zanichelli, Burden, Porto, Marchetti, Marinoni, Mellier, Moscardini, Nichol,
  Percival, Phleps, \& Wolk}]{Guzzo2014}
Guzzo, L., Scodeggio, M., Garilli, B., {et~al.} 2014, A\&A, 566, A108

\bibitem[{Haardt \& Maraschi(1991)}]{Haardt1991}
Haardt, F. \& Maraschi, L. 1991, ApJ, 380, L51

\bibitem[{Hambly {et~al.}(2008)Hambly, Collins, Cross, Mann, Read, Sutorius,
  Bond, Bryant, Emerson, Lawrence, Rimoldini, Stewart, Williams, Adamson,
  Hirst, Dye, \& Warren}]{Hambly2008}
Hambly, N.~C., Collins, R.~S., Cross, N. J.~G., {et~al.} 2008, MNRAS, 384, 637

\bibitem[{{Hickox} \& {Alexander}(2018)}]{Hickox2018}
{Hickox}, R. \& {Alexander}, D. 2018, ARAA, 56, 625

\bibitem[{Hickox {et~al.}(2007)Hickox, Jones, Forman, Murray, Brodwin, Brown,
  Eisenhardt, Stern, Kochanek, Eisenstein, Cool, Jannuzi, Dey, Brand, Gorjian,
  \& Caldwell}]{Hickox2007}
Hickox, R.~C., Jones, C., Forman, W.~R., {et~al.} 2007, ApJ, 671, 1365

\bibitem[{Hickox {et~al.}(2017)Hickox, Myers, Greene, Hainline, Zakamska, \&
  DiPompeo}]{Hickox2017}
Hickox, R.~C., Myers, A.~D., Greene, J.~E., {et~al.} 2017, ApJ, 849, 53

\bibitem[{Hönig \& Kishimoto(2010)}]{Hoenig2010}
Hönig, S.~F. \& Kishimoto, M. 2010, A\&A, 523, A27

\bibitem[{Hönig \& Kishimoto(2017)}]{Hoenig2017}
Hönig, S.~F. \& Kishimoto, M. 2017, ApJ, 838, L20

\bibitem[{Hönig {et~al.}(2010)Hönig, Kishimoto, Gandhi, Smette, Asmus,
  Duschl, Polletta, \& Weigelt}]{Hoenig2010a}
Hönig, S.~F., Kishimoto, M., Gandhi, P., {et~al.} 2010, A\&A, 515, A23

\bibitem[{Irwin {et~al.}(2004)Irwin, Lewis, Hodgkin, Bunclark, Evans, McMahon,
  Emerson, Stewart, \& Beard}]{Irwin2004}
Irwin, M.~J., Lewis, J., Hodgkin, S., {et~al.} 2004, in Optimizing Scientific
  Return for Astronomy through Information Technologies ({SPIE})

\bibitem[{Isobe {et~al.}(1986)Isobe, Feigelson, \& Nelson}]{Isobe1986}
Isobe, T., Feigelson, E.~D., \& Nelson, P.~I. 1986, ApJ, 306, 490

\bibitem[{Jaffarian \& Gaskell(2020)}]{Jaffarian2020}
Jaffarian, G.~W. \& Gaskell, C.~M. 2020, MNRAS
  [\eprint{http://arxiv.org/abs/2001.08900v1}]

\bibitem[{{Kalberla} {et~al.}(2005){Kalberla}, {Burton}, {Hartmann}, {Arnal},
  {Bajaja}, {Morras}, \& {P{\"o}ppel}}]{Kalberla2005}
{Kalberla}, P. M.~W., {Burton}, W.~B., {Hartmann}, D., {et~al.} 2005, A\&A,
  440, 775

\bibitem[{Kormendy {et~al.}(1996{\natexlab{a}})Kormendy, Bender, Ajhar,
  Dressler, Faber, Gebhardt, Grillmair, Lauer, Richstone, \&
  Tremaine}]{Kormendy1996}
Kormendy, J., Bender, R., Ajhar, E.~A., {et~al.} 1996{\natexlab{a}}, ApJ, 473,
  L91

\bibitem[{Kormendy {et~al.}(1996{\natexlab{b}})Kormendy, Bender, Richstone,
  Ajhar, Dressler, Faber, Gebhardt, Grillmair, Lauer, \&
  Tremaine}]{johnkormendy1996}
Kormendy, J., Bender, R., Richstone, D., {et~al.} 1996{\natexlab{b}}, ApJ, 459

\bibitem[{Koutoulidis {et~al.}(2018)Koutoulidis, Georgantopoulos, Mountrichas,
  Plionis, Georgakakis, Akylas, \& Rovilos}]{koutoulidis2018}
Koutoulidis, L., Georgantopoulos, I., Mountrichas, G., {et~al.} 2018, MNRAS,
  481, 3063

\bibitem[{Krumpe {et~al.}(2008)Krumpe, Lamer, Corral, Schwope, Carrera,
  Barcons, Page, Mateos, Tedds, \& Watson}]{Krumpe2008}
Krumpe, M., Lamer, G., Corral, A., {et~al.} 2008, Astronomy {\&} Astrophysics,
  483, 415

\bibitem[{LaMassa {et~al.}(2016)LaMassa, Civano, Brusa, Stern, Glikman,
  Gallagher, Urry, Cales, Cappelluti, Cardamone, Comastri, Farrah, Greene,
  Komossa, Merloni, Mroczkowski, Natarajan, Richards, Salvato, Schawinski, \&
  Treister}]{LaMassa2016a}
LaMassa, S.~M., Civano, F., Brusa, M., {et~al.} 2016, ApJ, 818, 88

\bibitem[{Lavalley {et~al.}(1992)Lavalley, Isobe, \& Feigelson}]{Lavalley1992}
Lavalley, M., Isobe, T., \& Feigelson, E. 1992, in Astronomical Society of the
  Pacific Conference Series, Vol.~25, Astronomical Data Analysis Software and
  Systems I, ed. D.~M. {Worrall}, C.~{Biemesderfer}, \& J.~{Barnes}, 245

\bibitem[{Malkan {et~al.}(1998)Malkan, Gorjian, \& Tam}]{Malkan1998}
Malkan, M.~A., Gorjian, V., \& Tam, R. 1998, Astrophys. J., Suppl. Ser., 117,
  25

\bibitem[{Masoura {et~al.}(2018)Masoura, Mountrichas, Georgantopoulos, Ruiz,
  Magdis, \& Plionis}]{Masoura2018}
Masoura, V.~A., Mountrichas, G., Georgantopoulos, I., {et~al.} 2018, A\&A, 618,
  A31

\bibitem[{{Mateos} {et~al.}(2012){Mateos}, {Alonso-Herrero}, {Carrera},
  {Blain}, {Watson}, {Barcons}, {Braito}, {Severgnini}, {Donley}, \&
  {Stern}}]{Mateos2012}
{Mateos}, S., {Alonso-Herrero}, A., {Carrera}, F.~J., {et~al.} 2012, MNRAS,
  426, 3271

\bibitem[{Merloni {et~al.}(2013)Merloni, Bongiorno, Brusa, Iwasawa, Mainieri,
  Magnelli, Salvato, Berta, Cappelluti, Comastri, Fiore, Gilli, Koekemoer,
  Floc, Lusso, Lutz, Miyaji, Pozzi, Riguccini, Rosario, Silverman, Symeonidis,
  Treister, Vignali, \& Zamorani}]{Merloni2013}
Merloni, A., Bongiorno, A., Brusa, M., {et~al.} 2013, MNRAS, 437, 3550

\bibitem[{Mountrichas {et~al.}(2017)Mountrichas, Georgantopoulos, Secrest,
  Ordov{\'{a}}s-Pascual, Corral, Akylas, Mateos, Carrera, \&
  Batziou}]{Mountrichas2017}
Mountrichas, G., Georgantopoulos, I., Secrest, N.~J., {et~al.} 2017, MNRAS,
  468, 3042

\bibitem[{Nandra \& Pounds(1994)}]{Nandra1994}
Nandra, K. \& Pounds, K.~A. 1994, MNRAS, 268, 405

\bibitem[{Nenkova {et~al.}(2008{\natexlab{a}})Nenkova, Sirocky, Nikutta,
  Ivezi{\'{c}}, \& Elitzur}]{Nenkova2008}
Nenkova, M., Sirocky, M.~M., Nikutta, R., Ivezi{\'{c}}, {\v{Z}}., \& Elitzur,
  M. 2008{\natexlab{a}}, ApJ, 685, 160

\bibitem[{Nenkova {et~al.}(2008{\natexlab{b}})Nenkova, Sirocky, \&
  Elitzur}]{Nenkova2008a}
Nenkova, M., Sirocky, M.~M.;~Ivezi{\'c}, {\v Z}., \& Elitzur, M.
  2008{\natexlab{b}}, ApJ, 685, 147

\bibitem[{Netzer(2015)}]{Netzer2015}
Netzer, H. 2015, ARA\&A, 53, 365

\bibitem[{Noll {et~al.}(2009)Noll, Burgarella, Giovannoli, Buat, Marcillac, \&
  Mu{\~{n}}oz-Mateos}]{Noll2009}
Noll, S., Burgarella, D., Giovannoli, E., {et~al.} 2009, A\&A, 507, 1793

\bibitem[{{Oliver} {et~al.}(2012)}]{Oliver2012}
{Oliver}, S.~J. {et~al.} 2012, MNRAS, 424, 1614

\bibitem[{Pierre {et~al.}(2016)Pierre, Pacaud, Adami, Alis, Altieri, Baran,
  Benoist, Birkinshaw, Bongiorno, Bremer, Brusa, Butler, Ciliegi, Chiappetti,
  Clerc, Corasaniti, Coupon, Breuck, Democles, Desai, Delhaize, Devriendt,
  Dubois, Eckert, Elyiv, Ettori, Evrard, Faccioli, Farahi, Ferrari, Finet,
  Fotopoulou, Fourmanoit, Gandhi, Gastaldello, Gastaud, Georgantopoulos, Giles,
  Guennou, Guglielmo, Horellou, Husband, Huynh, Iovino, Kilbinger, Koulouridis,
  Lavoie, Brun, Fevre, Lidman, Lieu, Lin, Mantz, Maughan, Maurogordato,
  McCarthy, McGee, Melin, Melnyk, Menanteau, Novak, Paltani, Plionis,
  Poggianti, Pomarede, Pompei, Ponman, Ramos-Ceja, Ranalli, Rapetti,
  Raychaudury, Reiprich, Rottgering, Rozo, Rykoff, Sadibekova, Santos,
  Sauvageot, Schimd, Sereno, Smith, Smol{\v{c}}i{\'{c}}, Snowden, Spergel,
  Stanford, Surdej, Valageas, Valotti, Valtchanov, Vignali, Willis, \&
  Ziparo}]{Pierre2016}
Pierre, M., Pacaud, F., Adami, C., {et~al.} 2016, A\&A, 592, A1

\bibitem[{Pineau {et~al.}(2017)Pineau, Derriere, Motch, Carrera, Genova,
  Michel, Mingo, Mints, G{\'{o}}mez-Mor{\'{a}}n, Rosen, \&
  Camu{\~{n}}as}]{Pineau2017}
Pineau, F.-X., Derriere, S., Motch, C., {et~al.} 2017, A\&A, 597, A89

\bibitem[{Scodeggio {et~al.}(2018)Scodeggio, Guzzo, Garilli, Granett,
  Bolzonella, de~la Torre, Abbas, Adami, Arnouts, Bottini, Cappi, Coupon,
  Cucciati, Davidzon, Franzetti, Fritz, Iovino, Krywult, Brun, F{\`{e}}vre,
  Maccagni, Ma{\l}ek, Marchetti, Marulli, Polletta, Pollo, Tasca, Tojeiro,
  Vergani, Zanichelli, Bel, Branchini, Lucia, Ilbert, McCracken, Moutard,
  Peacock, Zamorani, Burden, Fumana, Jullo, Marinoni, Mellier, Moscardini, \&
  Percival}]{Scodeggio2018}
Scodeggio, M., Guzzo, L., Garilli, B., {et~al.} 2018, A\&A, 609, A84

\bibitem[{Siebenmorgen {et~al.}(2015)Siebenmorgen, Heymann, \&
  Efstathiou}]{Siebenmorgen2015}
Siebenmorgen, R., Heymann, F., \& Efstathiou, A. 2015, A\&A, 583, A120

\bibitem[{Stalevski {et~al.}(2016)Stalevski, Ricci, Ueda, Lira, Fritz, \&
  Baes}]{Stalevski2016}
Stalevski, M., Ricci, C., Ueda, Y., {et~al.} 2016, MNRAS, 458, 2288

\bibitem[{{Stern} {et~al.}(2012){Stern}, {Assef}, {Benford}, {Blain}, {Cutri},
  {Dey}, {Eisenhardt}, {Griffith}, {Jarrett}, {Lake}, {Masci}, {Petty},
  {Stanford}, {Tsai}, {Wright}, {Yan}, {Harrison}, \& {Madsen}}]{Stern2012}
{Stern}, D., {Assef}, R.~J., {Benford}, D.~J., {et~al.} 2012, ApJ, 753, 30

\bibitem[{{Ueda} {et~al.}(2014){Ueda}, {Akiyama}, {Hasinger}, {Miyaji}, \&
  {Watson}}]{Ueda2014}
{Ueda}, Y., {Akiyama}, M., {Hasinger}, G., {Miyaji}, T., \& {Watson}, M.~G.
  2014, ApJ, 786, 104

\bibitem[{Ulrich {et~al.}(1997)Ulrich, Maraschi, \& Urry}]{Ulrich1997}
Ulrich, M.-H., Maraschi, L., \& Urry, C.~M. 1997, ARA\&A, 35, 445

\bibitem[{Wilkins \& Gallo(2015)}]{Wilkins2015}
Wilkins, D.~R. \& Gallo, L.~C. 2015, MNRAS, 448, 703

\bibitem[{{Wright} {et~al.}(2010){Wright}, {Eisenhardt}, {Mainzer}, {Ressler},
  {Cutri}, {Jarrett}, {Kirkpatrick}, {Padgett}, {McMillan}, {Skrutskie},
  {Stanford}, {Cohen}, {Walker}, {Mather}, {Leisawitz}, {Gautier}, {McLean},
  {Benford}, {Lonsdale}, {Blain}, {Mendez}, {Irace}, {Duval}, {Liu}, {Royer},
  {Heinrichsen}, {Howard}, {Shannon}, {Kendall}, {Walsh}, {Larsen}, {Cardon},
  {Schick}, {Schwalm}, {Abid}, {Fabinsky}, {Naes}, \& {Tsai}}]{Wright2010}
{Wright}, E.~L., {Eisenhardt}, P.~R.~M., {Mainzer}, A.~K., {et~al.} 2010, AJ,
  140, 1868

\bibitem[{Yan {et~al.}(2013)Yan, Donoso, Tsai, Stern, Assef, Eisenhardt, Blain,
  Cutri, Jarrett, Stanford, Wright, Bridge, \& Riechers}]{Yan2013}
Yan, L., Donoso, E., Tsai, C.-W., {et~al.} 2013, AJ, 145, 55

\end{thebibliography}
\bibliographystyle{aa}

\appendix

\section{X-ray, optical spectra, and SEDs}
\label{appendix_figs}

In this section, we present the X-ray, SEDs, and the optical spectra (when available) for each one of the 47 optically red sources in our sample. Numbers 1 and 2 are used to denote the classification of the sources based on the various criteria. Specifically, 1 refers to unobscured and 2 to obscured sources. Number 0 is used in those cases where the optical spectrum is too noisy to extract any useful information. Number 3 is used to denote IMD type of classification. The numbers in parentheses refer to the classification based on the X-ray spectrum, the SED, and the optical spectrum, respectively. We use three decimal points for spectroscopic redshifts and two decimal points for photometric redshifts.

\clearpage

\begin{figure}
\includegraphics[height=0.87\columnwidth]{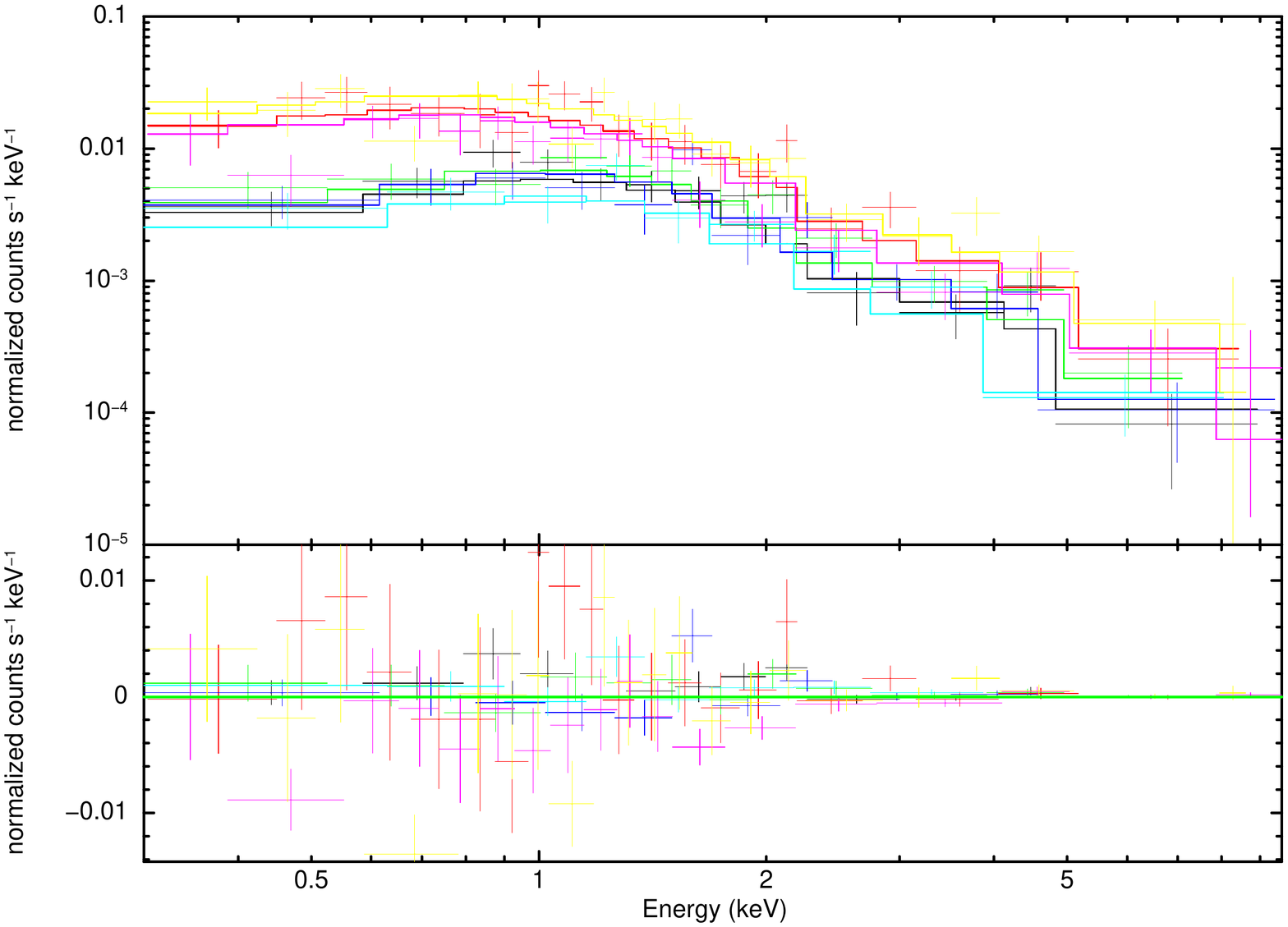}
\includegraphics[height=0.86\columnwidth]{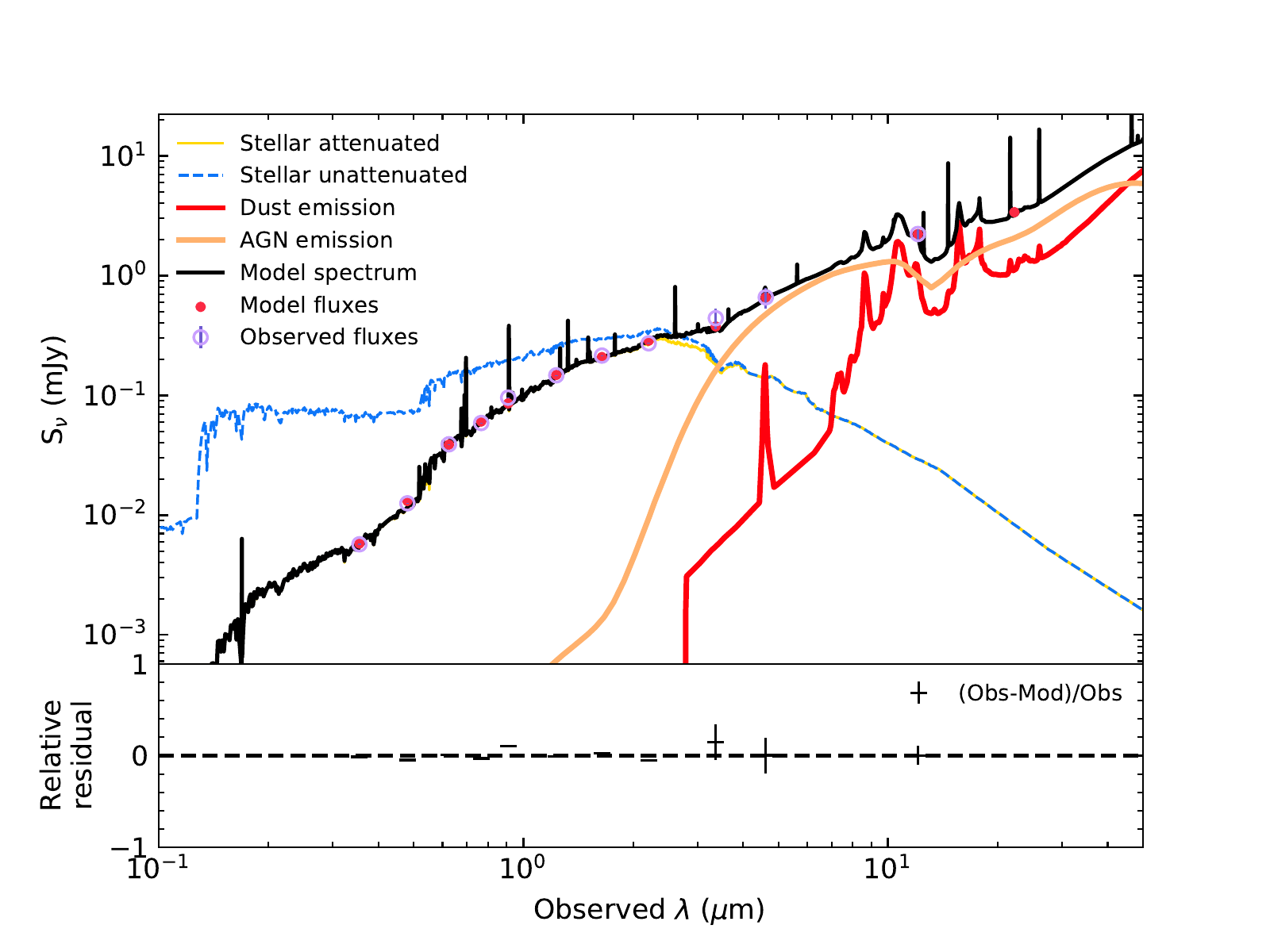}
\includegraphics[height=0.74\columnwidth]{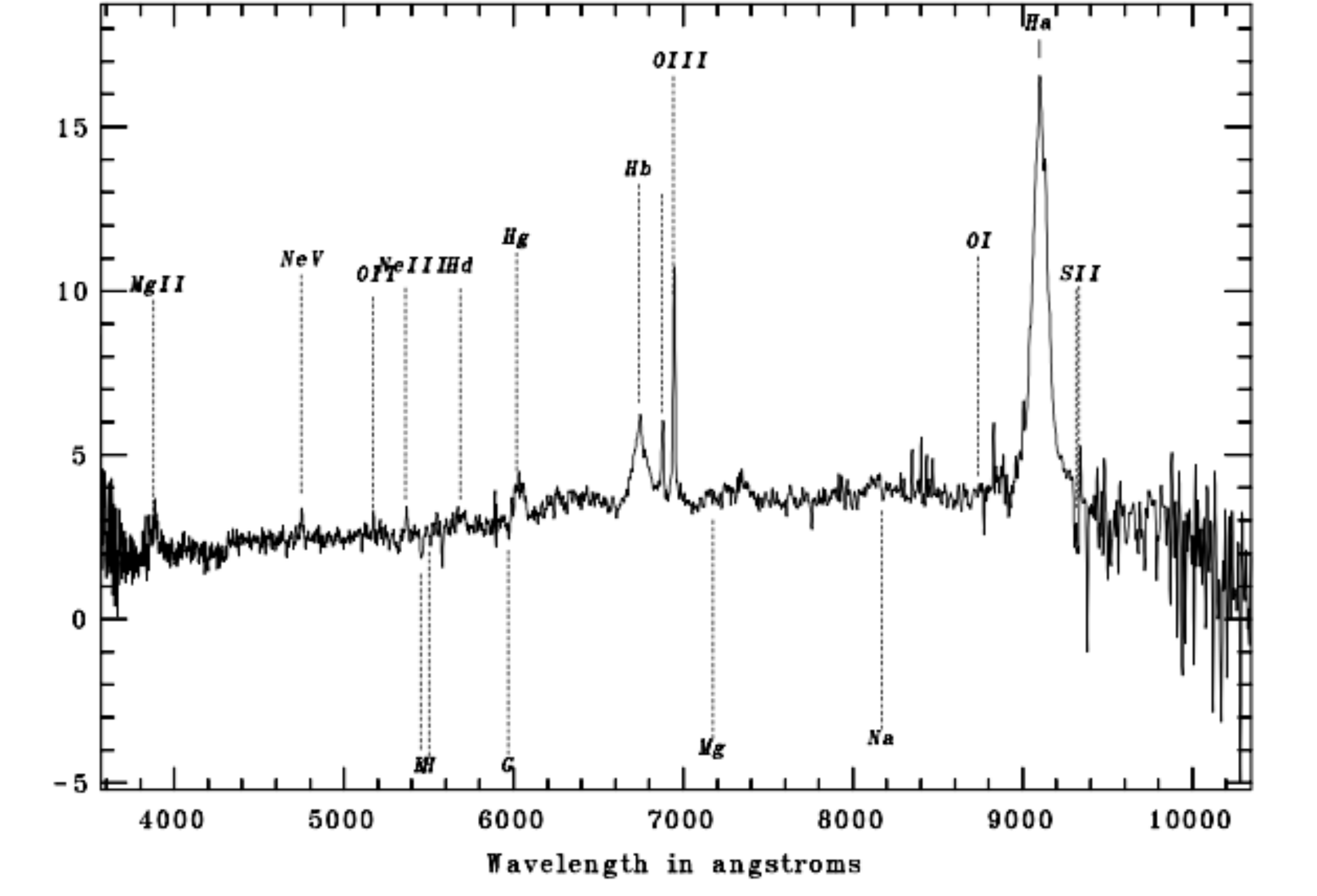}
\caption{J021835.7-053758~(1,2,1), z=0.387\\}
\label{4}
\end{figure}
\begin{figure}
\includegraphics[height=0.87\columnwidth]{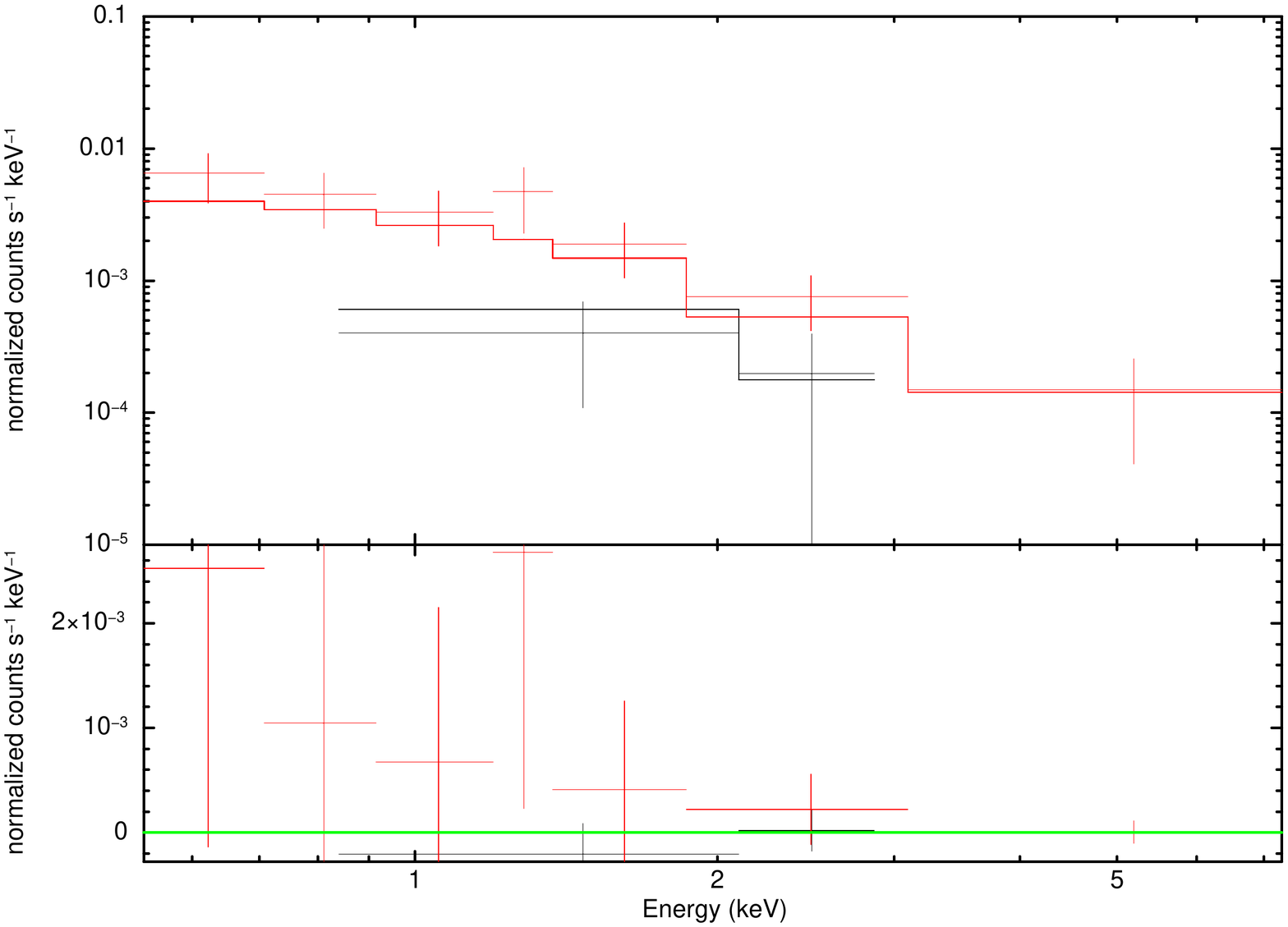}
\includegraphics[height=0.86\columnwidth]{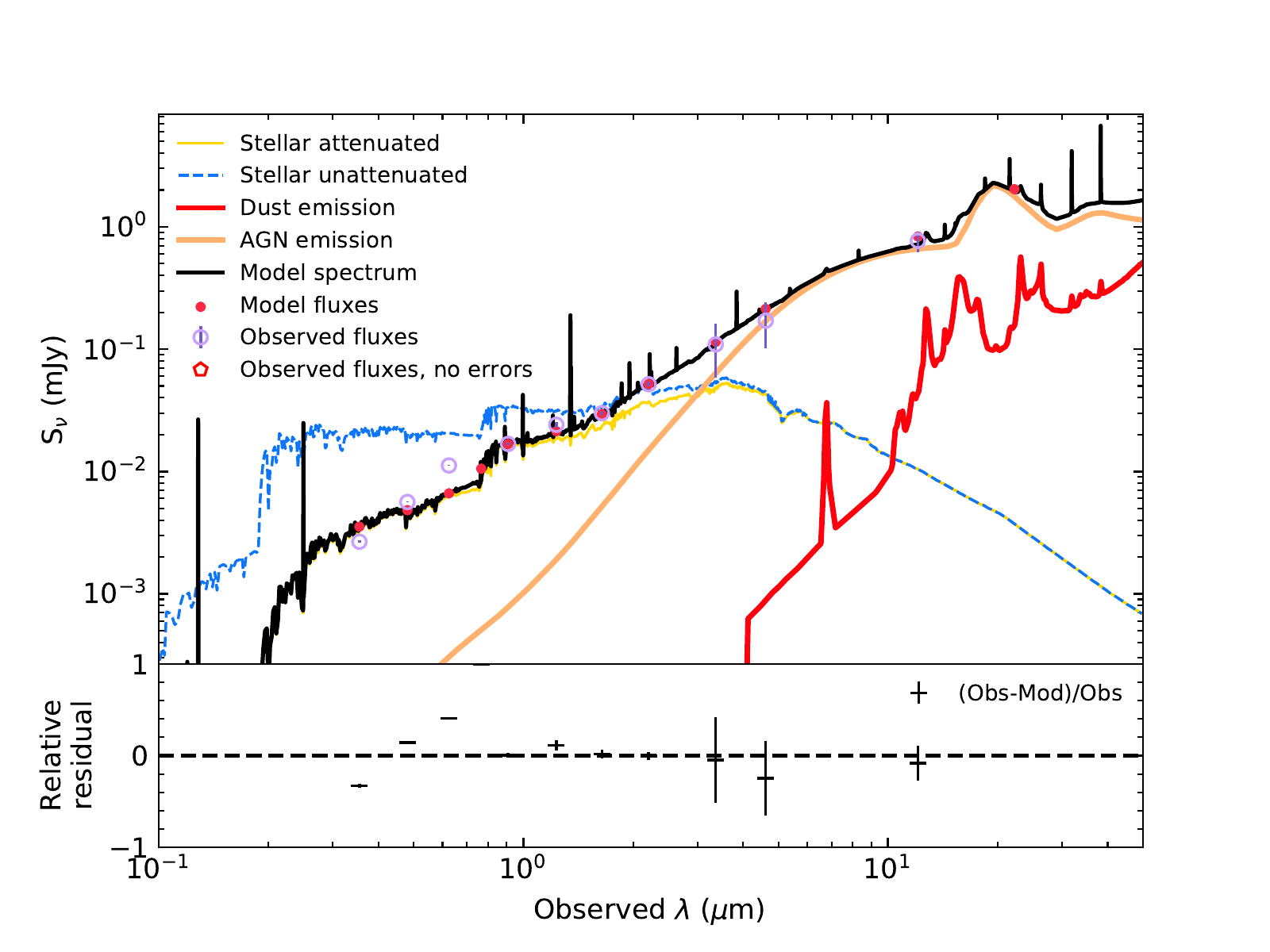}
\includegraphics[height=0.74\columnwidth]{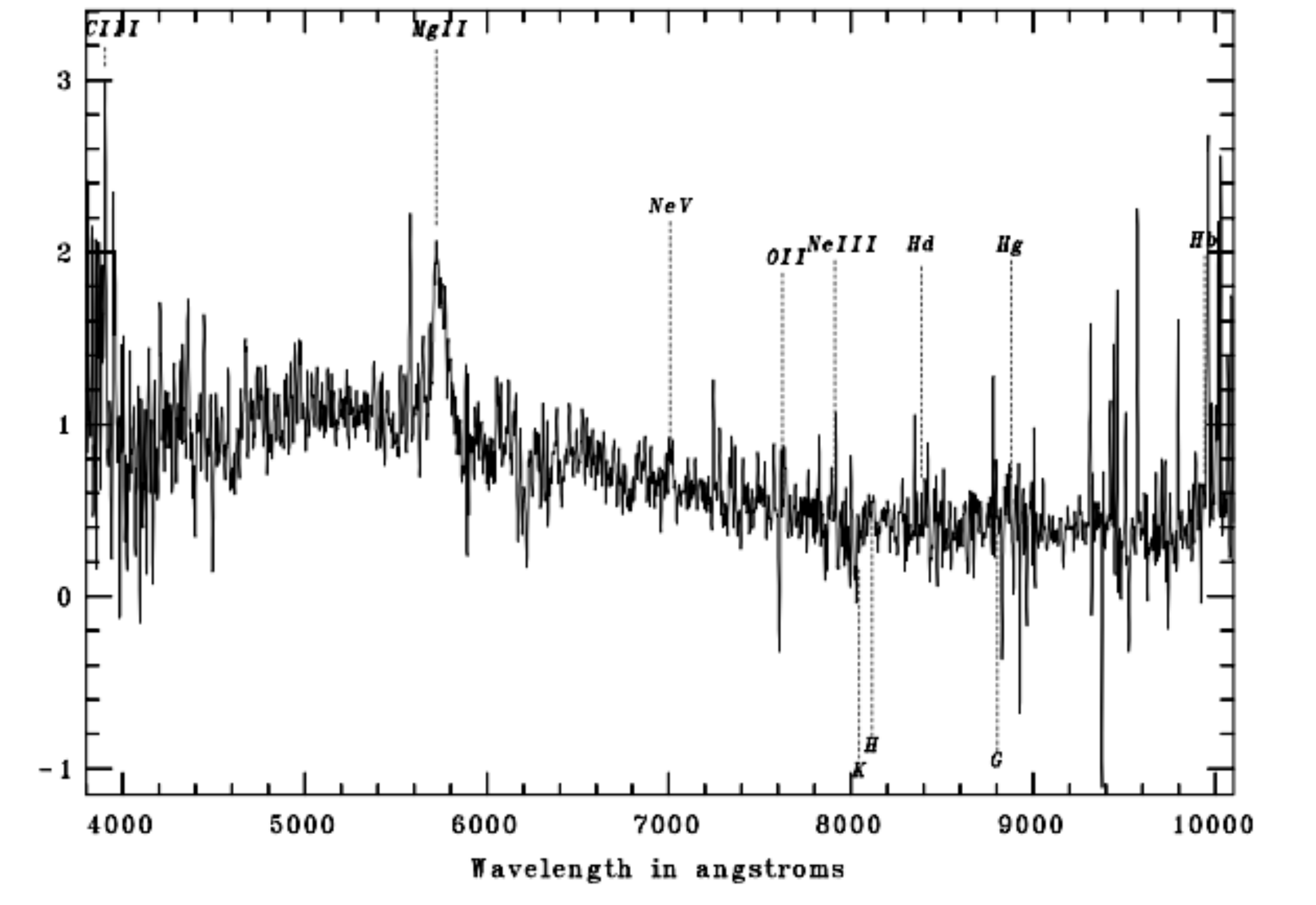}
\caption{J022848.4-044426~(2,2,1), z=1.046}
\label{}
\end{figure}
\begin{figure}
\includegraphics[height=0.87\columnwidth]{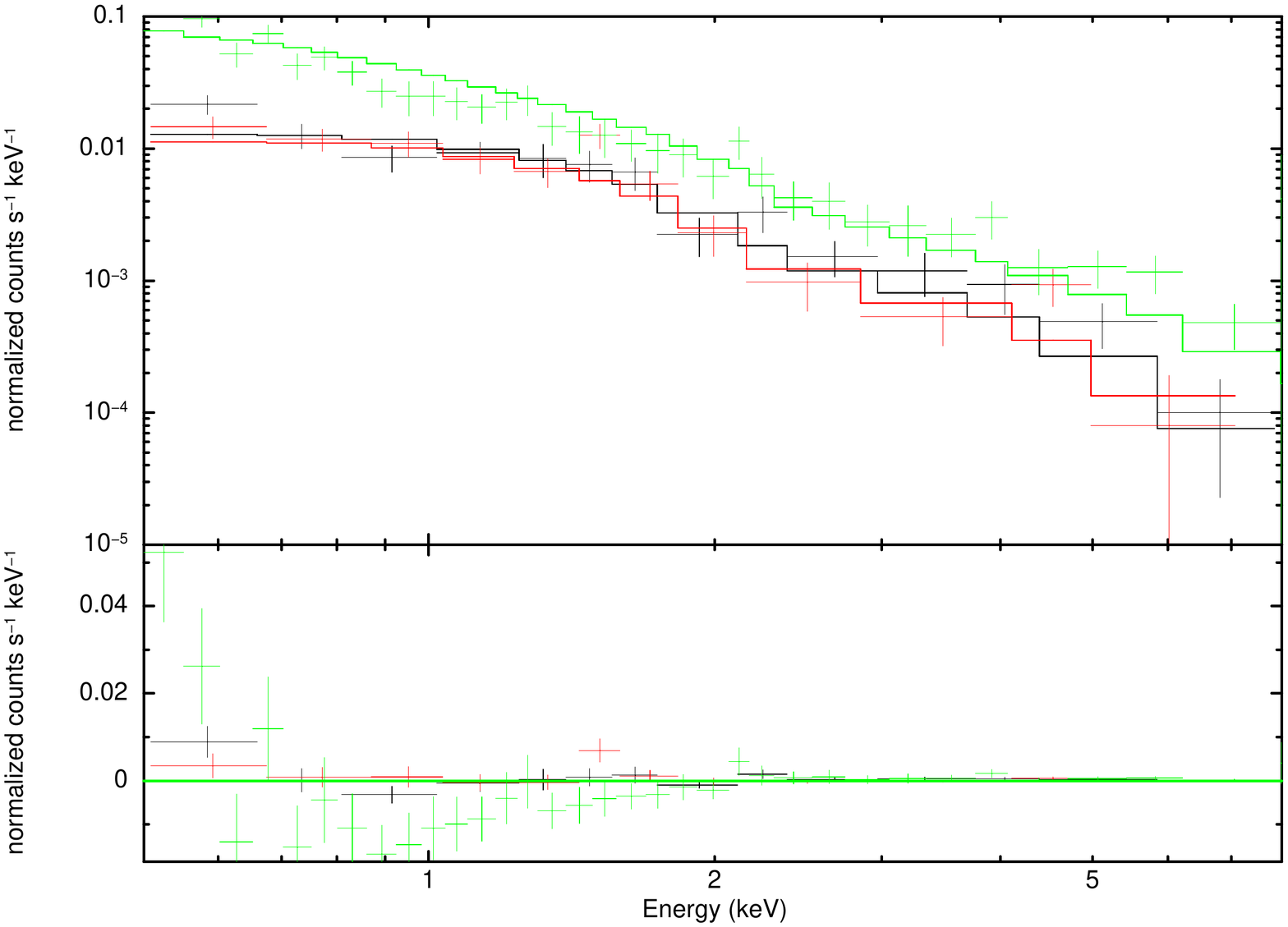}
\includegraphics[height=0.86\columnwidth]{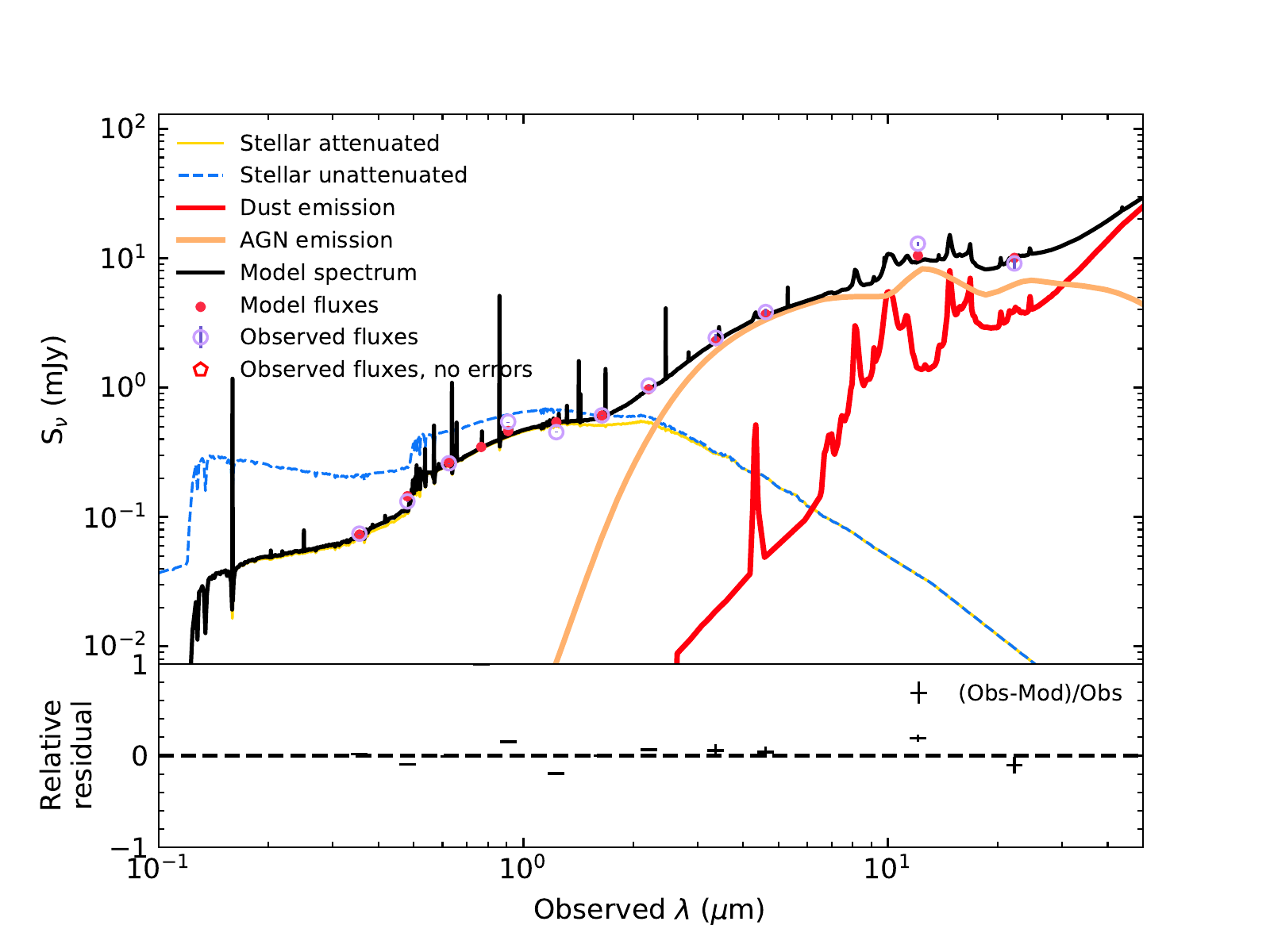}
\includegraphics[height=0.74\columnwidth]{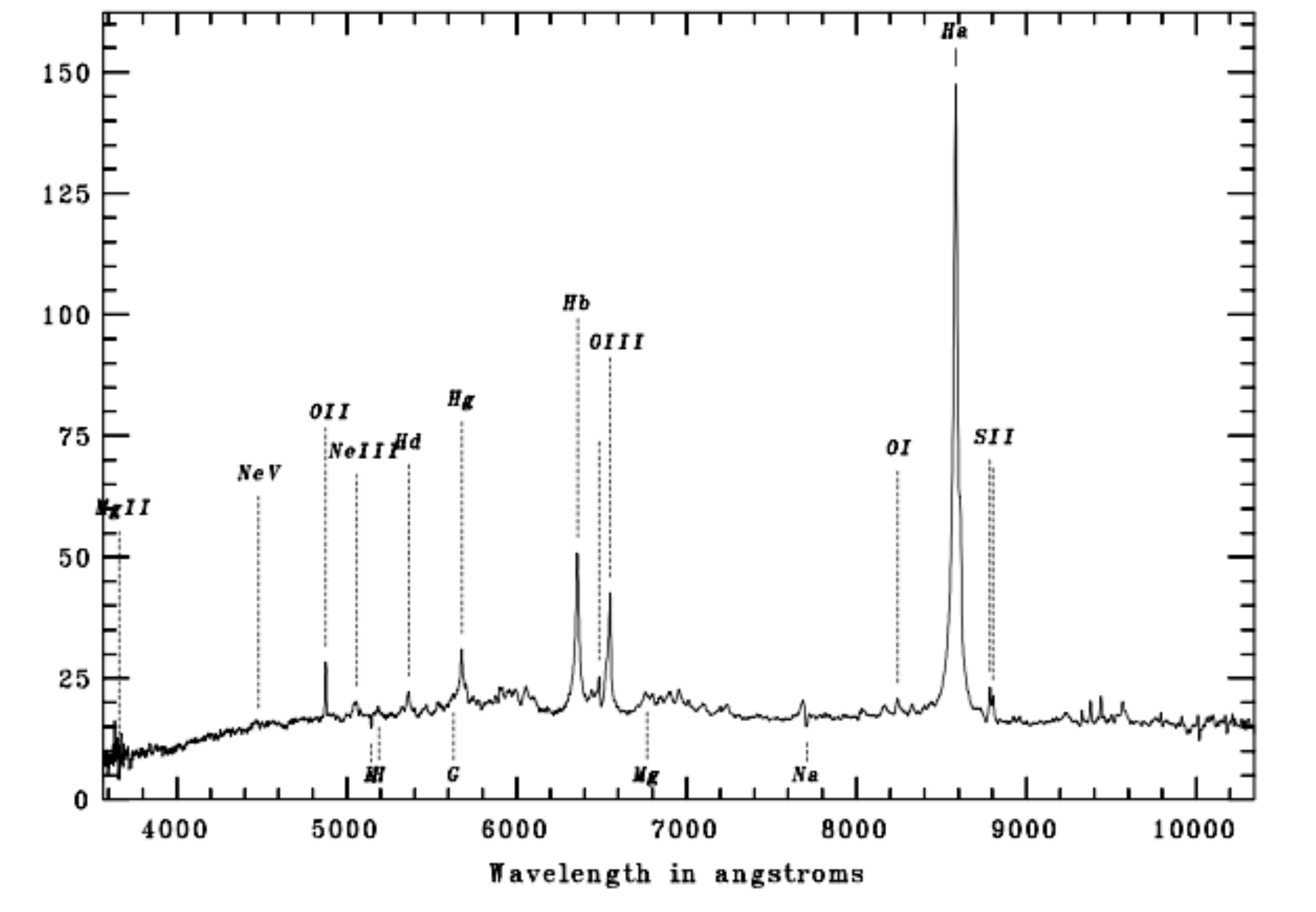}
\caption{J022928.4-051124~(2,2,1), z=0.307}
\label{}
\end{figure}
\begin{figure}
\includegraphics[height=0.87\columnwidth]{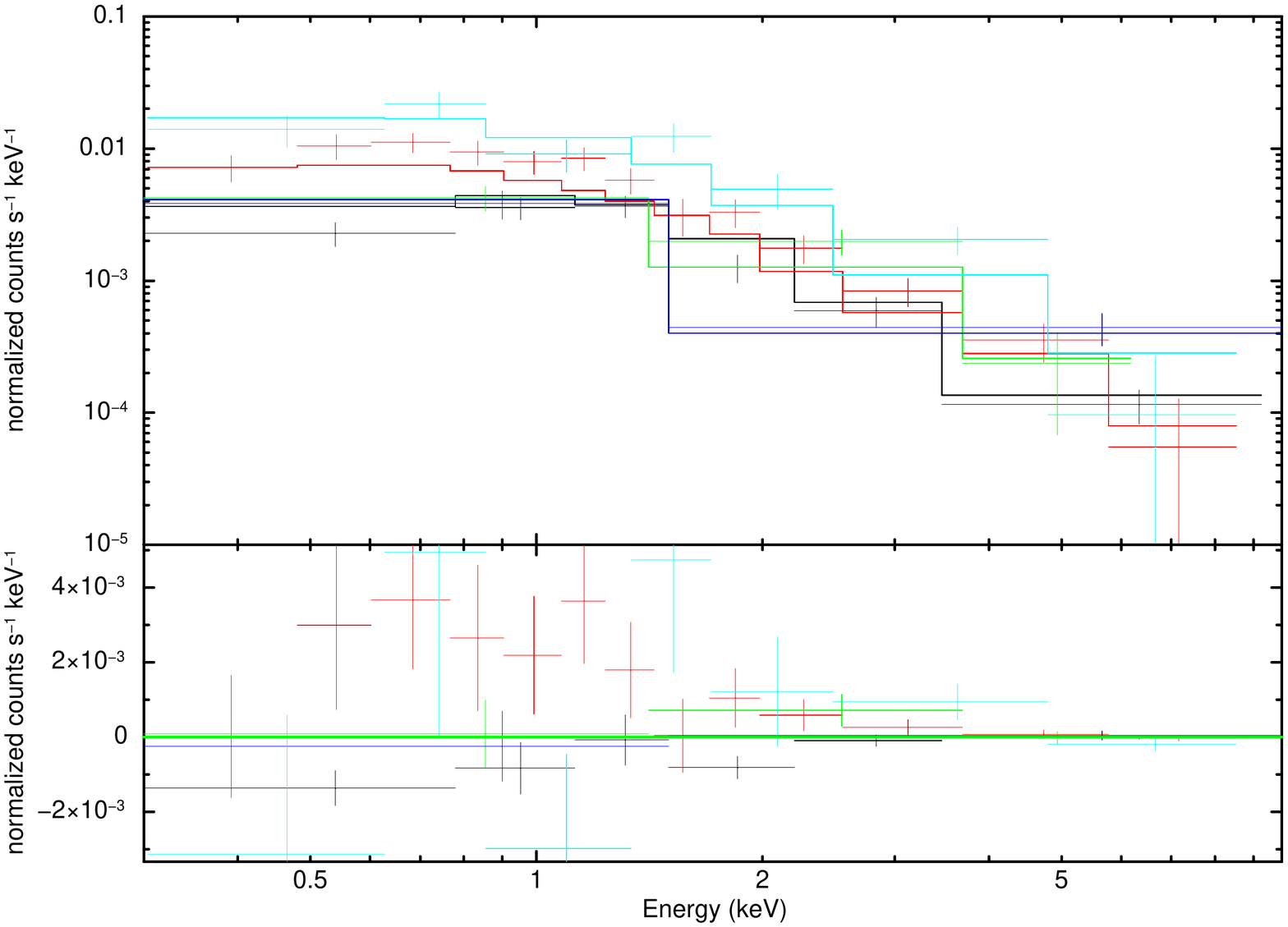}
\includegraphics[height=0.86\columnwidth]{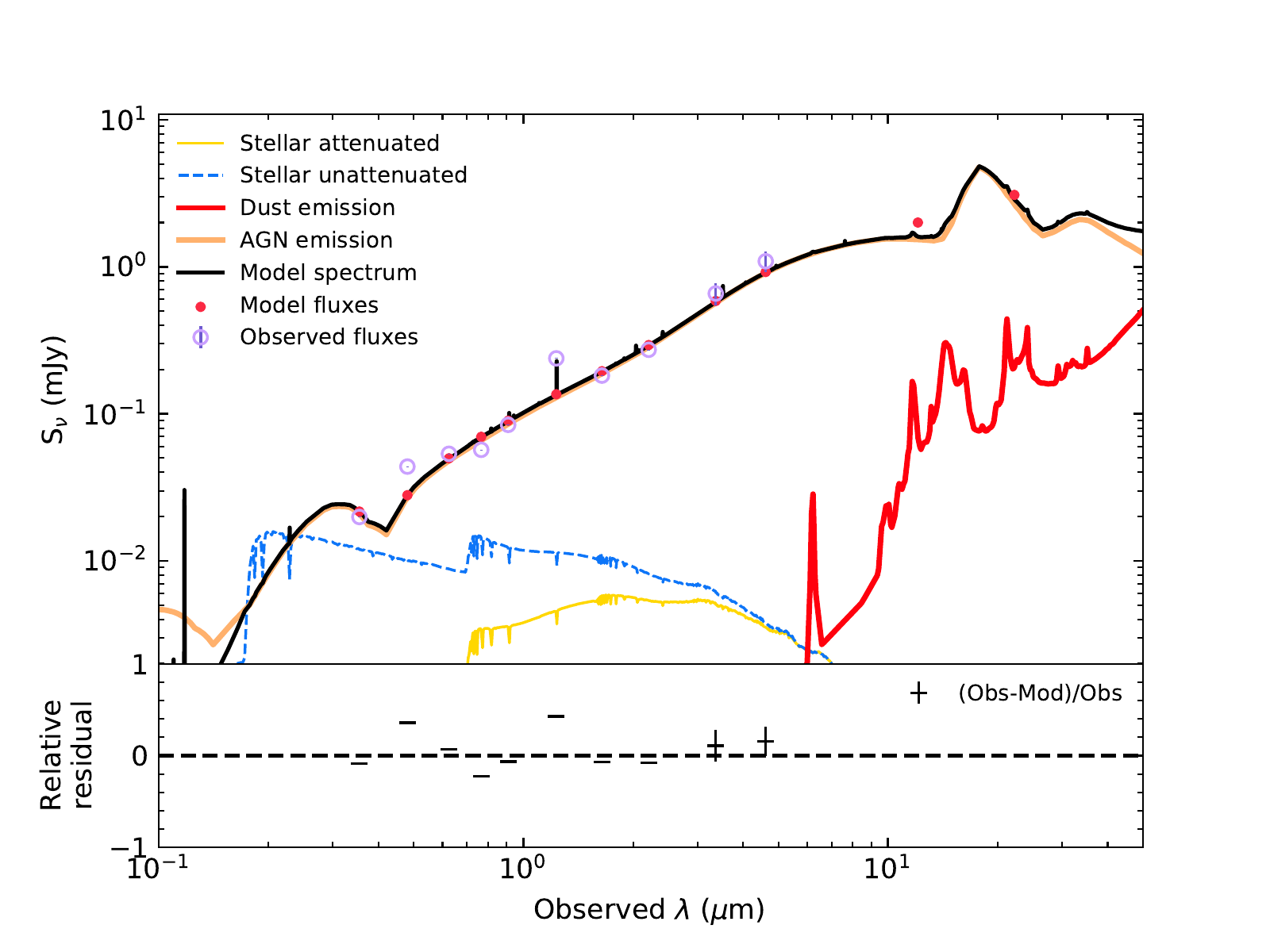}
\includegraphics[height=0.74\columnwidth]{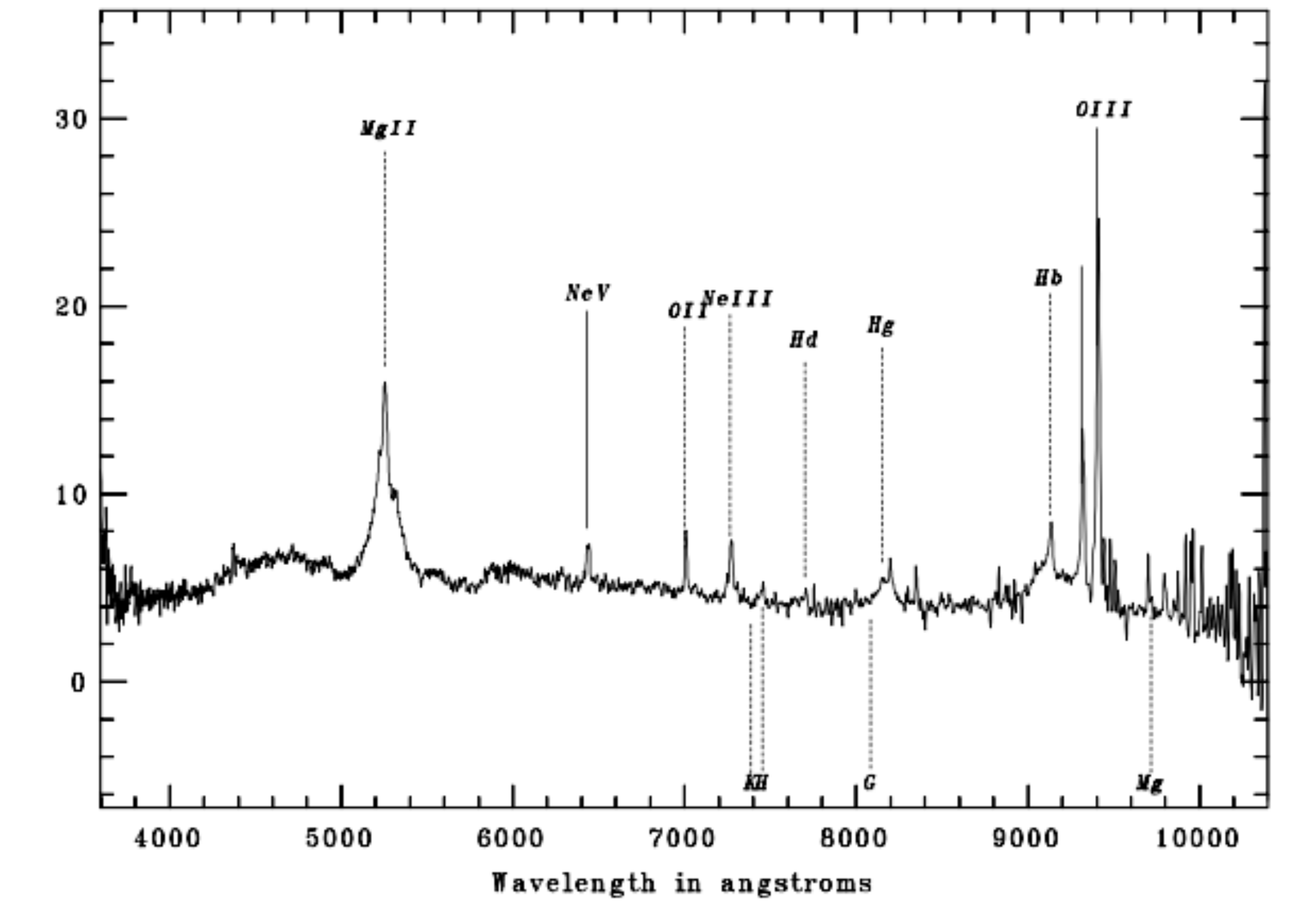}
\caption{J022809.0-041235~(1,1,1), z=0.879}
\label{}
\end{figure}
\begin{figure}
\includegraphics[height=0.87\columnwidth]{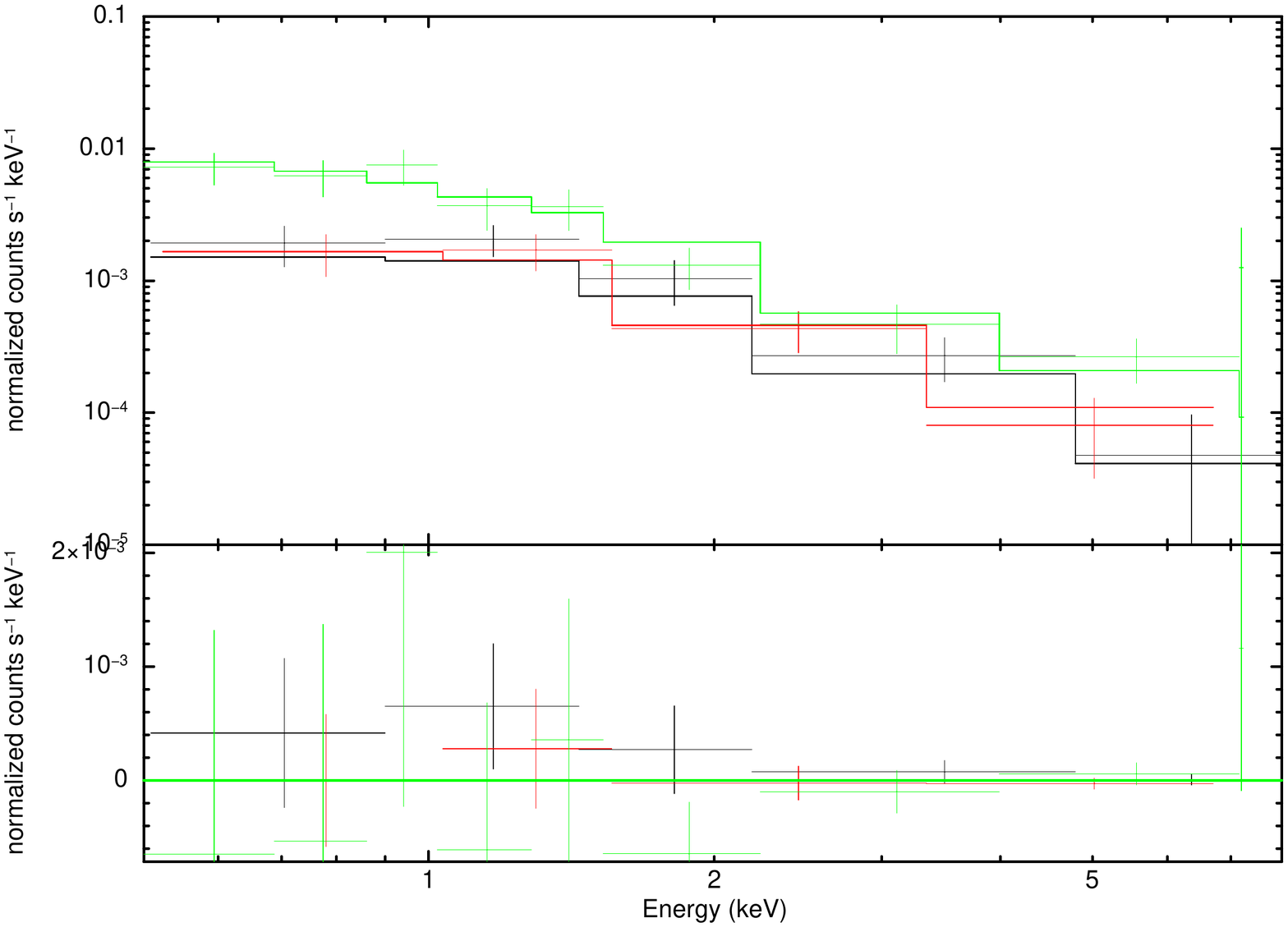}
\includegraphics[height=0.86\columnwidth]{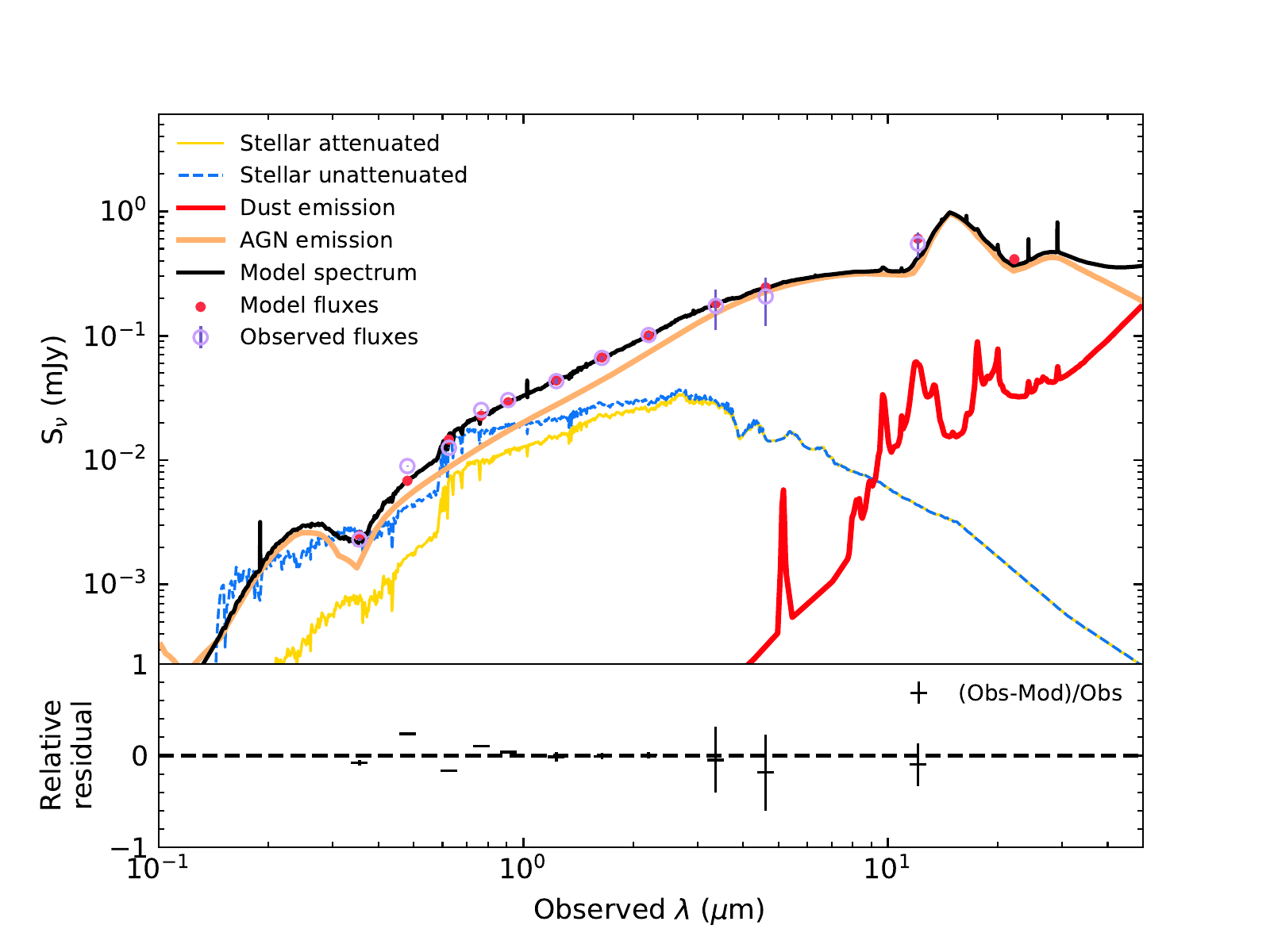}
\includegraphics[height=0.74\columnwidth]{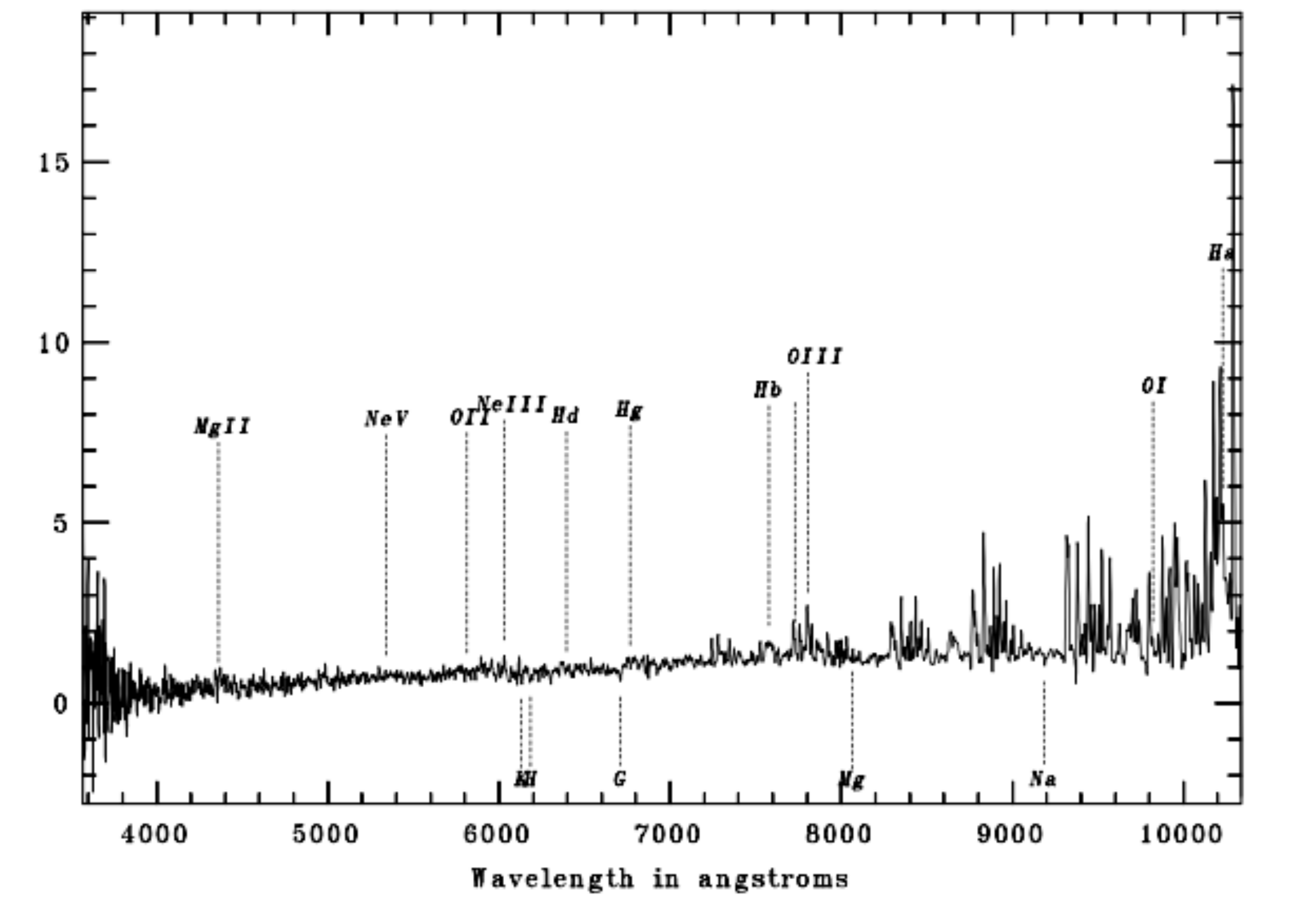}
\caption{J020654.9-064552~(1,2,2), z=1.412}
\label{}
\end{figure}
\begin{figure}
\includegraphics[height=0.87\columnwidth]{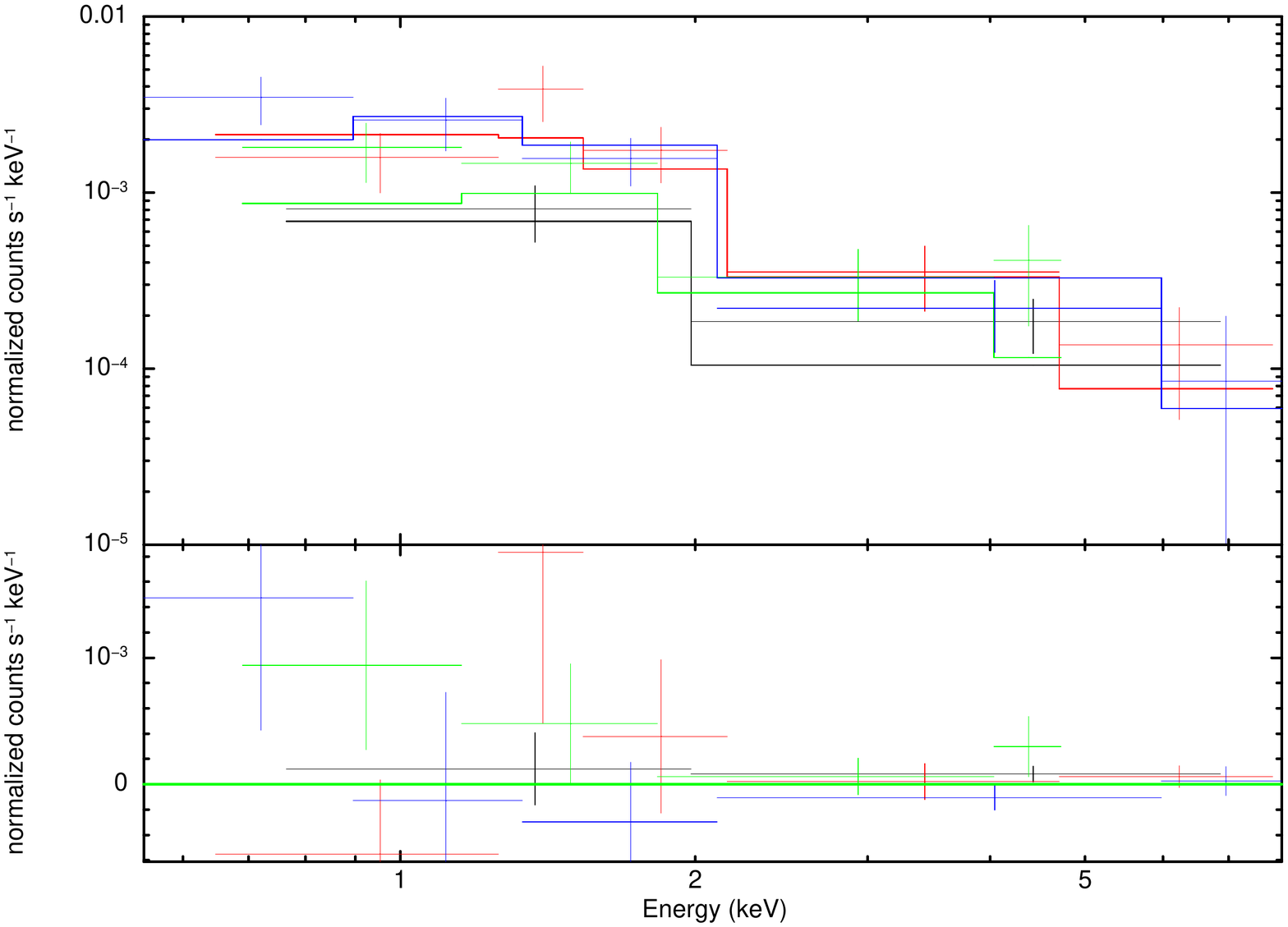}
\includegraphics[height=0.86\columnwidth]{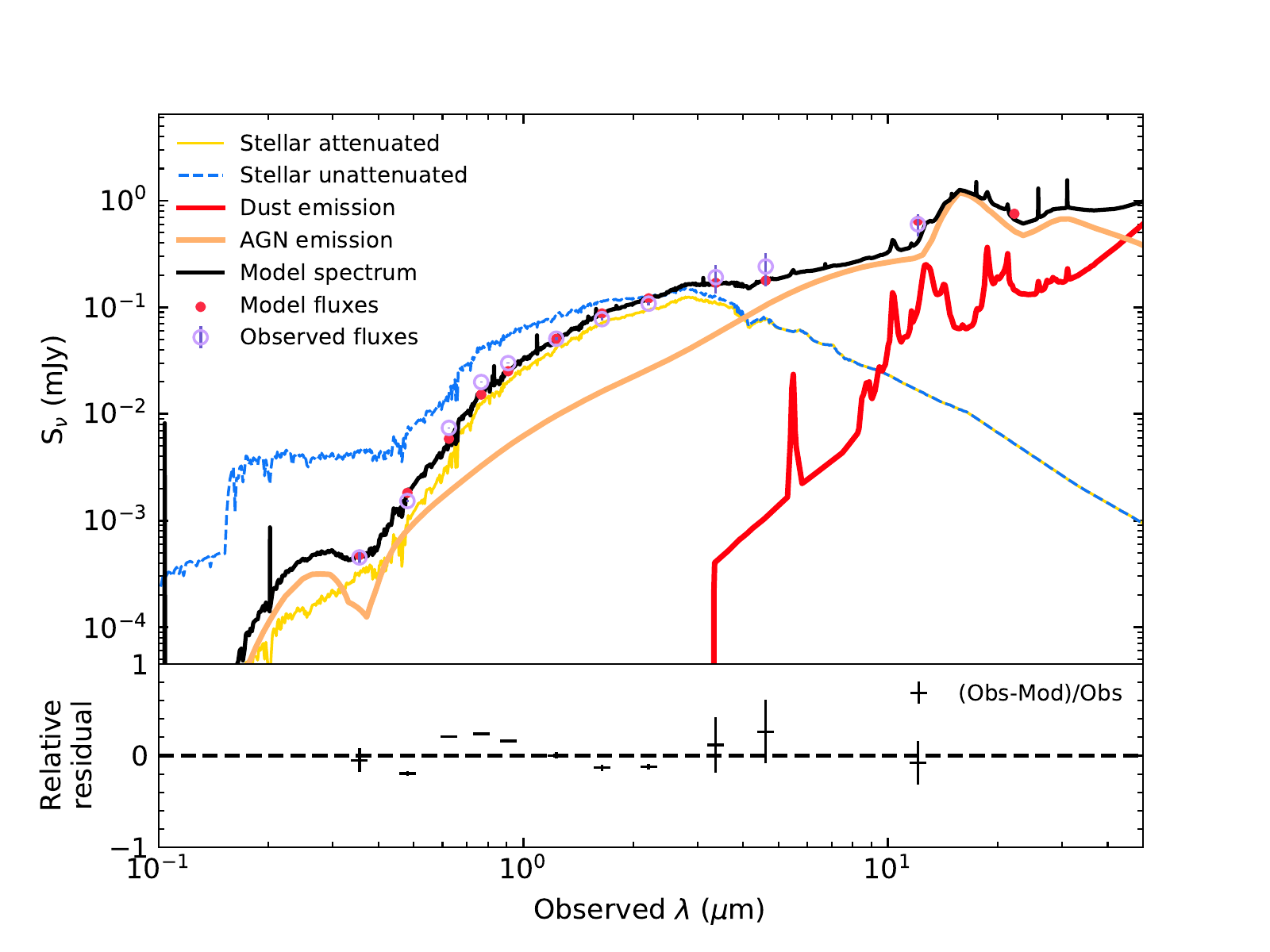}
\includegraphics[height=0.74\columnwidth]{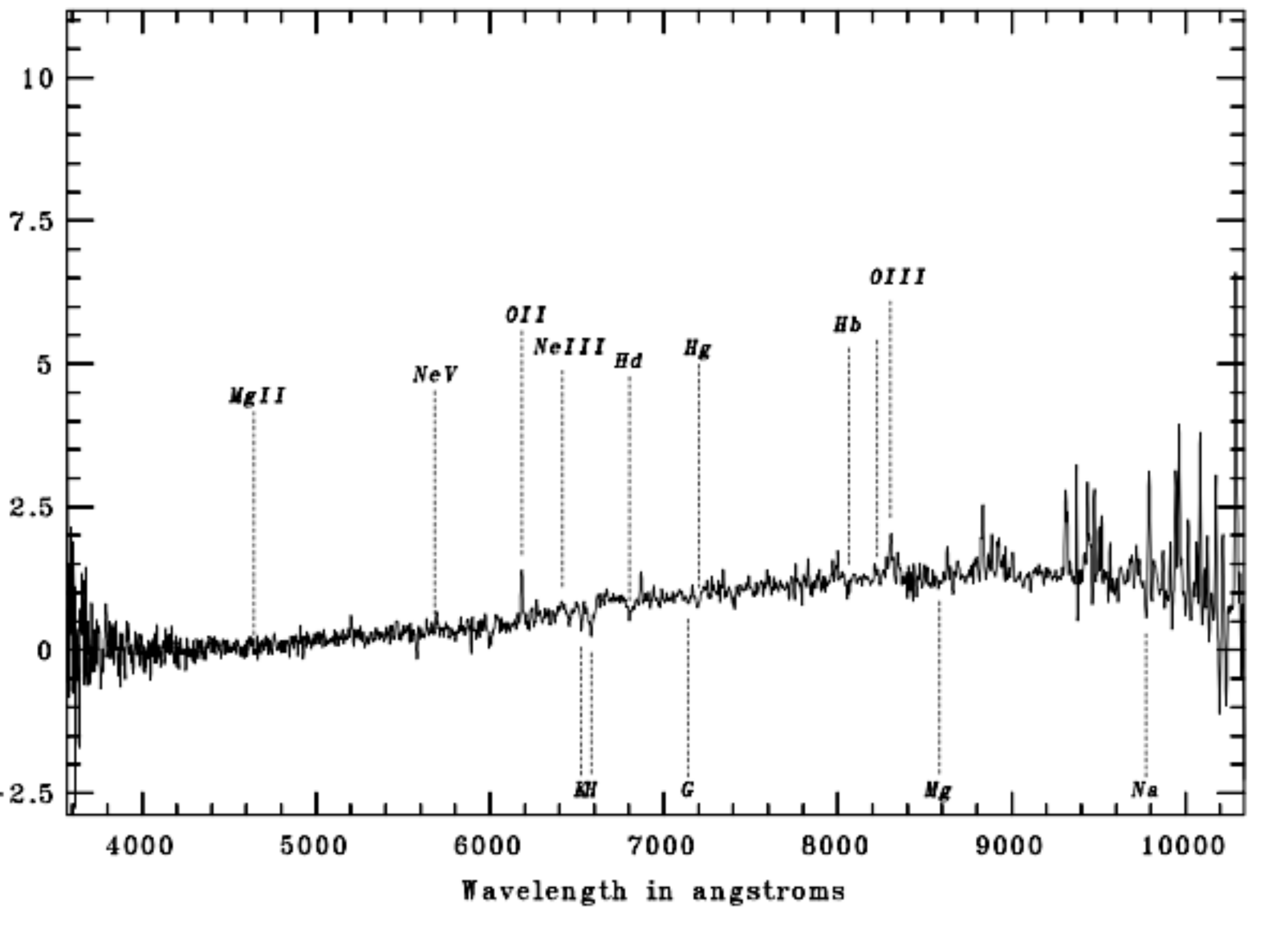}
\caption{J020410.4-063924~(2,2,2), z=0.414}
\label{}
\end{figure}
\begin{figure}
\includegraphics[height=0.87\columnwidth]{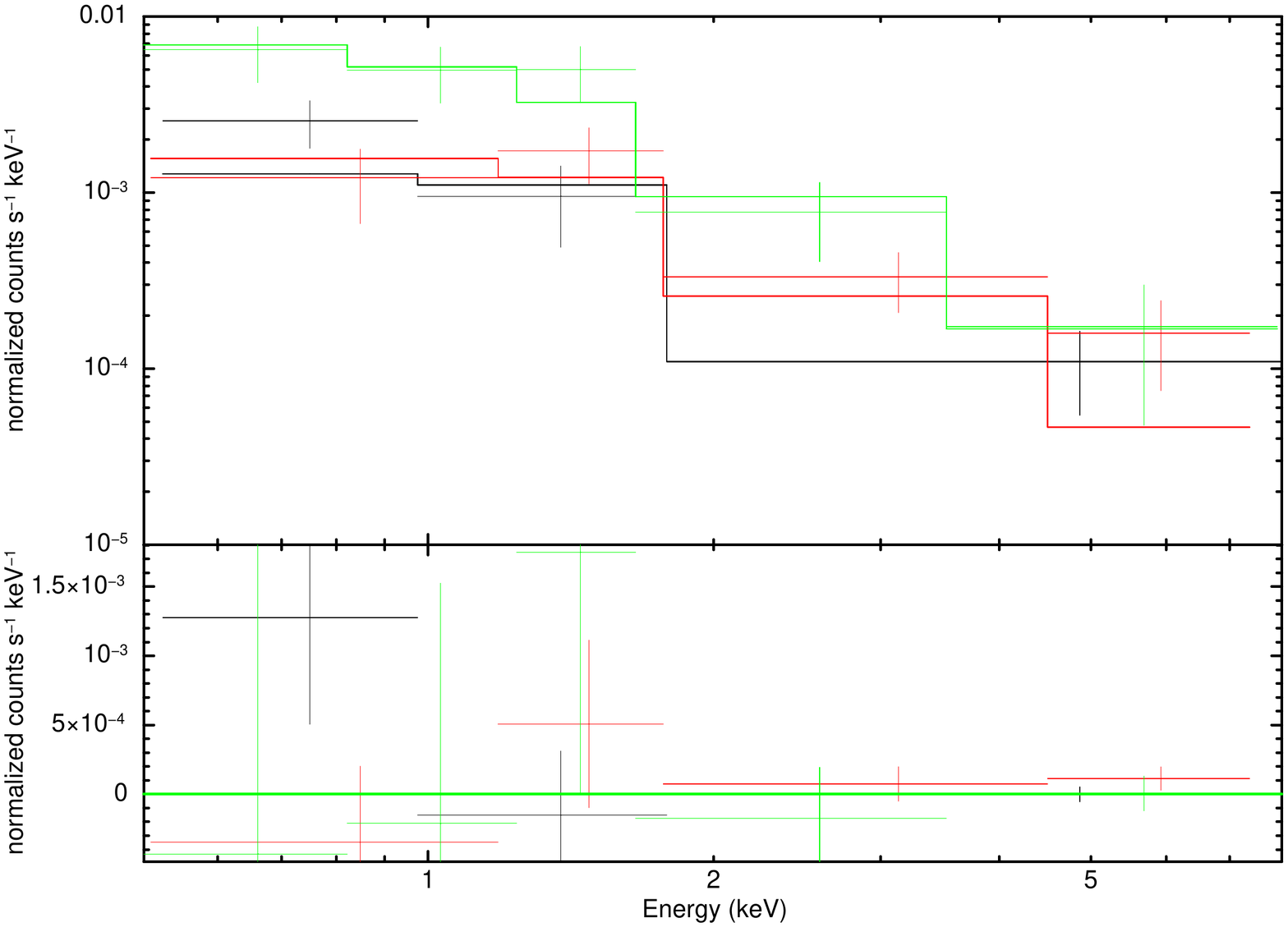}
\includegraphics[height=0.86\columnwidth]{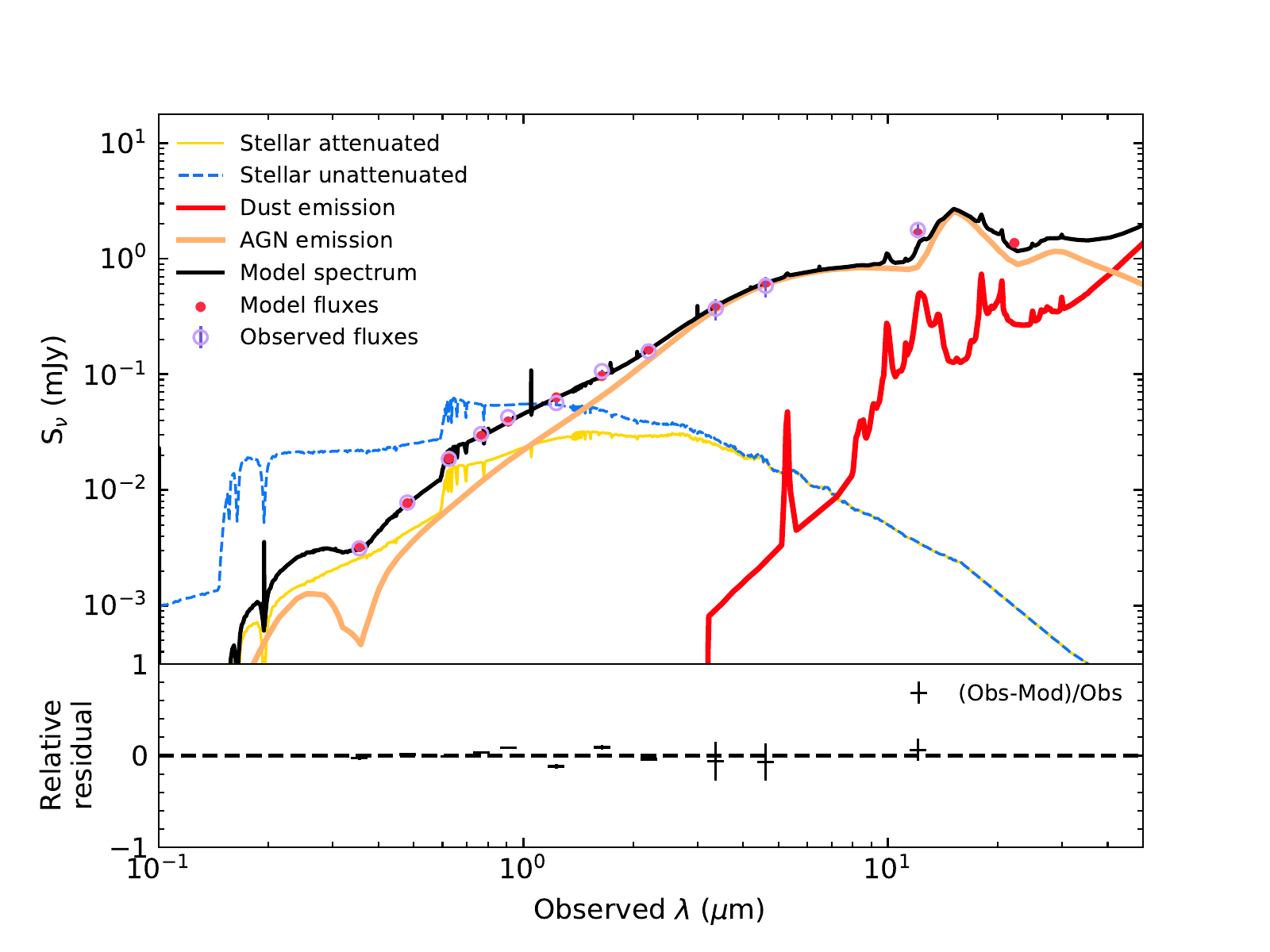}
\includegraphics[height=0.74\columnwidth]{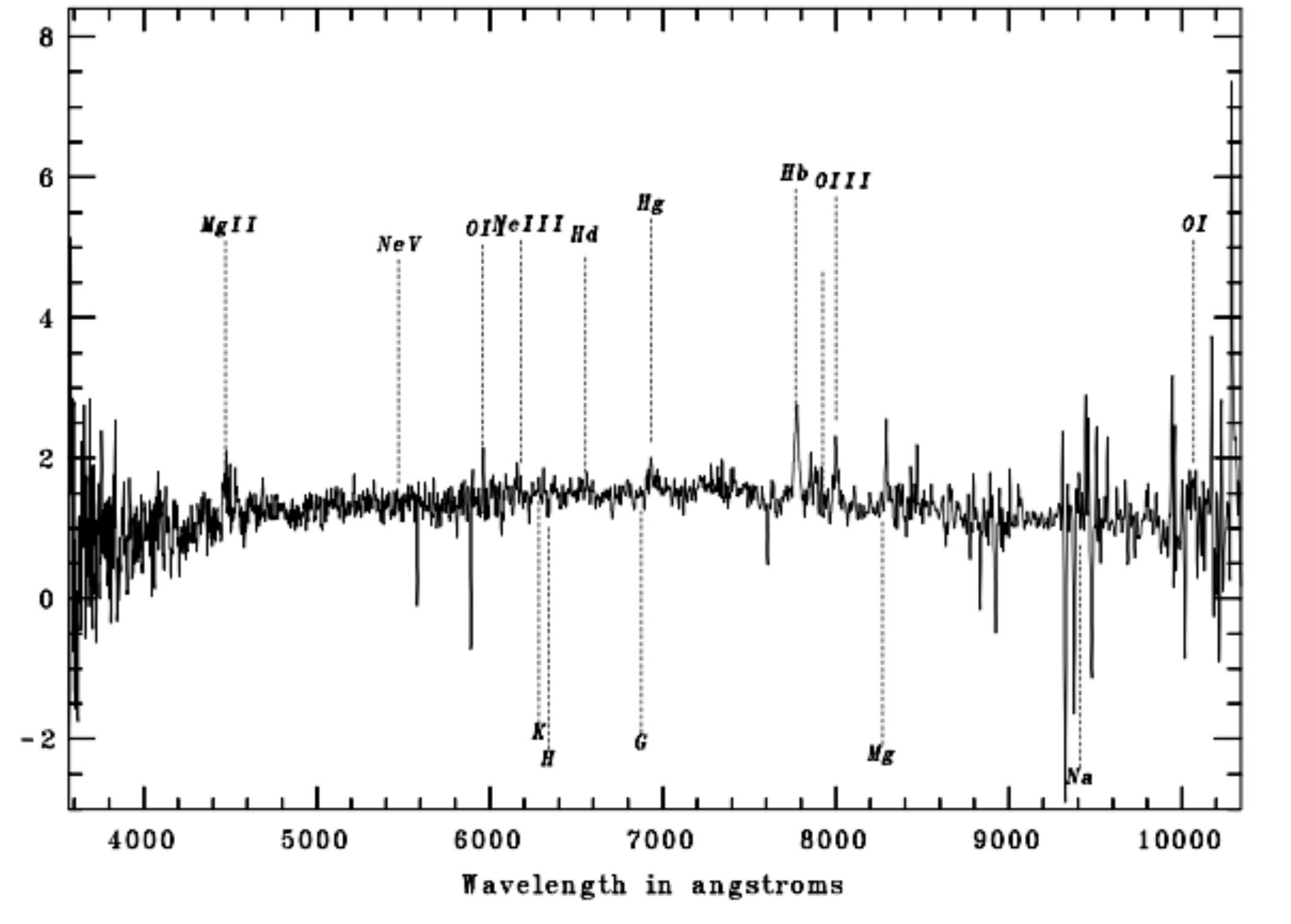}
\caption{J023315.5-054747~(1,2,2), z=0.598}
\label{}
\end{figure}
\begin{figure}
\includegraphics[height=0.87\columnwidth]{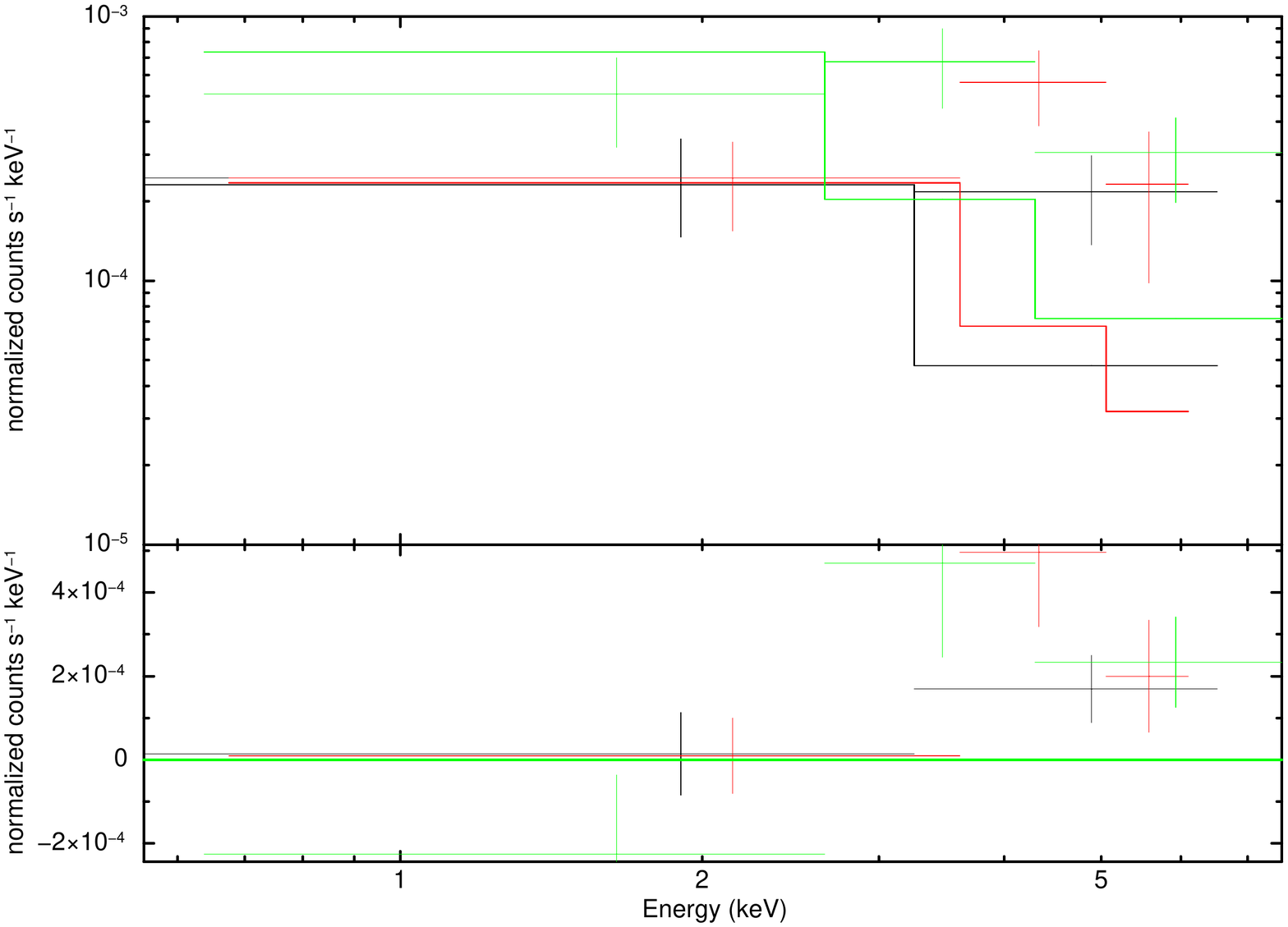}
\includegraphics[height=0.86\columnwidth]{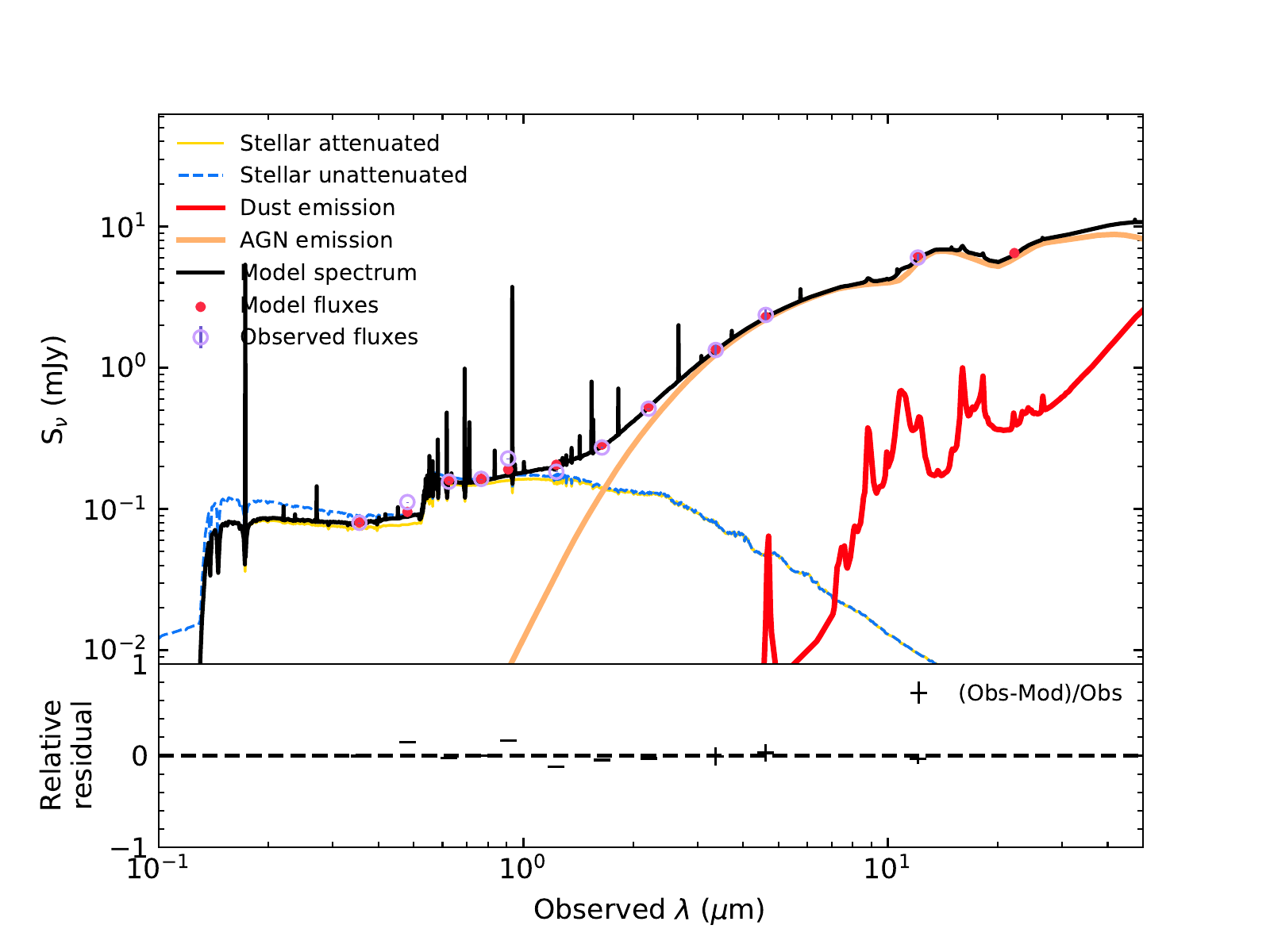}
\includegraphics[height=0.74\columnwidth]{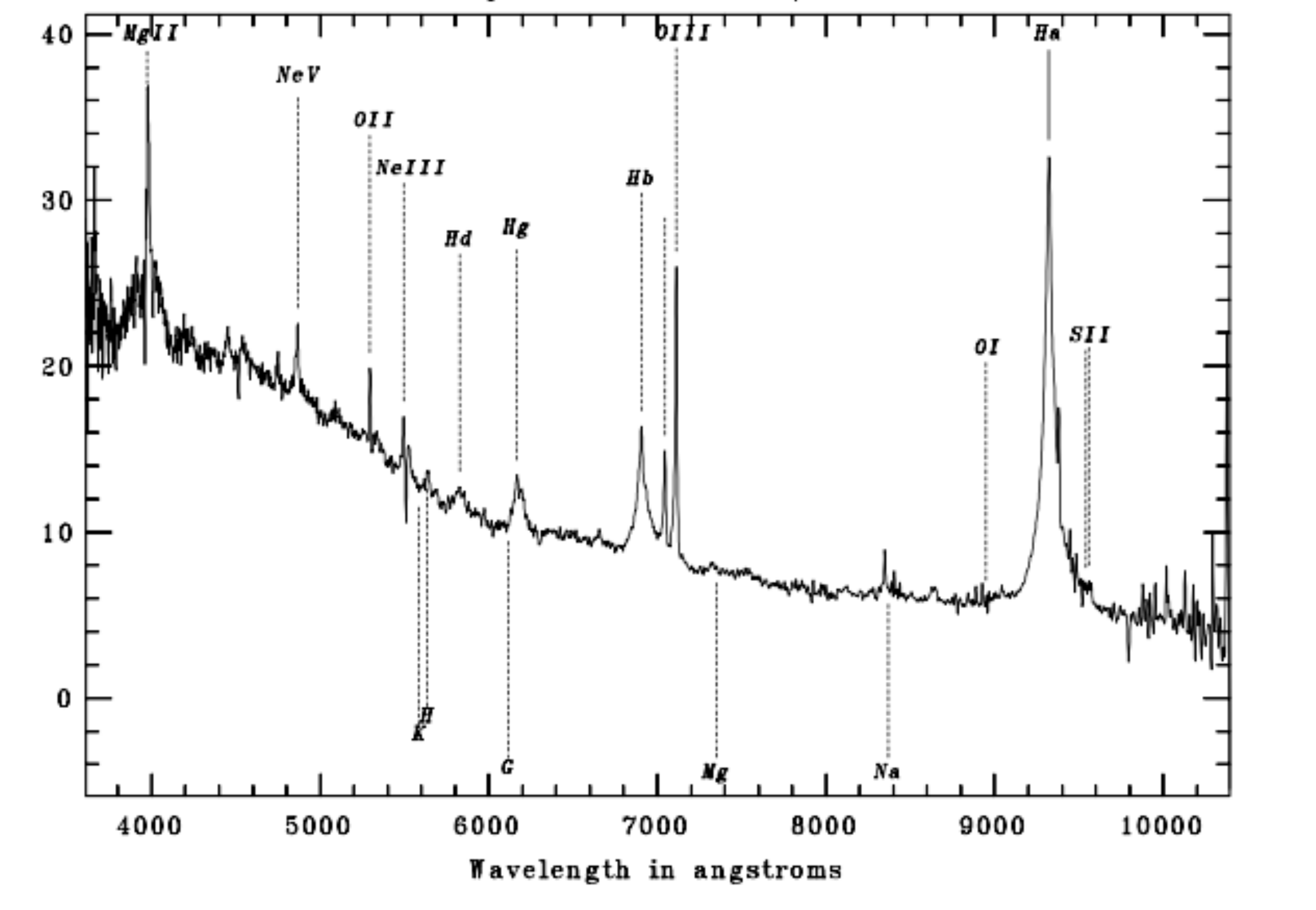}
\caption{J021337.9-042814~(2,2,1), z=0.419}
\label{}
\end{figure}
\begin{figure}
\includegraphics[height=0.87\columnwidth]{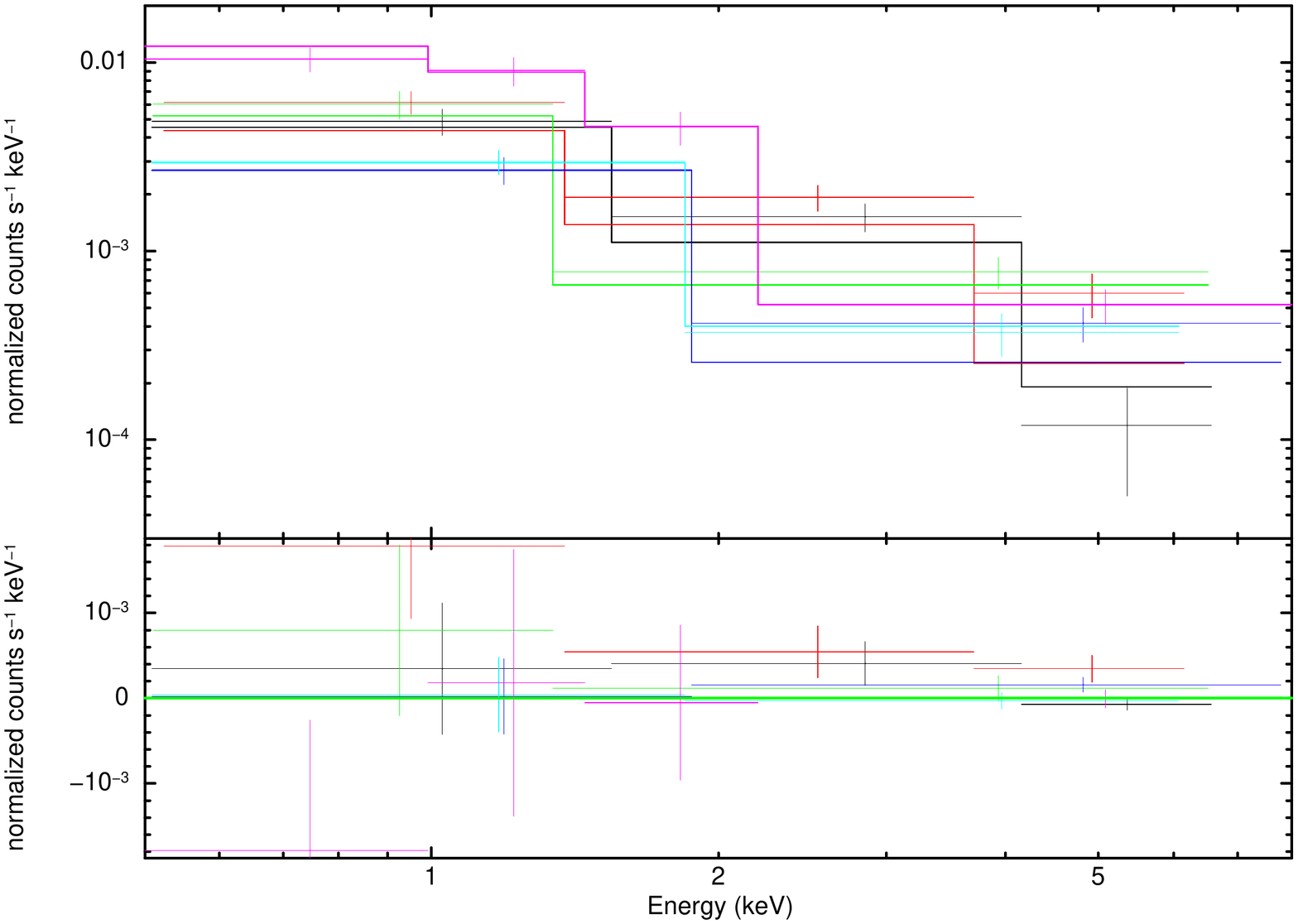}
\includegraphics[height=0.86\columnwidth]{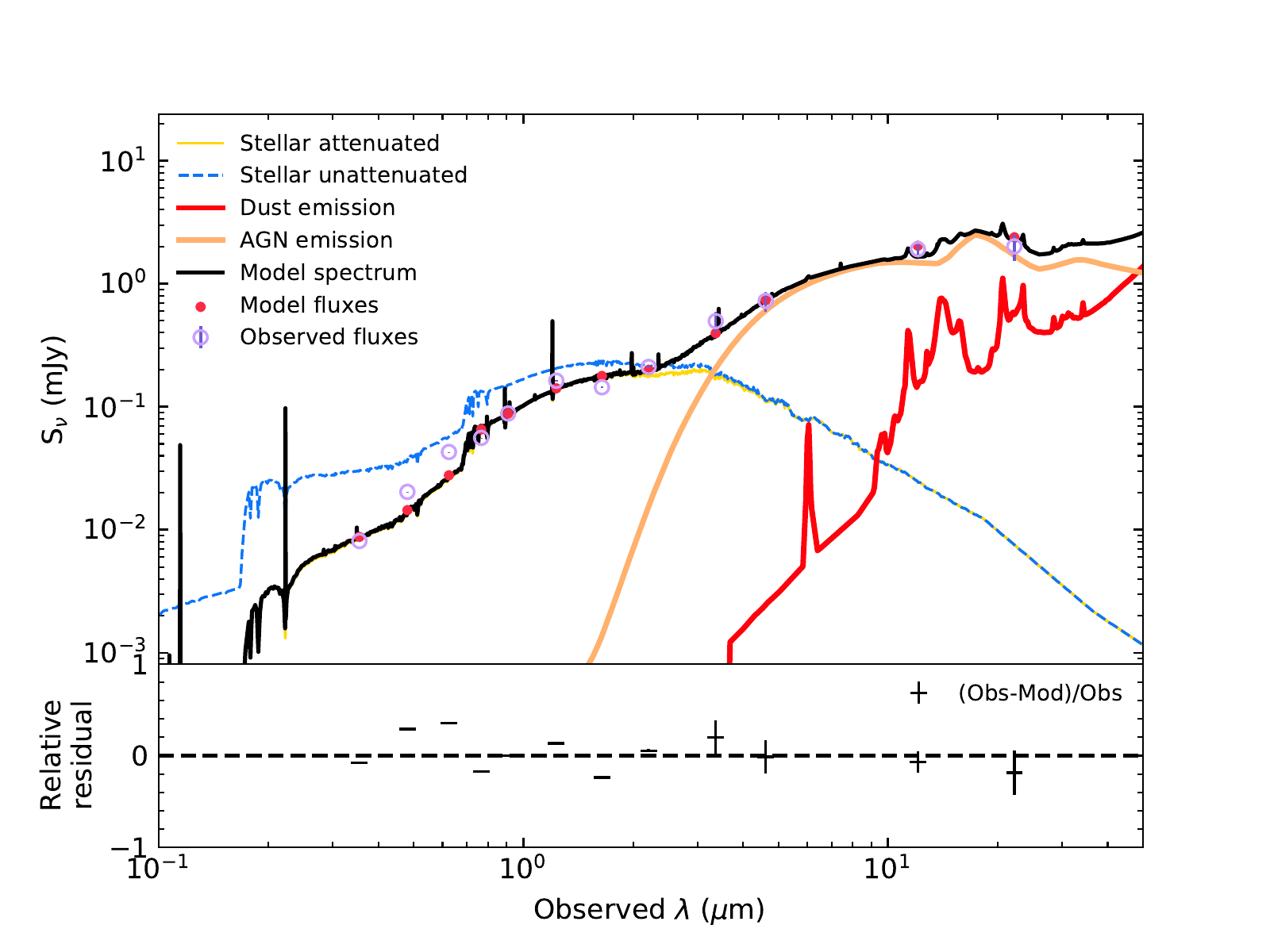}
\includegraphics[height=0.74\columnwidth]{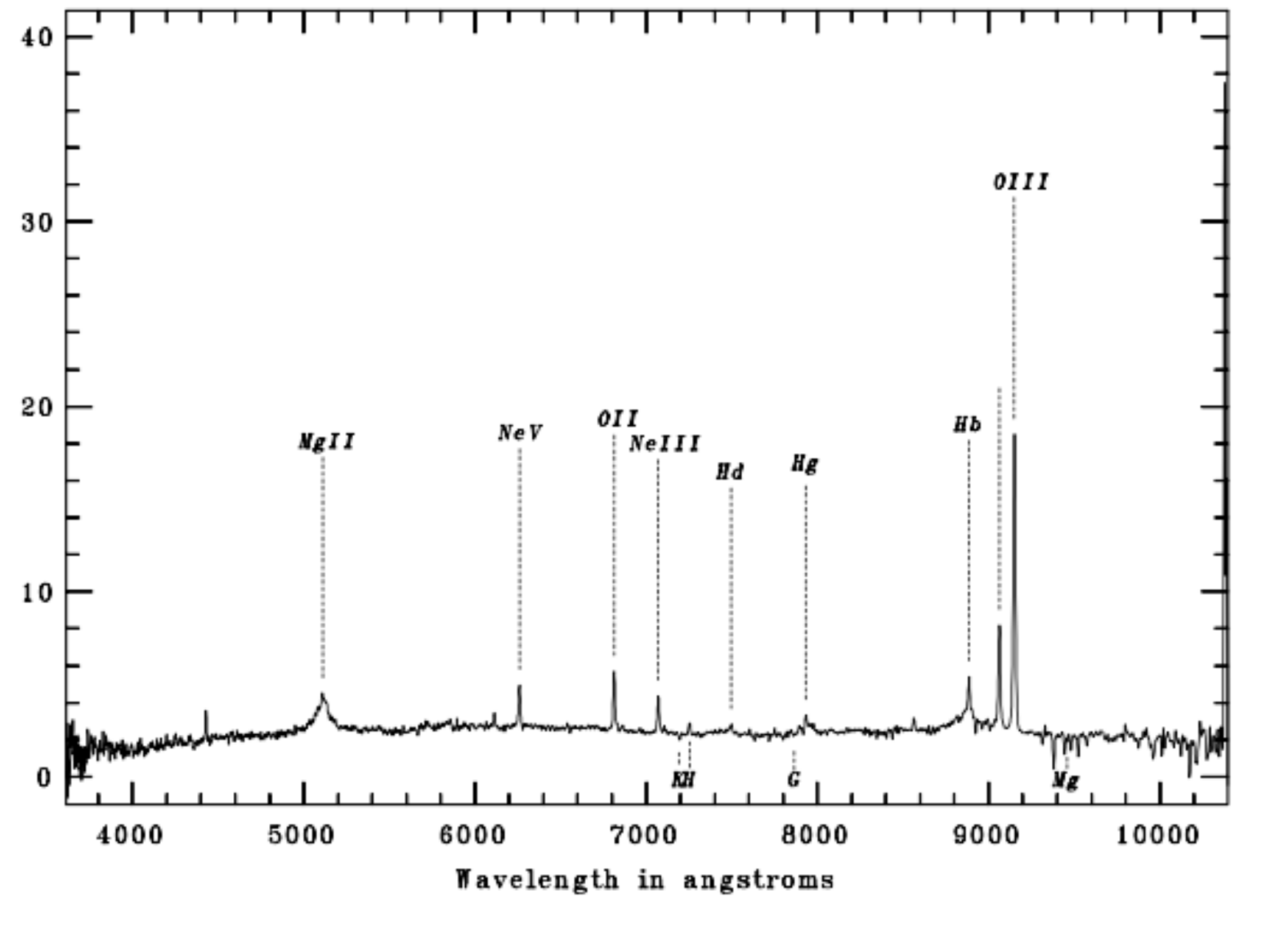}
\caption{J020436.4-042833~(1,2,1), z=0.827}
\label{}
\end{figure}
\begin{figure}
\includegraphics[height=0.87\columnwidth]{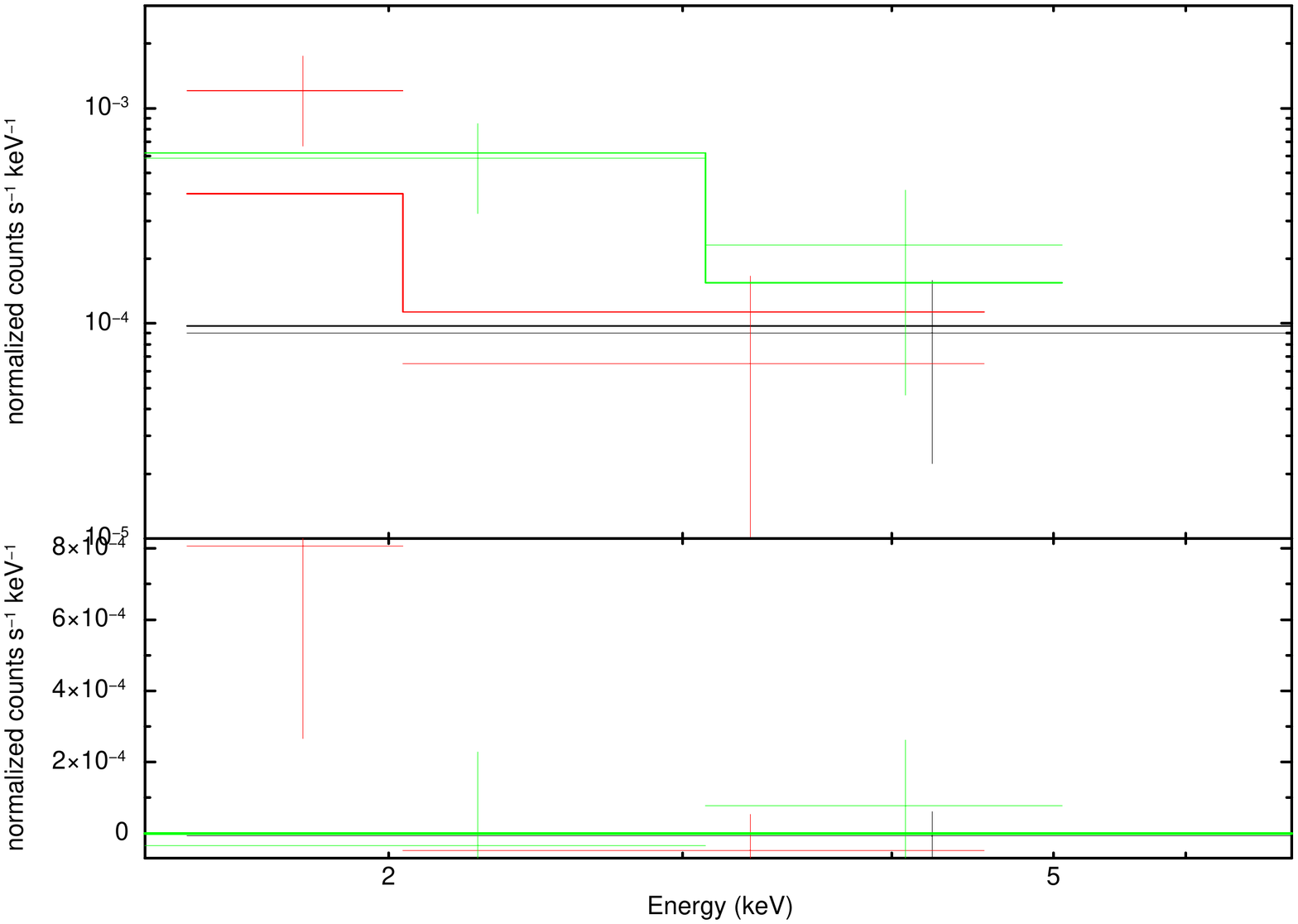}
\includegraphics[height=0.86\columnwidth]{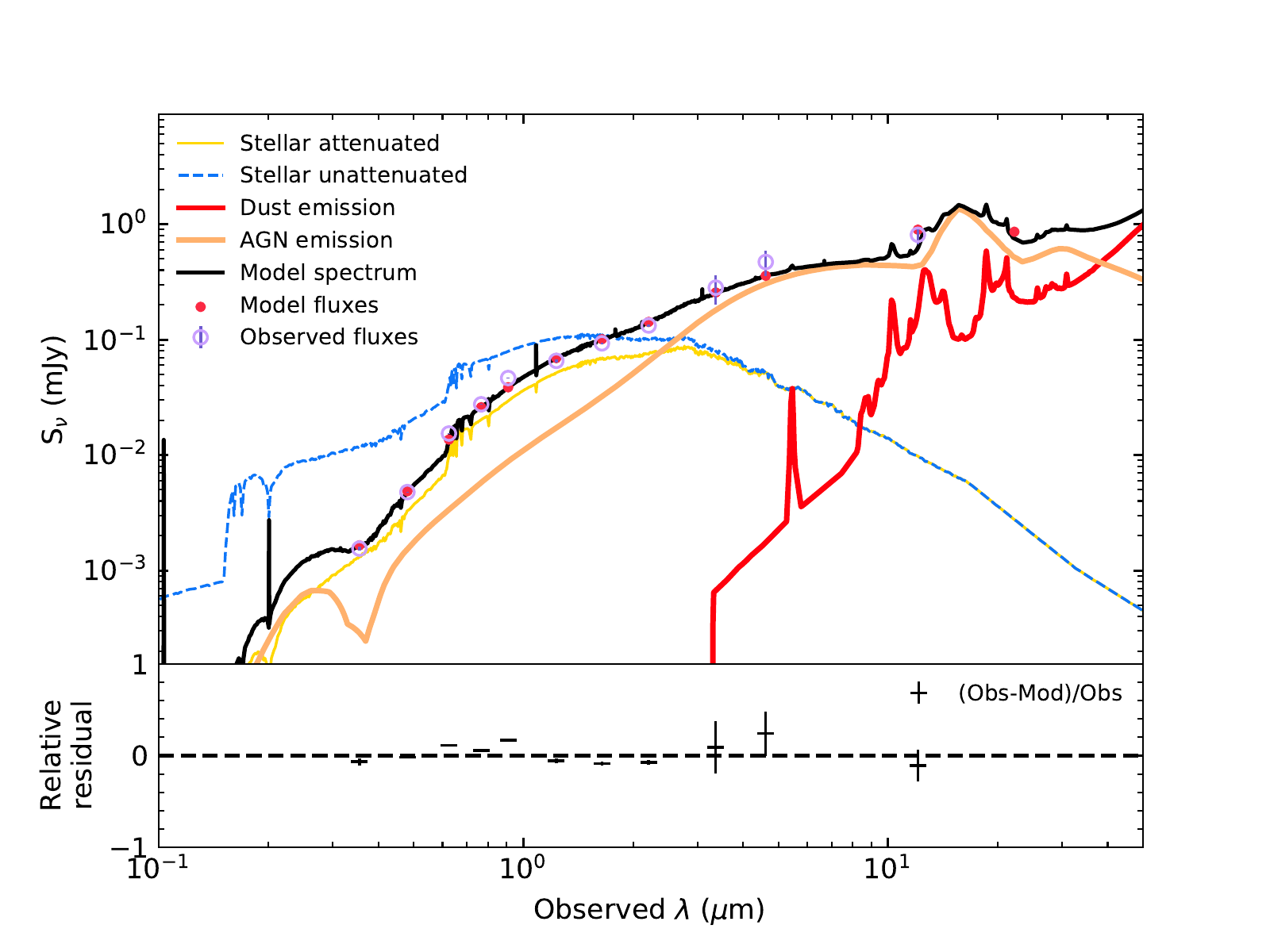}
\includegraphics[height=0.74\columnwidth]{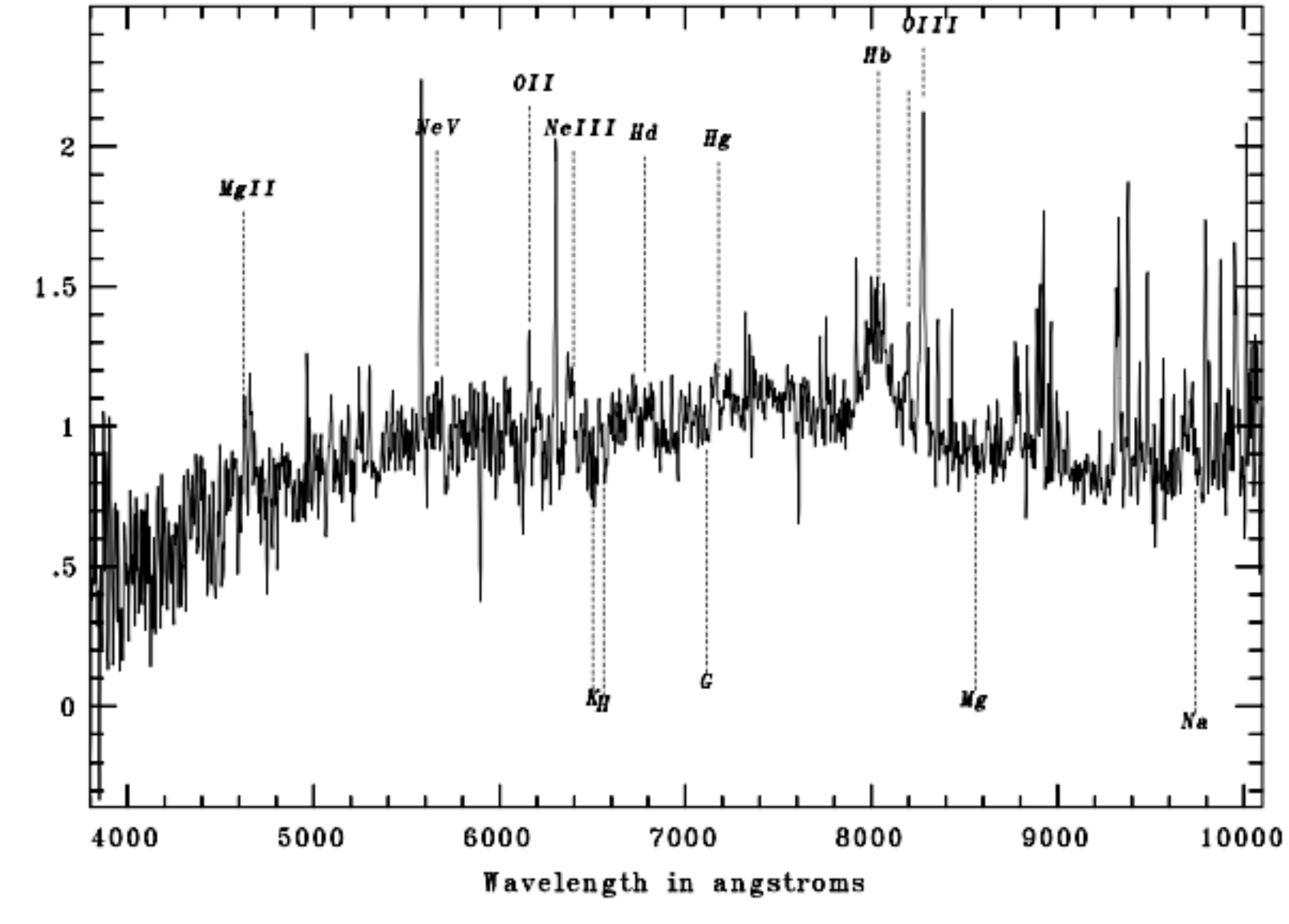}
\caption{J020543.0-051656~(2,2,1), z=0.653}
\label{}
\end{figure}
\begin{figure}
\includegraphics[height=0.87\columnwidth]{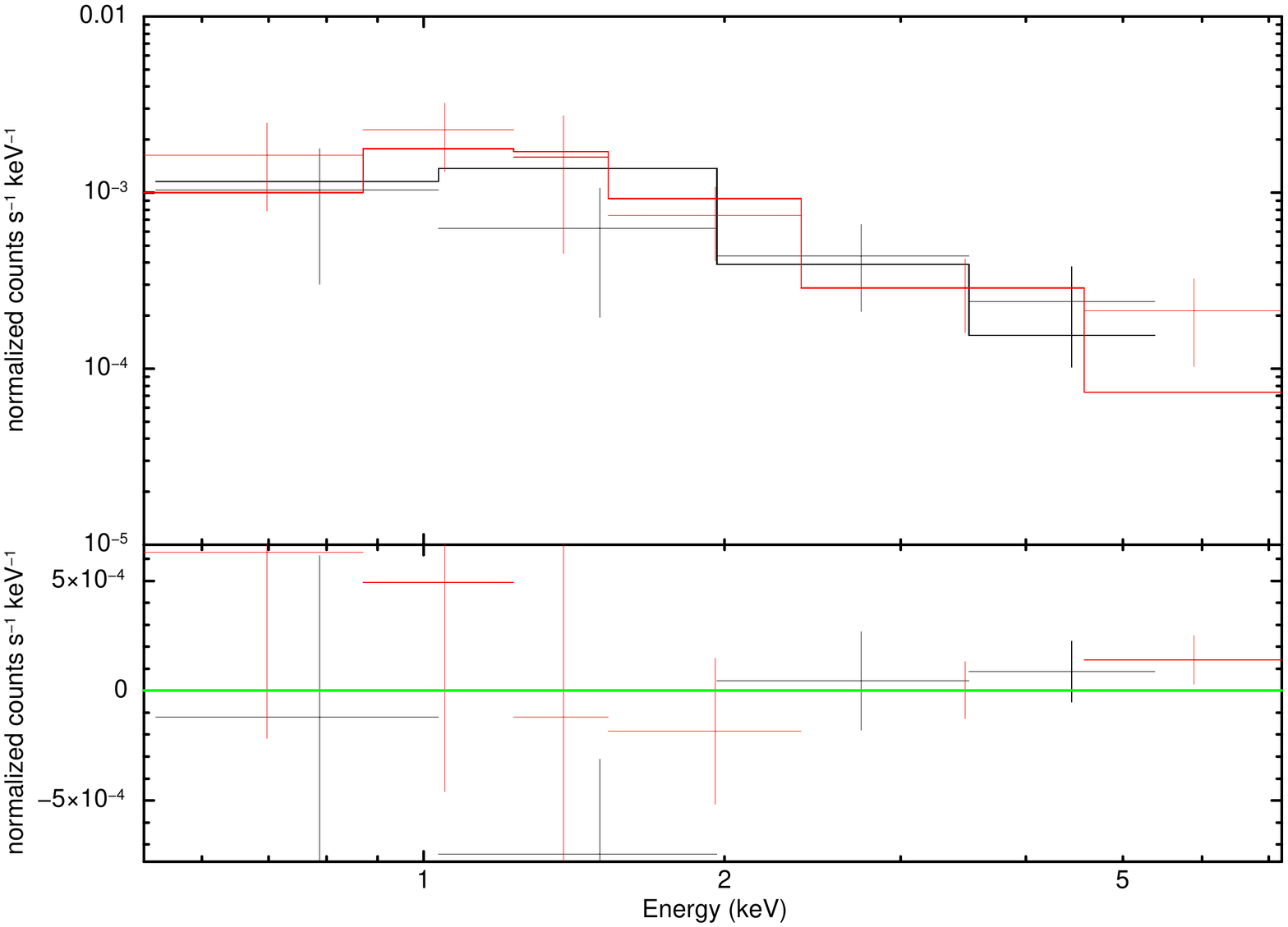}
\includegraphics[height=0.86\columnwidth]{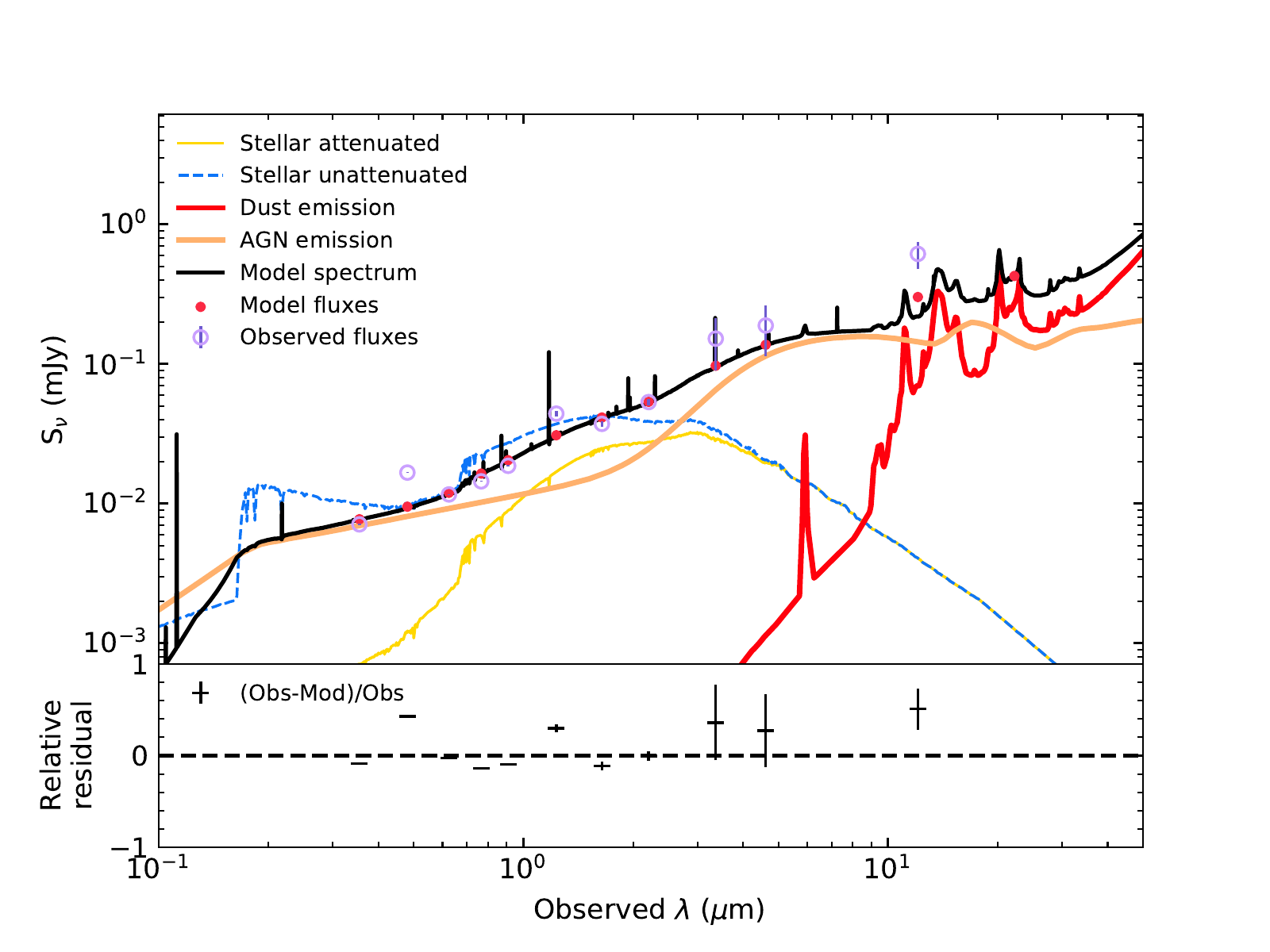}
\includegraphics[height=0.74\columnwidth]{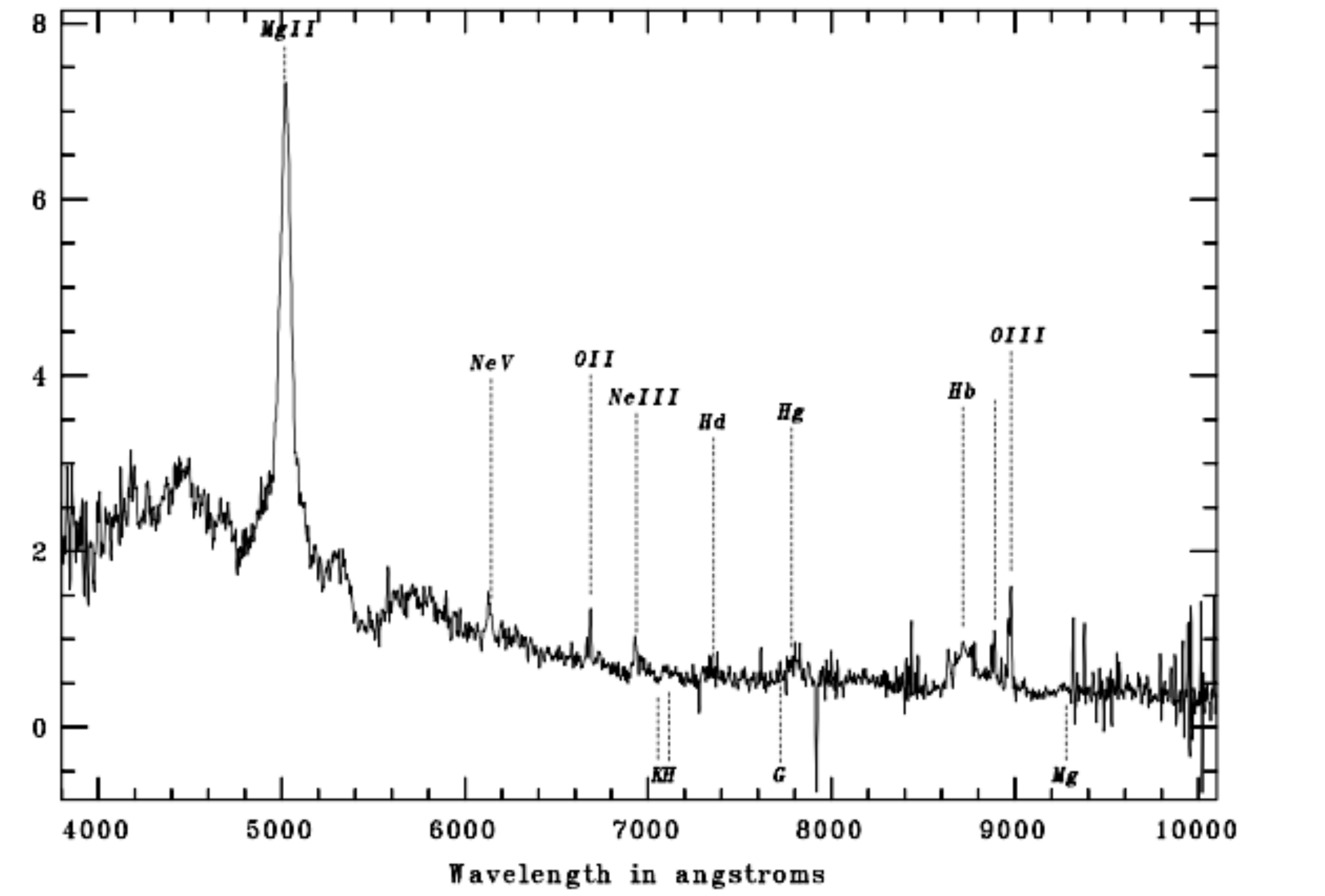}
\caption{J020517.3-051024~(2,0,1), z=0.792}
\label{}
\end{figure}
\begin{figure}
\includegraphics[height=0.87\columnwidth]{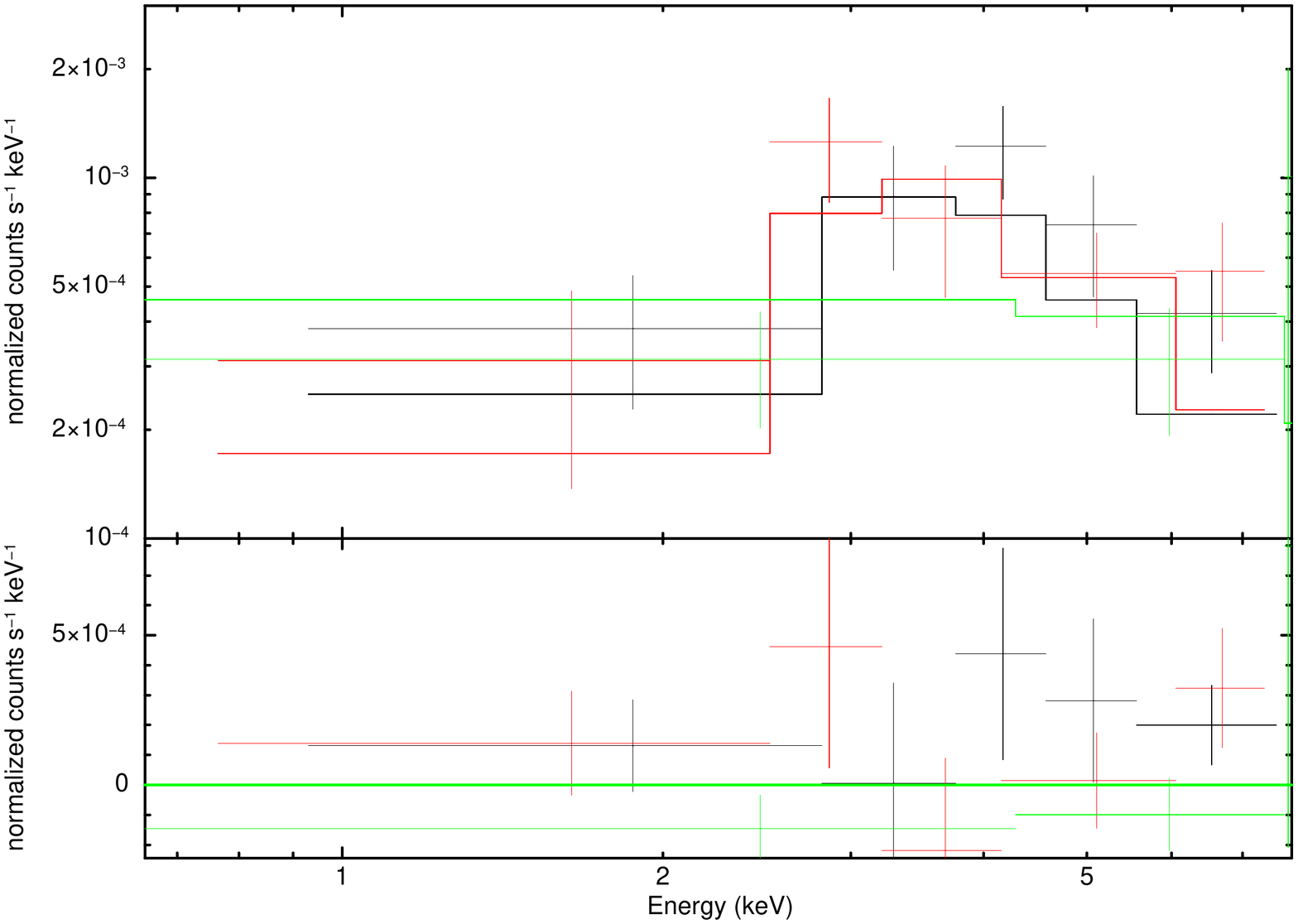}
\includegraphics[height=0.86\columnwidth]{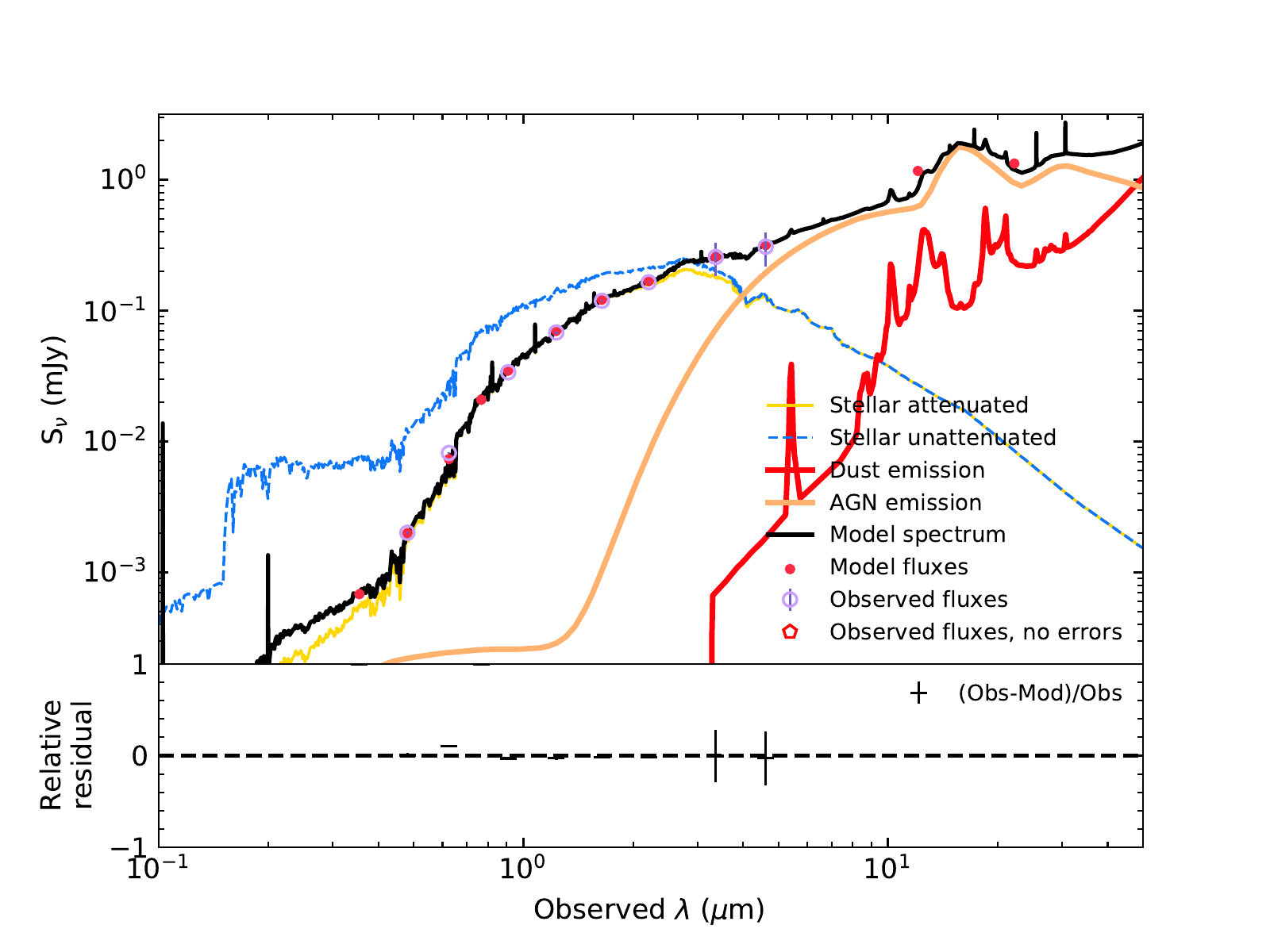}
\includegraphics[height=0.74\columnwidth]{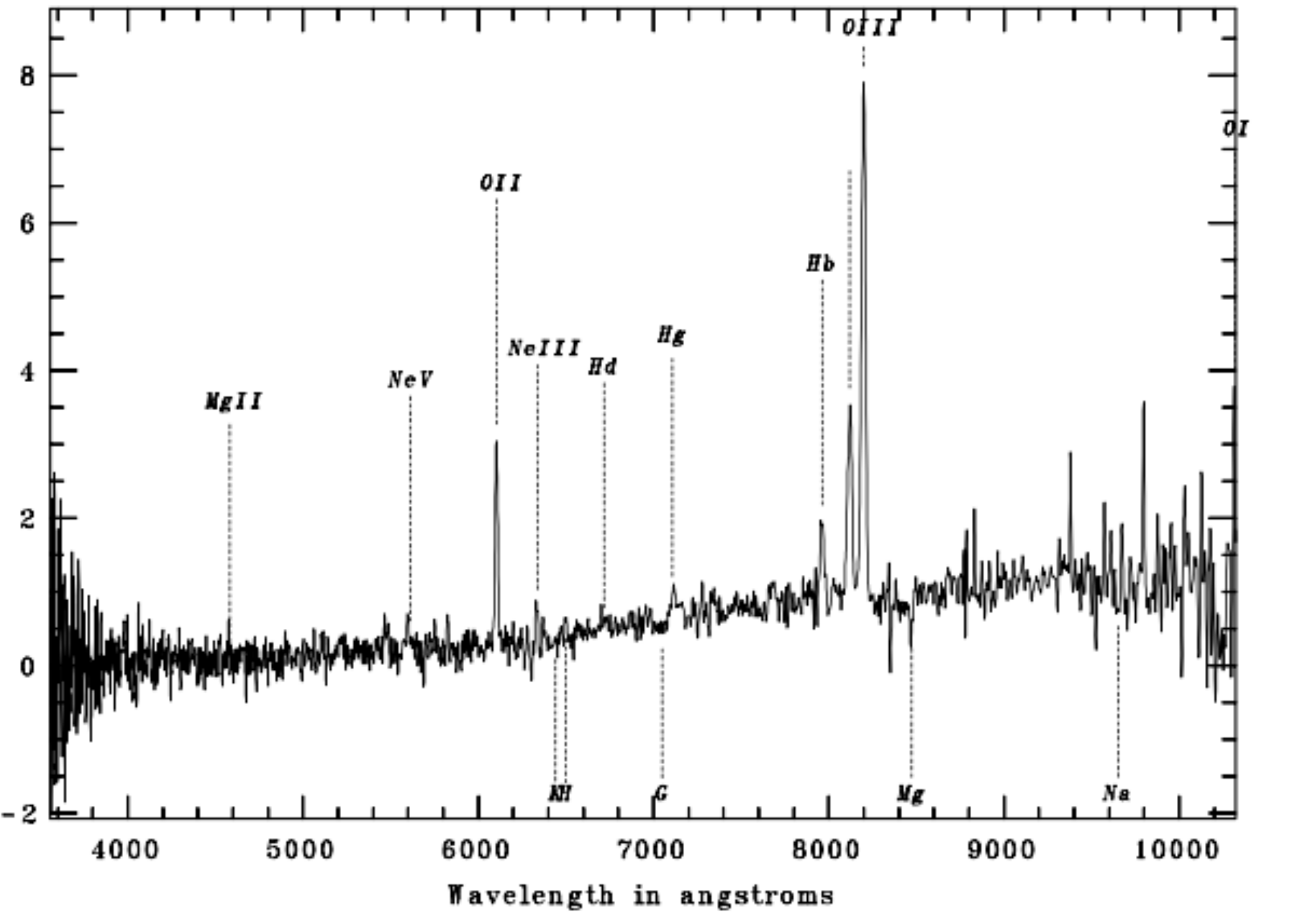}
\caption{J022244.3-030525~(2,2,2), z=0.637}
\label{}
\end{figure}
\begin{figure}
\includegraphics[height=0.87\columnwidth]{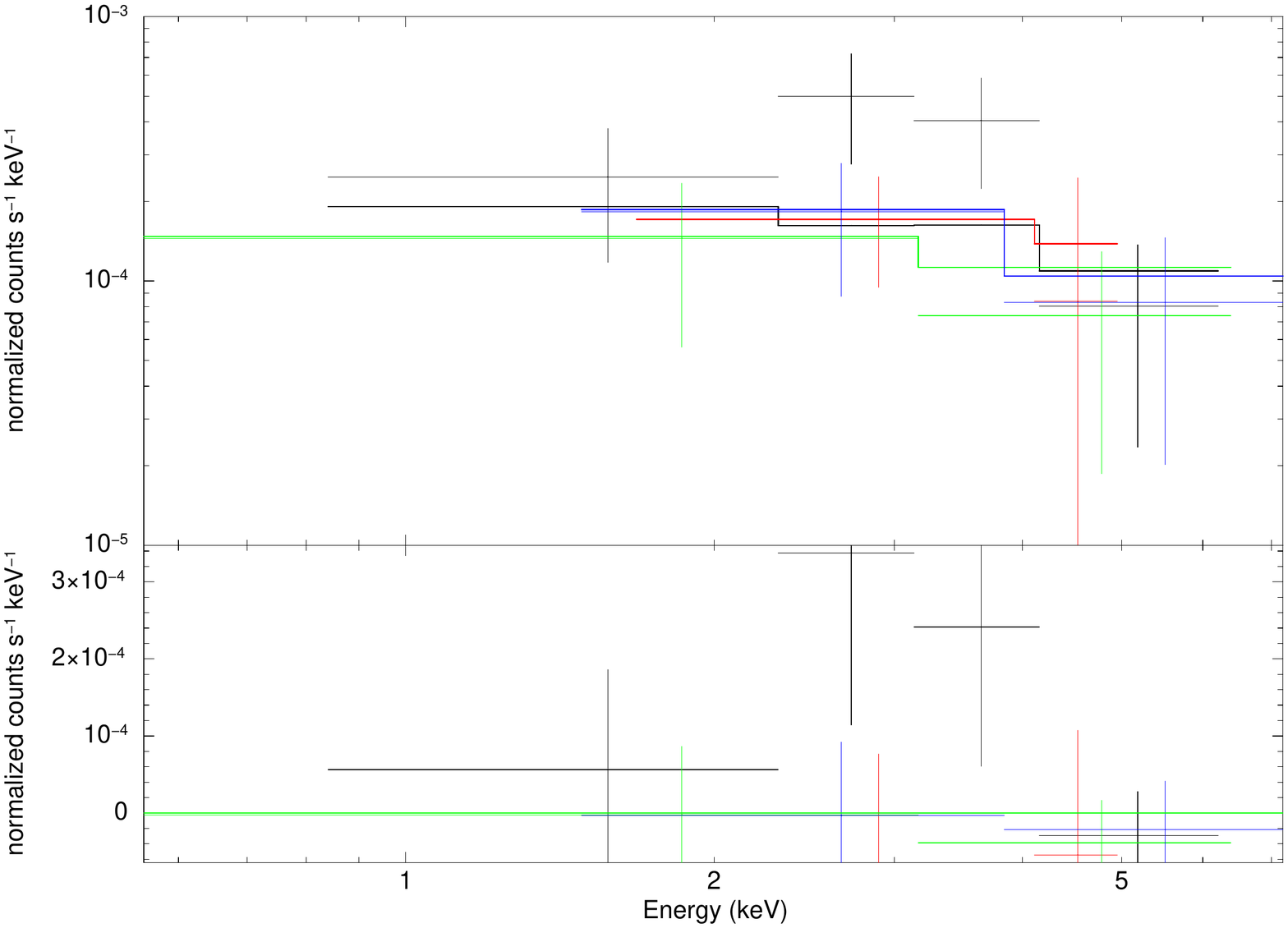}
\includegraphics[height=0.86\columnwidth]{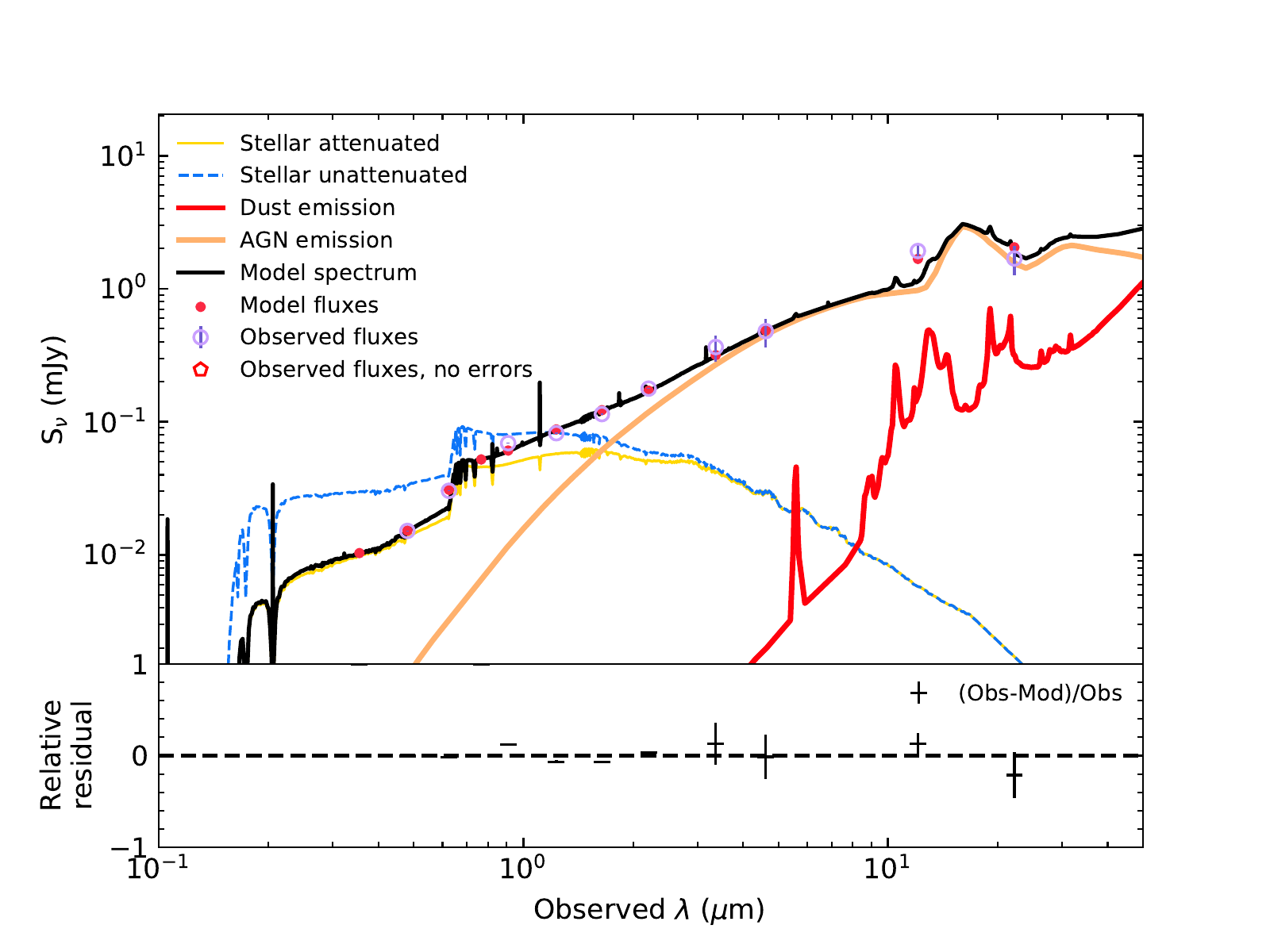}
\includegraphics[height=0.74\columnwidth]{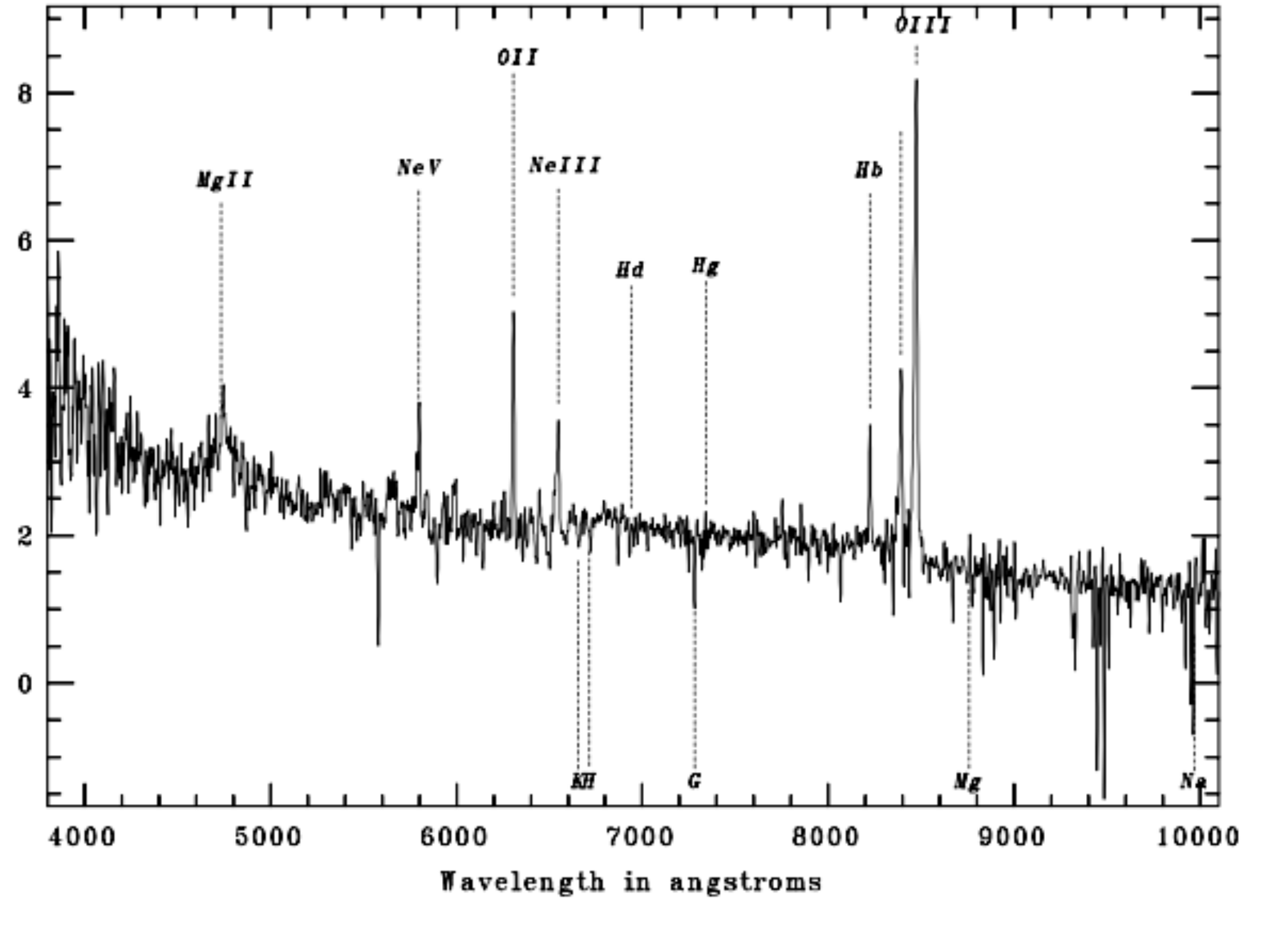}
\caption{J022323.4-031157~(2,2,3), z=0.691}
\label{}
\end{figure}
\begin{figure}
\includegraphics[height=0.87\columnwidth]{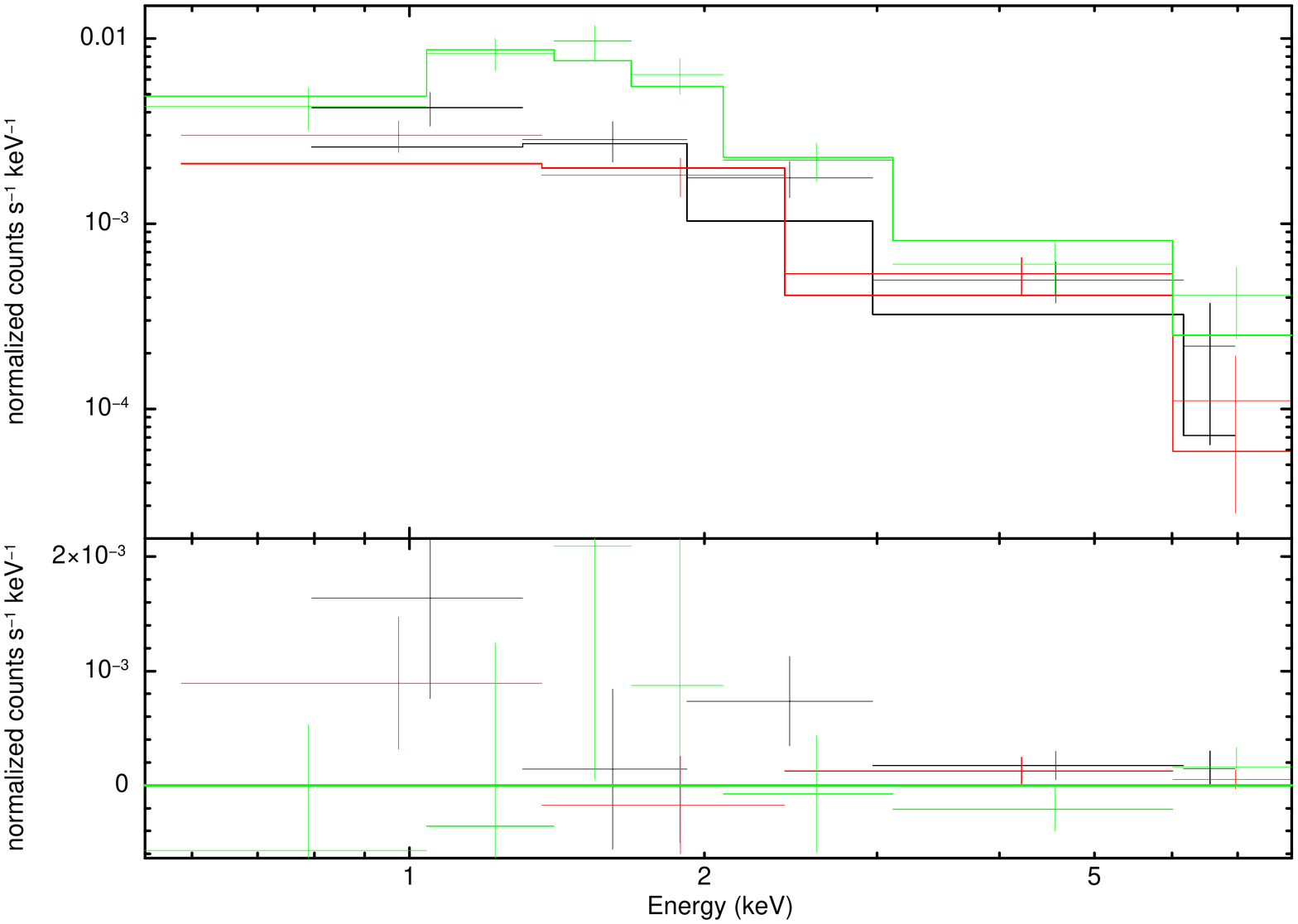}
\includegraphics[height=0.86\columnwidth]{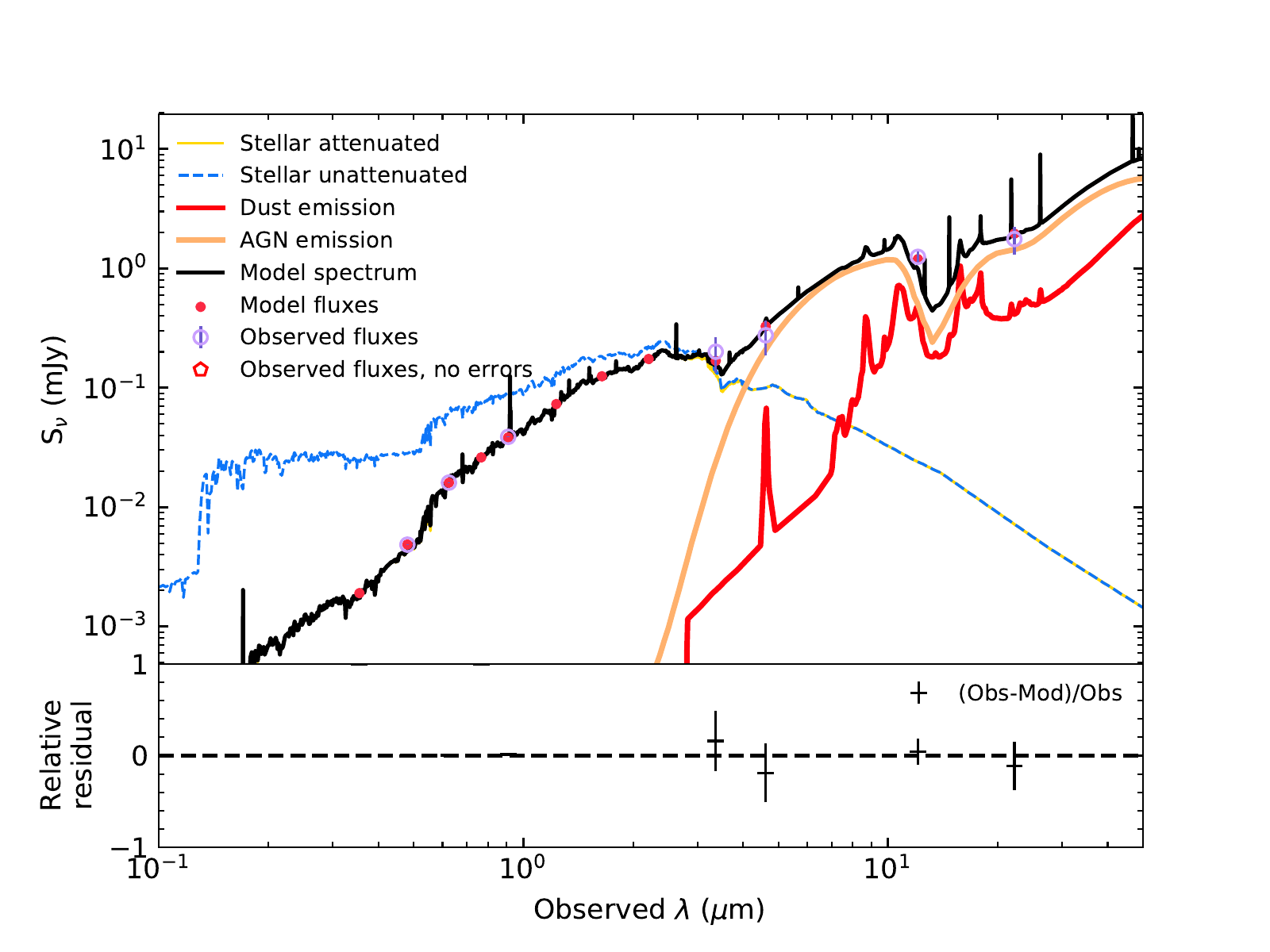}
\includegraphics[height=0.74\columnwidth]{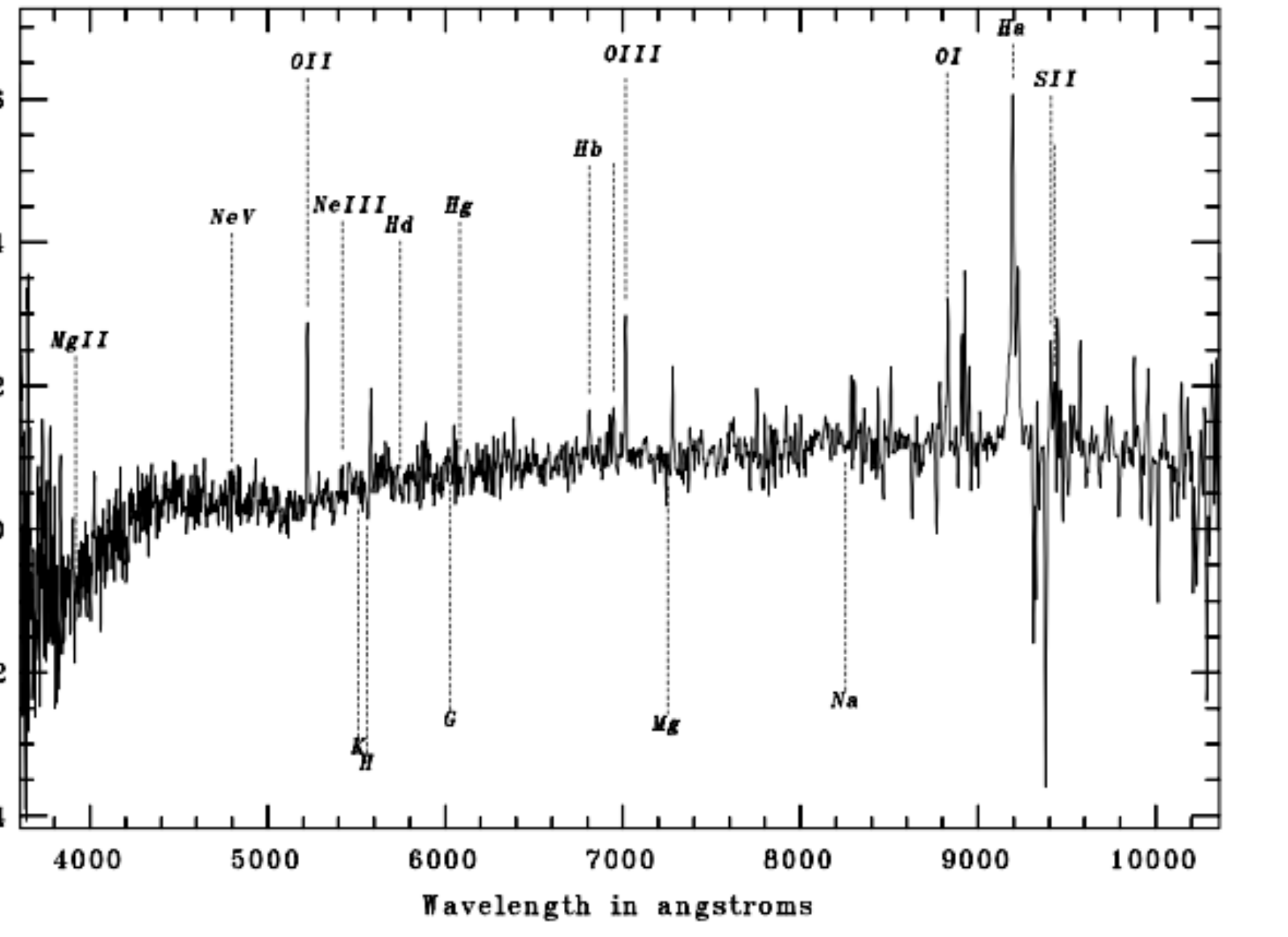}
\caption{J022209.6-025023~(2,2,2), z=0.400}
\label{}
\end{figure}
\begin{figure}
\includegraphics[height=0.87\columnwidth]{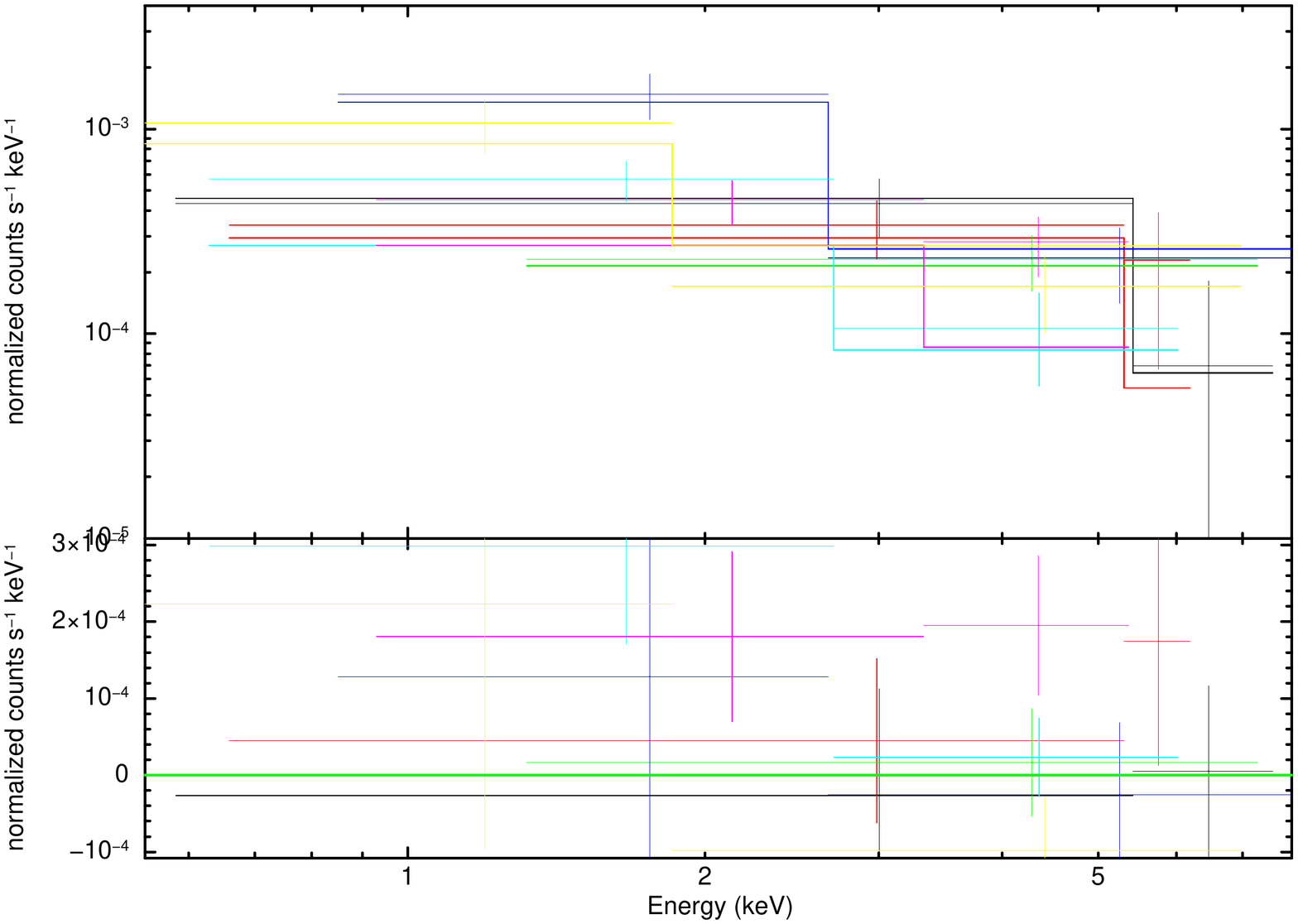}
\includegraphics[height=0.86\columnwidth]{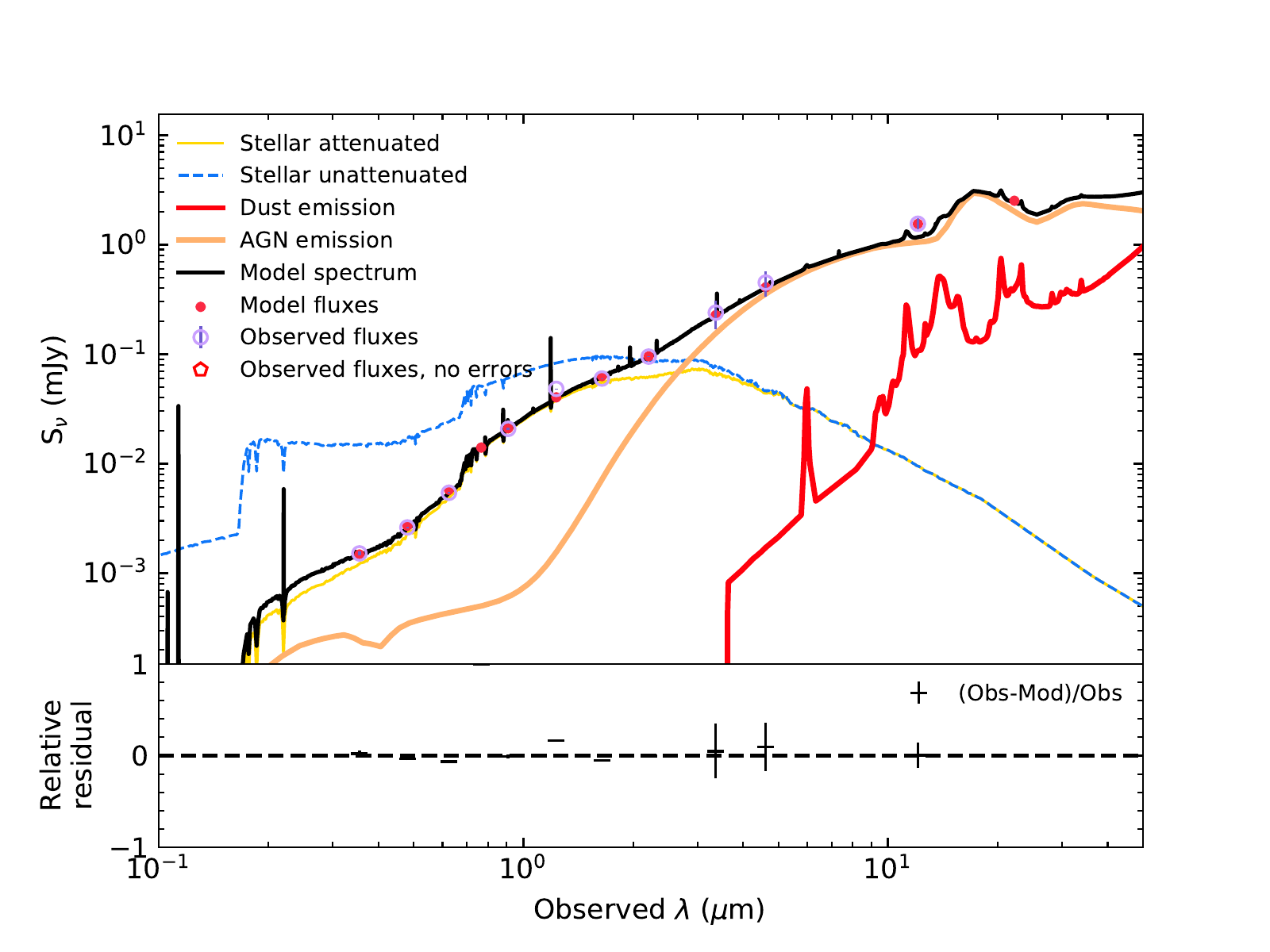}
\includegraphics[height=0.74\columnwidth]{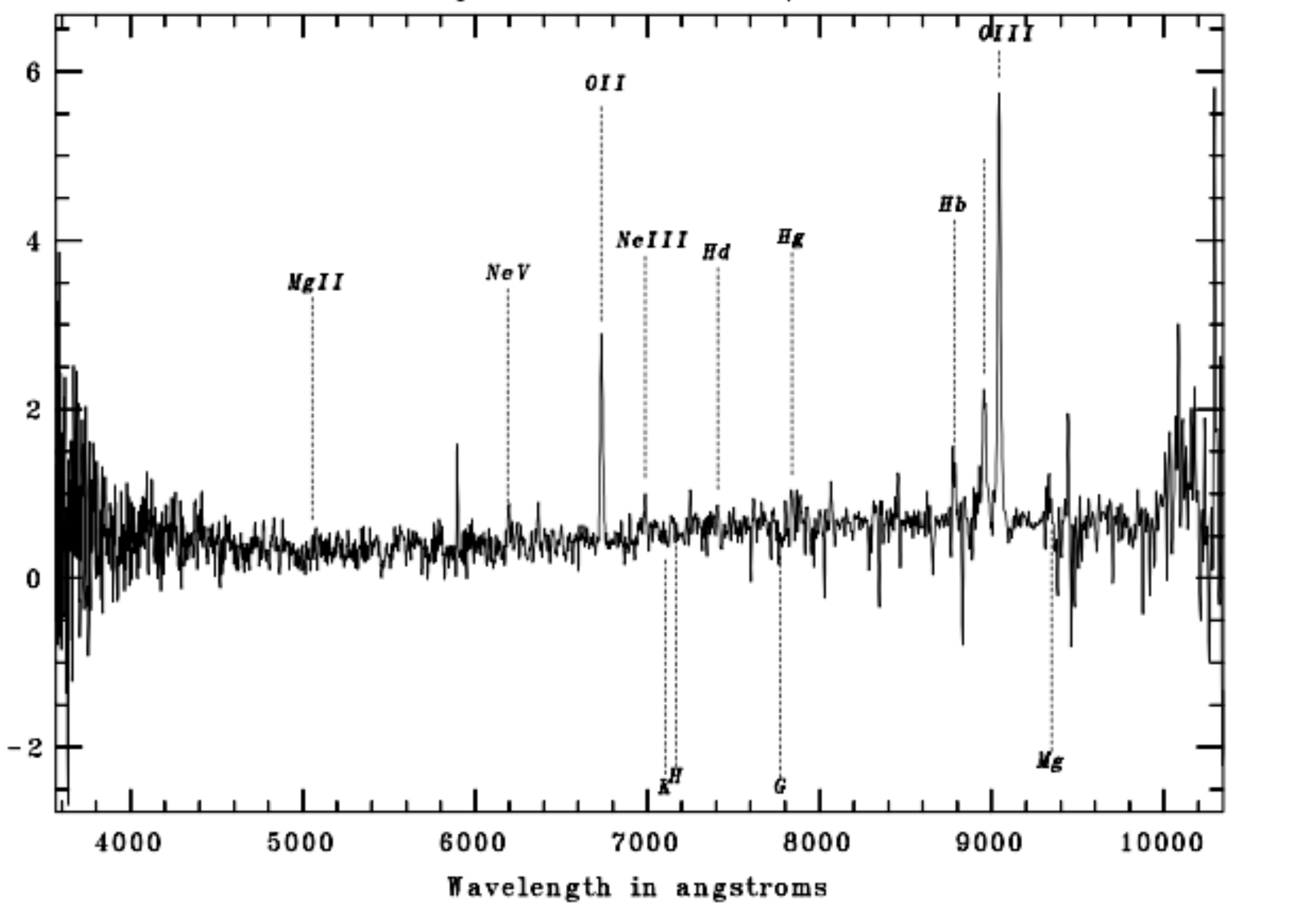}
\caption{J022750.7-052232~(2,2,3), z=0.804}
\label{}
\end{figure}
\begin{figure}
\includegraphics[height=0.87\columnwidth]{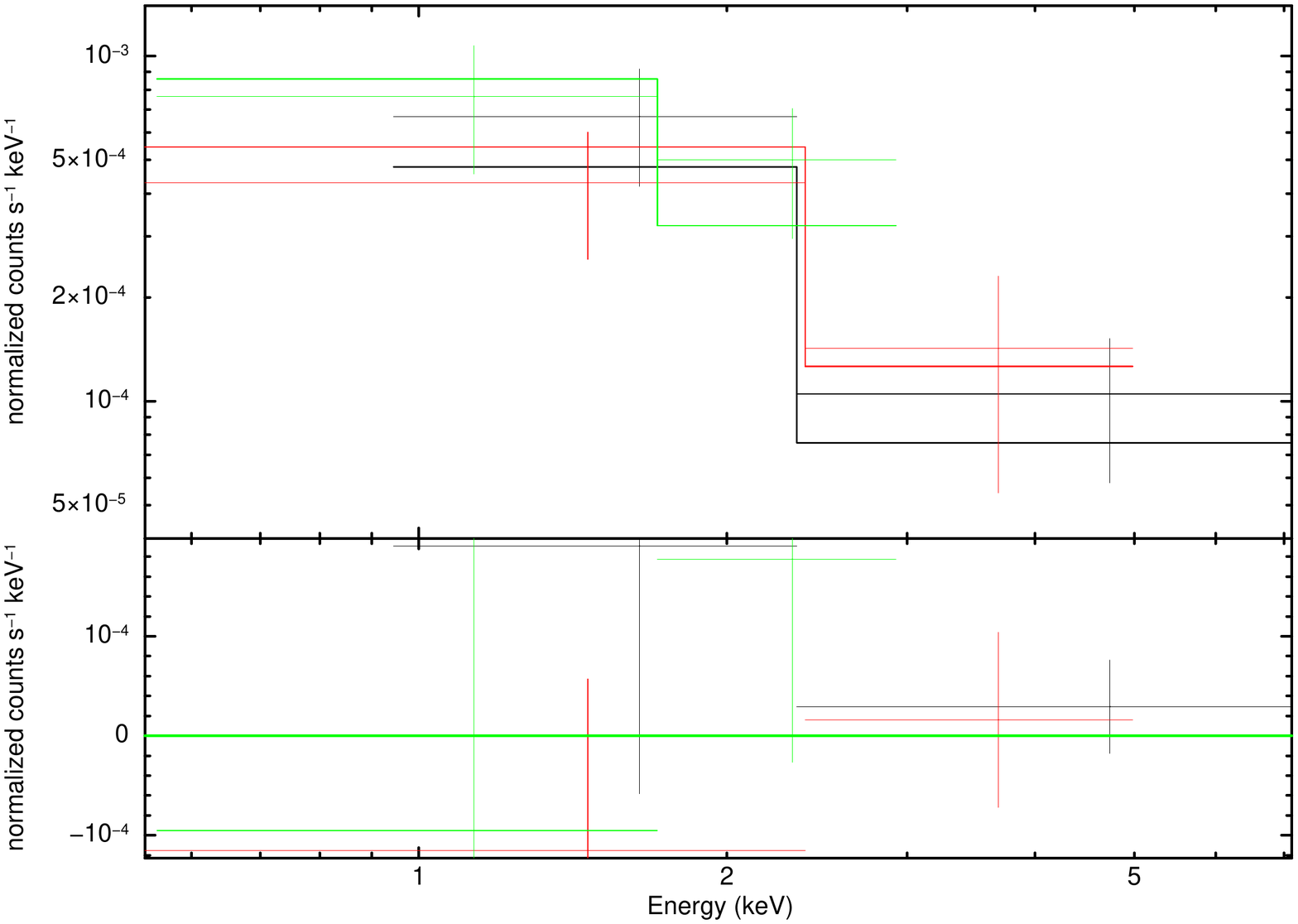}
\includegraphics[height=0.86\columnwidth]{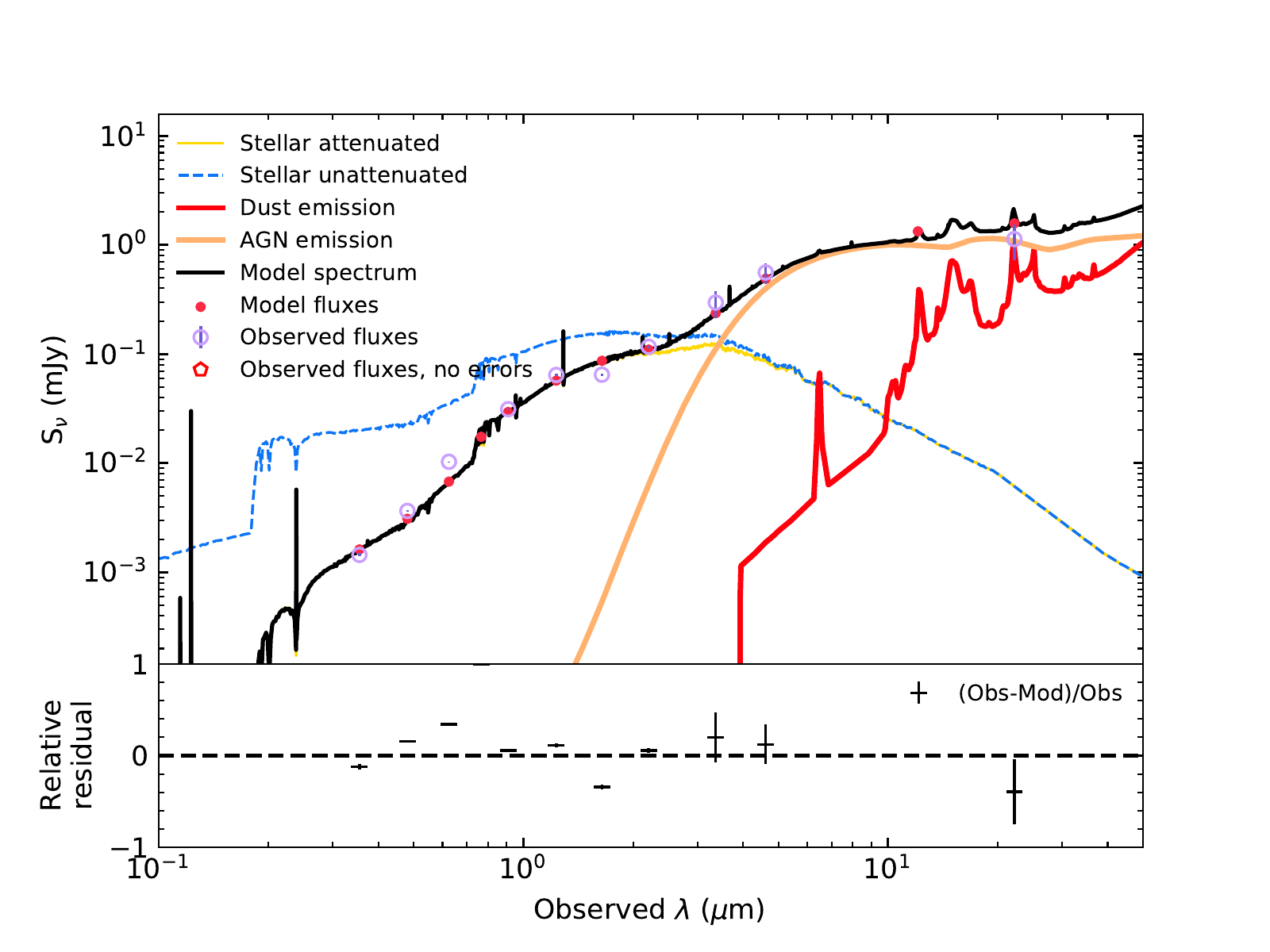}
\includegraphics[height=0.74\columnwidth]{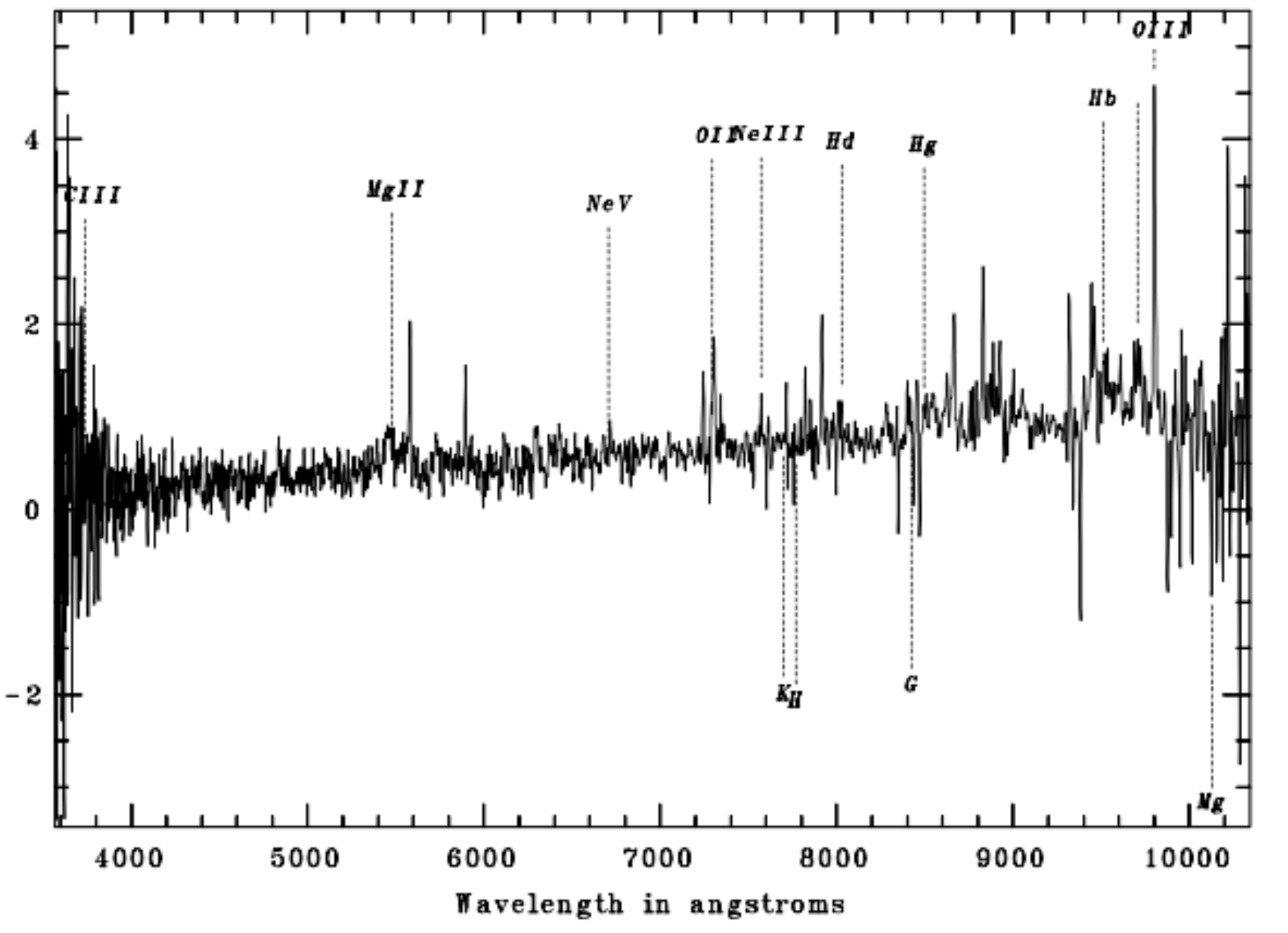}
\caption{J022758.4-053306~(1,2,2), z=0.956}
\label{}
\end{figure}
\clearpage

\begin{figure}
\includegraphics[height=0.87\columnwidth]{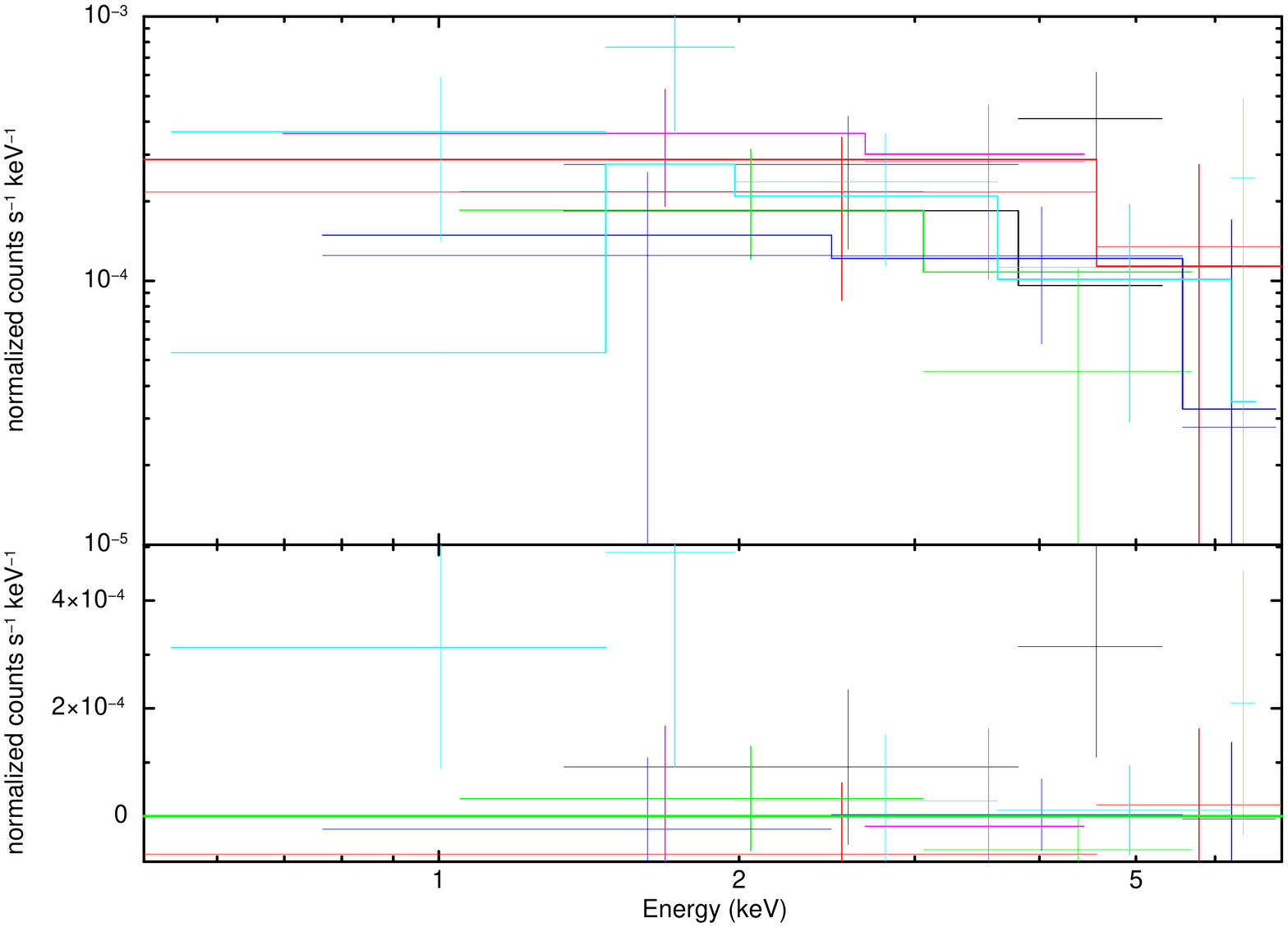}
\includegraphics[height=0.86\columnwidth]{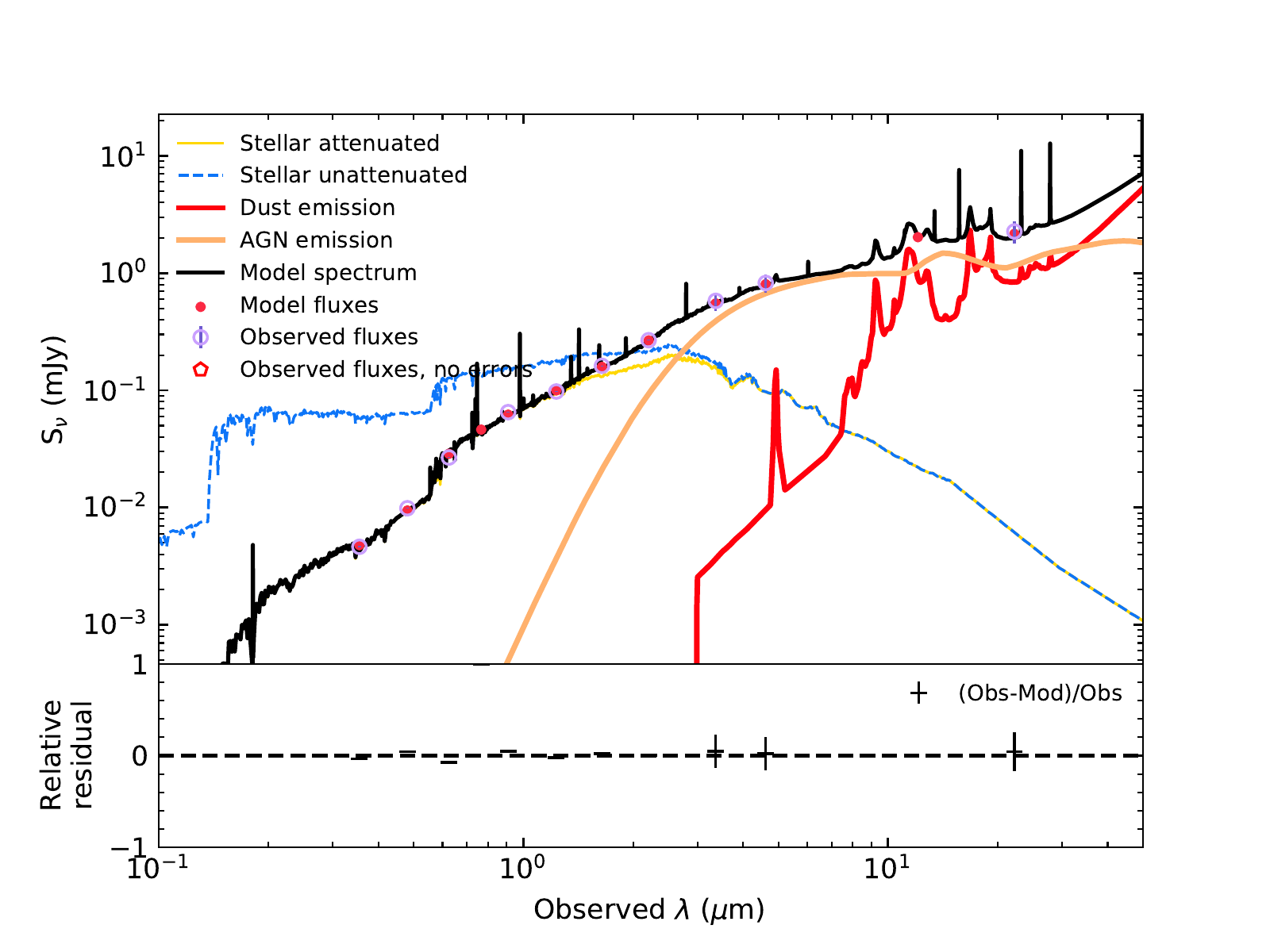}
\includegraphics[height=0.74\columnwidth]{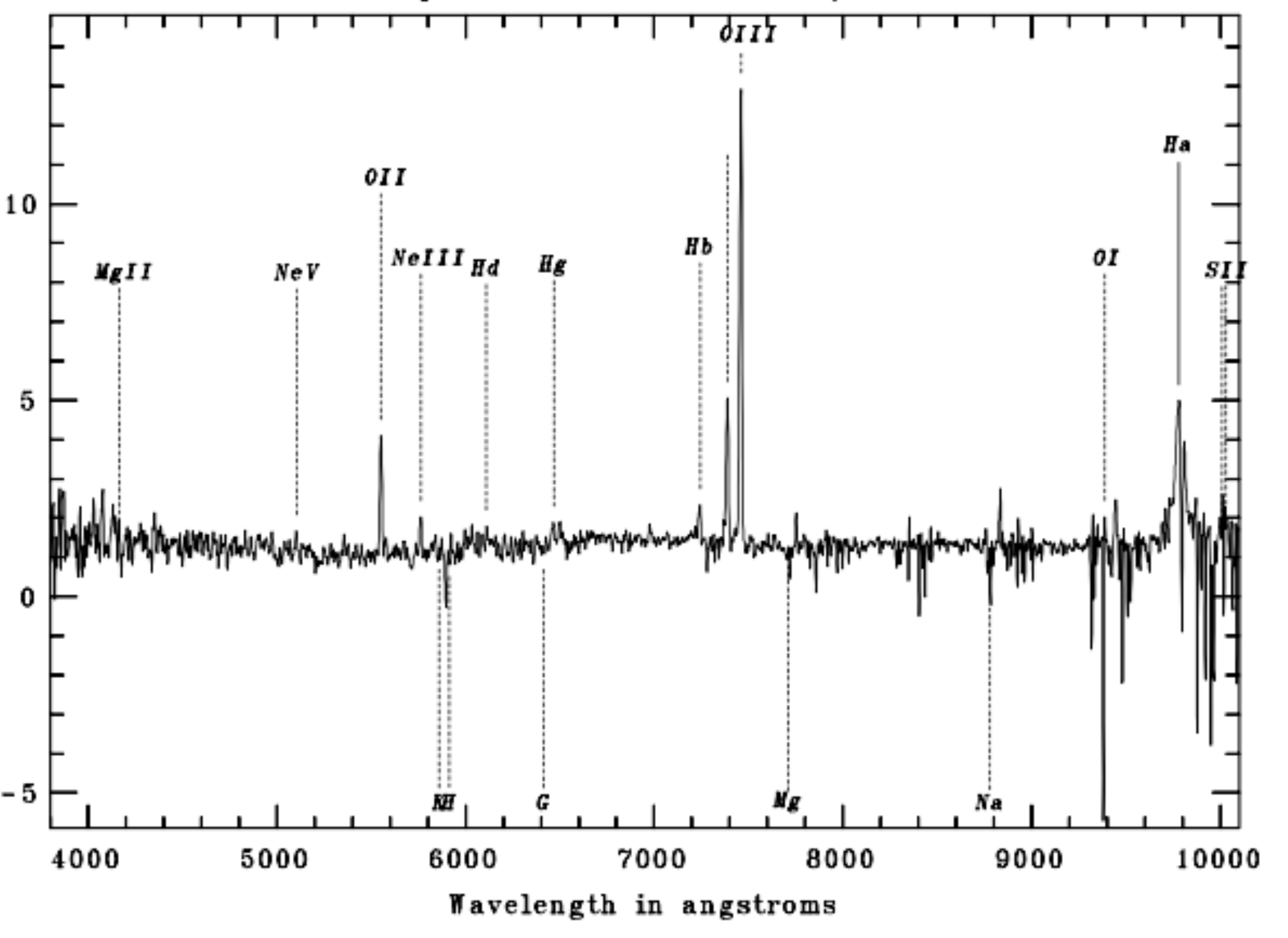}
\caption{J022453.2-054050~(2,2,3), z=0.488}
\label{}
\end{figure}
\begin{figure}
\includegraphics[height=0.87\columnwidth]{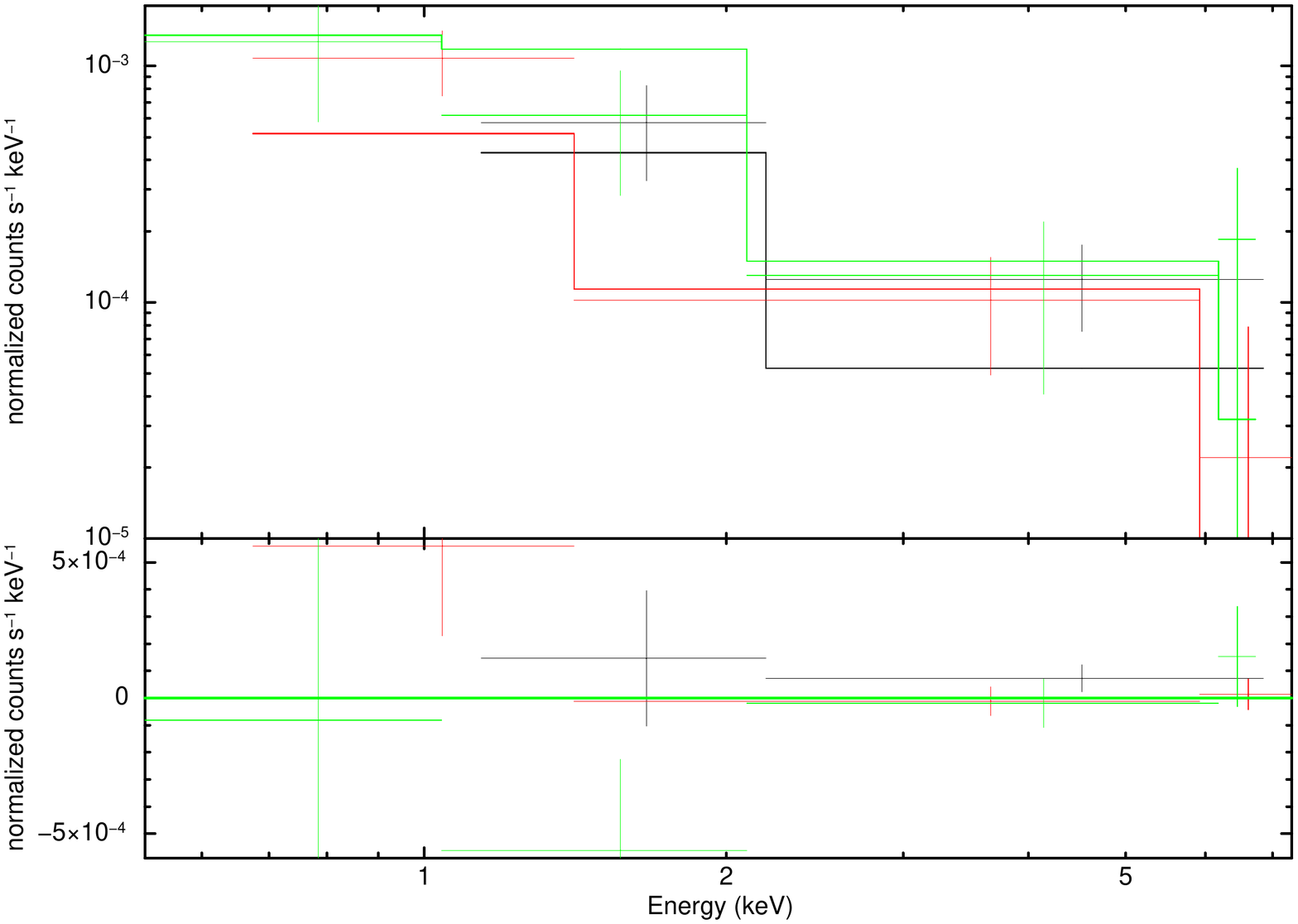}
\includegraphics[height=0.86\columnwidth]{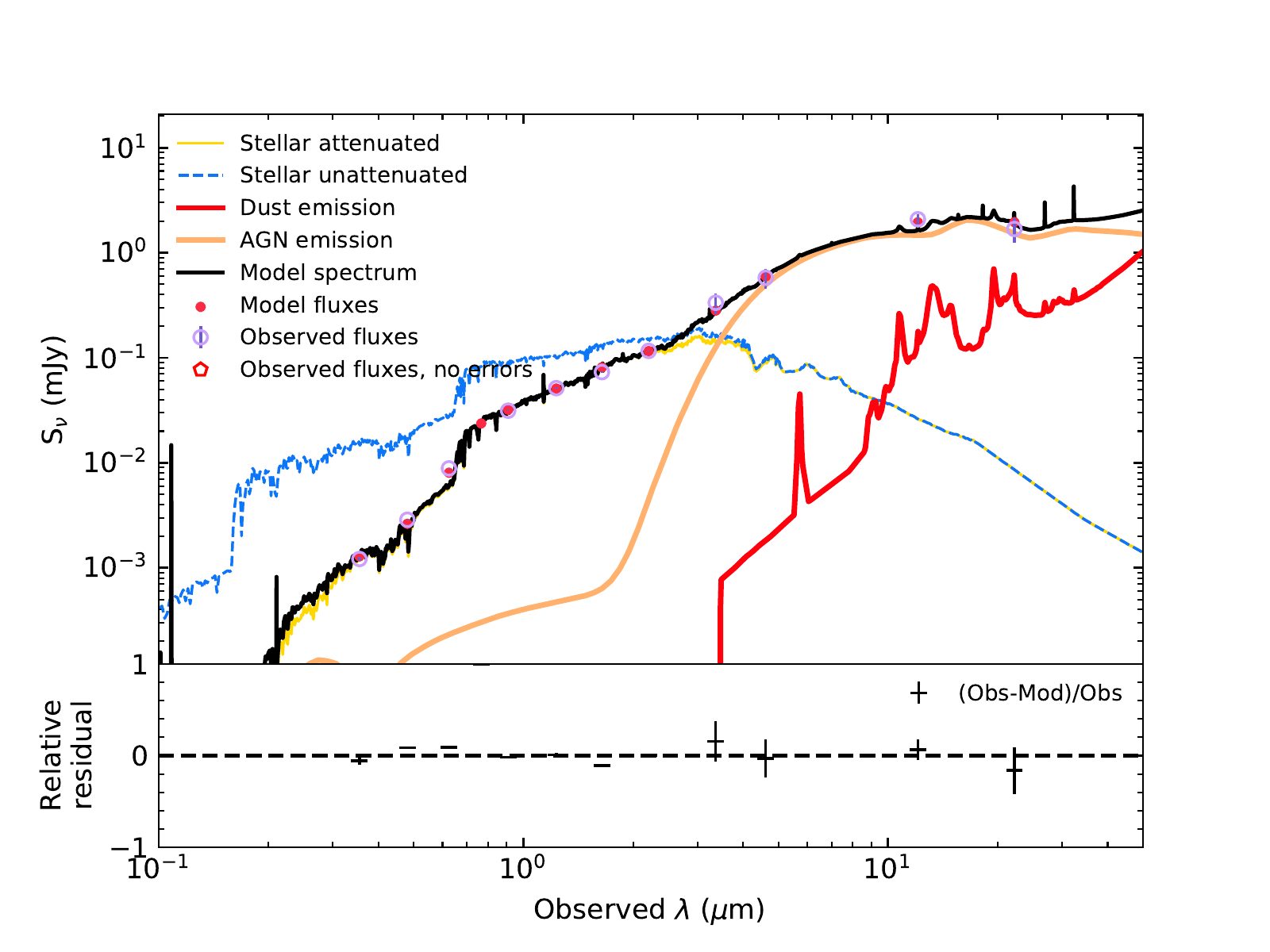}
\includegraphics[height=0.74\columnwidth]{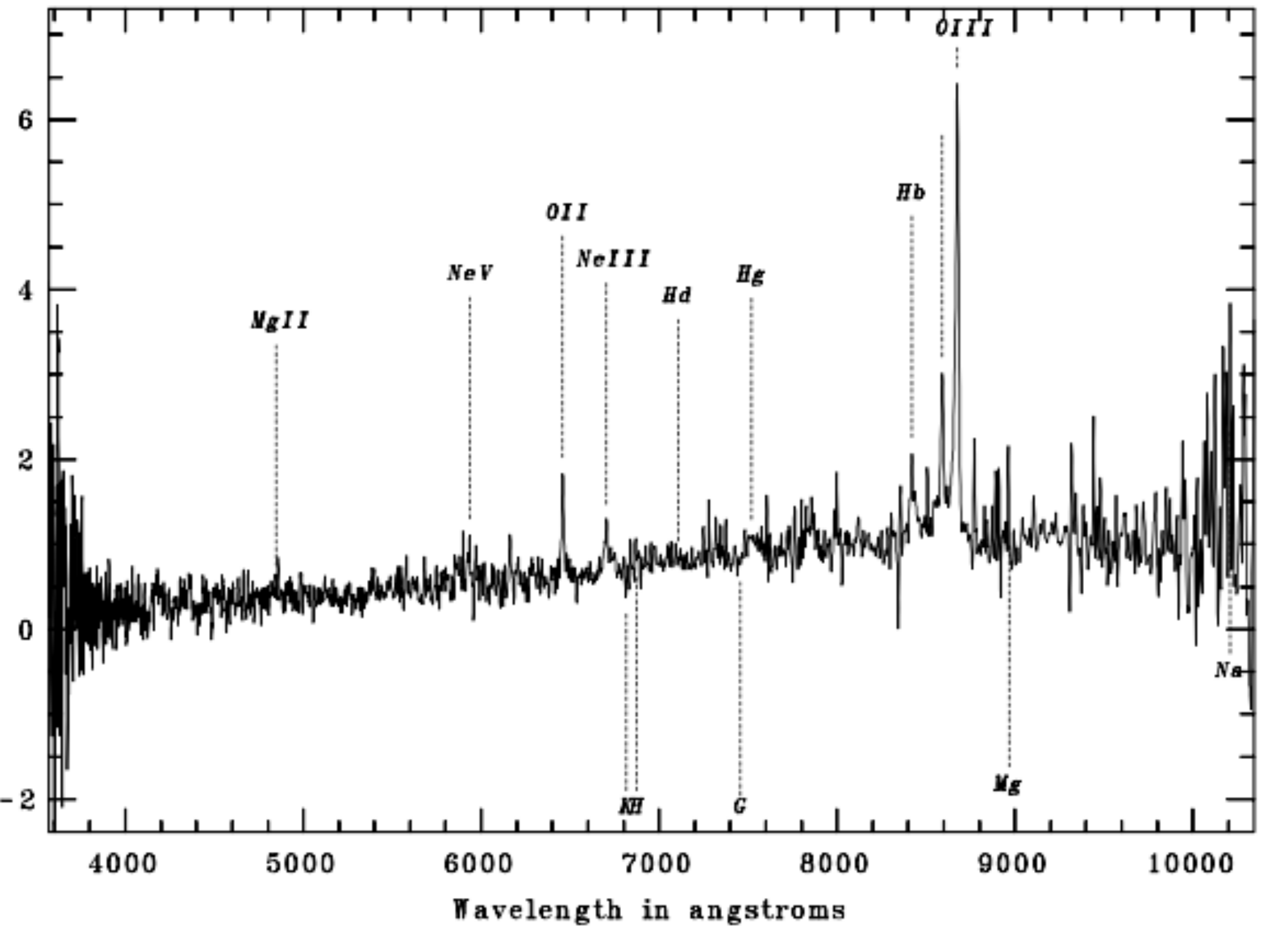}
\caption{J022258.8-055757~(1,2,3), z=0.732}
\label{}
\end{figure}
\begin{figure}
\includegraphics[height=0.87\columnwidth]{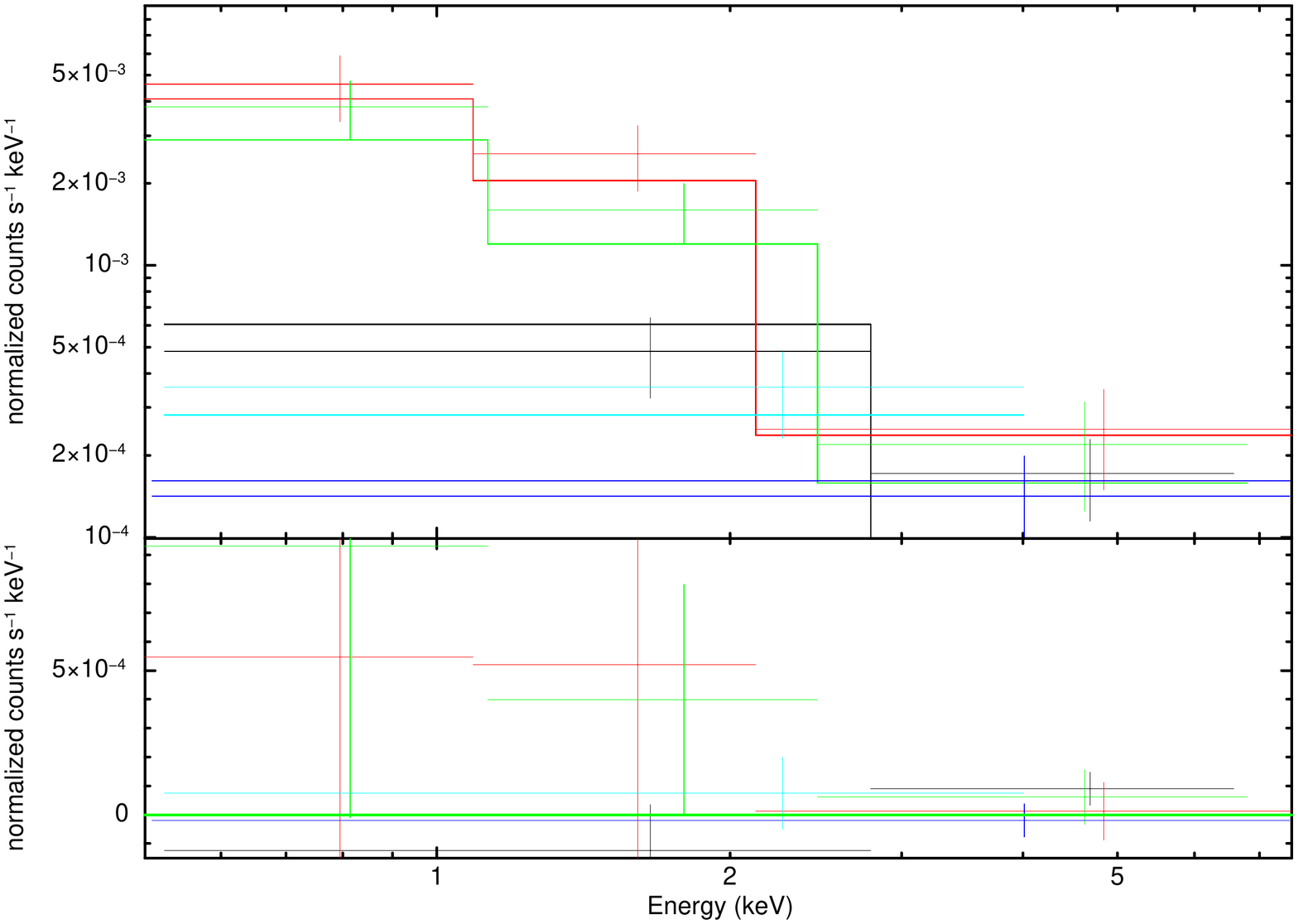}
\includegraphics[height=0.86\columnwidth]{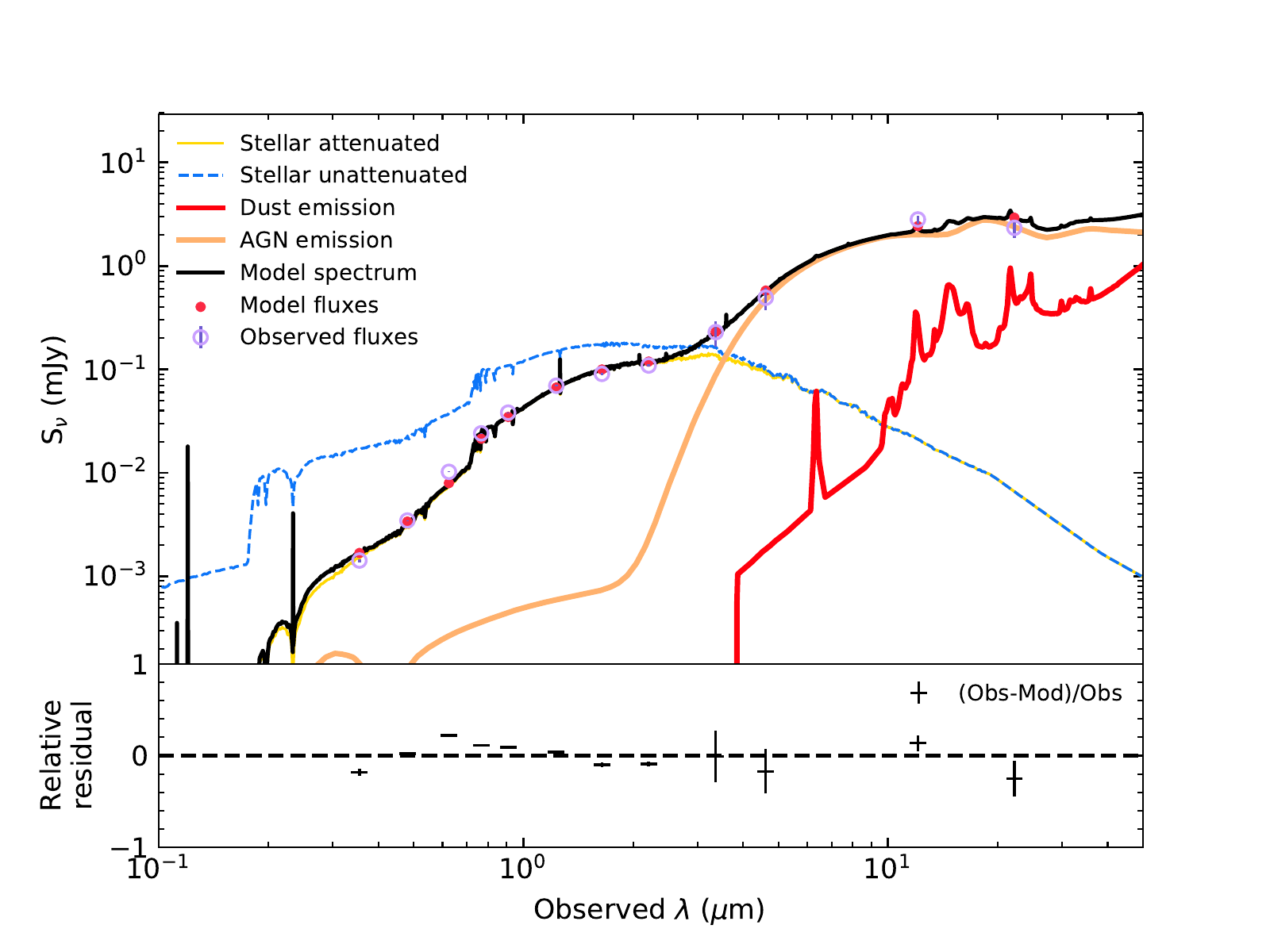}
\includegraphics[height=0.74\columnwidth]{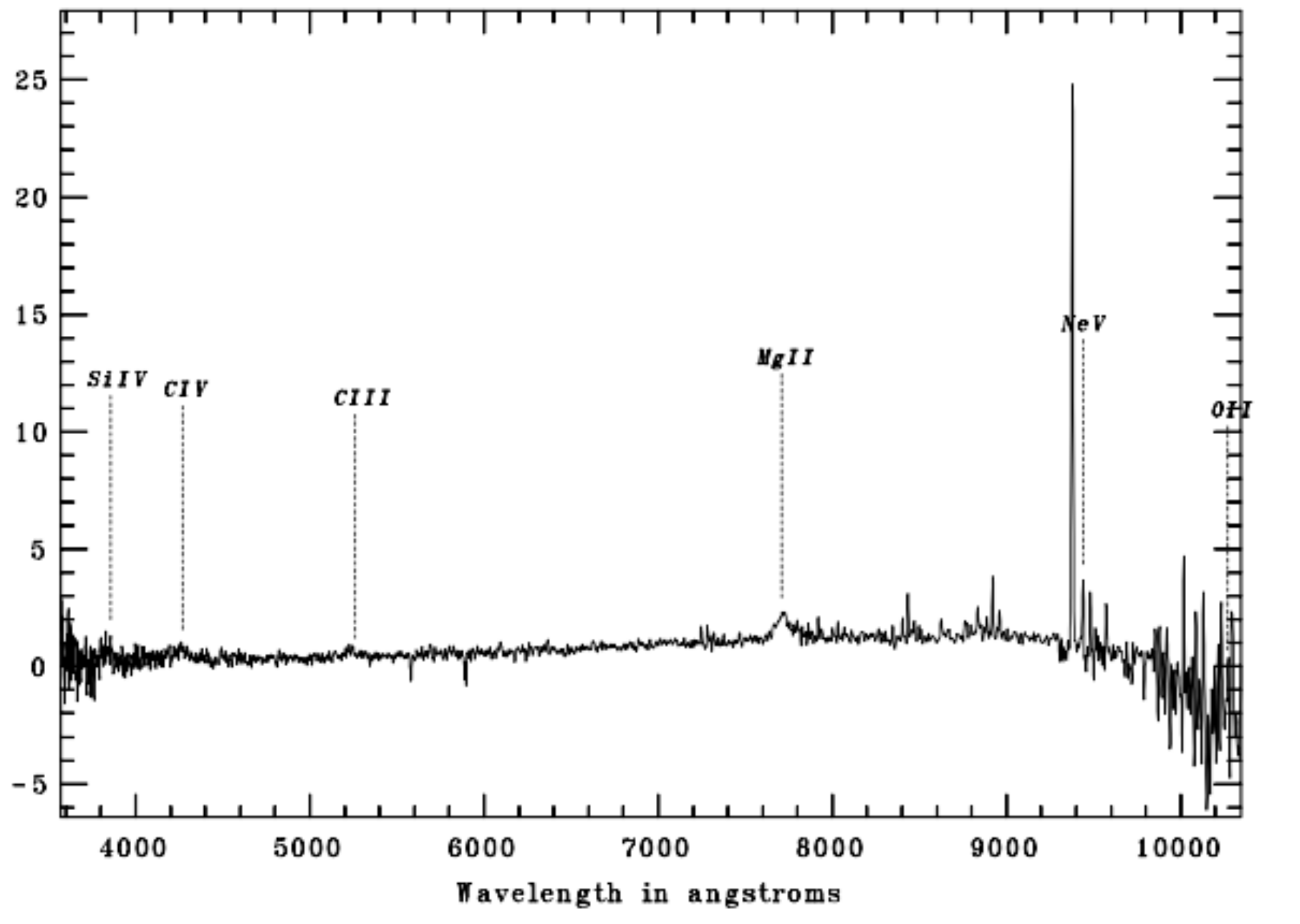}
\caption{J021844.6-054054~(1,2,3), z=0.671}
\label{}
\end{figure}
\begin{figure}
\includegraphics[height=0.87\columnwidth]{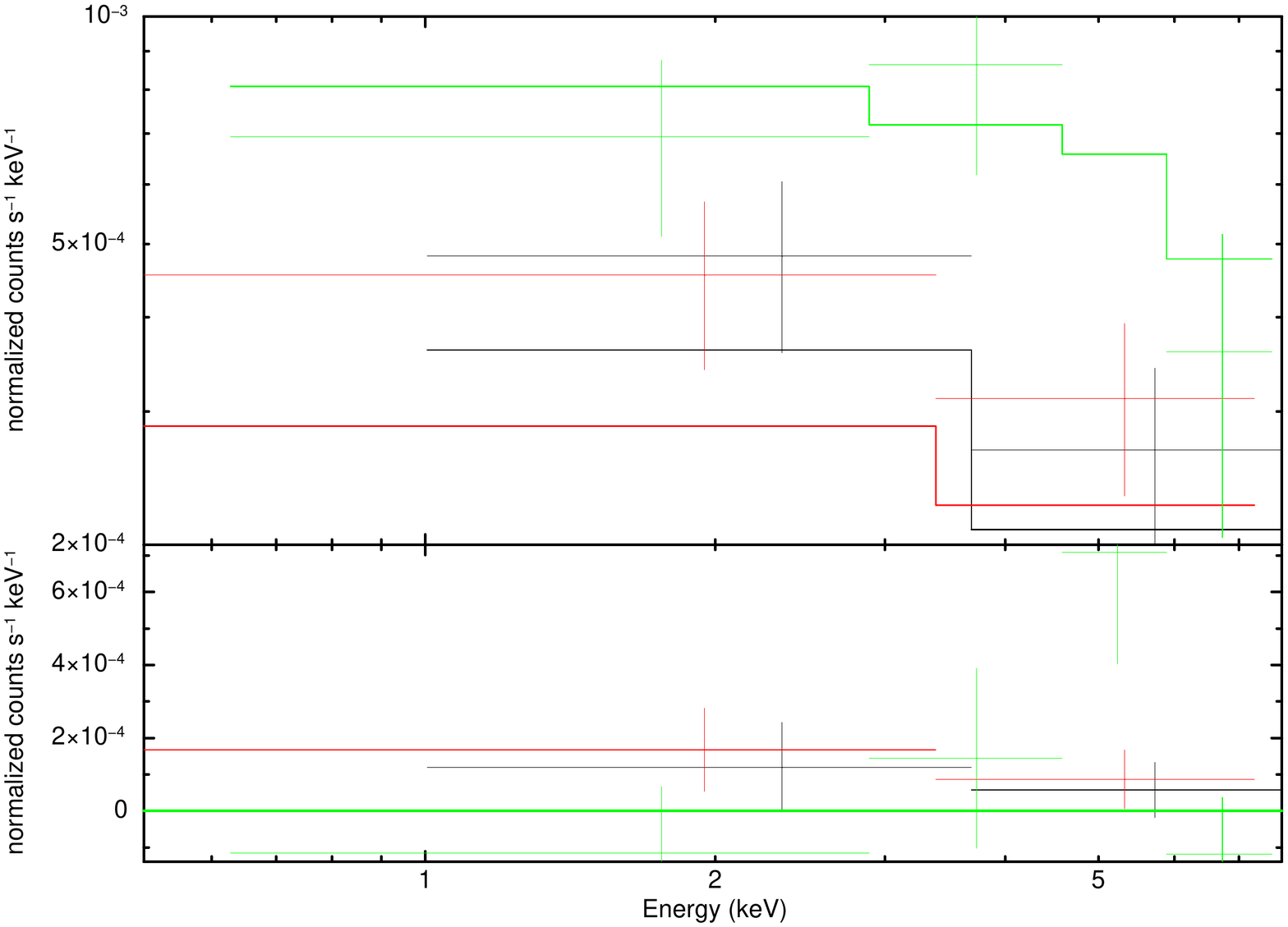}
\includegraphics[height=0.86\columnwidth]{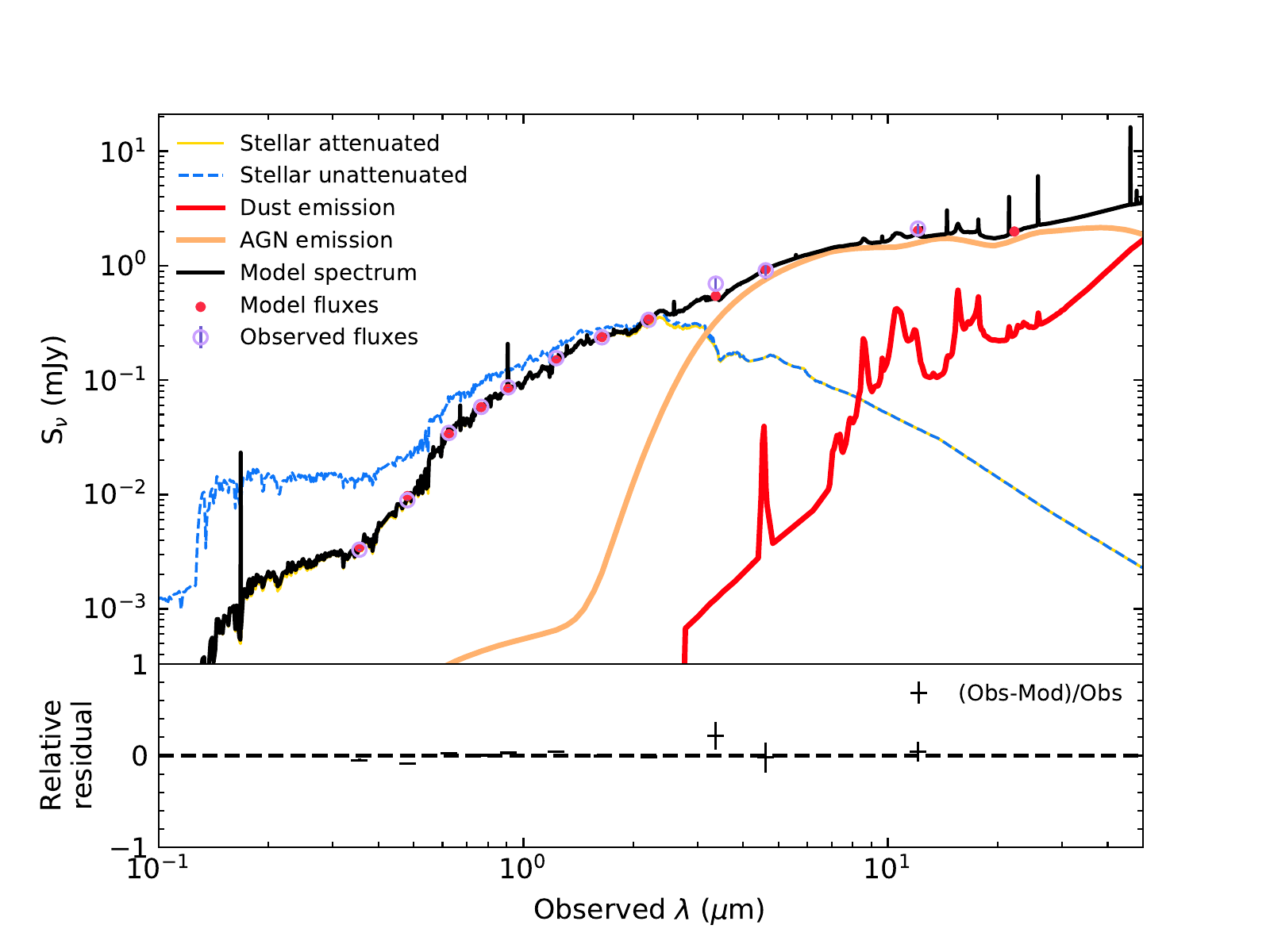}
\includegraphics[height=0.74\columnwidth]{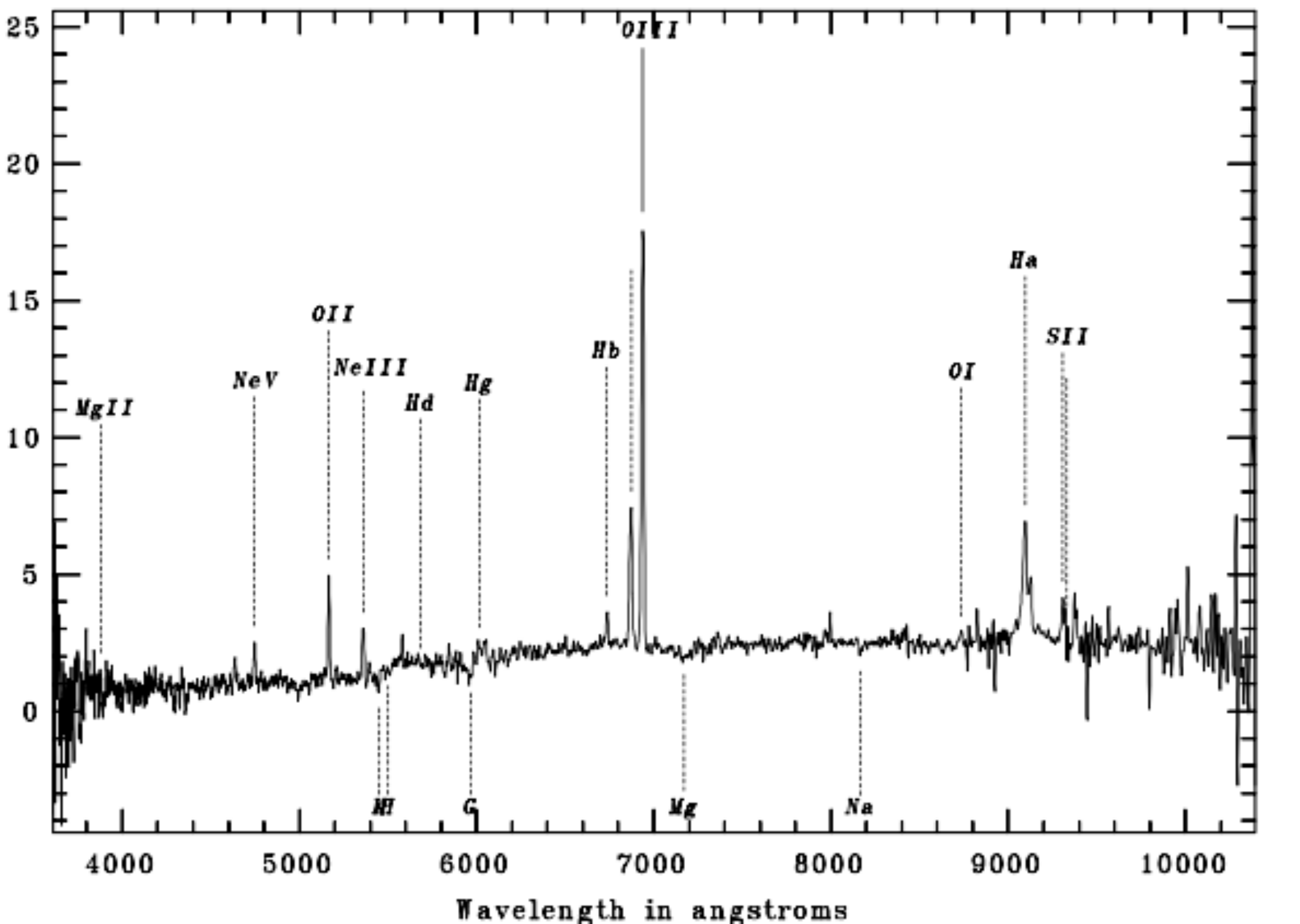}
\caption{J021523.2-044337~(1,2,2), z=0.860}
\label{}
\end{figure}
\begin{figure}
\includegraphics[height=0.87\columnwidth]{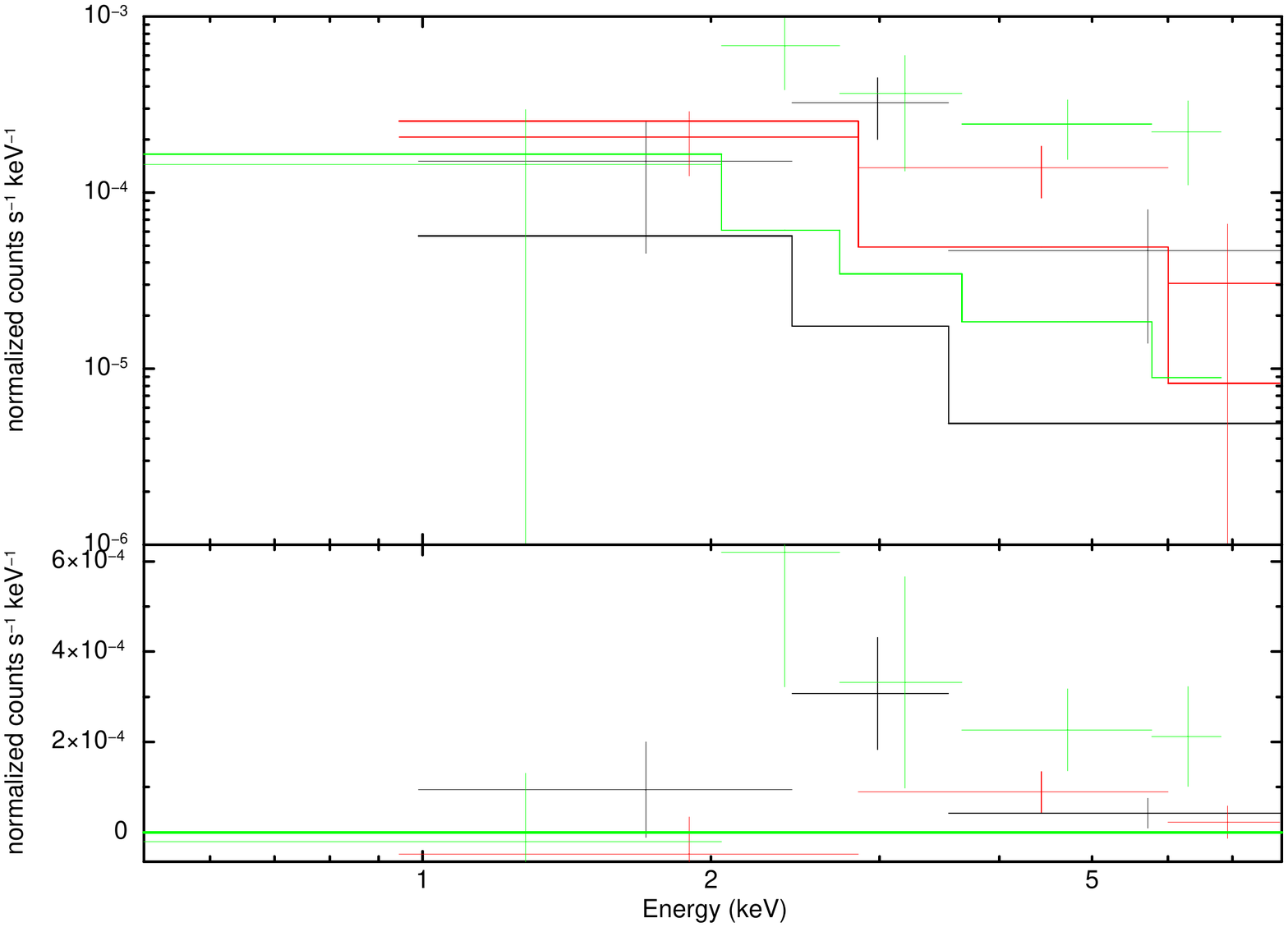}
\includegraphics[height=0.86\columnwidth]{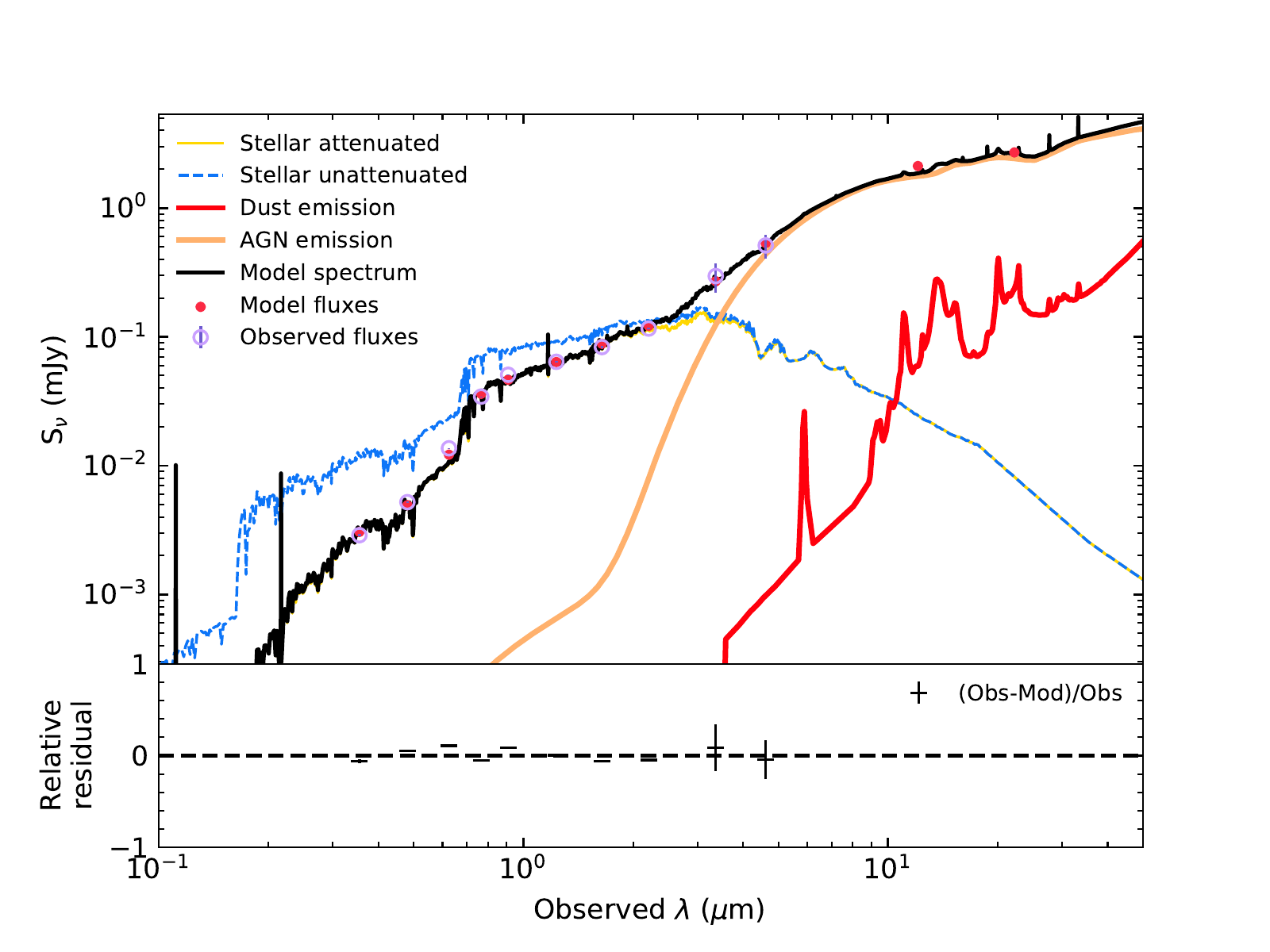}
\includegraphics[height=0.74\columnwidth]{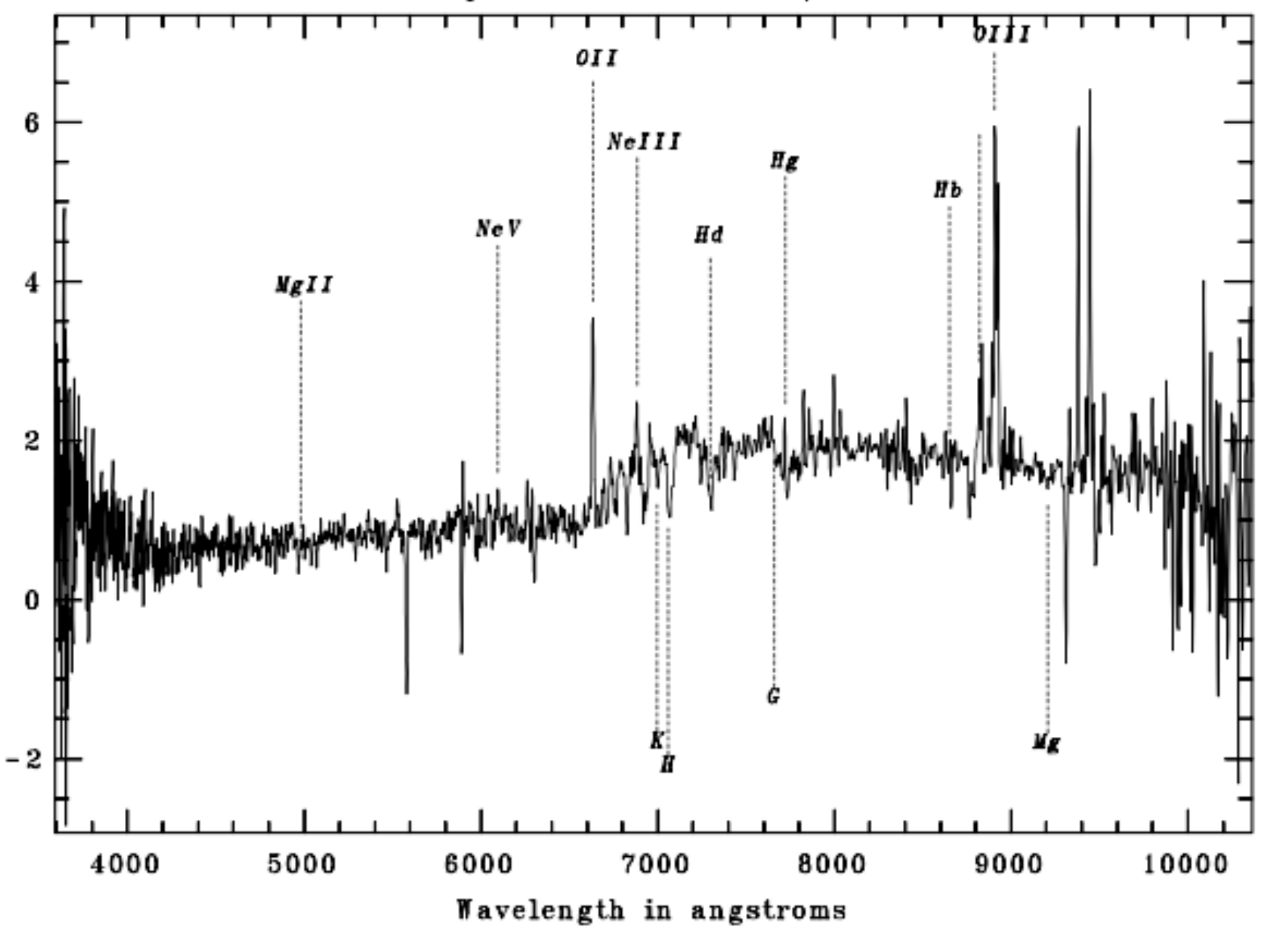}
\caption{J022321.9-045739~(2,2,2), z=0.779}
\label{}
\end{figure}
\begin{figure}
\includegraphics[height=0.87\columnwidth]{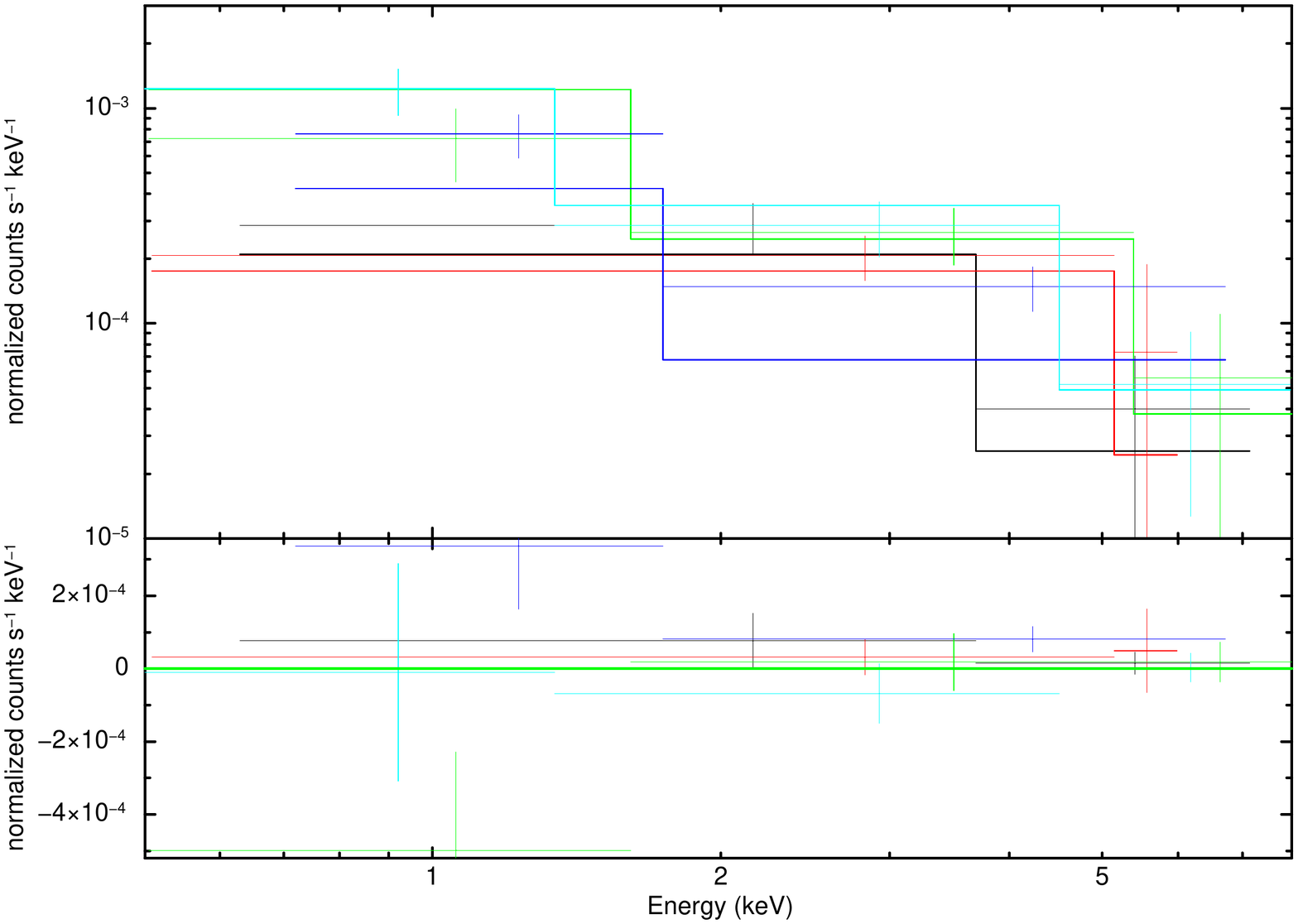}
\includegraphics[height=0.86\columnwidth]{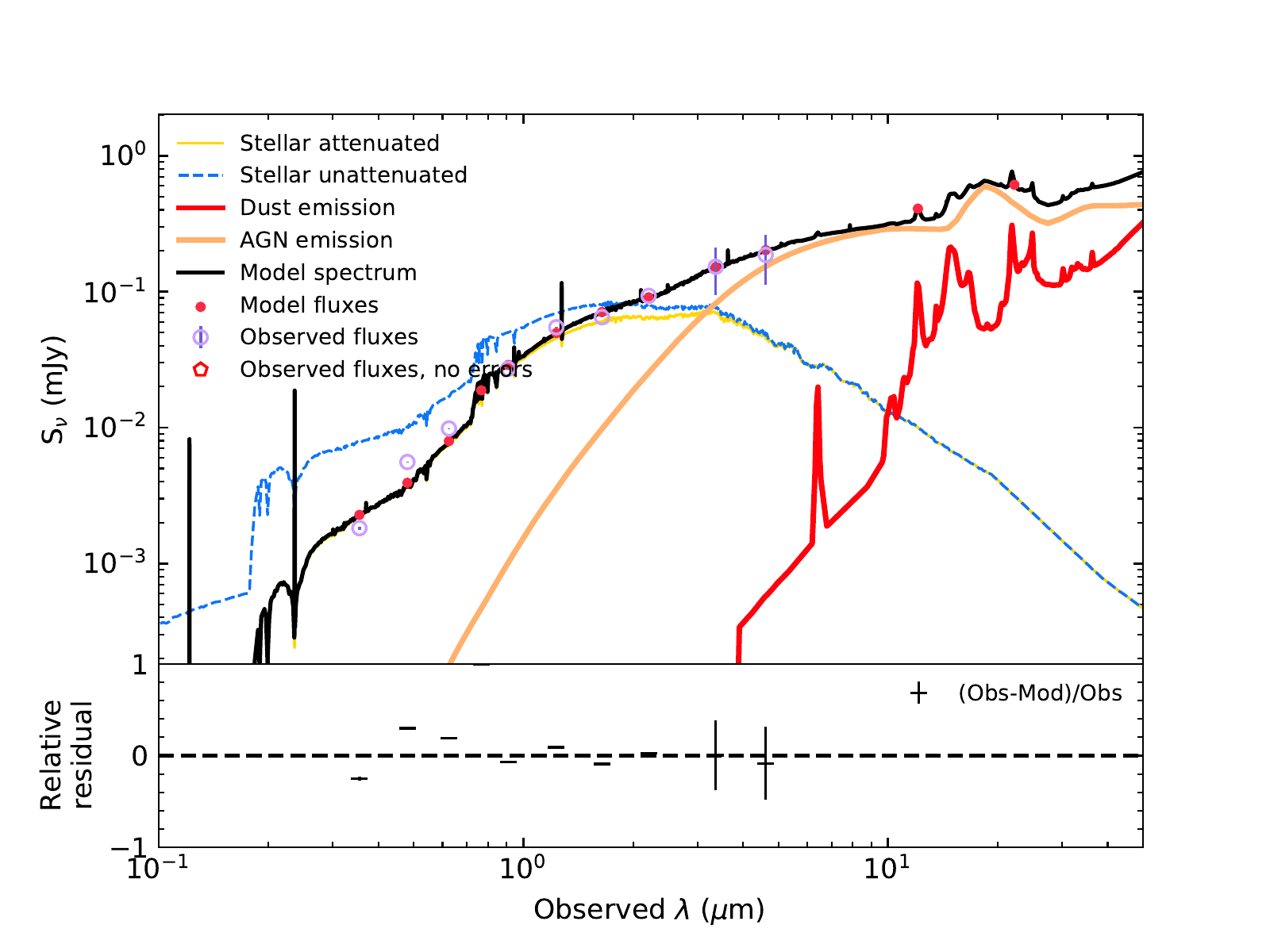}
\includegraphics[height=0.74\columnwidth]{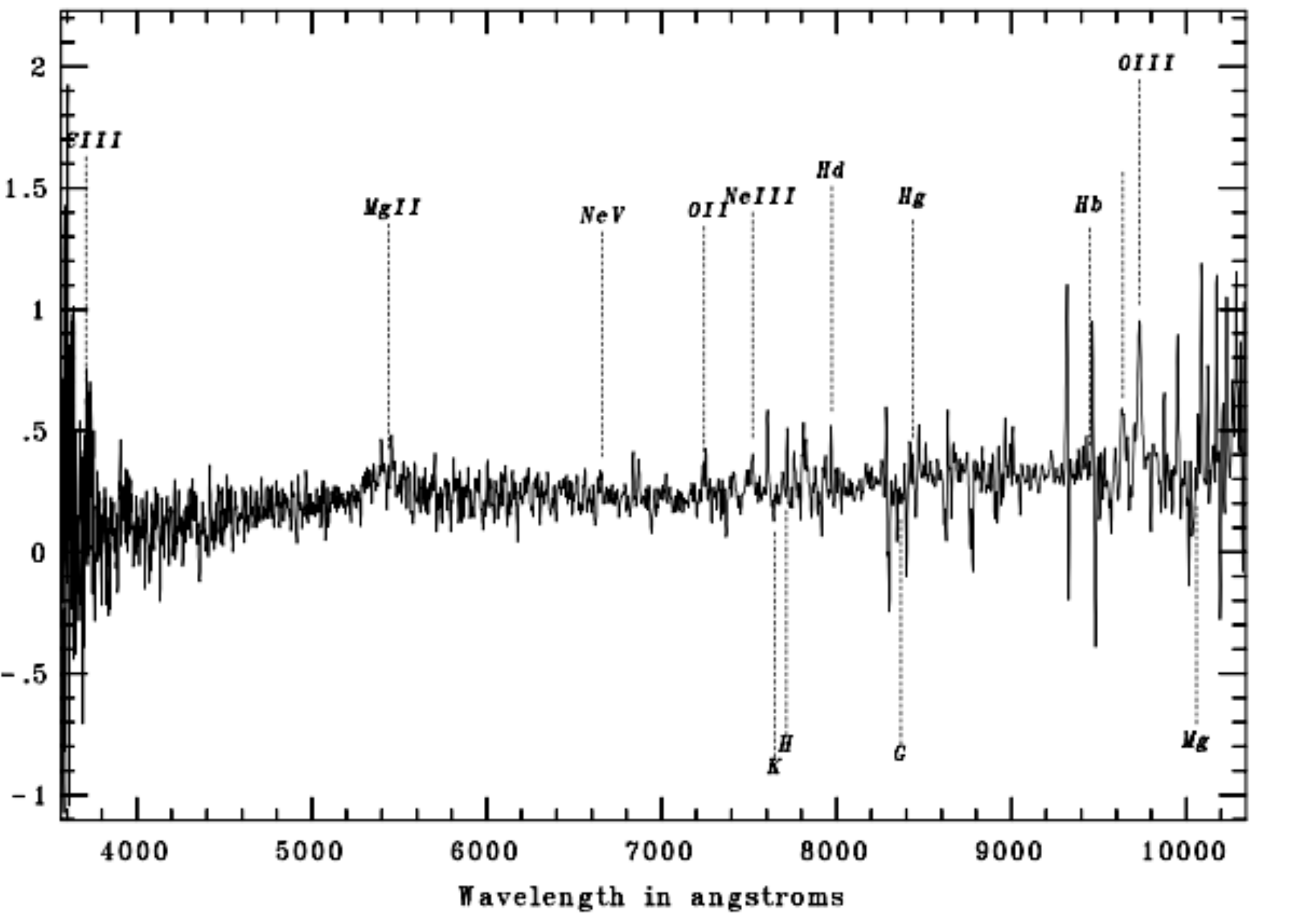}
\caption{J022443.6-050905~(2,2,3), z=0.943}
\label{}
\end{figure}
\begin{figure}
\includegraphics[height=0.87\columnwidth]{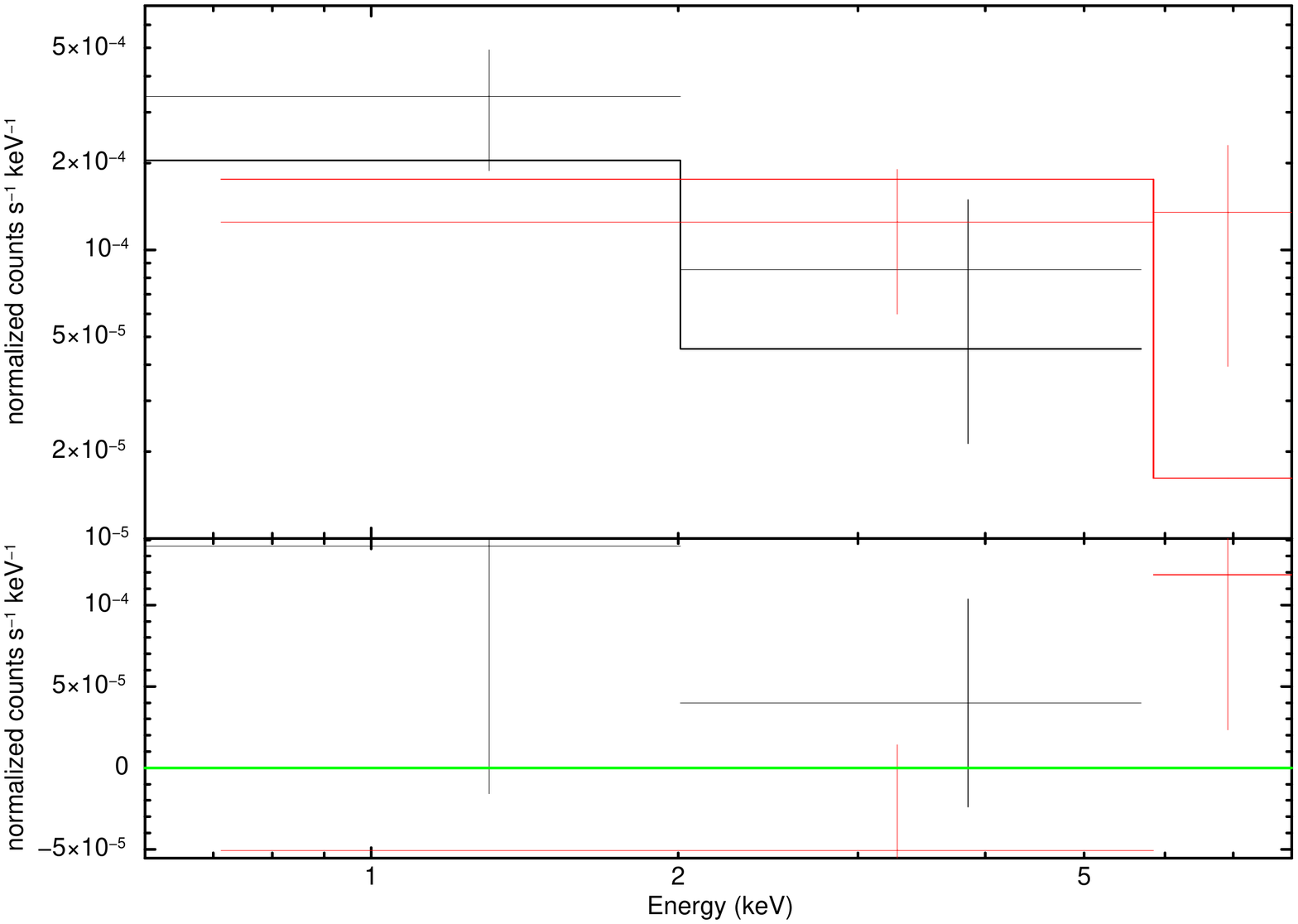}
\includegraphics[height=0.86\columnwidth]{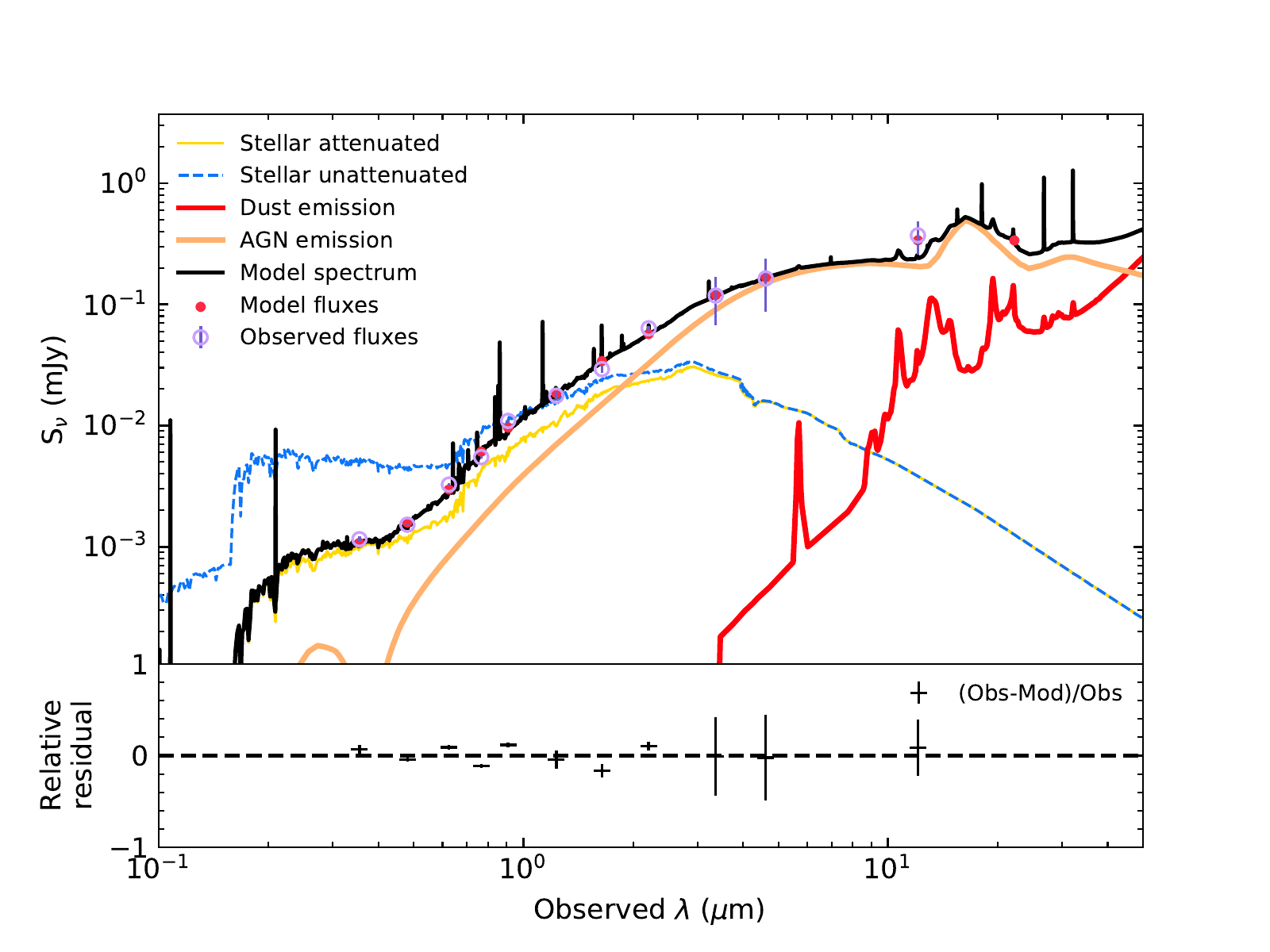}
\includegraphics[height=0.74\columnwidth]{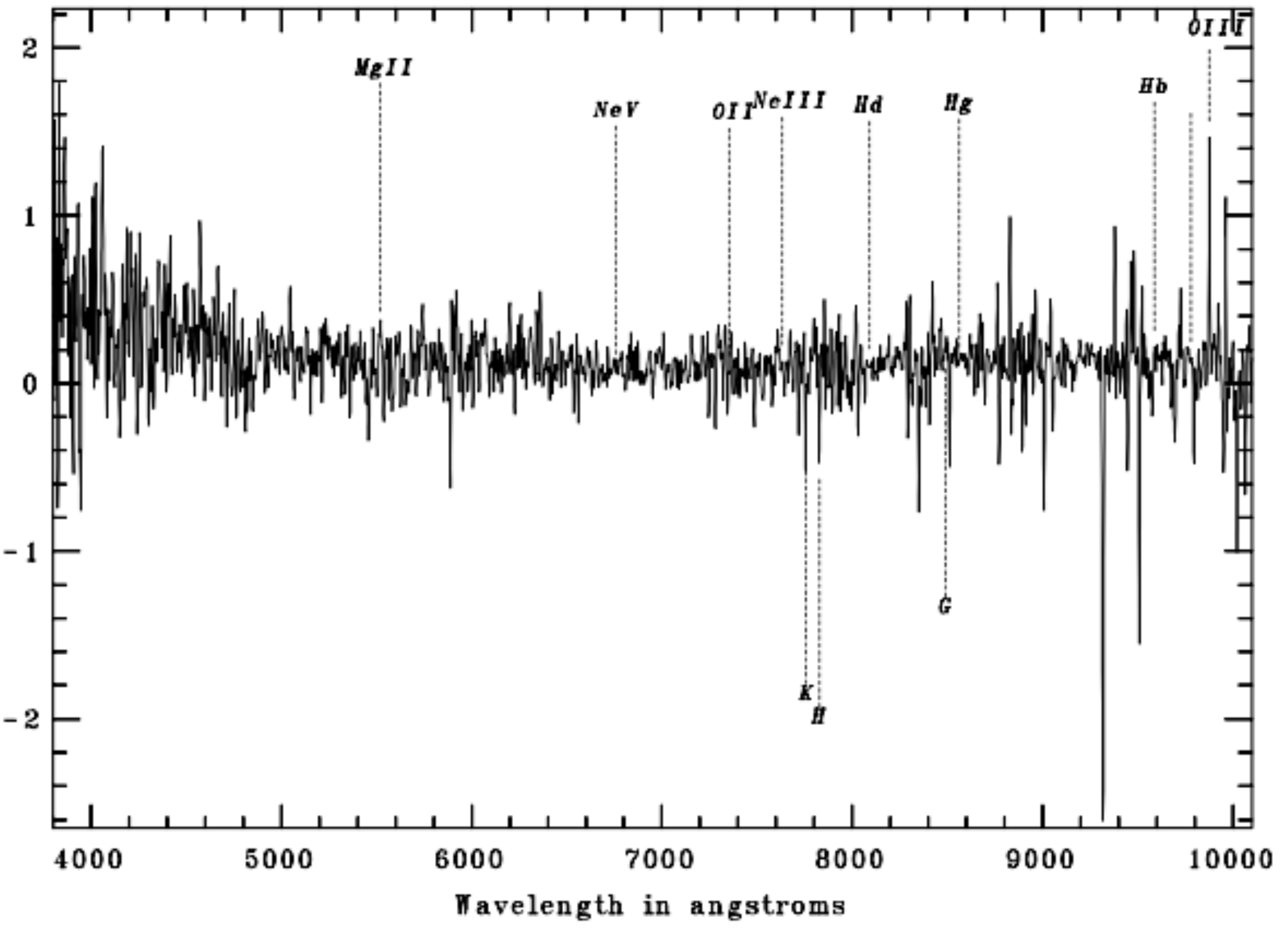}
\caption{J023418.0-041833~(1,2,0), z=0.582}
\label{}
\end{figure}
\begin{figure}
\includegraphics[height=0.87\columnwidth]{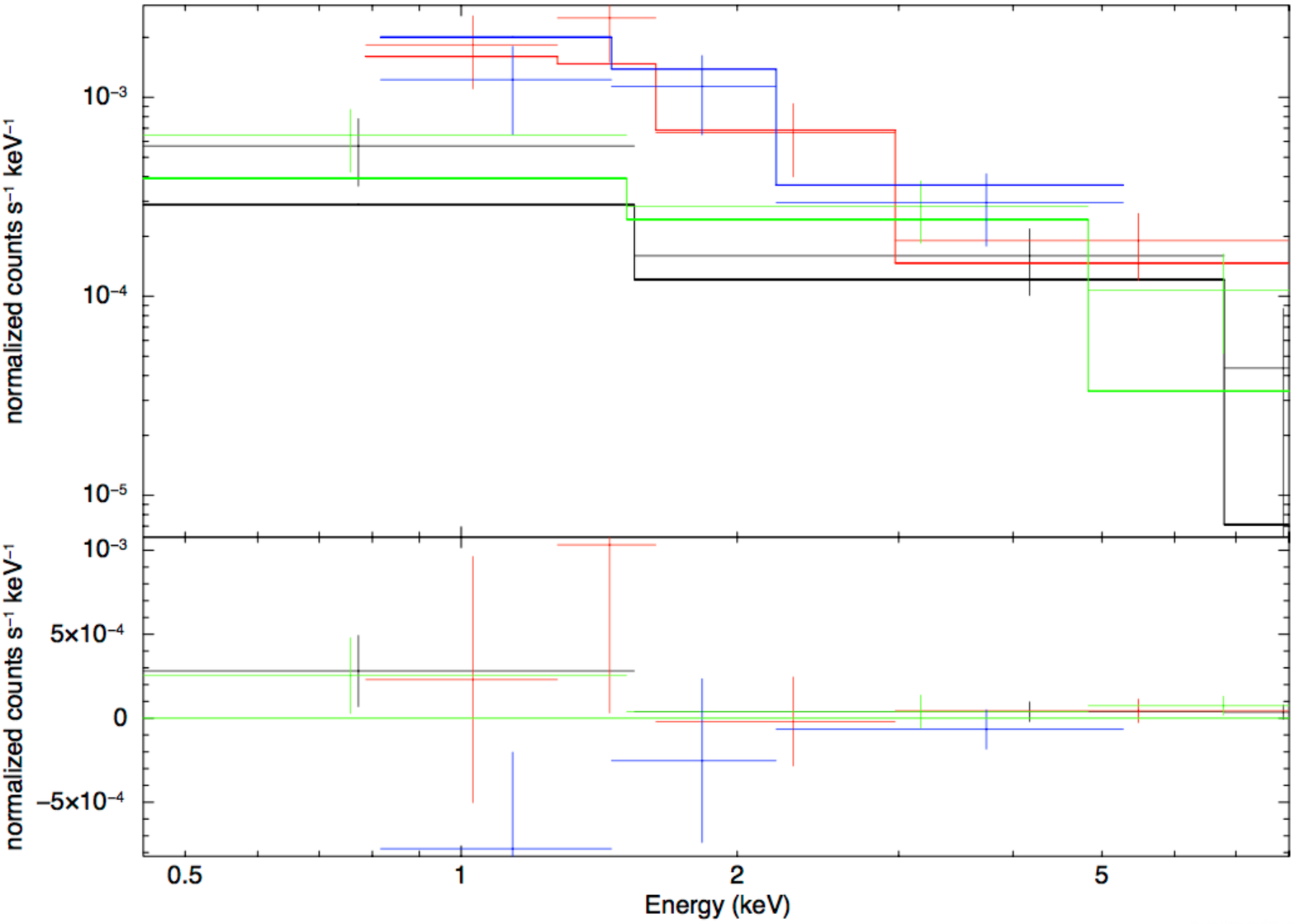}
\includegraphics[height=0.86\columnwidth]{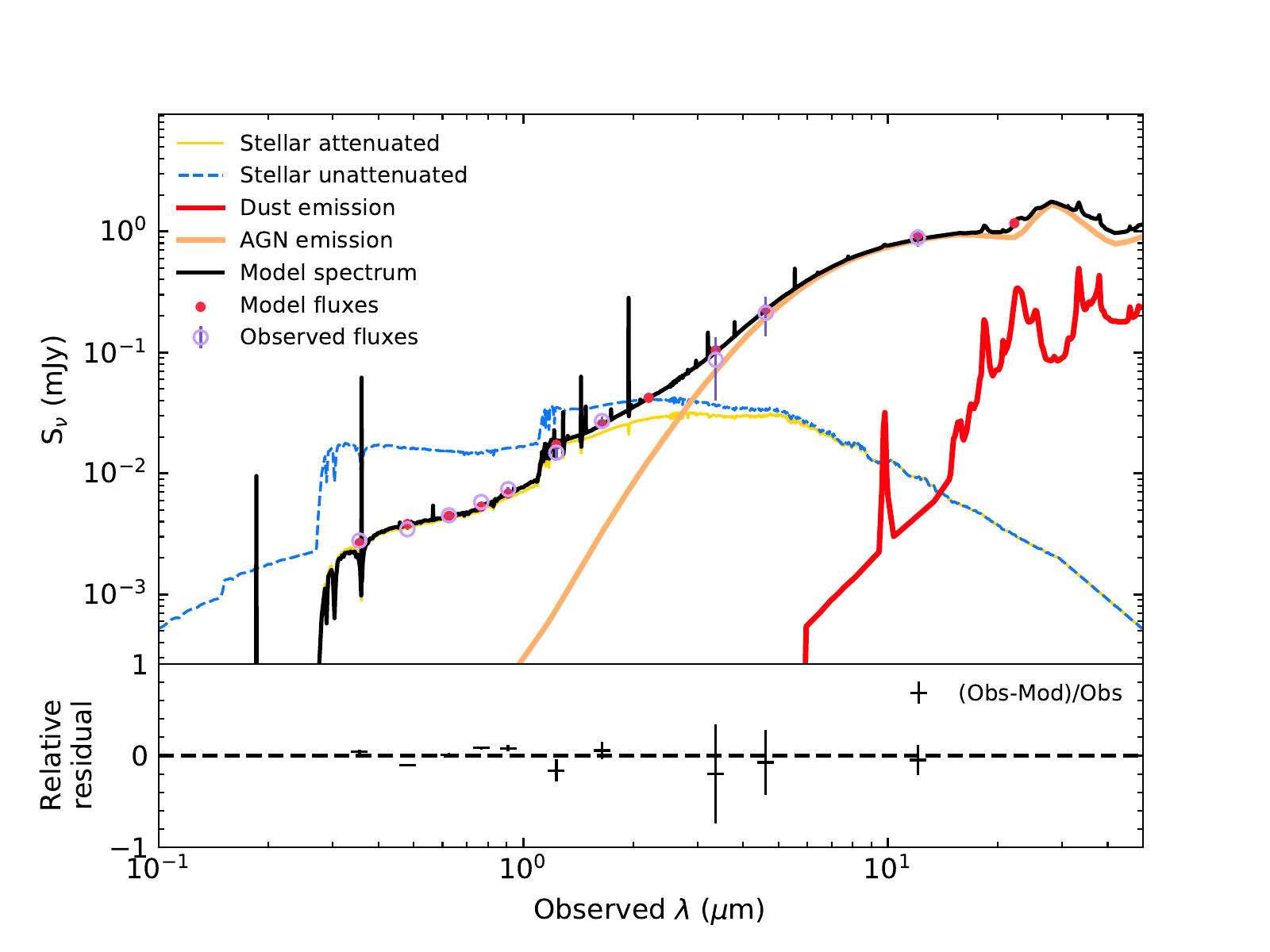}
\includegraphics[height=0.74\columnwidth]{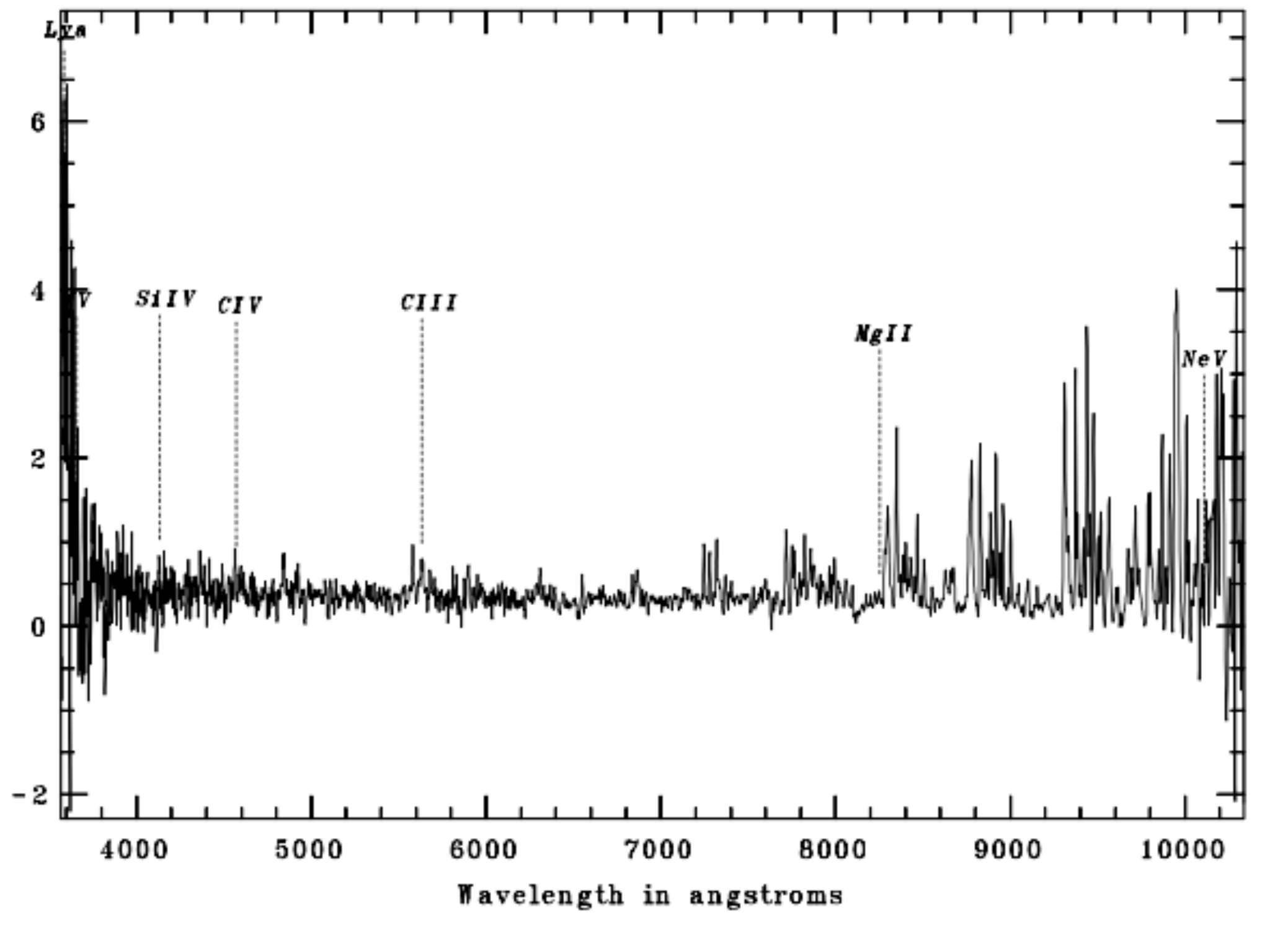}
\caption{J020543.7-063807~(1,2,0), z=0.772}
\label{}
\end{figure}
\clearpage
\begin{figure}
\includegraphics[height=0.87\columnwidth]{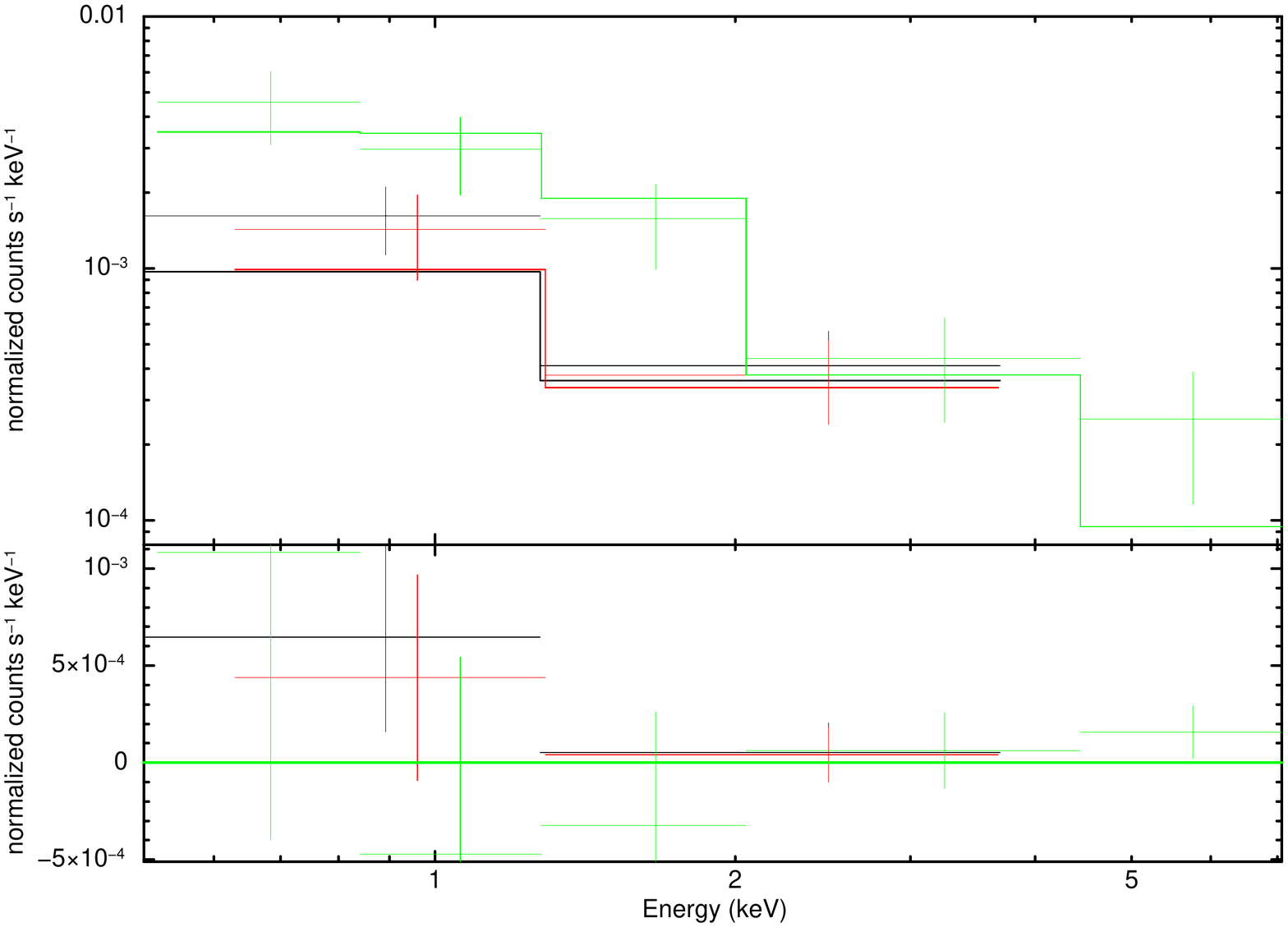}
\includegraphics[height=0.86\columnwidth]{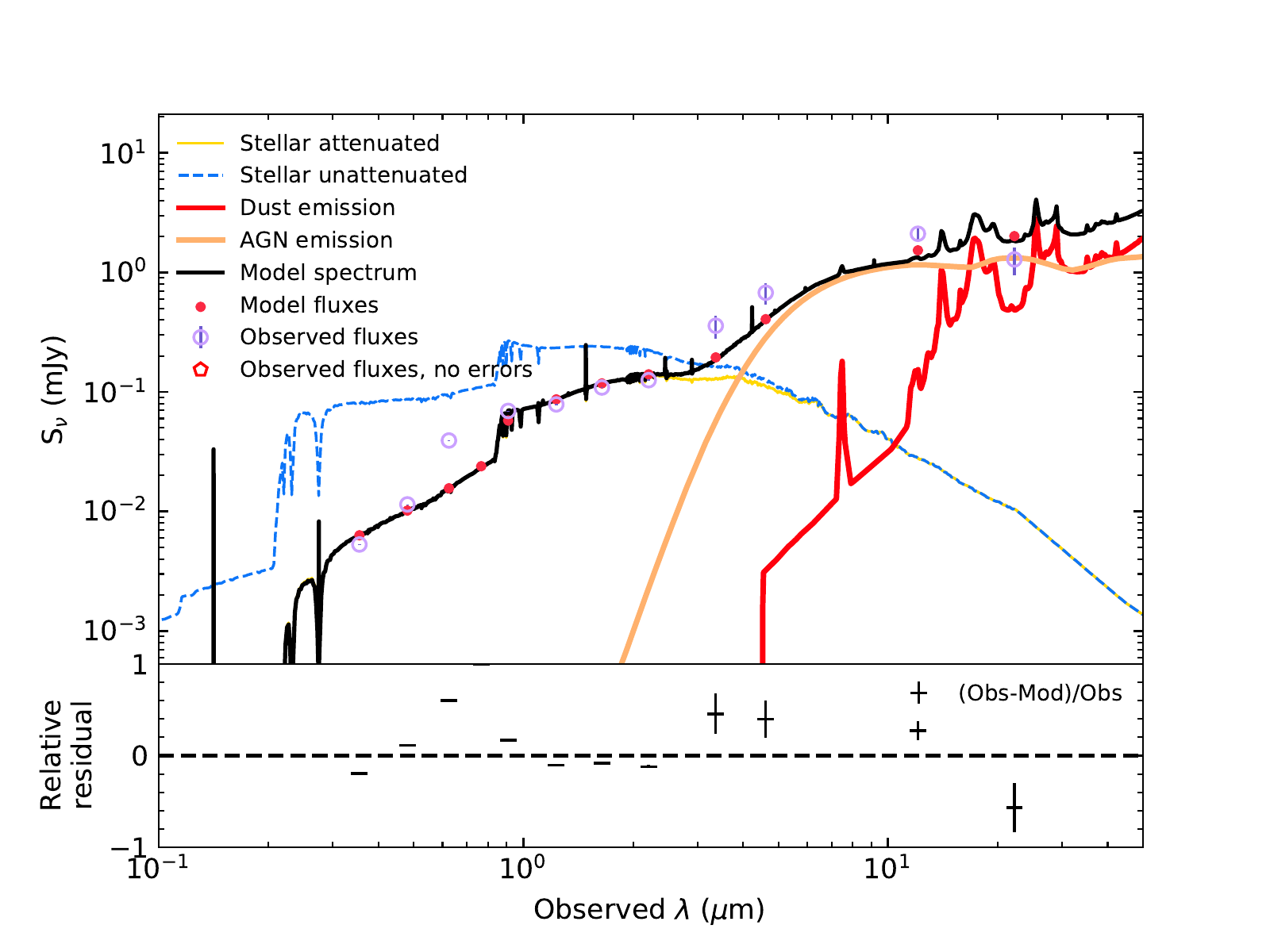}
\includegraphics[height=0.74\columnwidth]{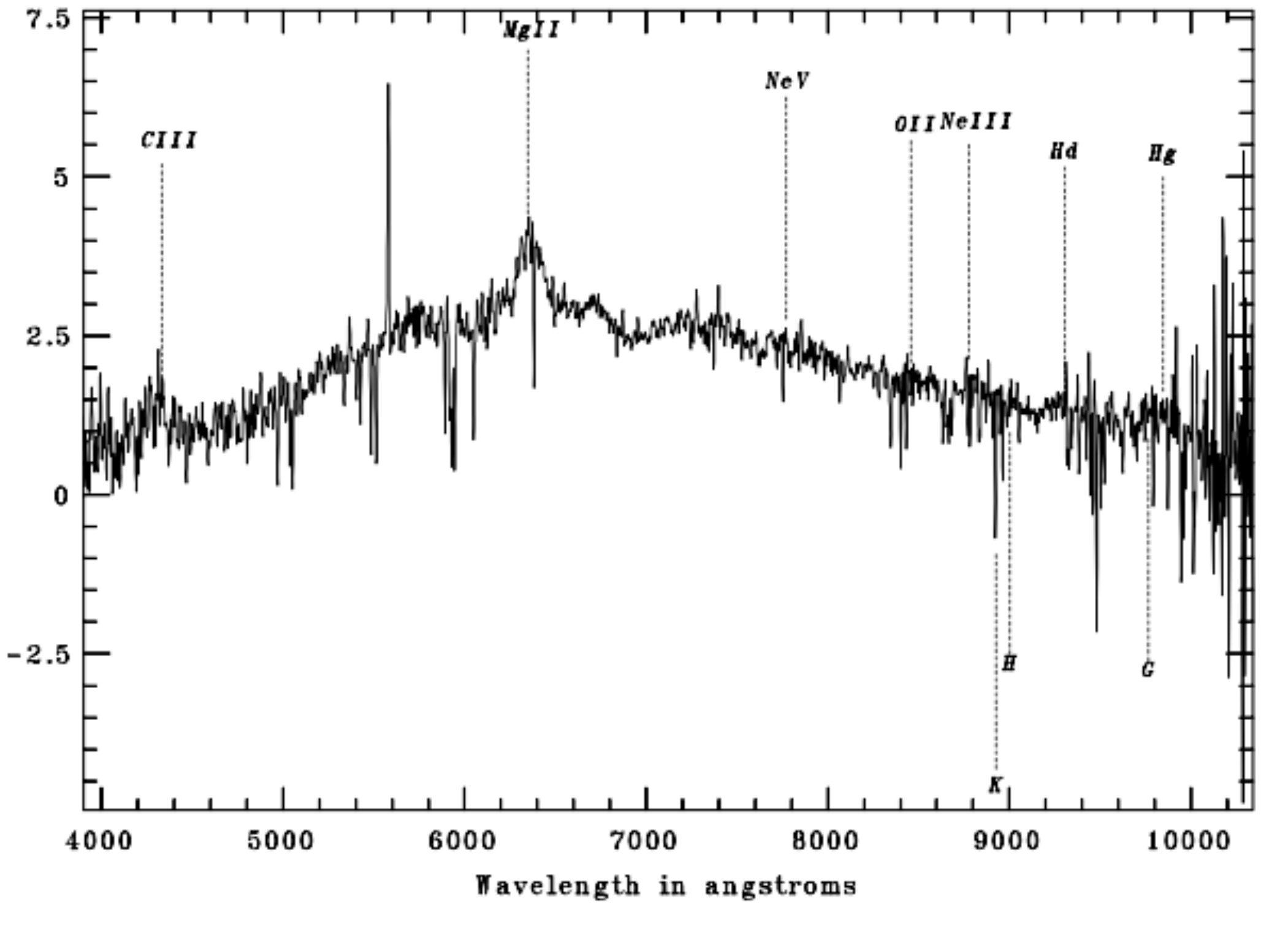}
\caption{J022932.6-055438~(1,2,1), z=1.263}
\label{}
\end{figure}
\begin{figure}
\includegraphics[height=0.87\columnwidth]{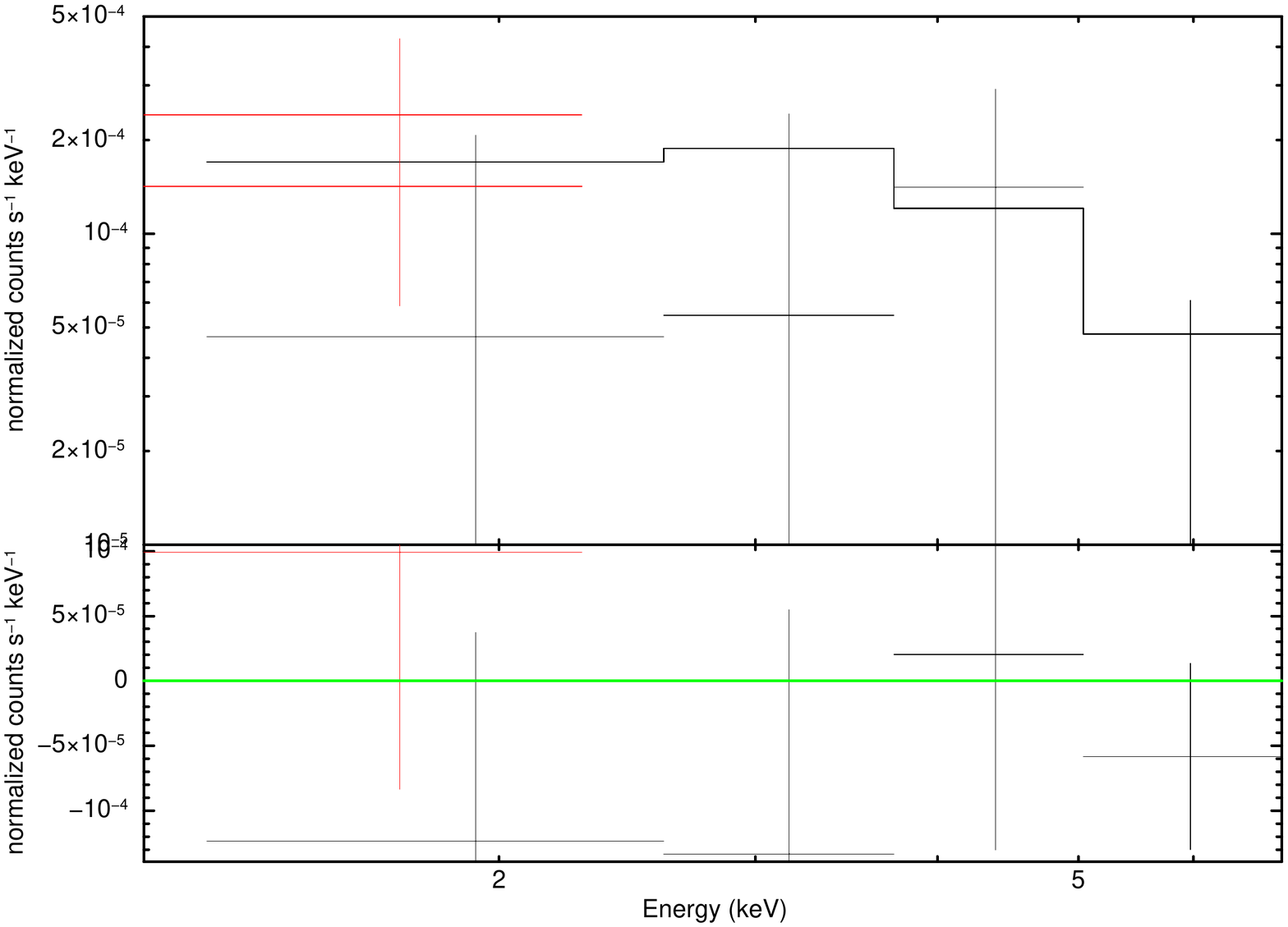}
\includegraphics[height=0.86\columnwidth]{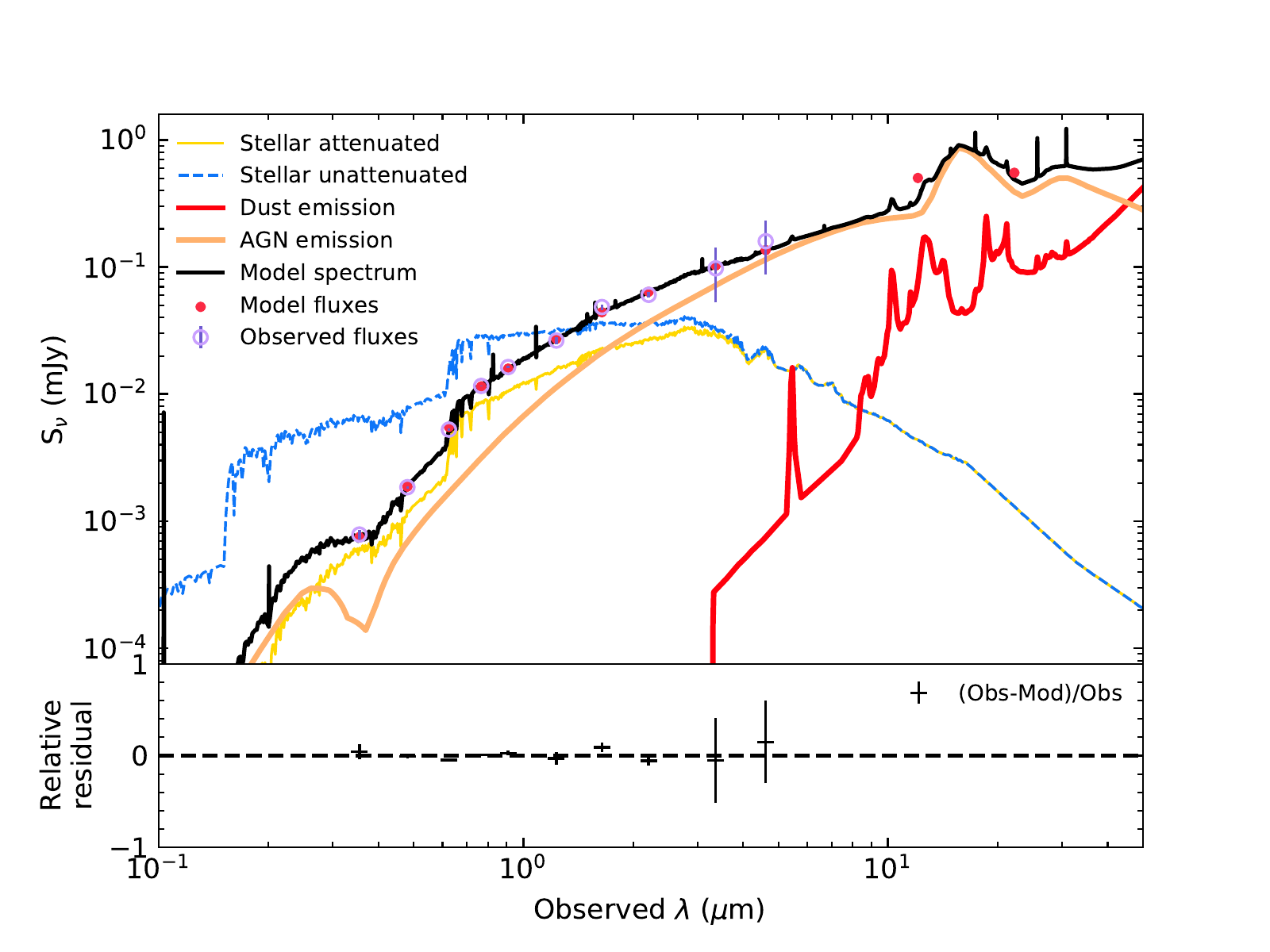}
\includegraphics[height=0.74\columnwidth]{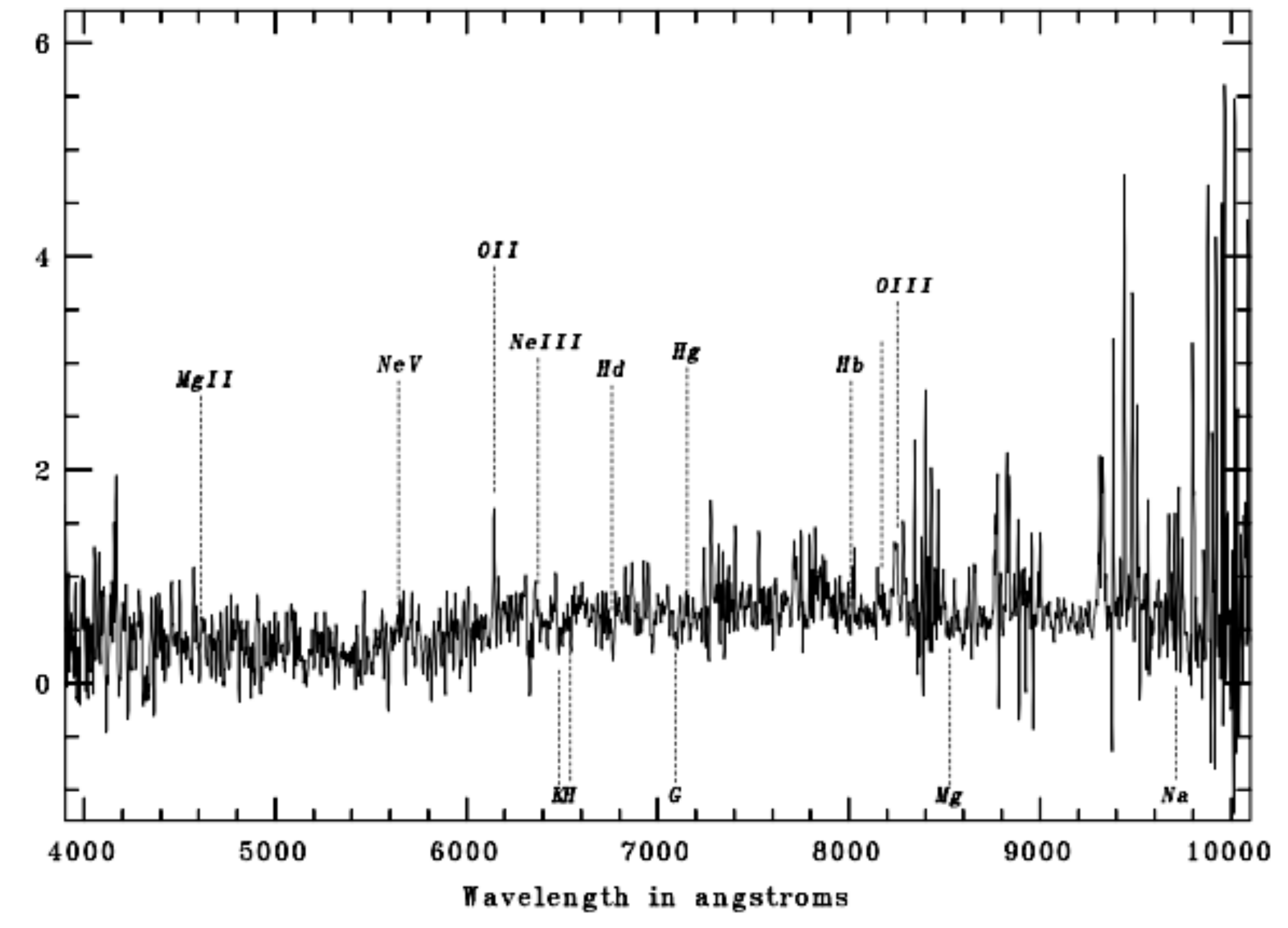}
\caption{J021239.2-054816~(2,2,0), z=0.711}
\label{}
\end{figure}
\begin{figure}
\includegraphics[height=0.87\columnwidth]{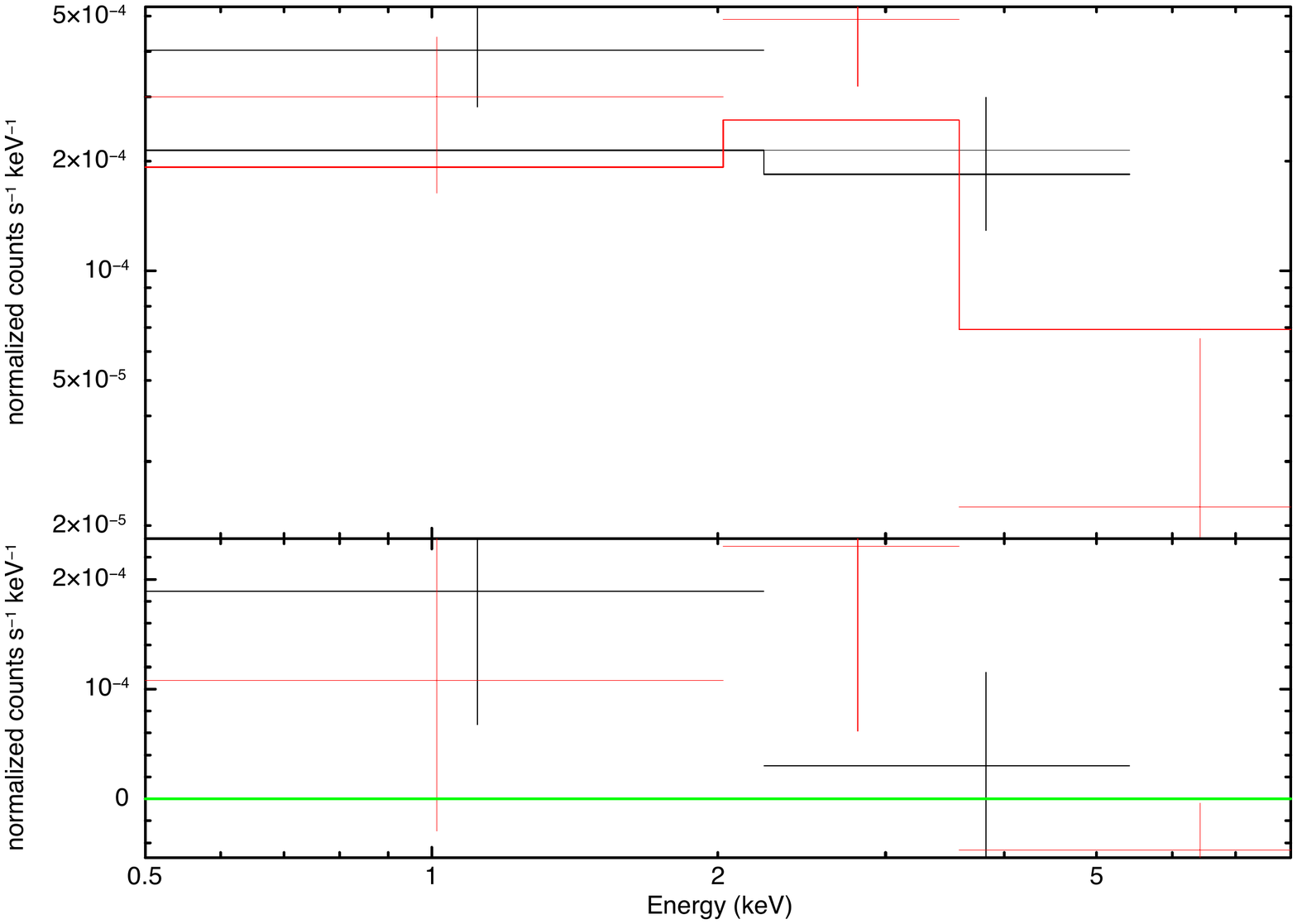}
\includegraphics[height=0.86\columnwidth]{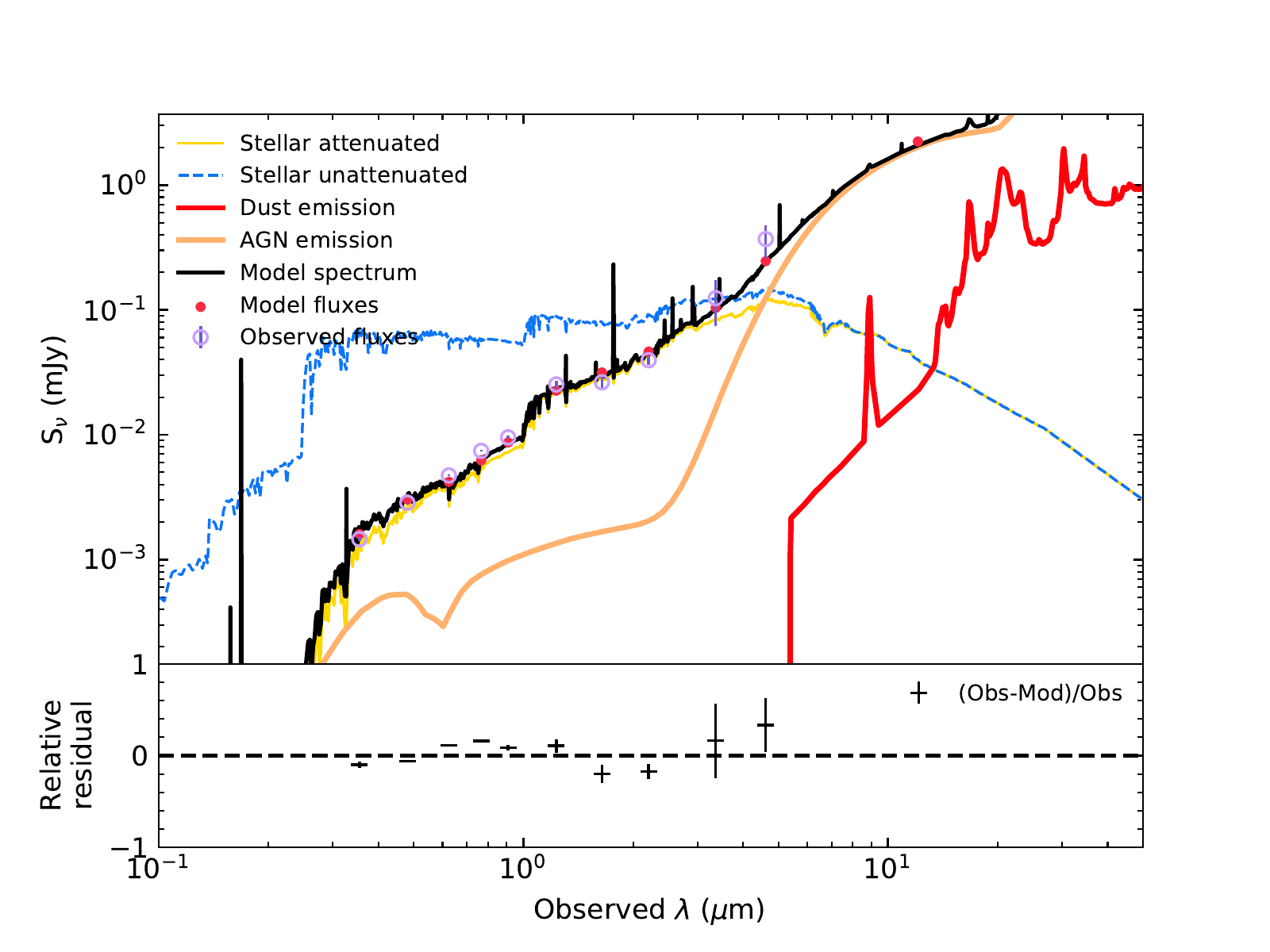}
\includegraphics[height=0.74\columnwidth]{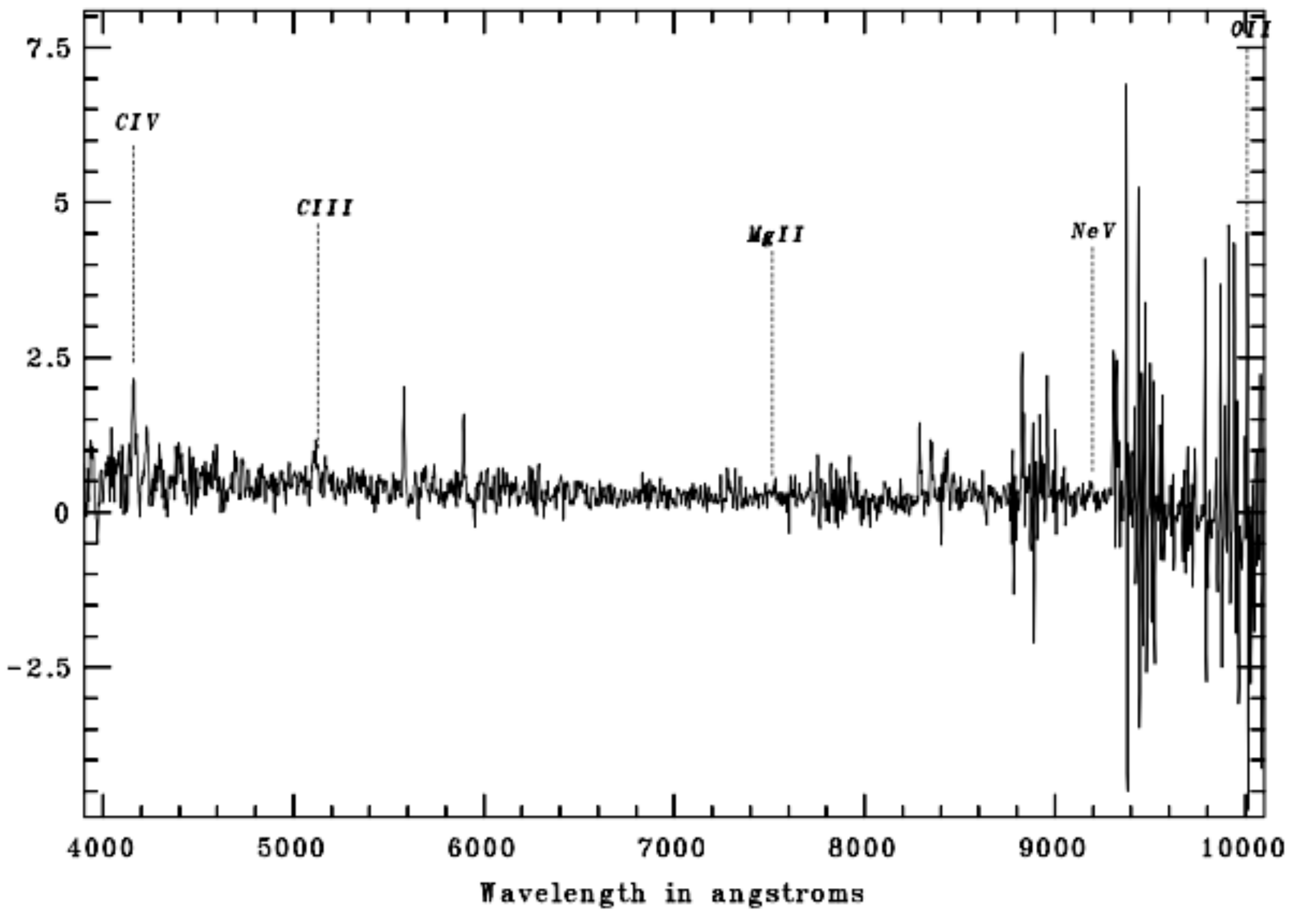}
\caption{J021511.4-060805~(2,2,1), z=0.772}
\label{}
\end{figure}
\begin{figure}
\includegraphics[height=0.87\columnwidth]{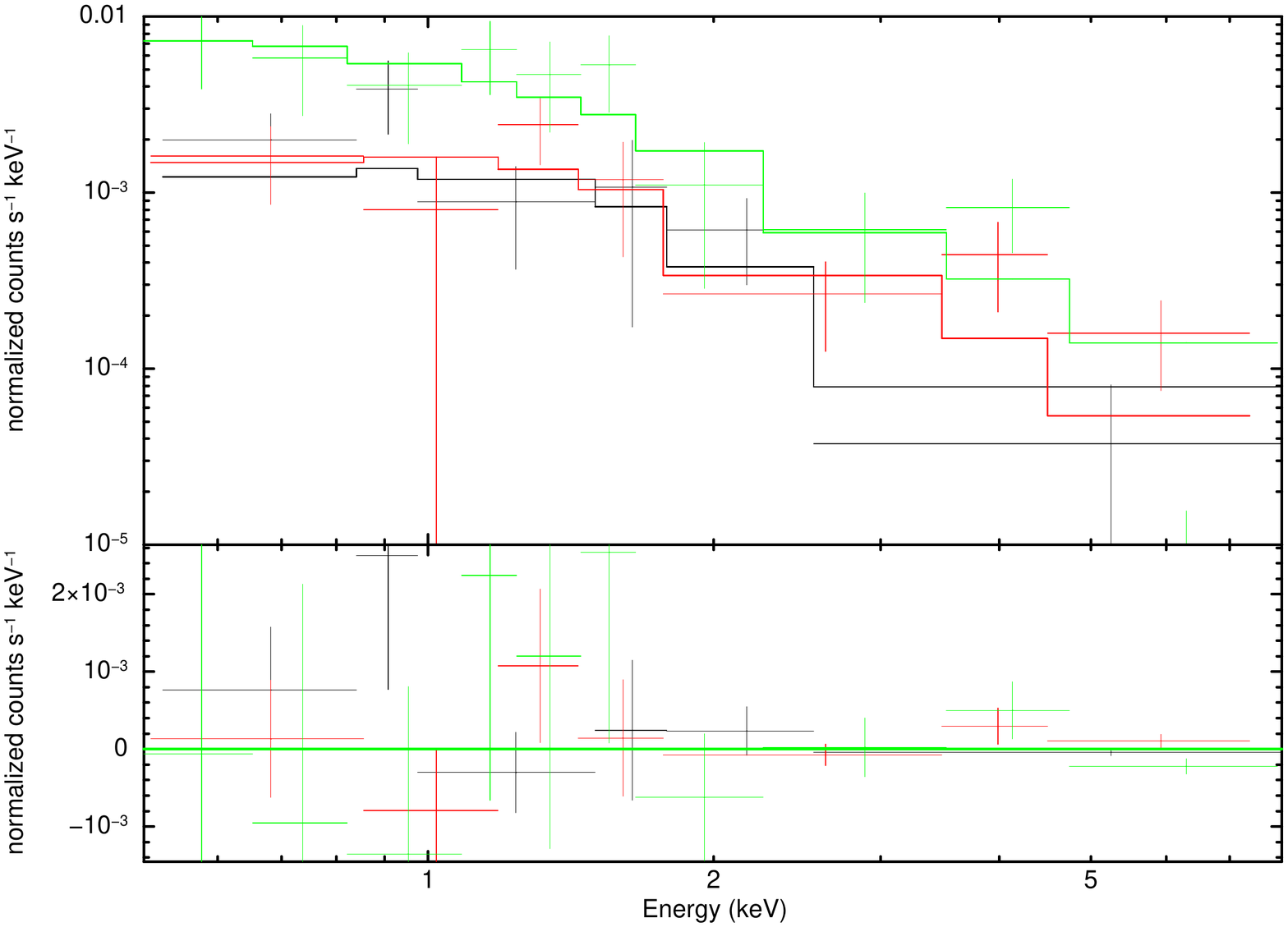}
\includegraphics[height=0.86\columnwidth]{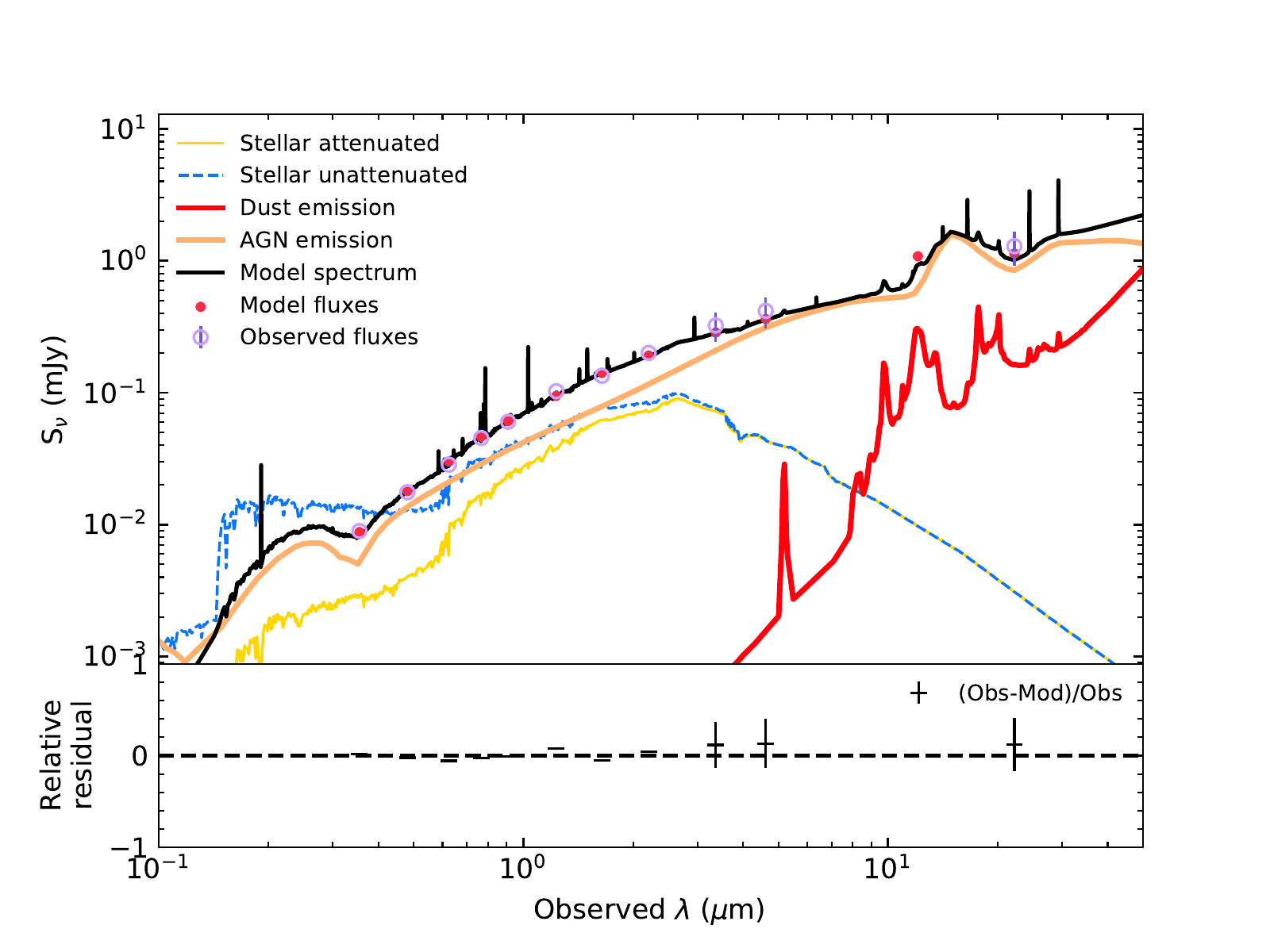}
\includegraphics[height=0.74\columnwidth]{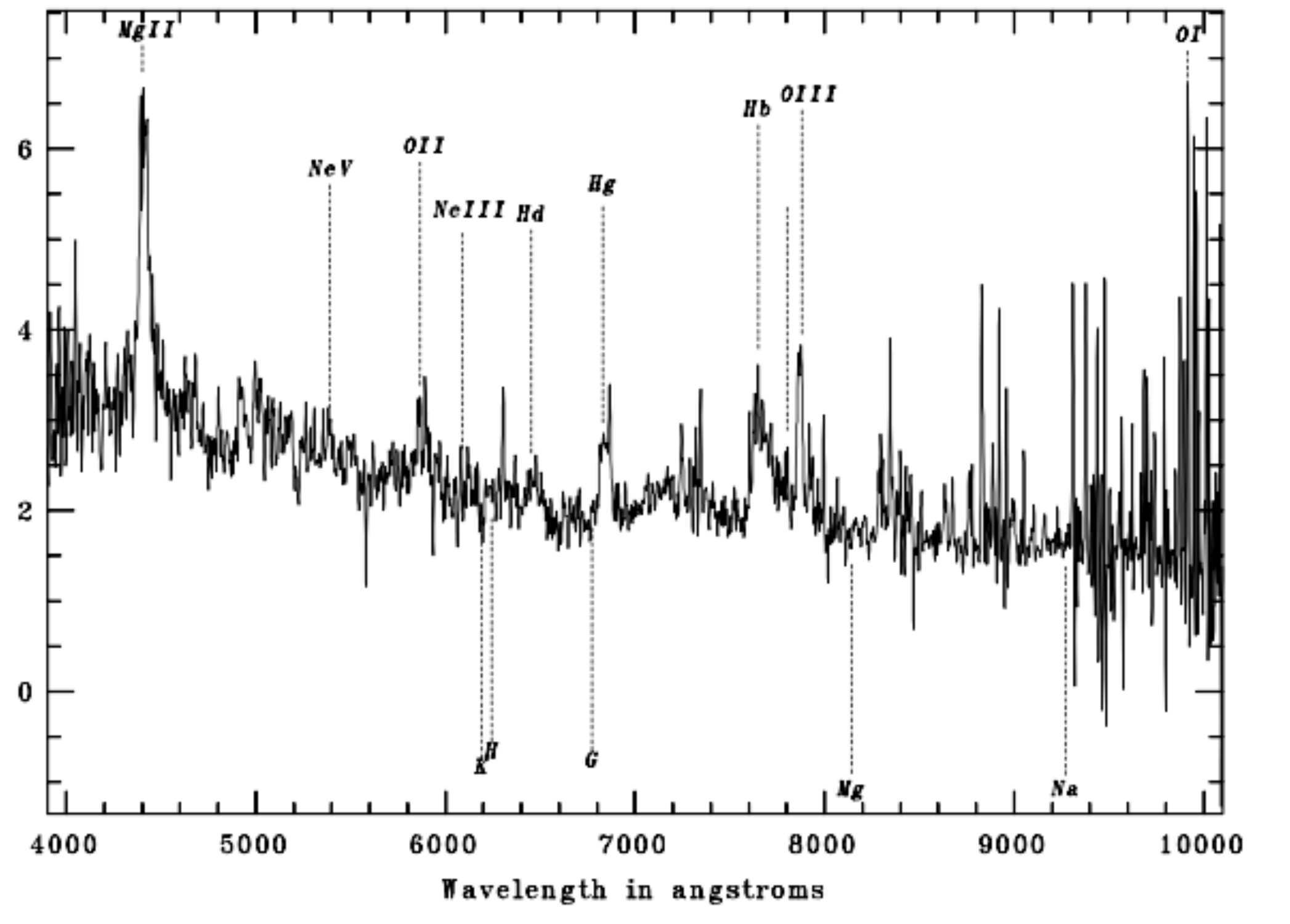}
\caption{J021808.8-055630~(1,0,1), z=0.572}
\label{}
\end{figure}
\begin{figure}
\includegraphics[height=0.87\columnwidth]{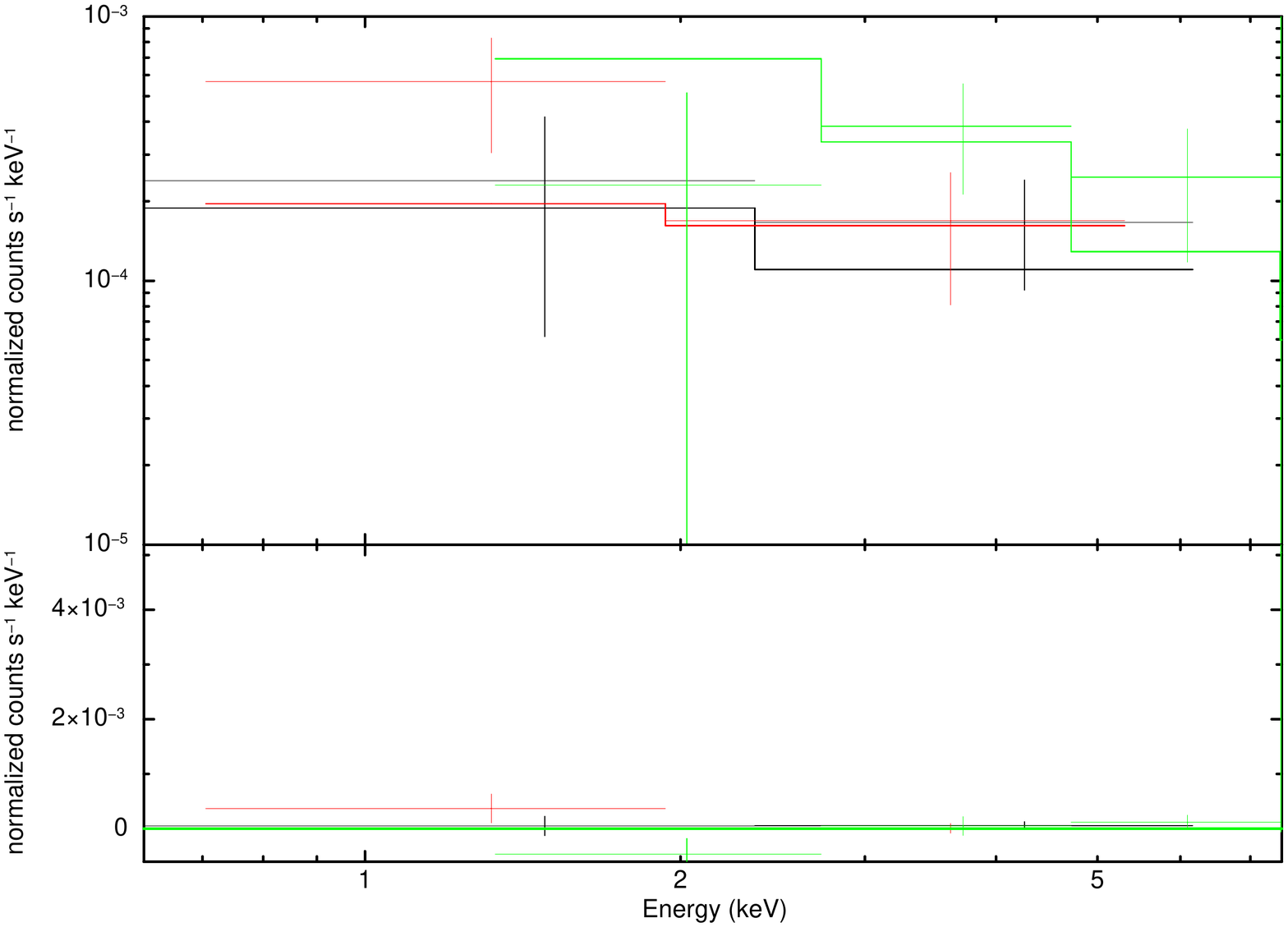}
\includegraphics[height=0.86\columnwidth]{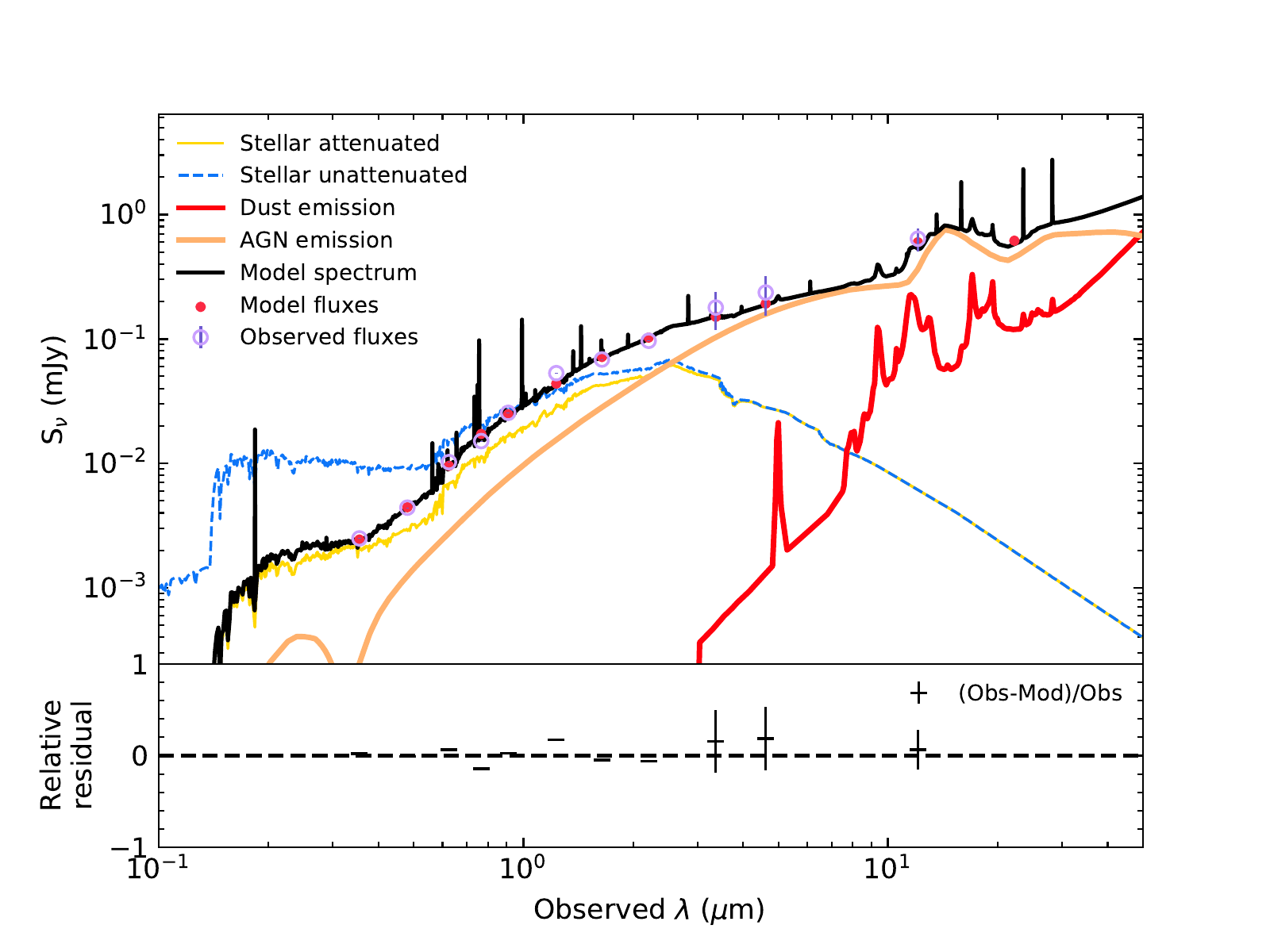}
\includegraphics[height=0.74\columnwidth]{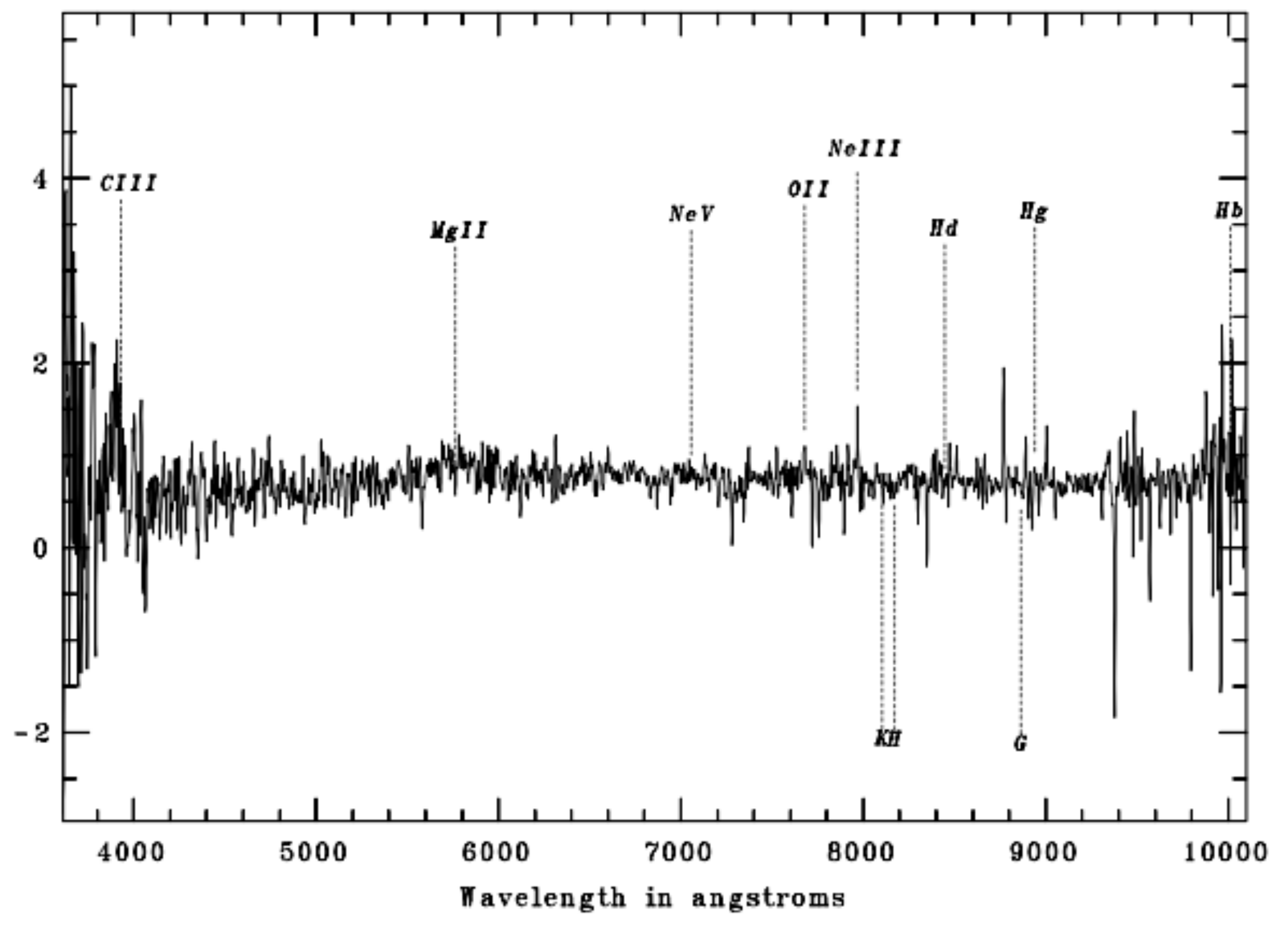}
\caption{J022538.9-040821~(2,2,1), z=0.733}
\label{}
\end{figure}
\begin{figure}
\includegraphics[height=0.87\columnwidth]{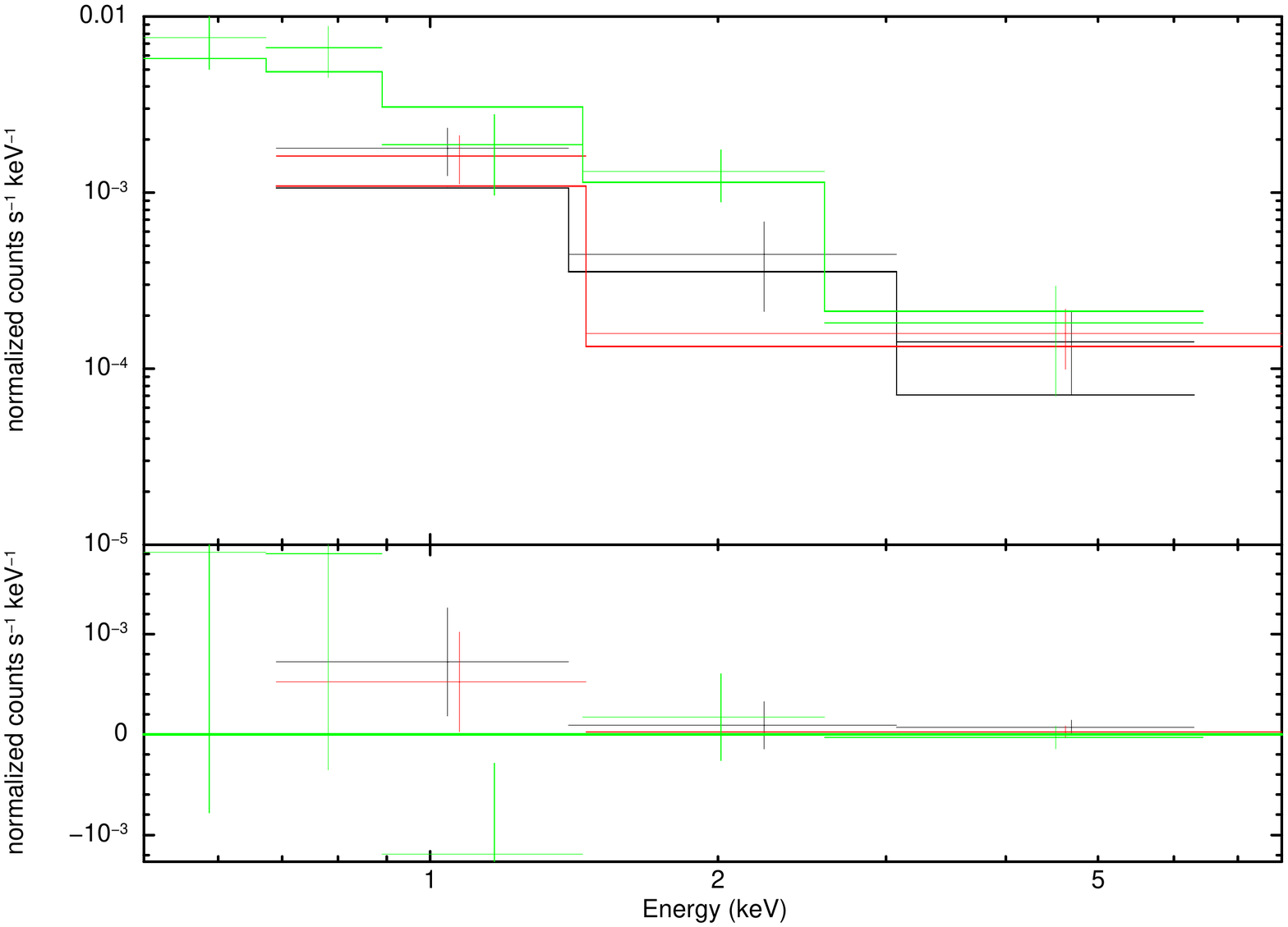}
\includegraphics[height=0.84\columnwidth]{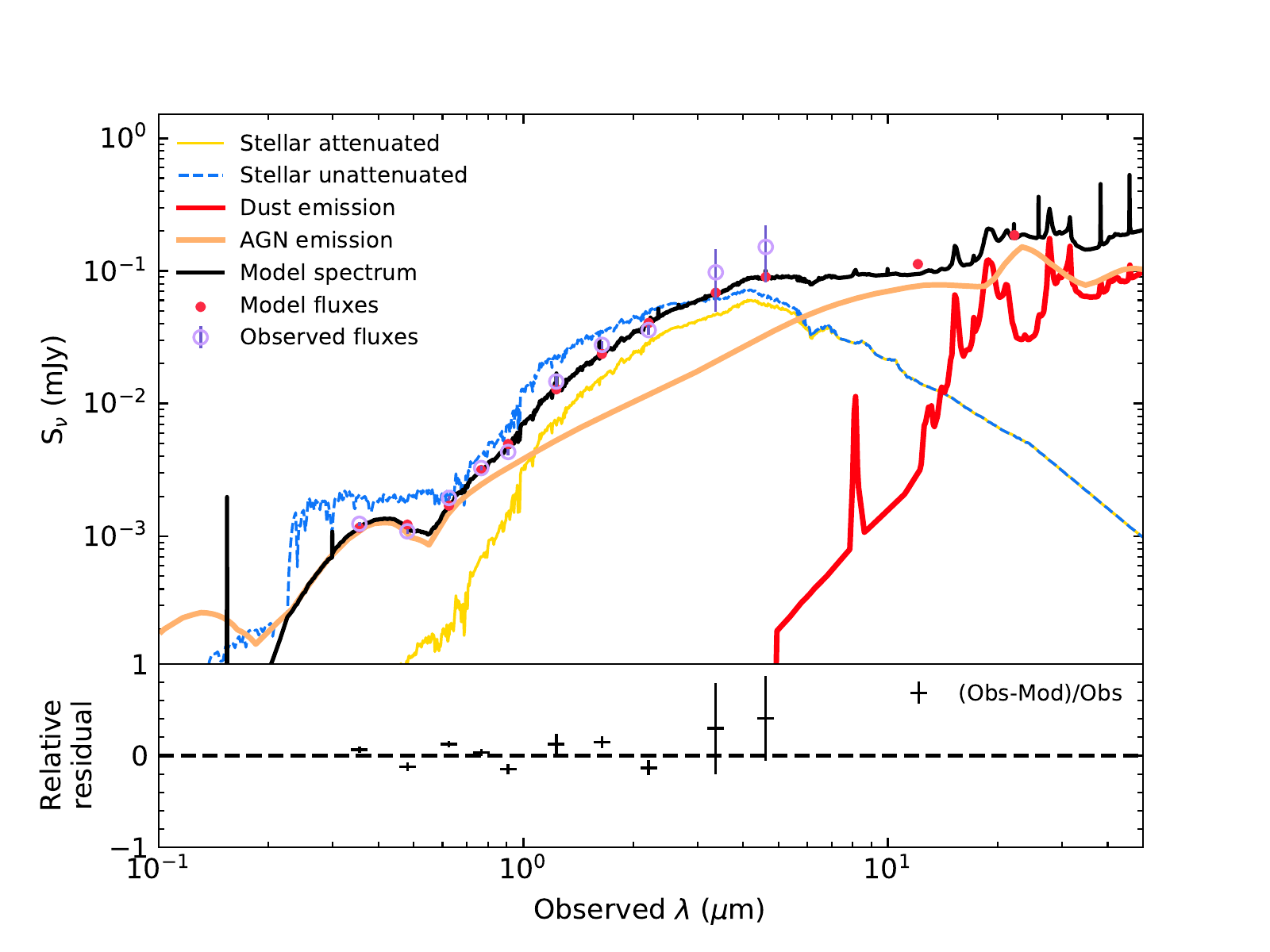}
\caption{J020845.1-051354~(2,2), z=1.46}
\label{}
\end{figure}
\begin{figure}
\includegraphics[height=0.87\columnwidth]{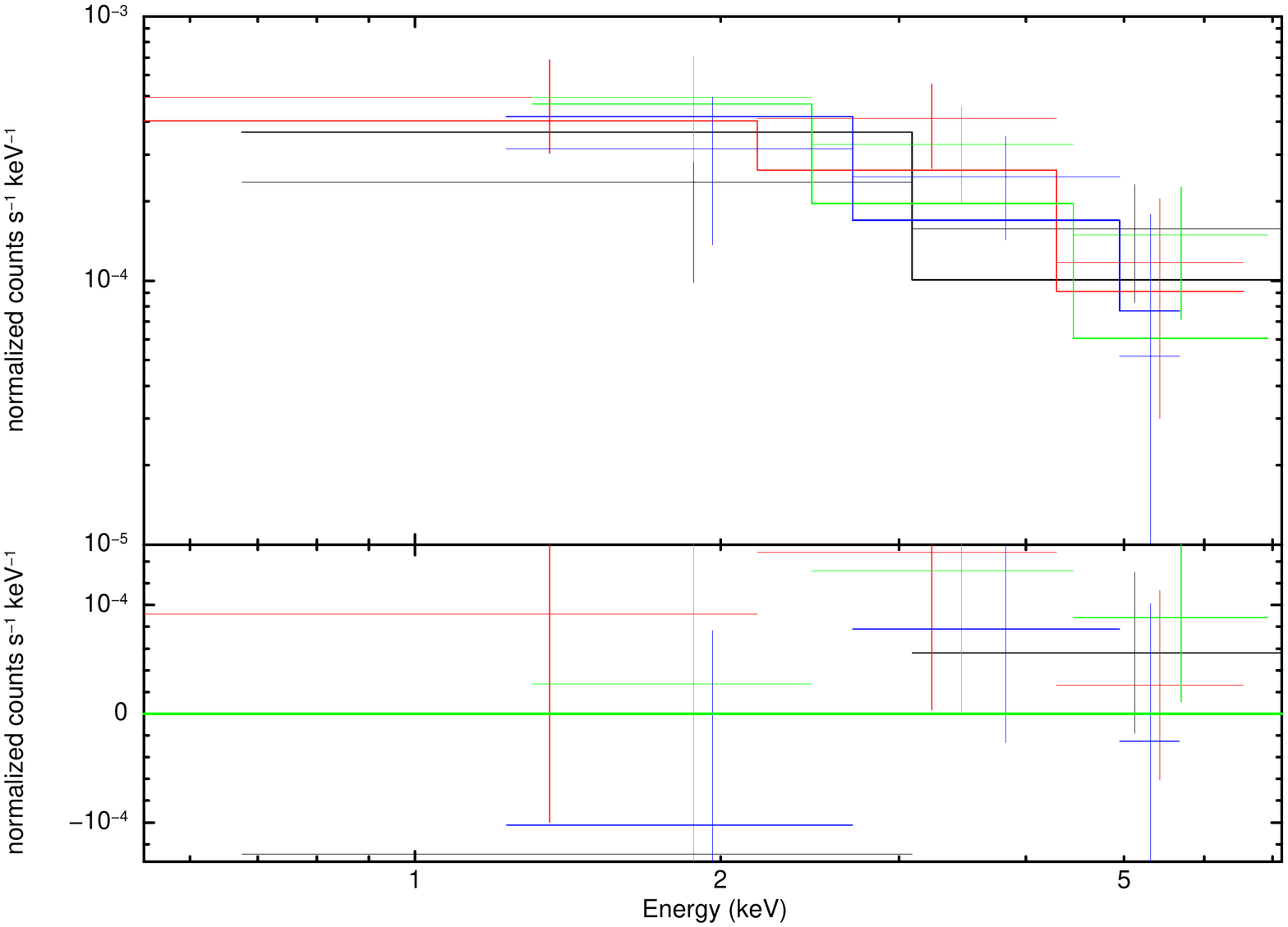}
\includegraphics[height=0.86\columnwidth]{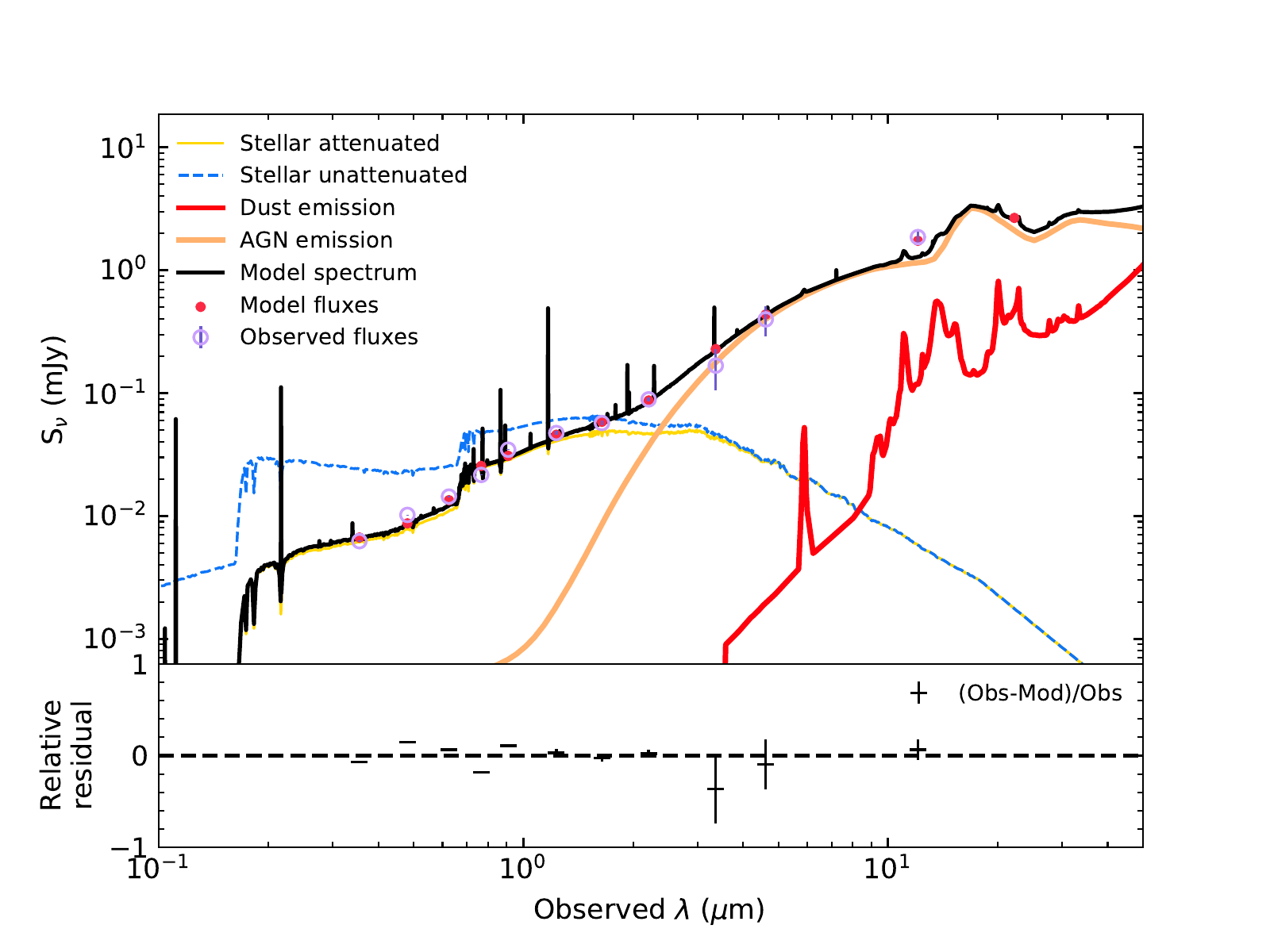}
\includegraphics[height=0.74\columnwidth]{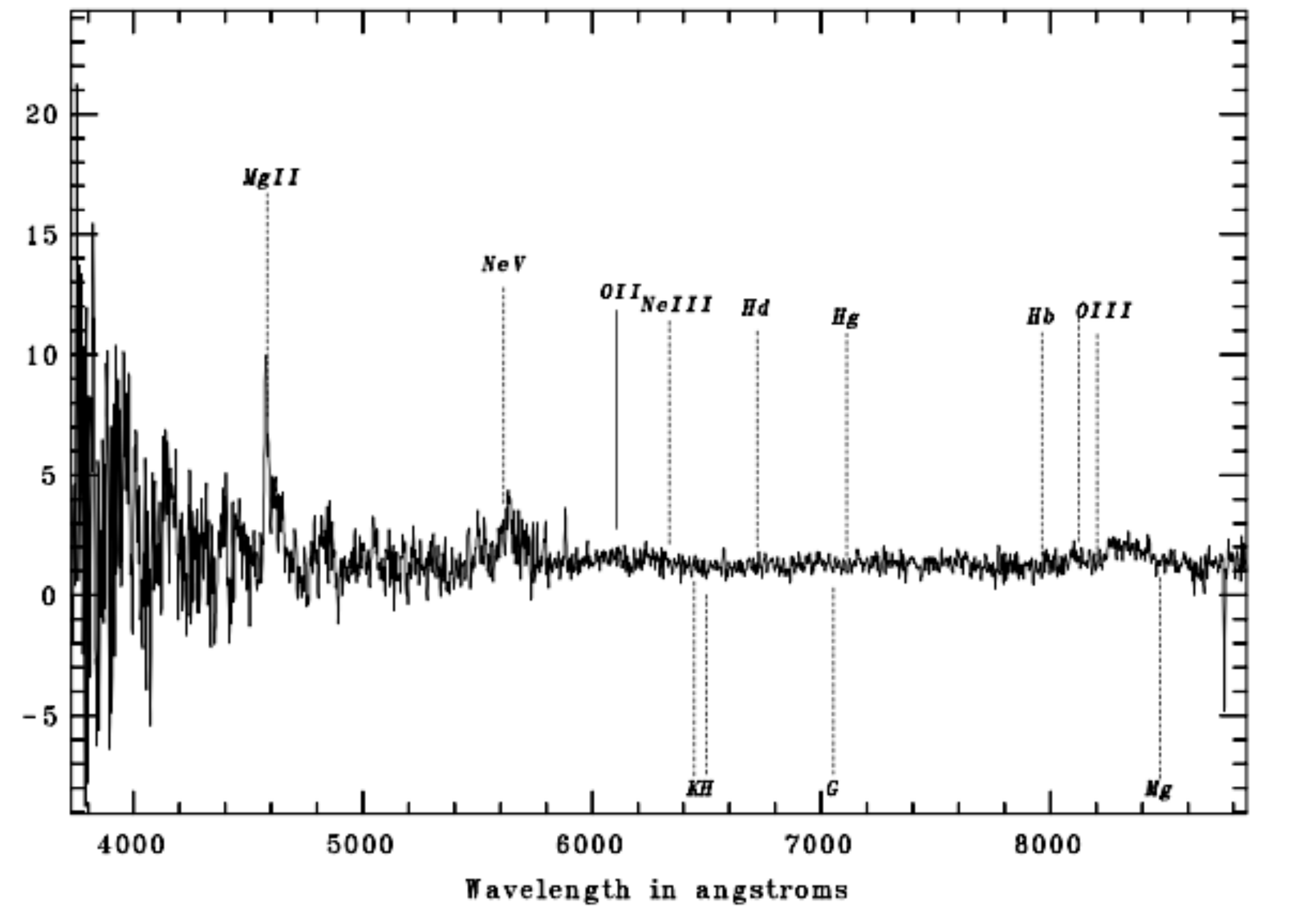}
\caption{J020953.9-055102~(2,2,1), z=0.642}
\label{}
\end{figure}
\begin{figure}
\includegraphics[height=0.87\columnwidth]{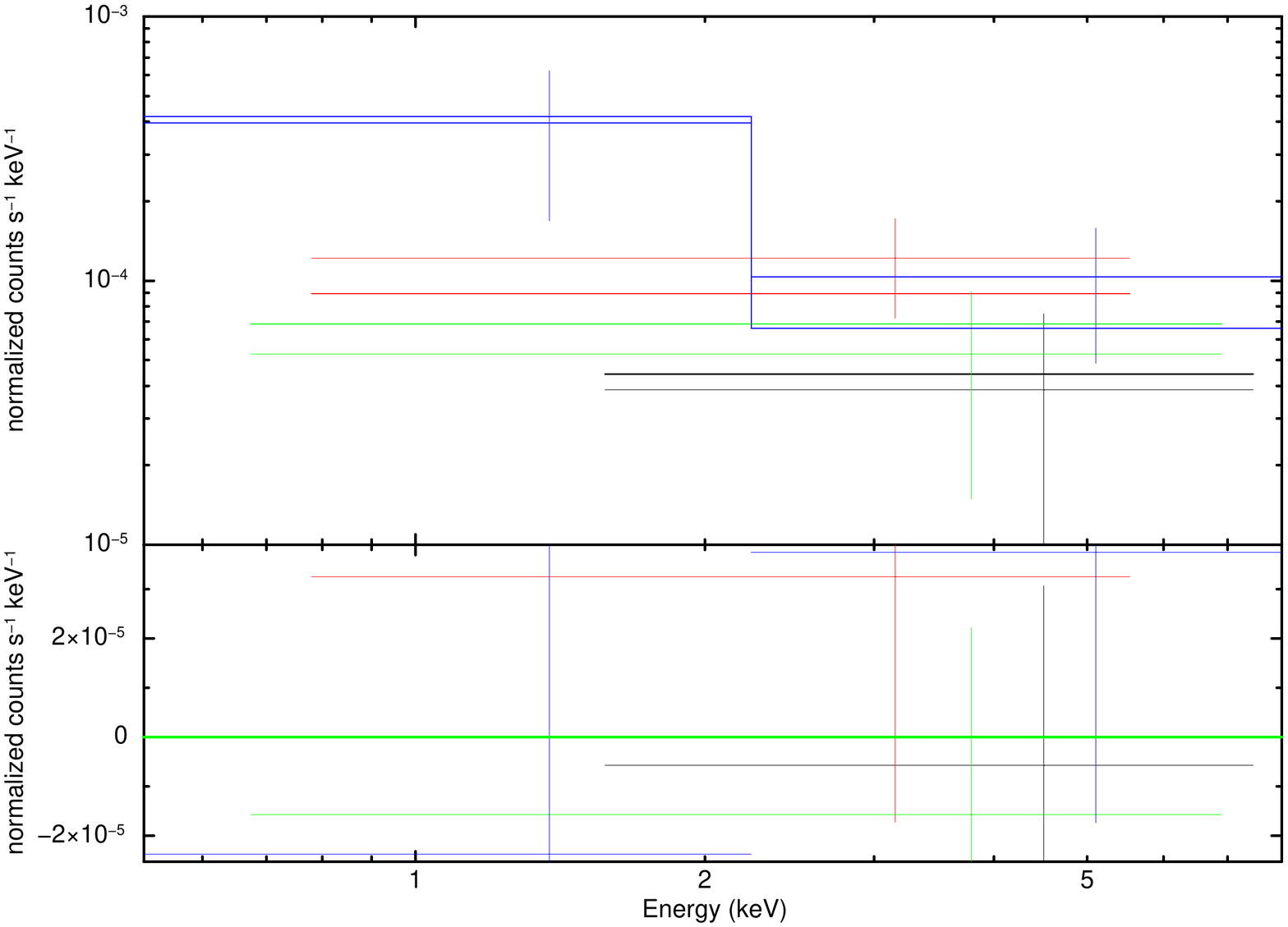}
\includegraphics[height=0.86\columnwidth]{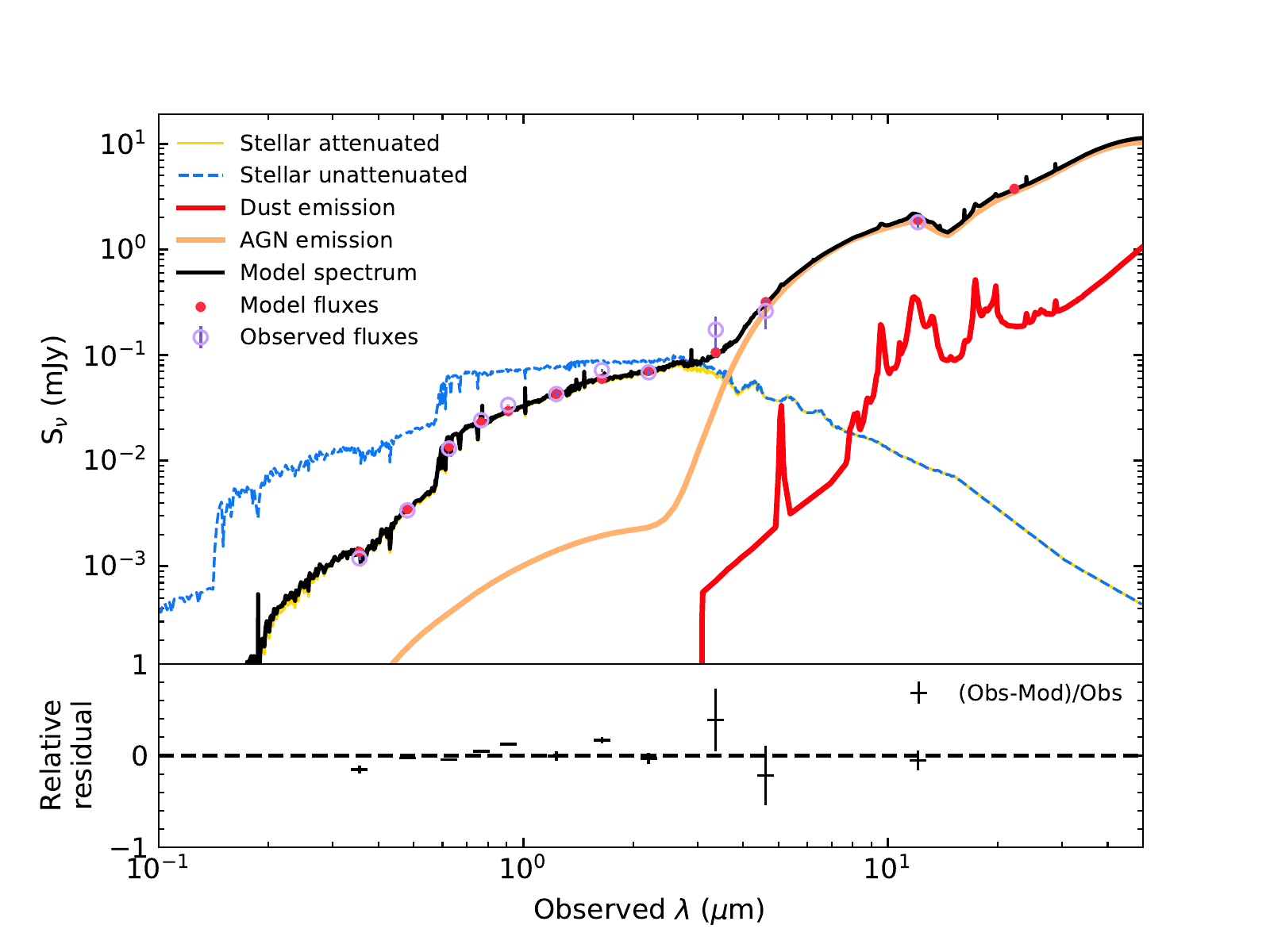}
\includegraphics[height=0.74\columnwidth]{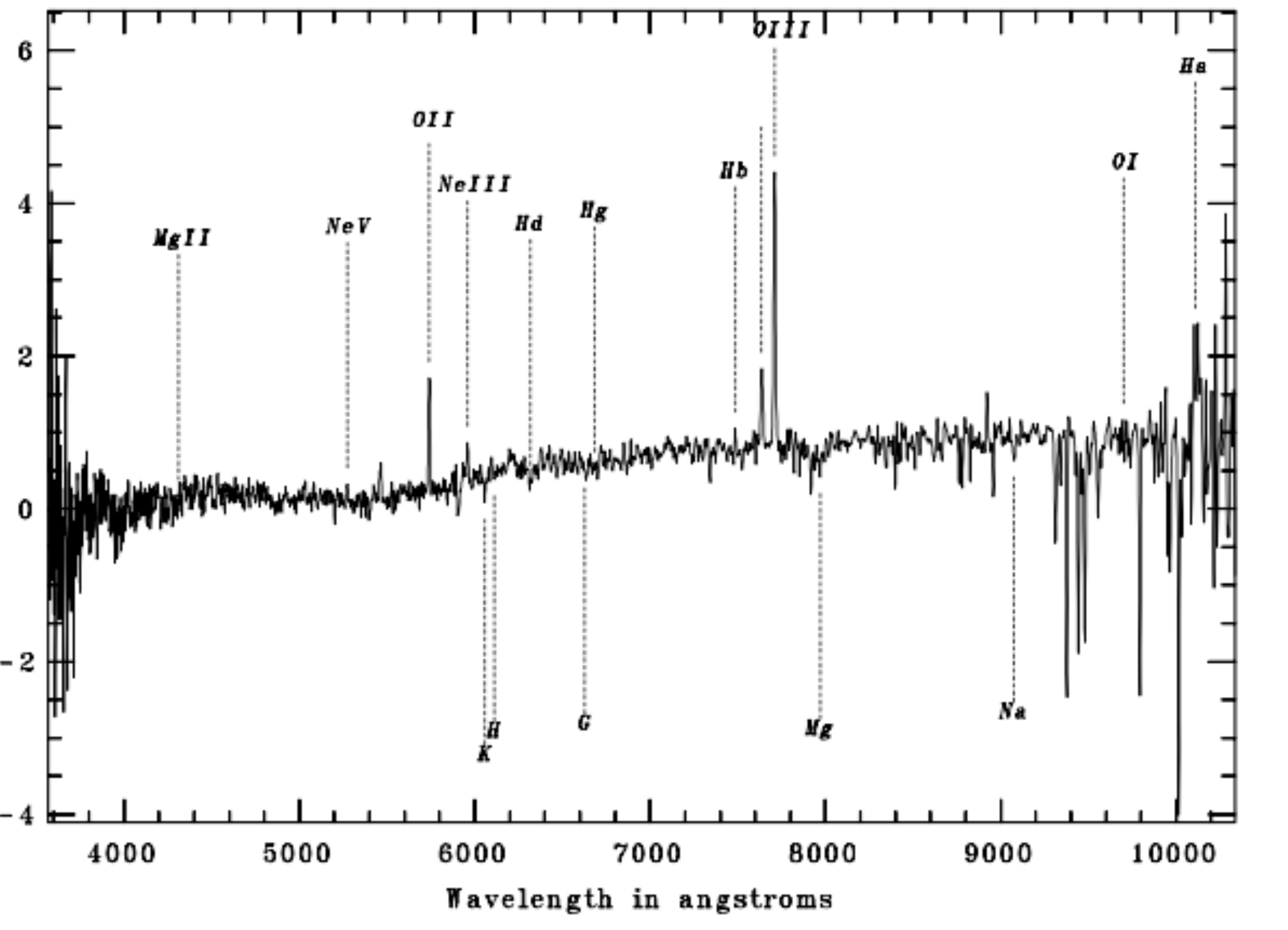}
\caption{J020806.6-055739~(2,2,2), z=0.539}
\label{}
\end{figure}
\begin{figure}
\includegraphics[height=0.85\columnwidth]{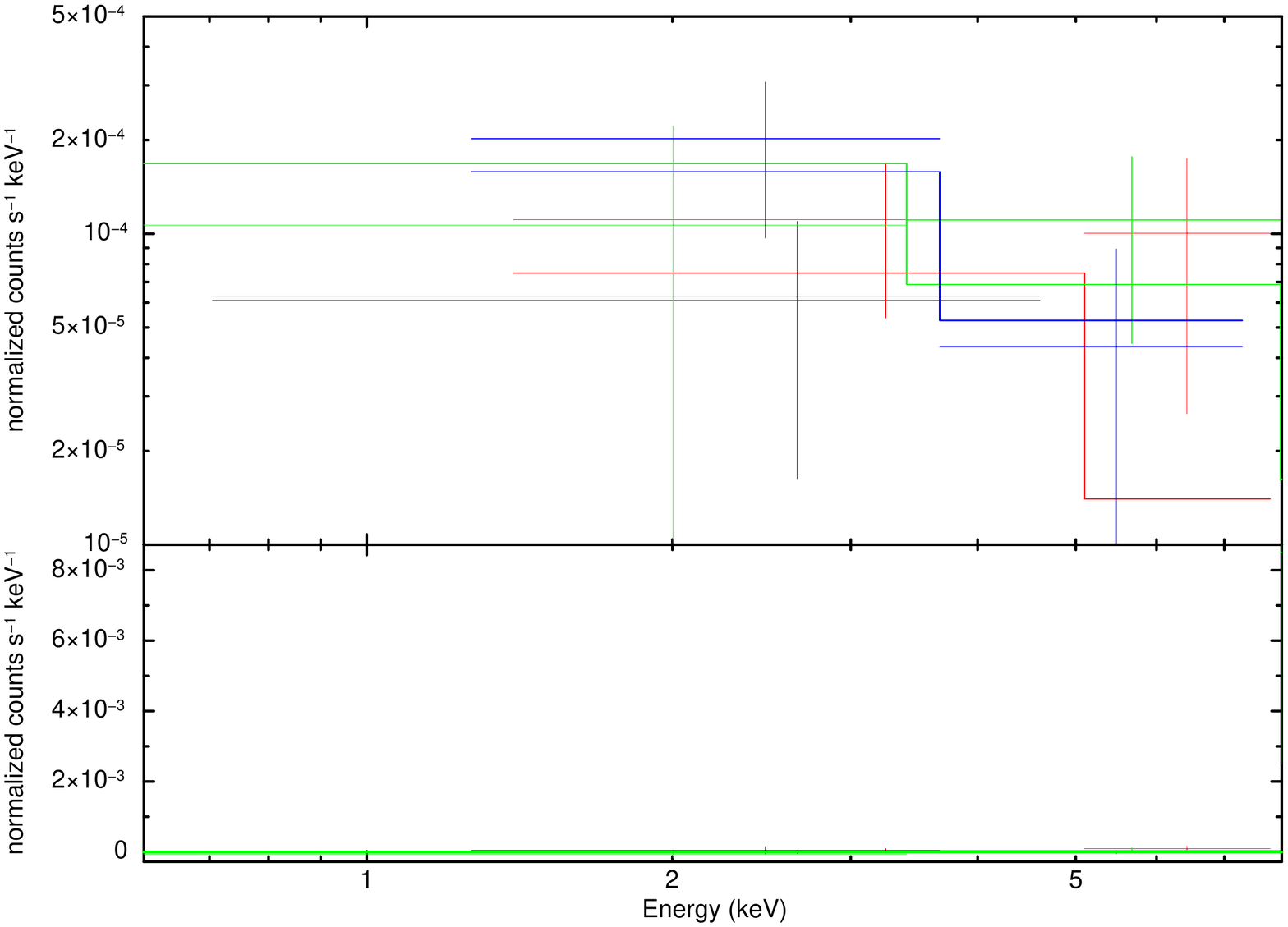}
\includegraphics[height=0.84\columnwidth]{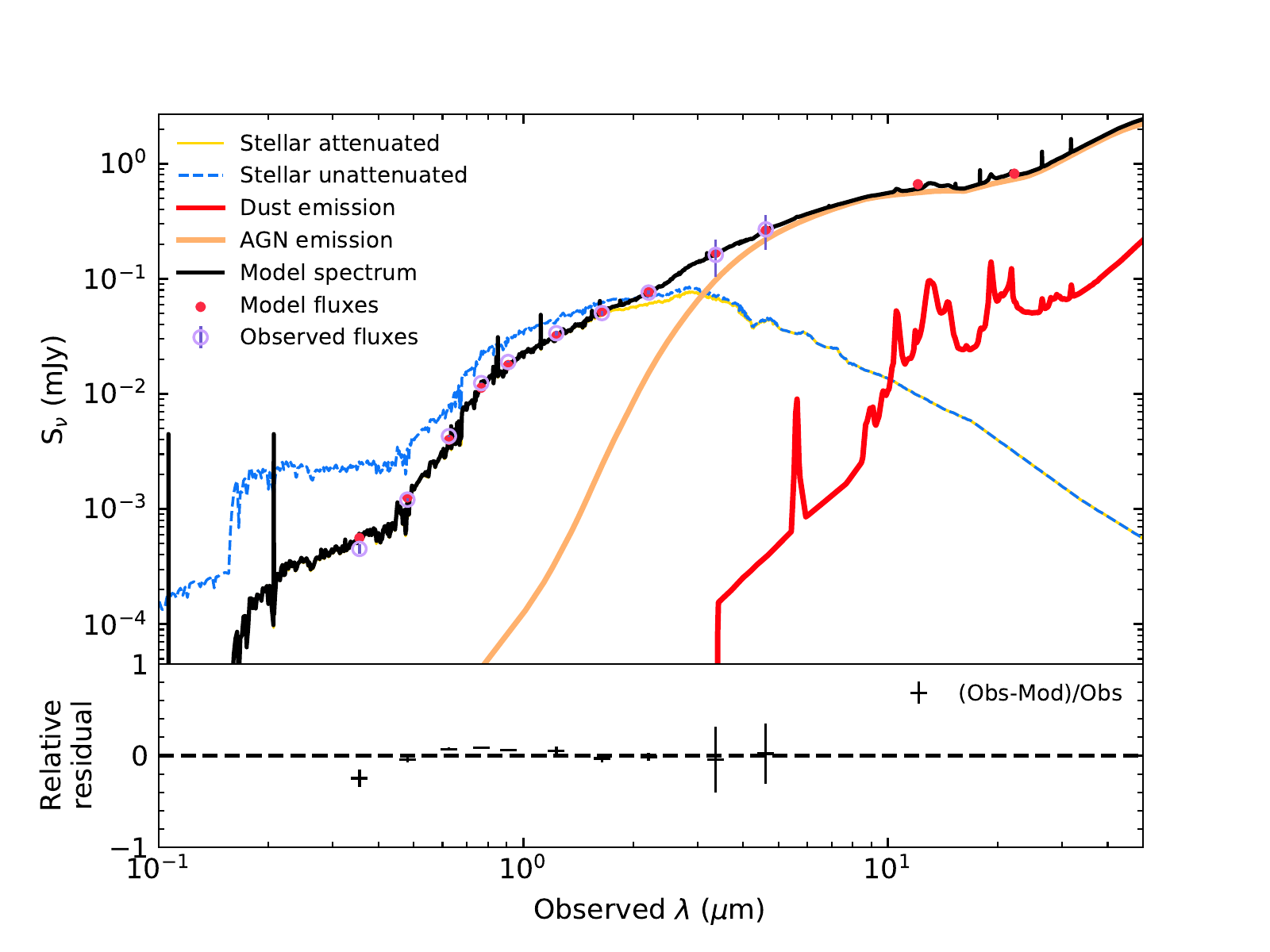}
\caption{J021509.0-054305~(2,2), z=0.70}
\label{}
\end{figure}
\begin{figure}
\includegraphics[height=0.85\columnwidth]{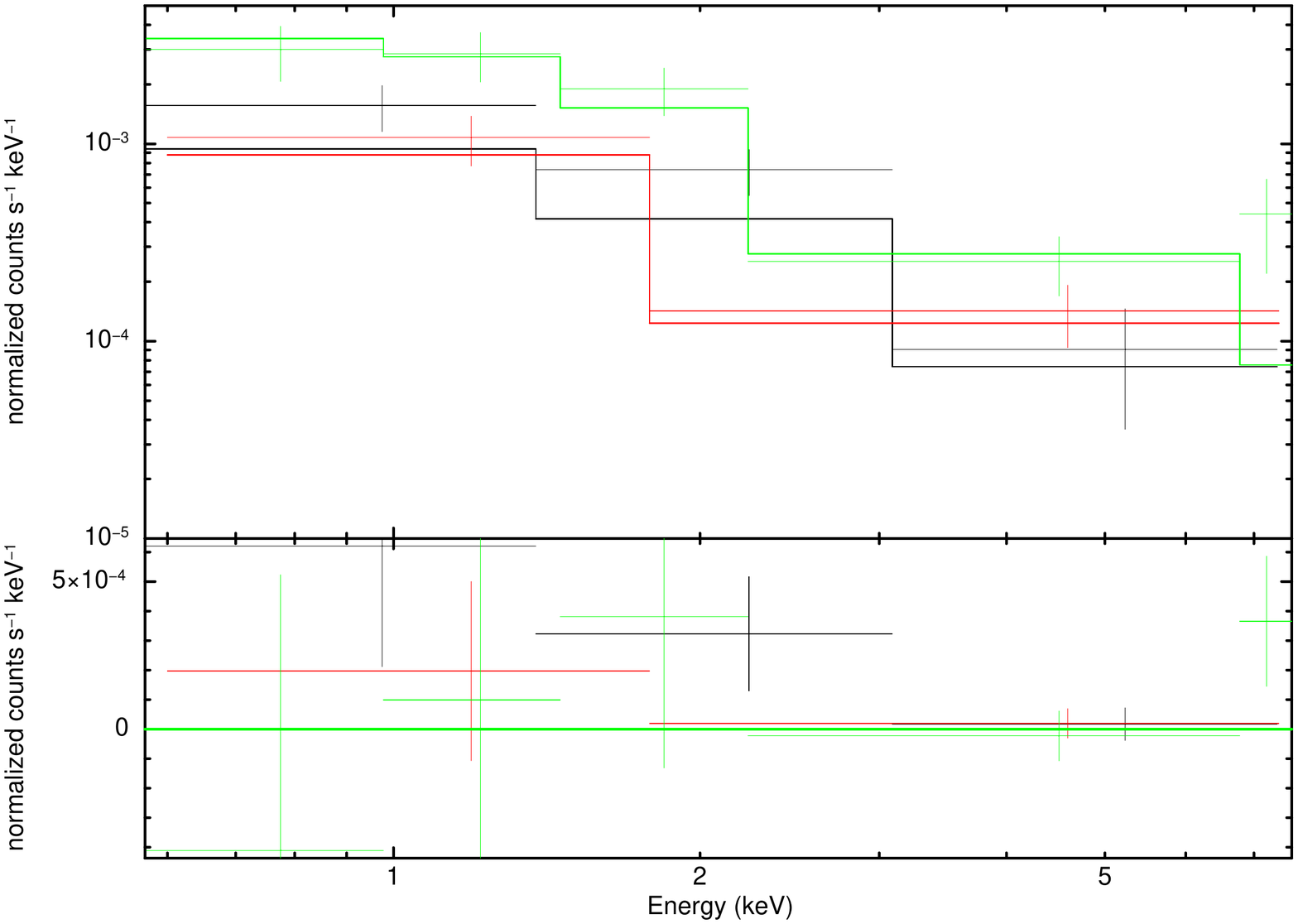}
\includegraphics[height=0.84\columnwidth]{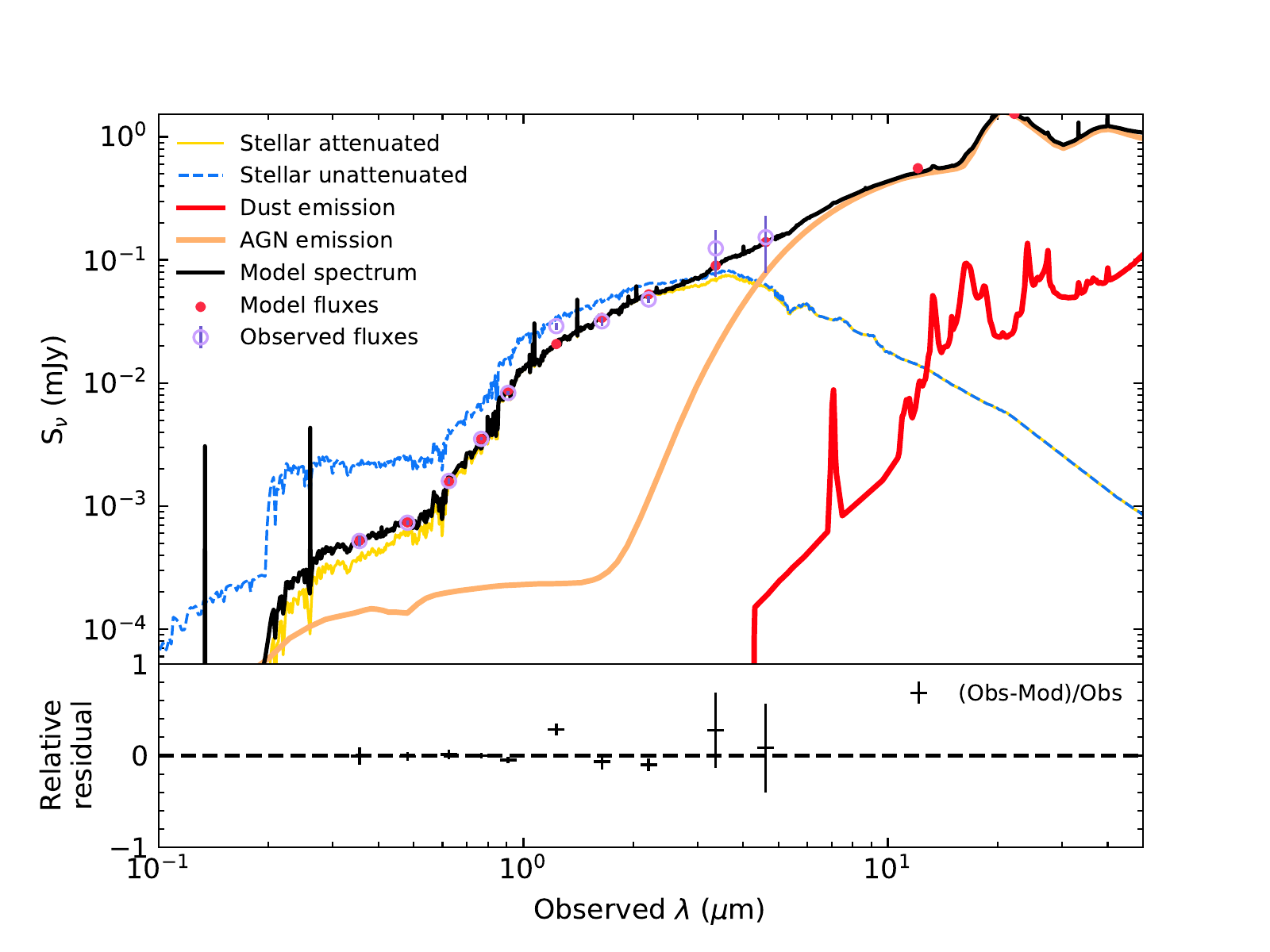}
\caption{J021512.9-060558~(2,2), z=1.14}
\label{}
\end{figure}
\begin{figure}
\includegraphics[height=0.85\columnwidth]{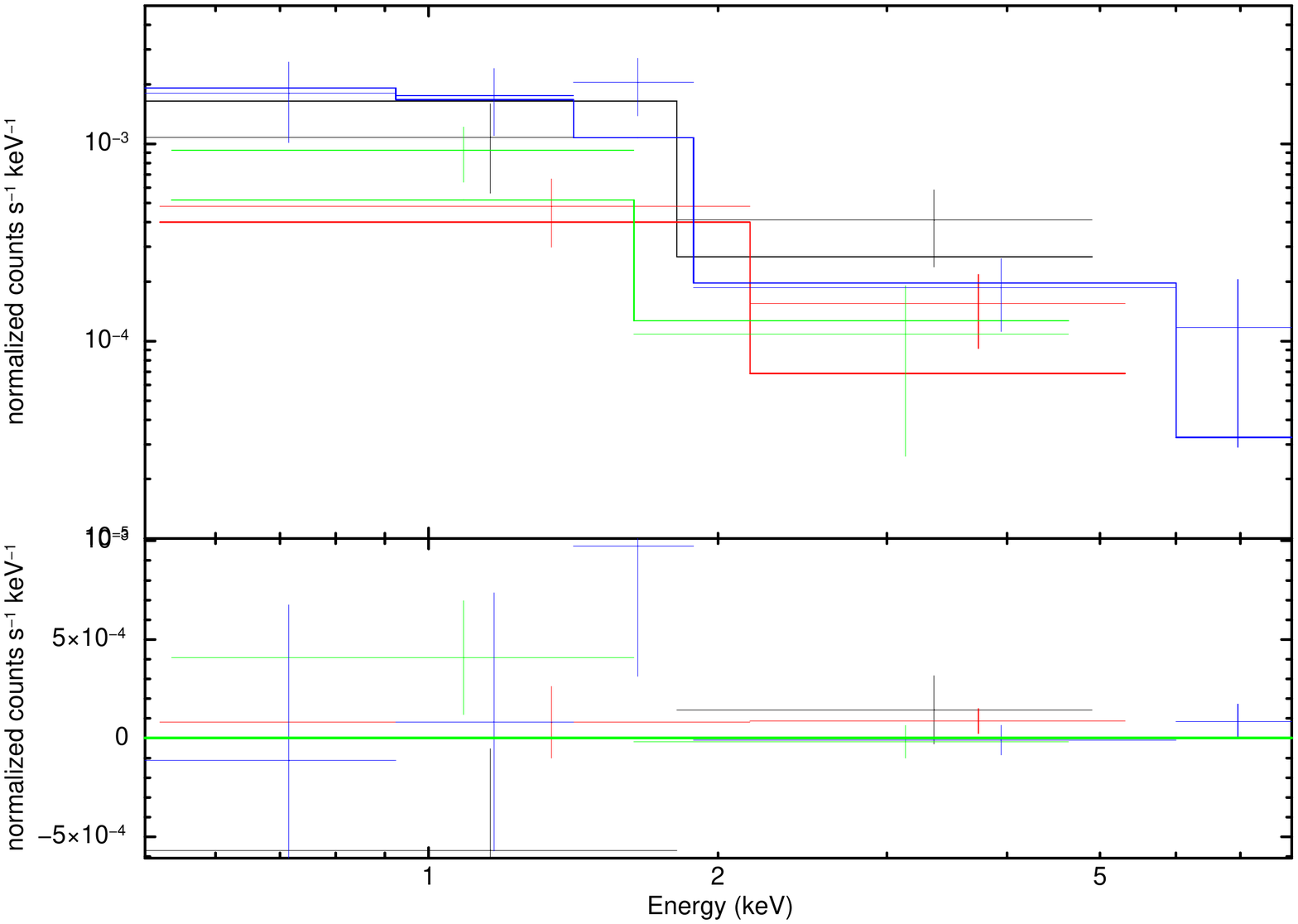}
\includegraphics[height=0.84\columnwidth]{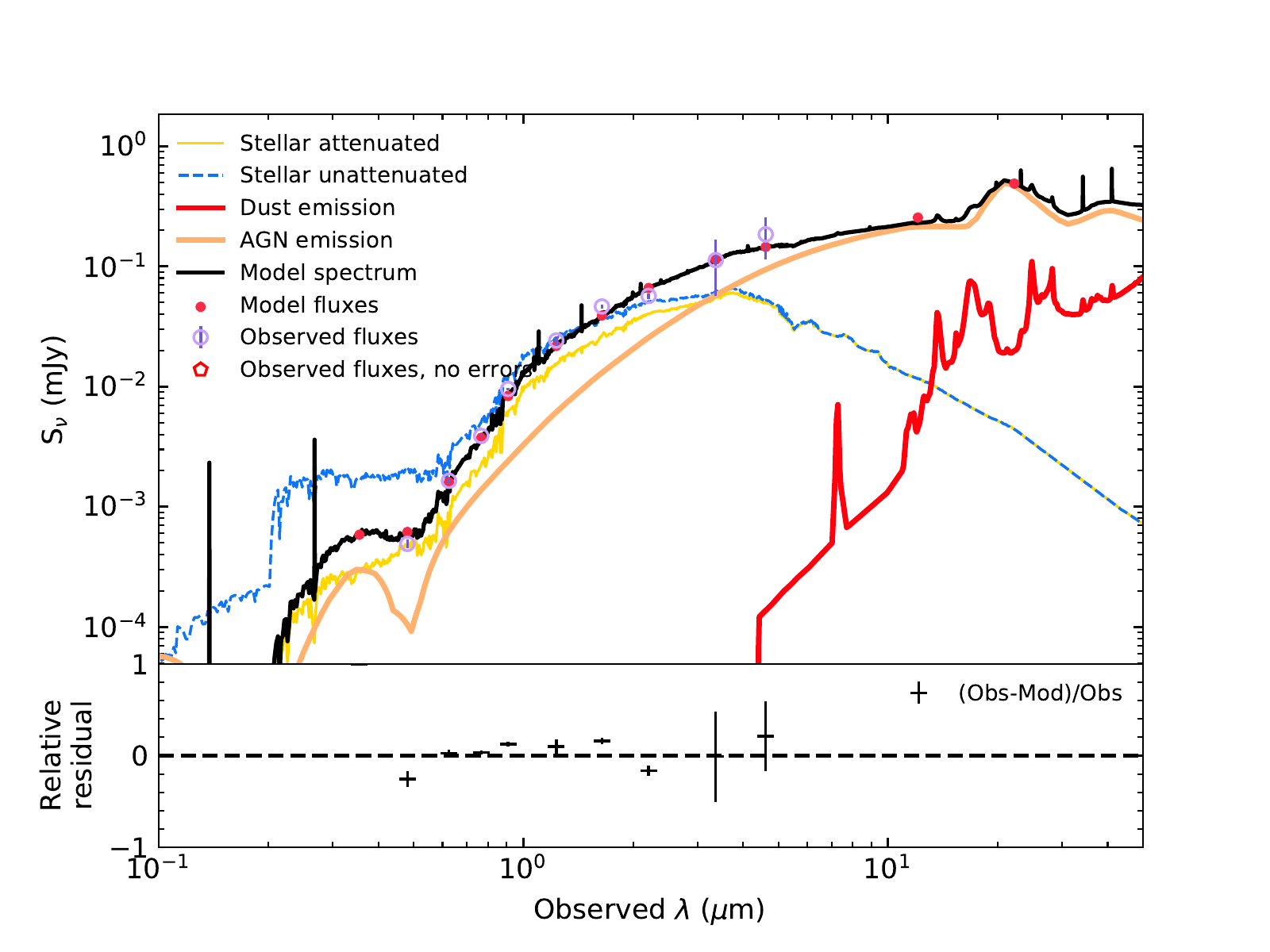}
\caption{J020529.5-051100~(2,2), z=1.20}
\label{}
\end{figure}
\begin{figure}
\includegraphics[height=0.85\columnwidth]{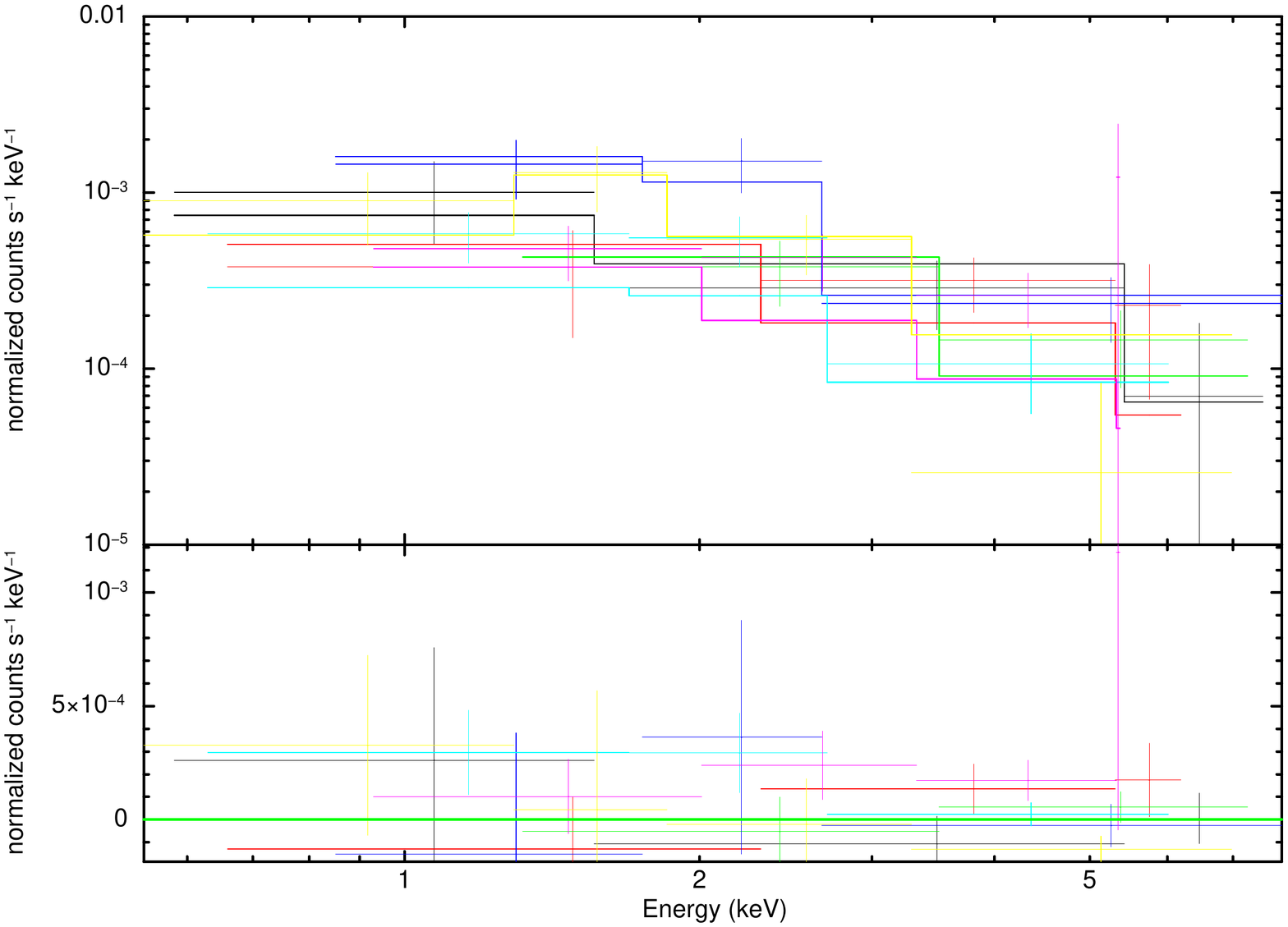}
\includegraphics[height=0.84\columnwidth]{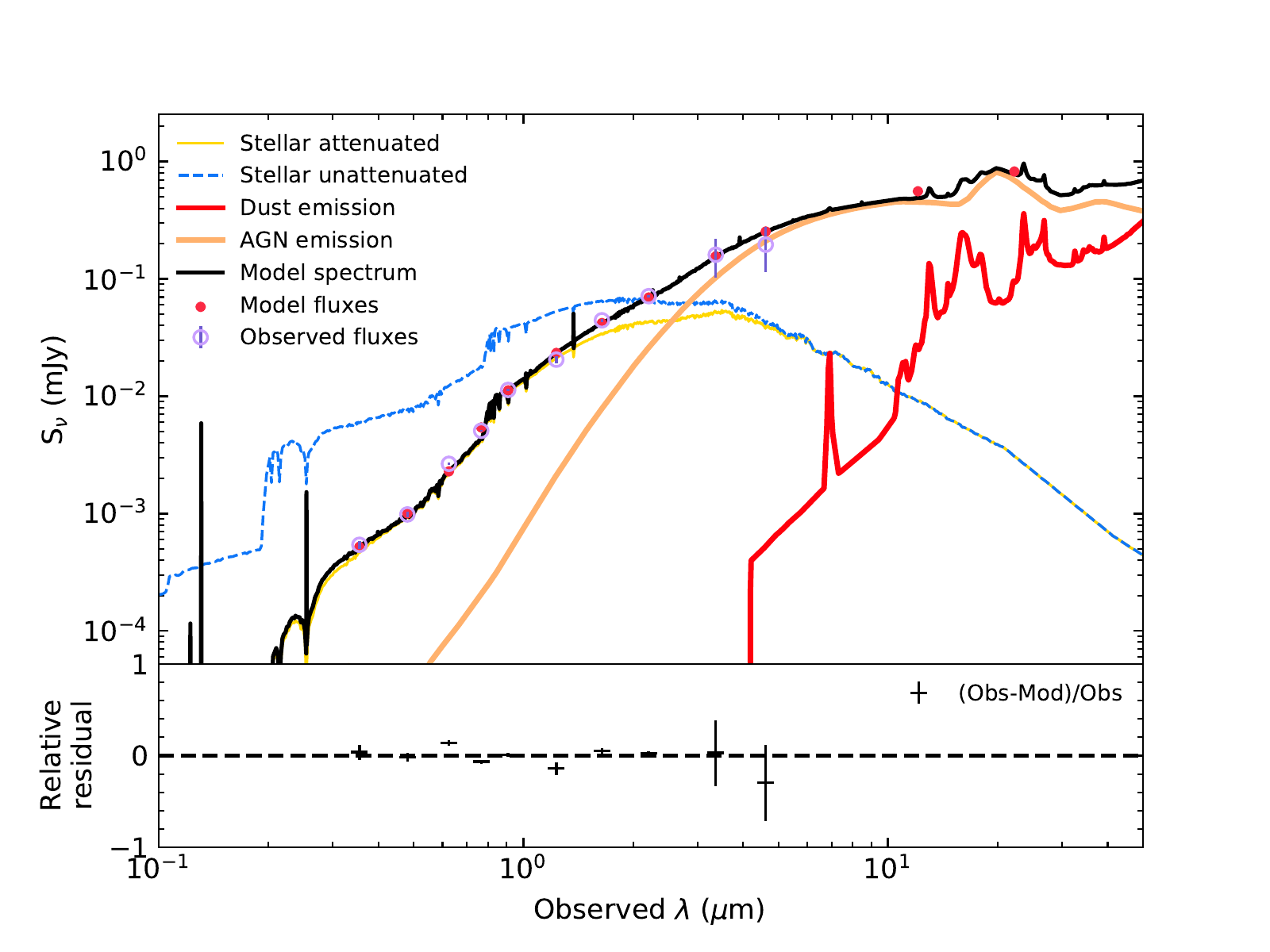}
\caption{J022404.0-035730~(2,2), z=0.17}
\label{}
\end{figure}
\begin{figure}
\includegraphics[height=0.85\columnwidth]{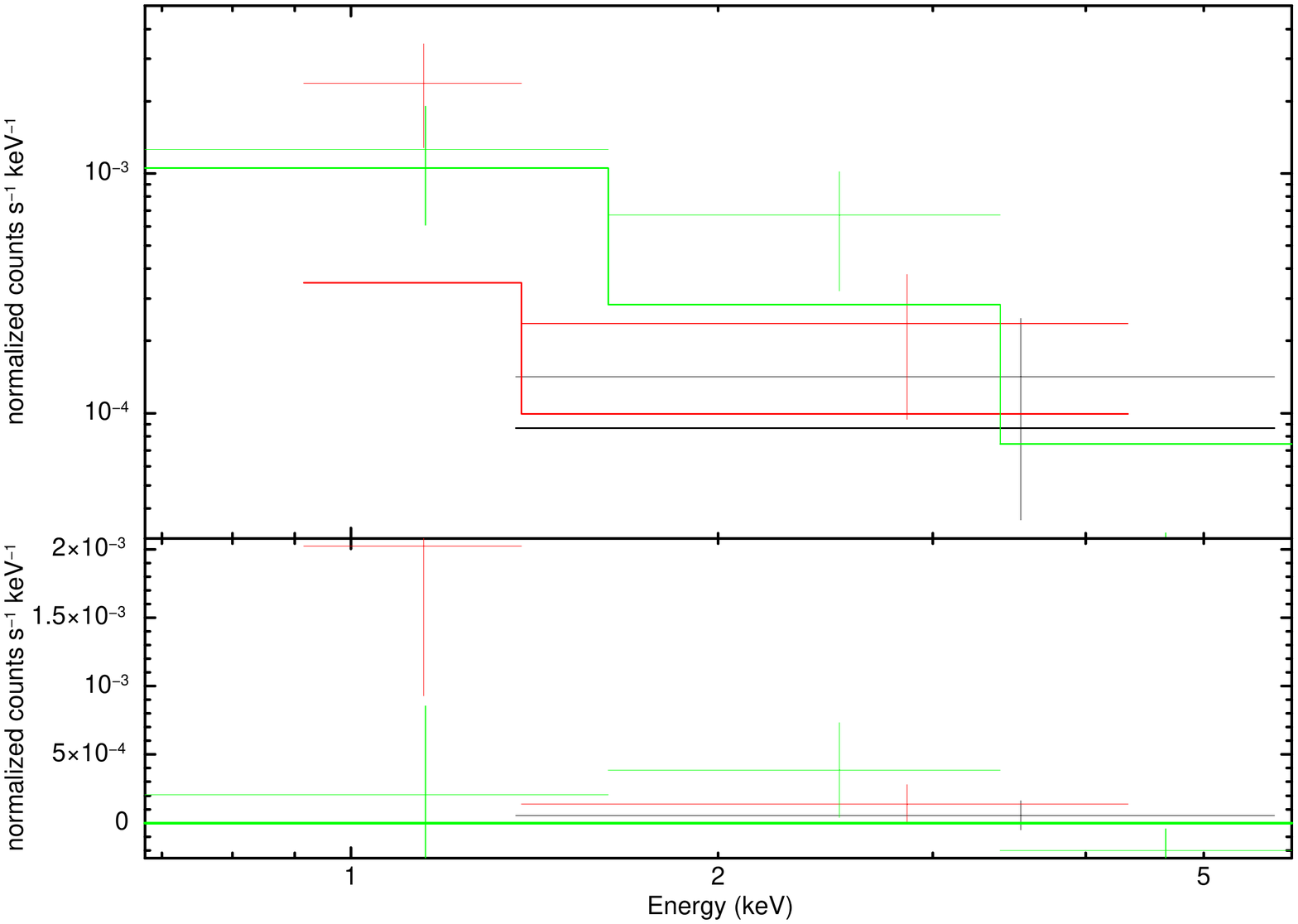}
\includegraphics[height=0.84\columnwidth]{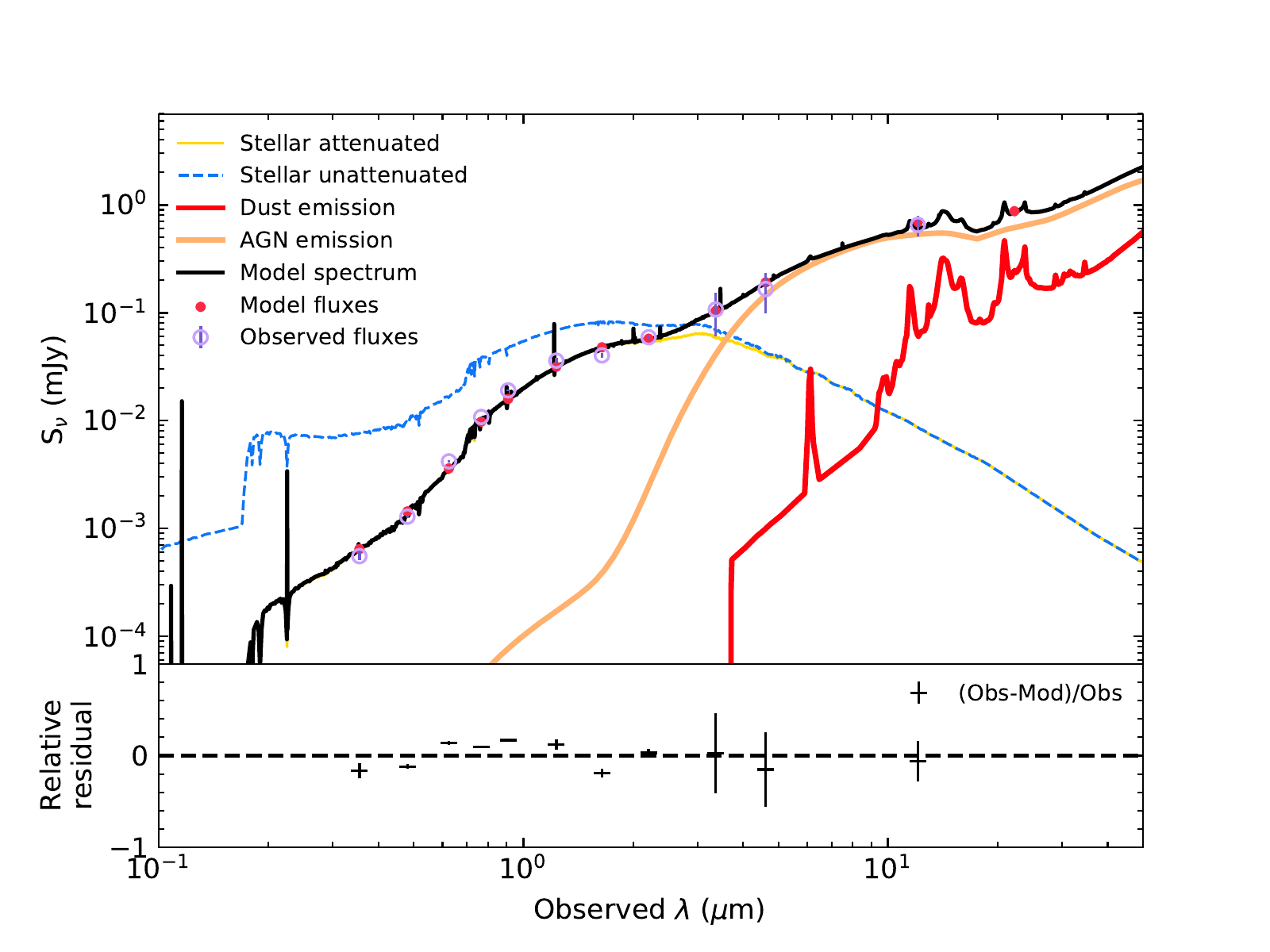}
\caption{J023501.0-055234~(2,2), z=0.85}
\label{}
\end{figure}
\begin{figure}
\includegraphics[height=0.85\columnwidth]{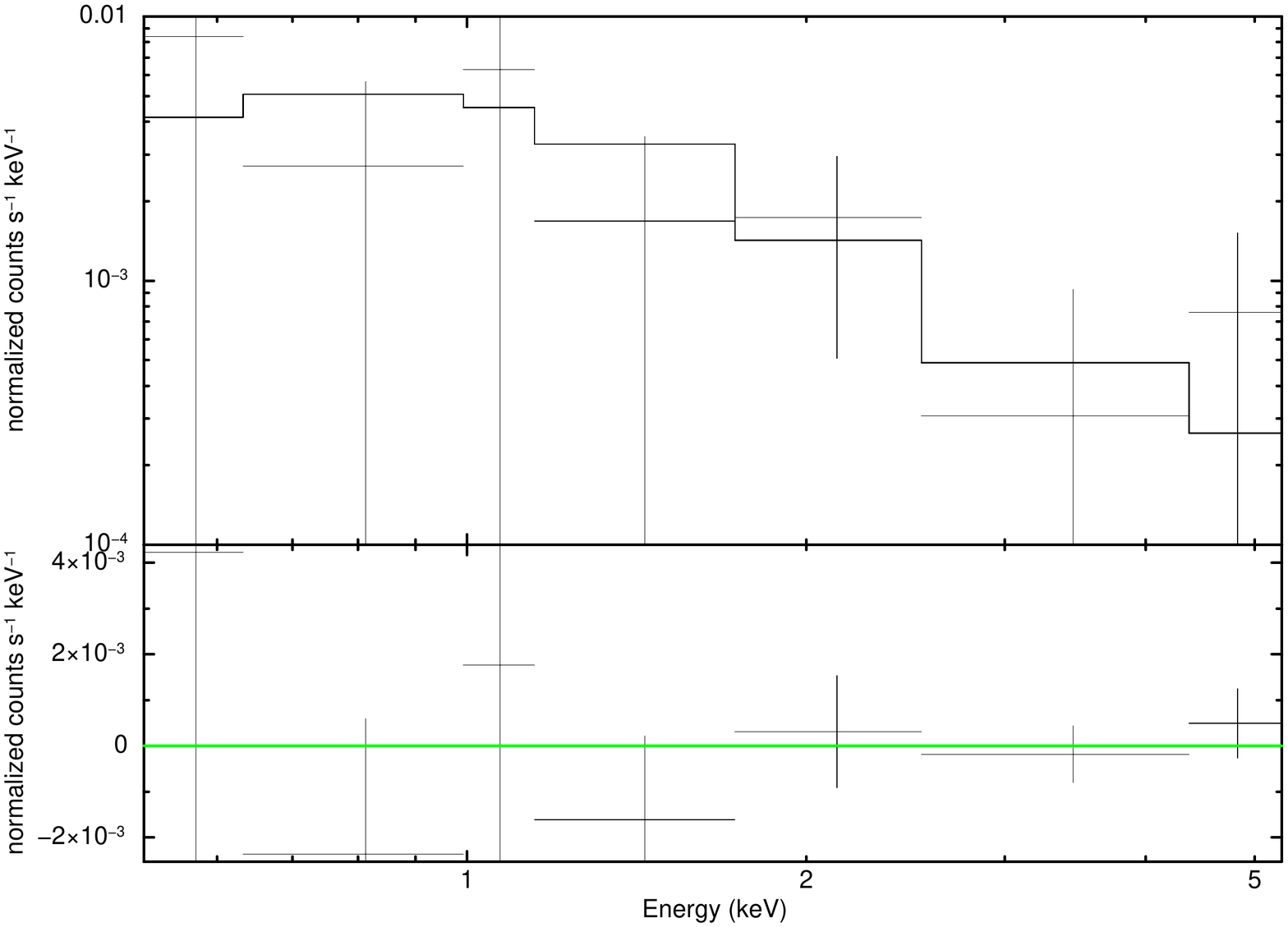}
\includegraphics[height=0.84\columnwidth]{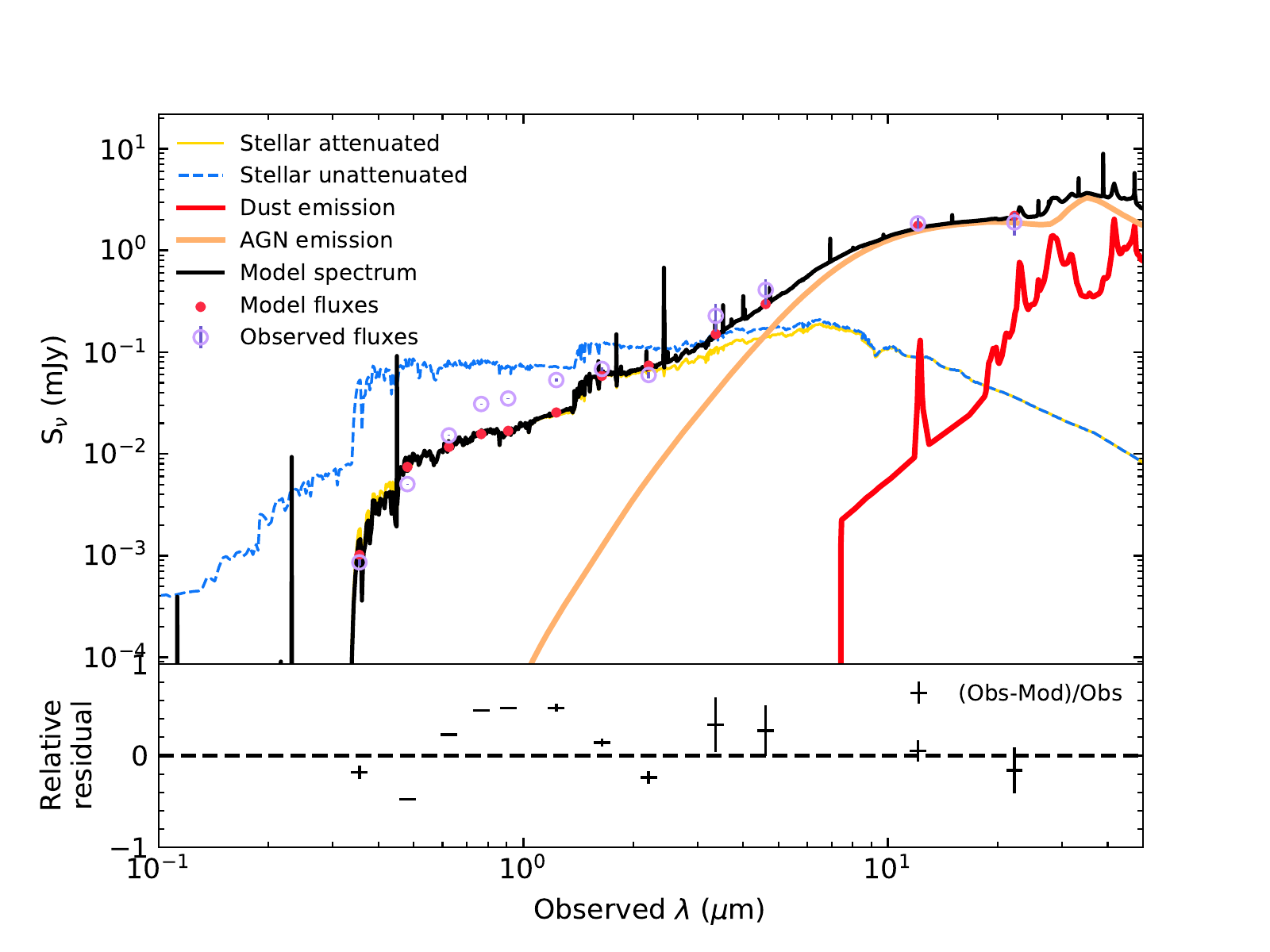}
\caption{J020210.6-041129~(2,2), z=2.71}
\label{}
\end{figure}
\clearpage
\begin{figure}
\includegraphics[height=0.85\columnwidth]{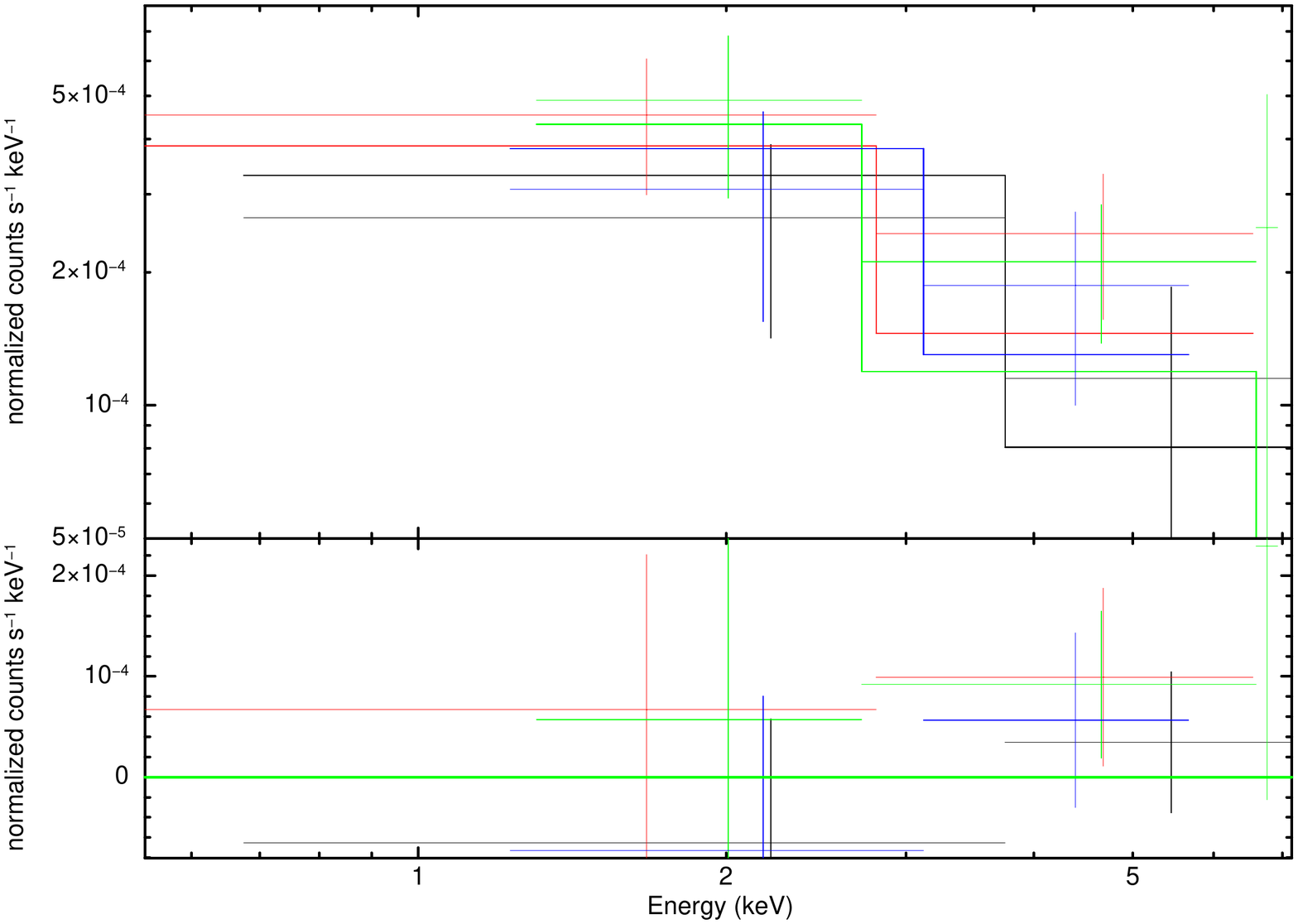}
\includegraphics[height=0.84\columnwidth]{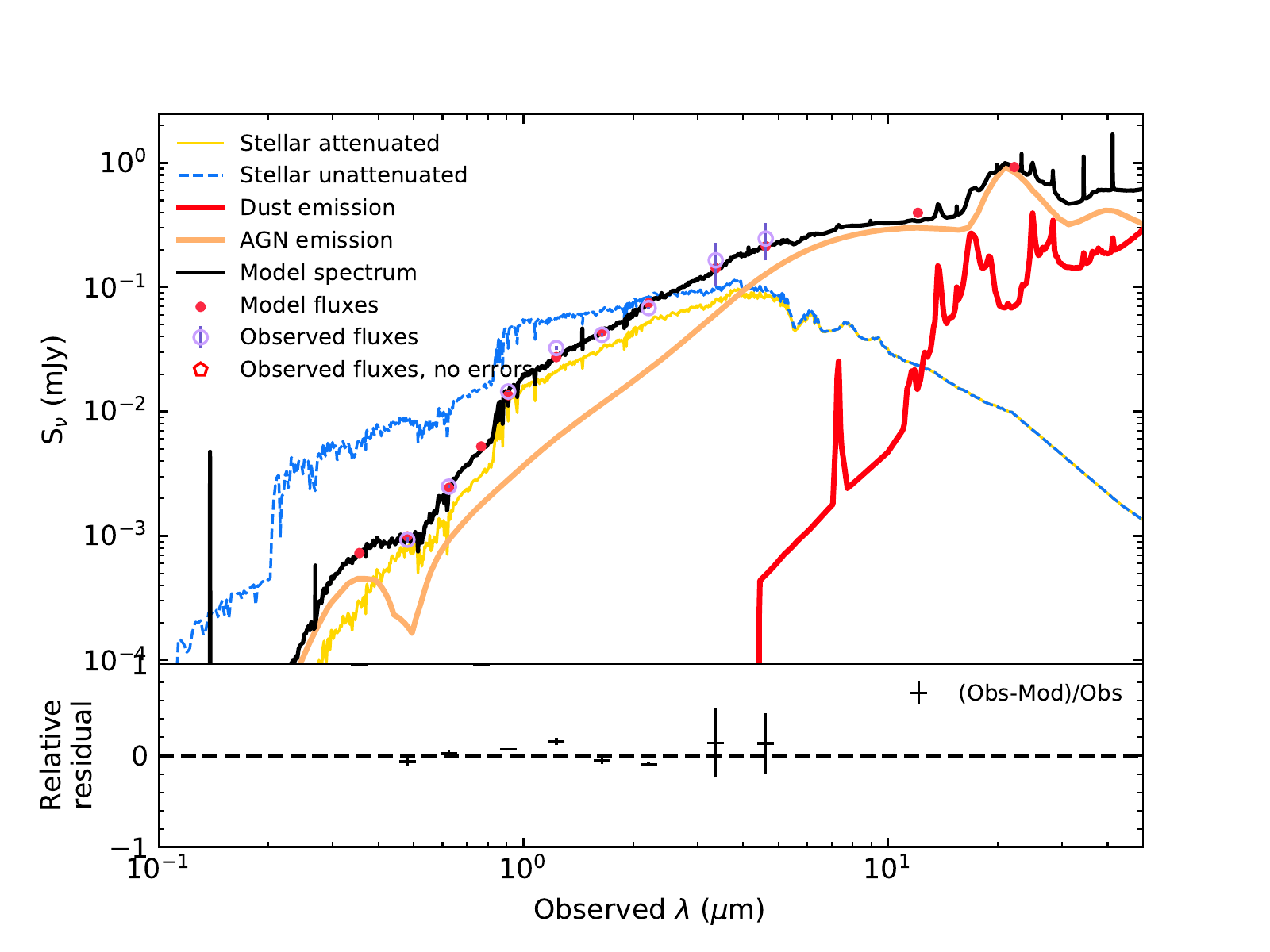}
\caption{J022650.3-025752~(2,2), z=1.21}
\label{}
\end{figure}
\begin{figure}
\includegraphics[height=0.85\columnwidth]{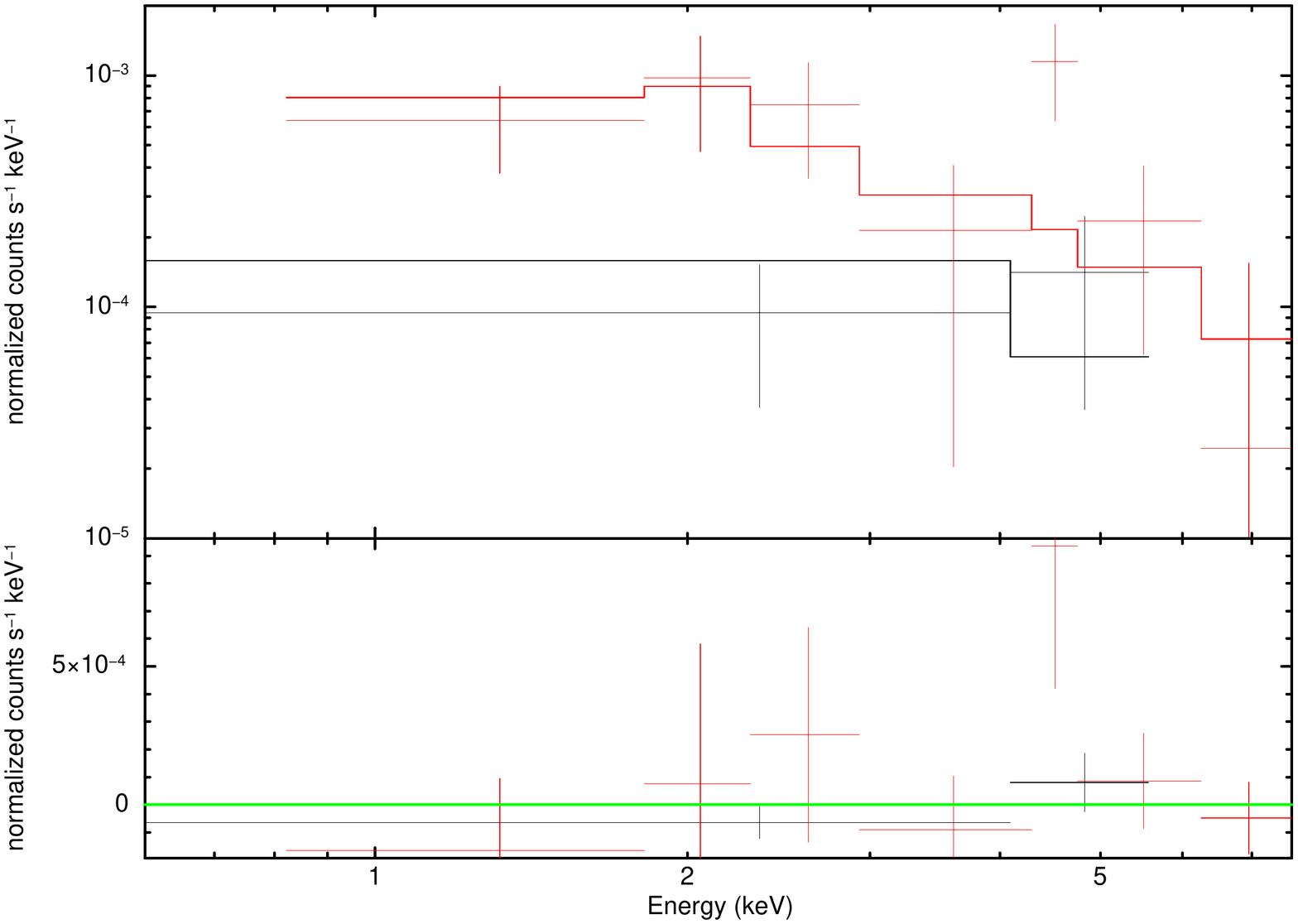}
\includegraphics[height=0.84\columnwidth]{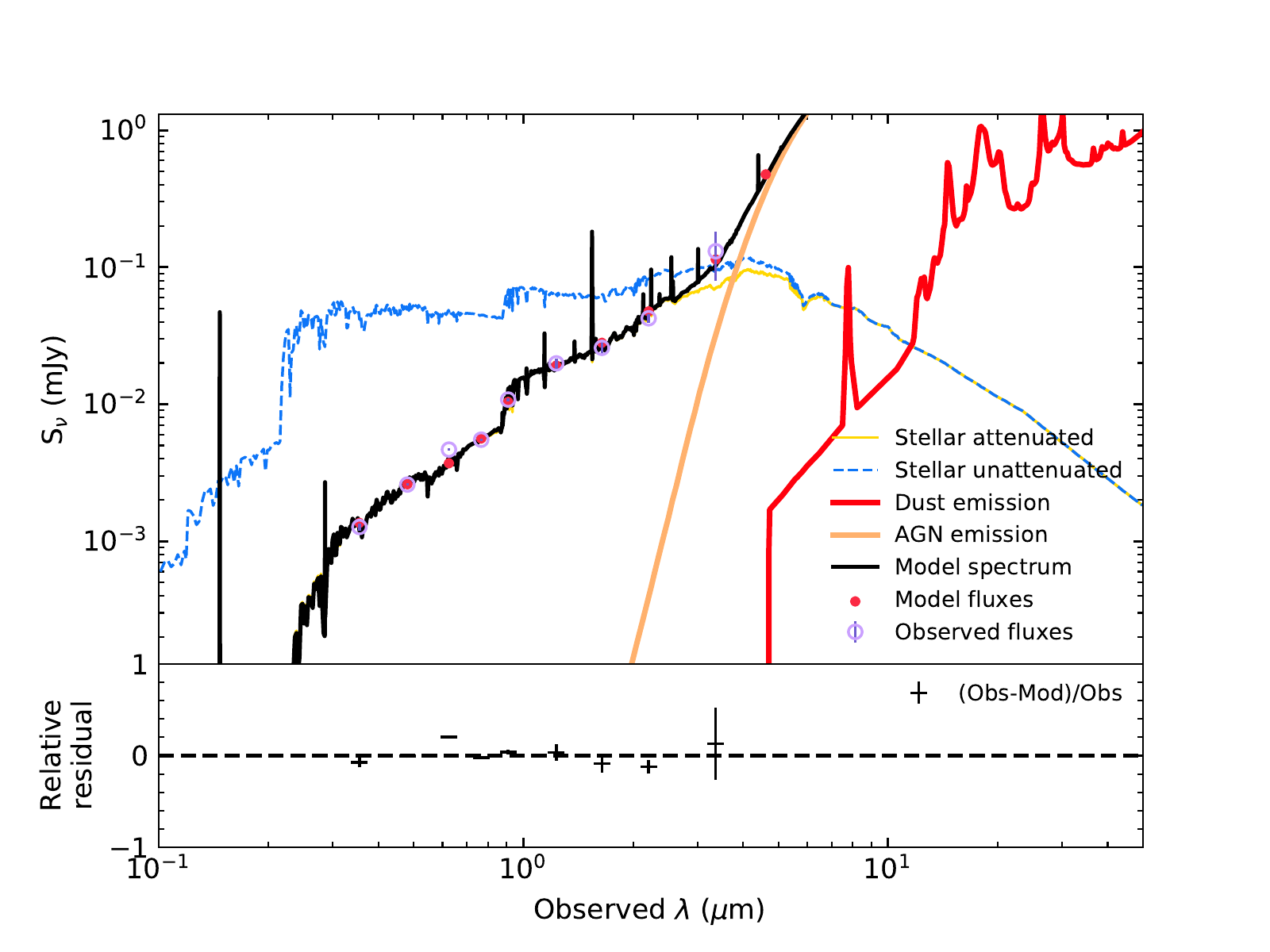}
\caption{J021303.7-040704~(2,2), z=1.35}
\label{}
\end{figure}
\begin{figure}
\includegraphics[height=0.85\columnwidth]{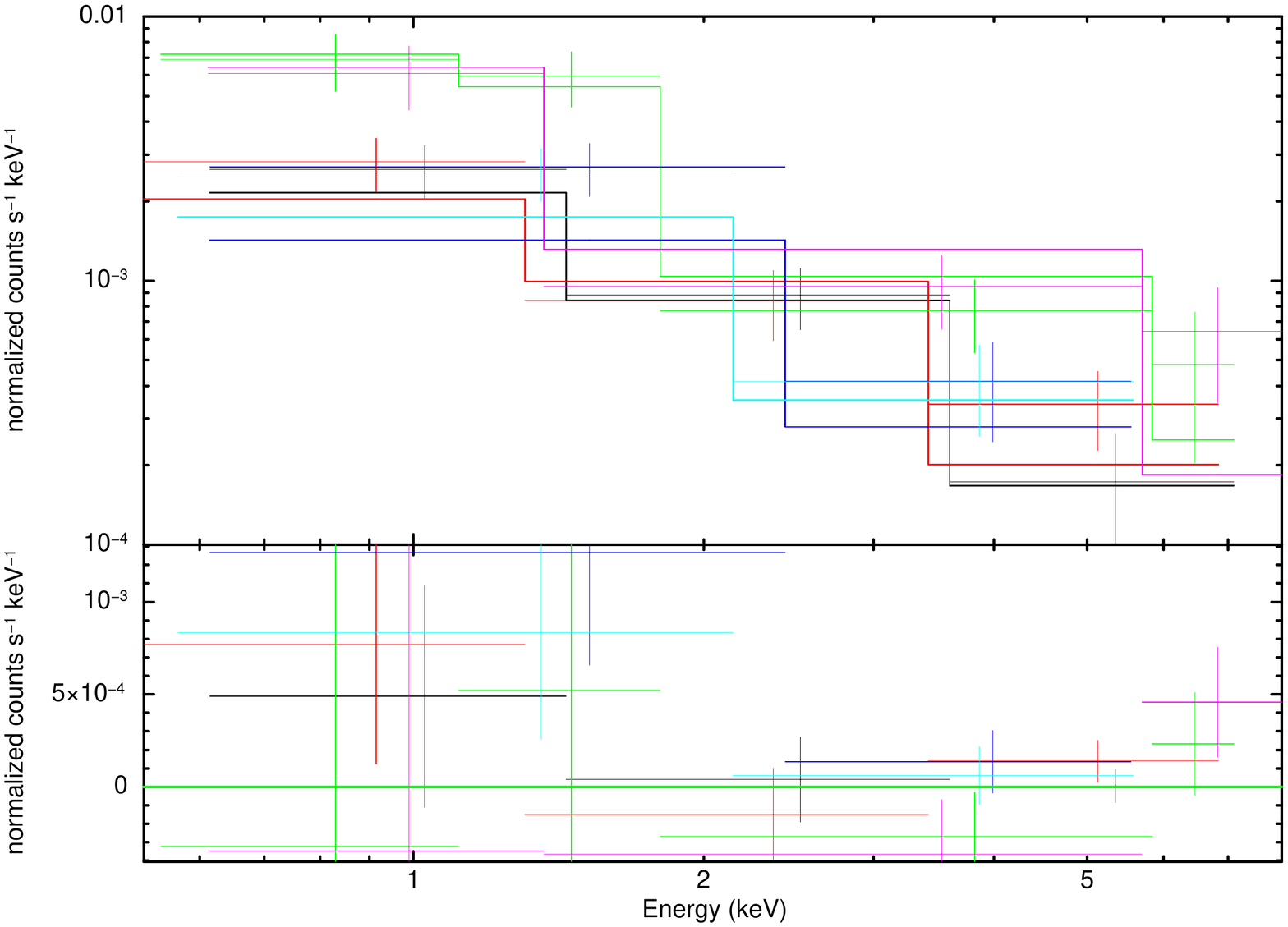}
\includegraphics[height=0.84\columnwidth]{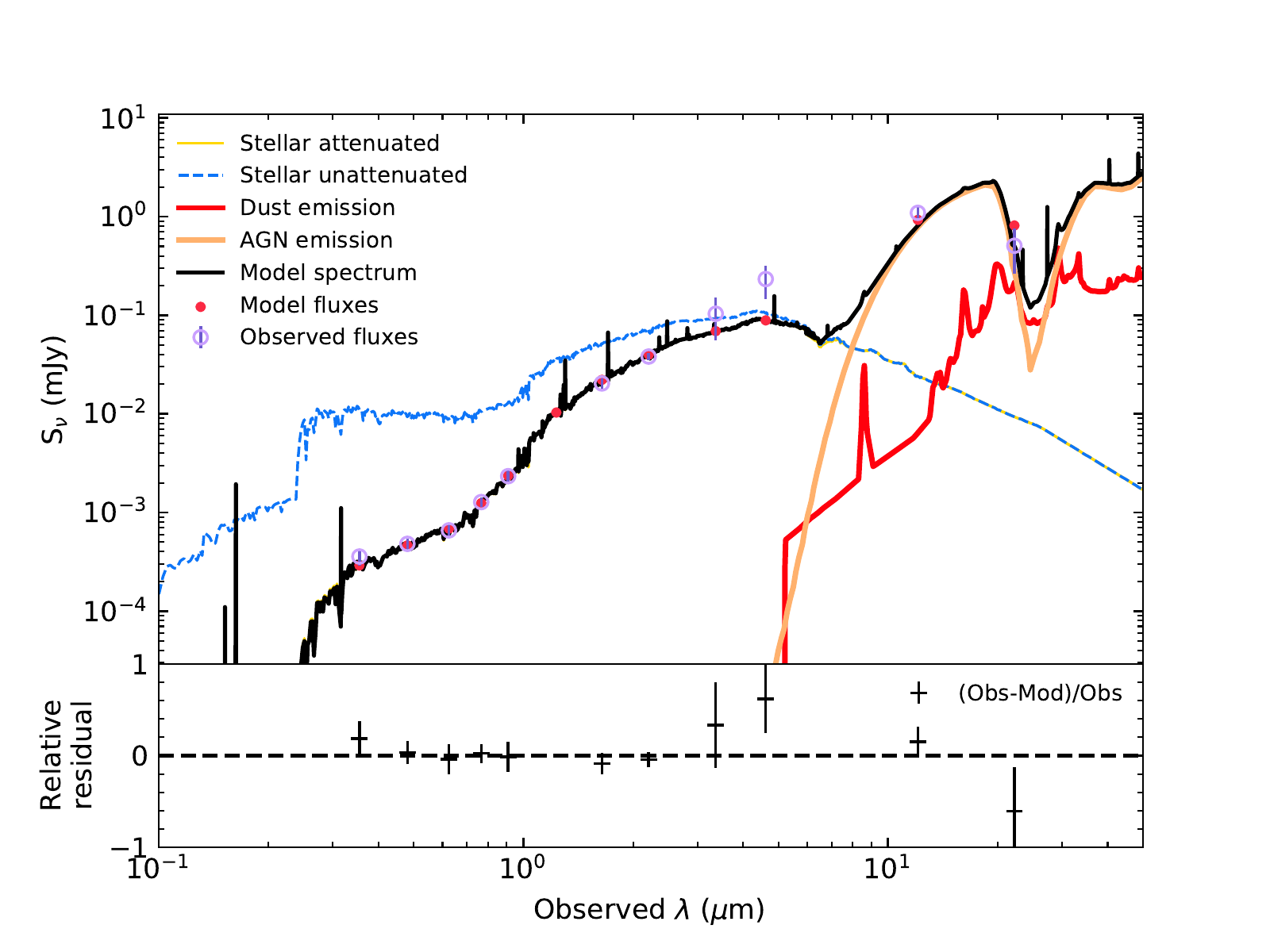}
\caption{J023357.7-054819~(2,2), z=1.60}
\label{}
\end{figure}
\begin{figure}
\includegraphics[height=0.85\columnwidth]{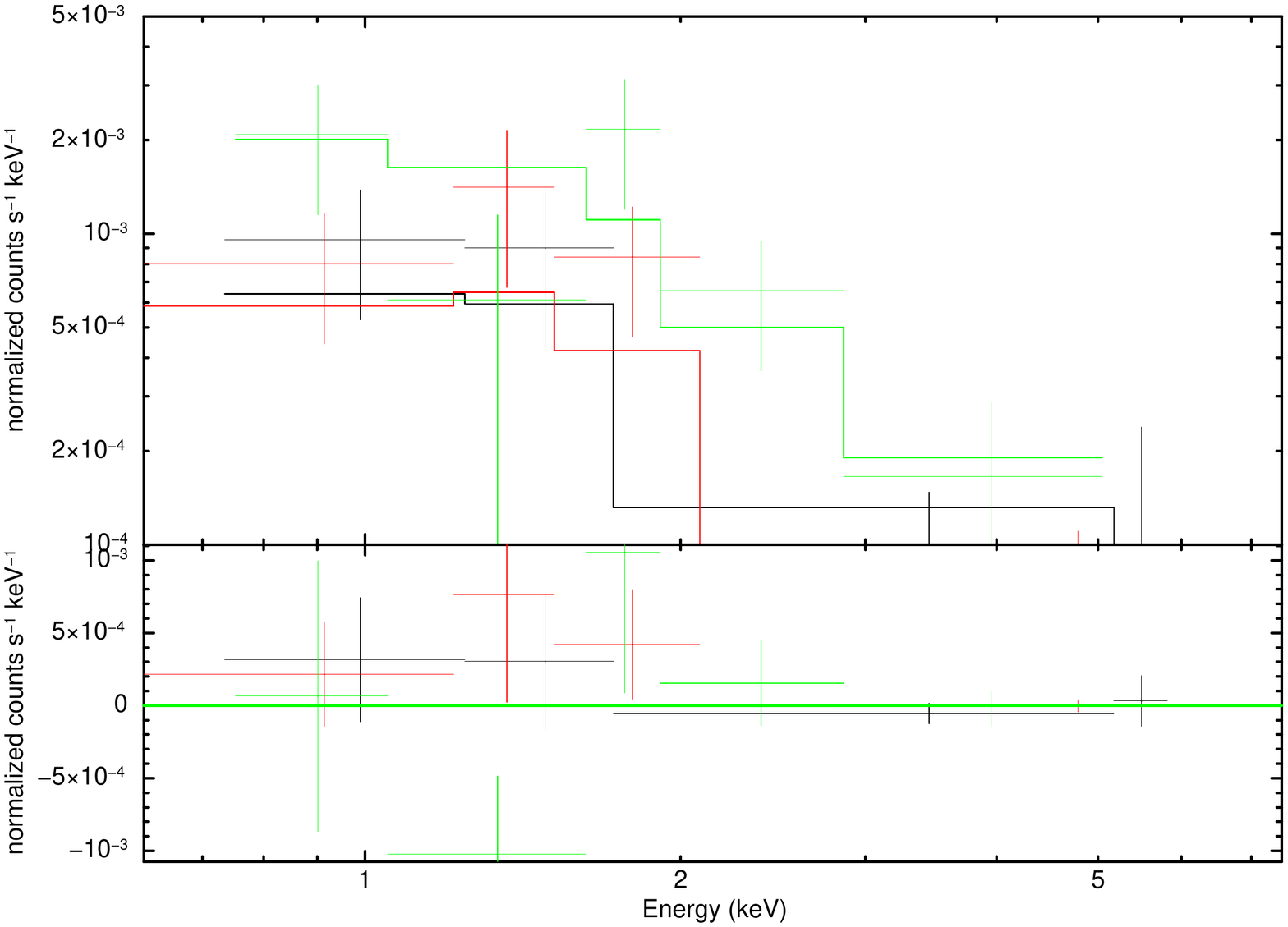}
\includegraphics[height=0.84\columnwidth]{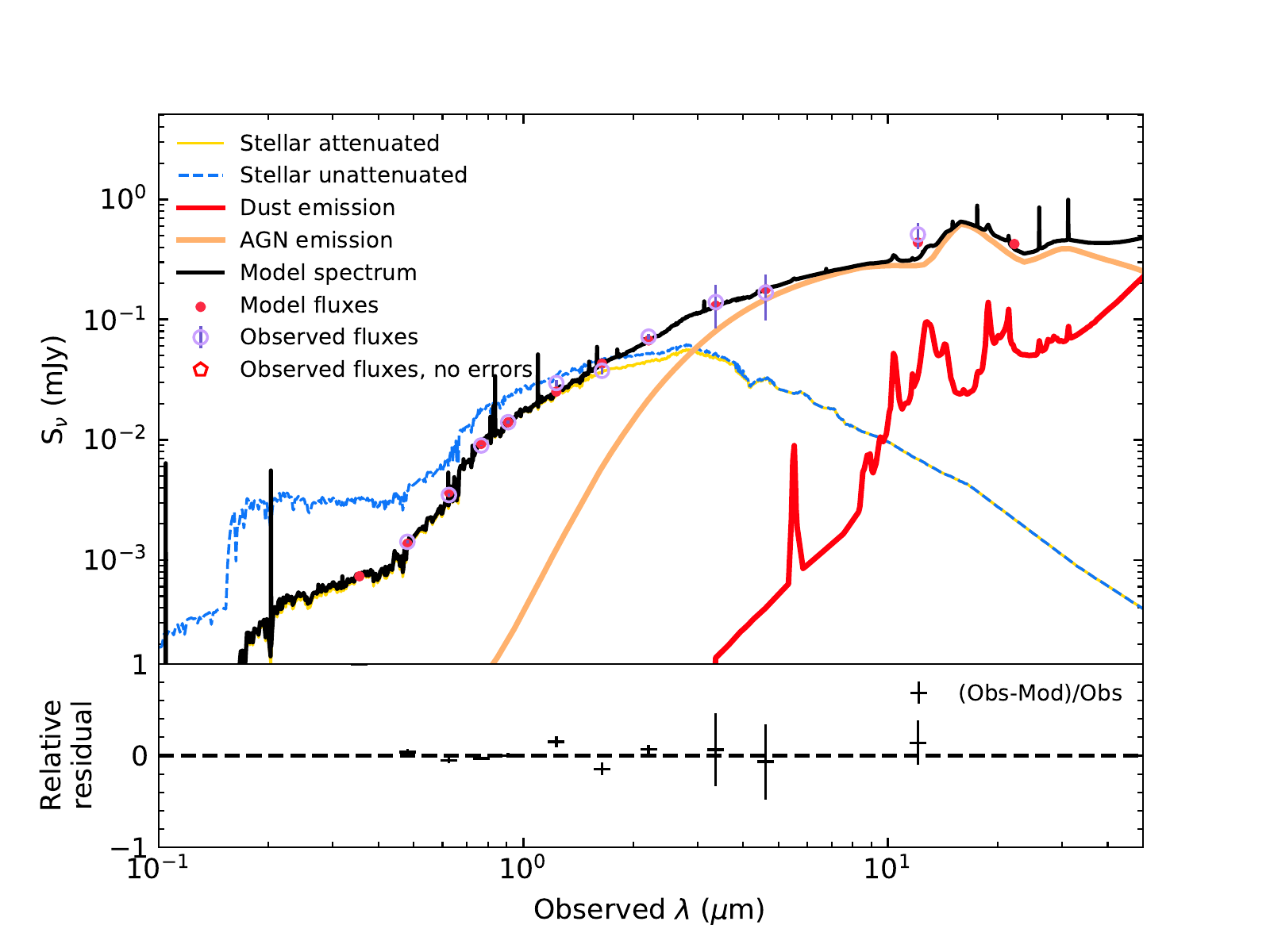}
\caption{J020335.0-064450~(2,2), z=0.67}
\label{}
\end{figure}
\begin{figure}
\includegraphics[height=0.85\columnwidth]{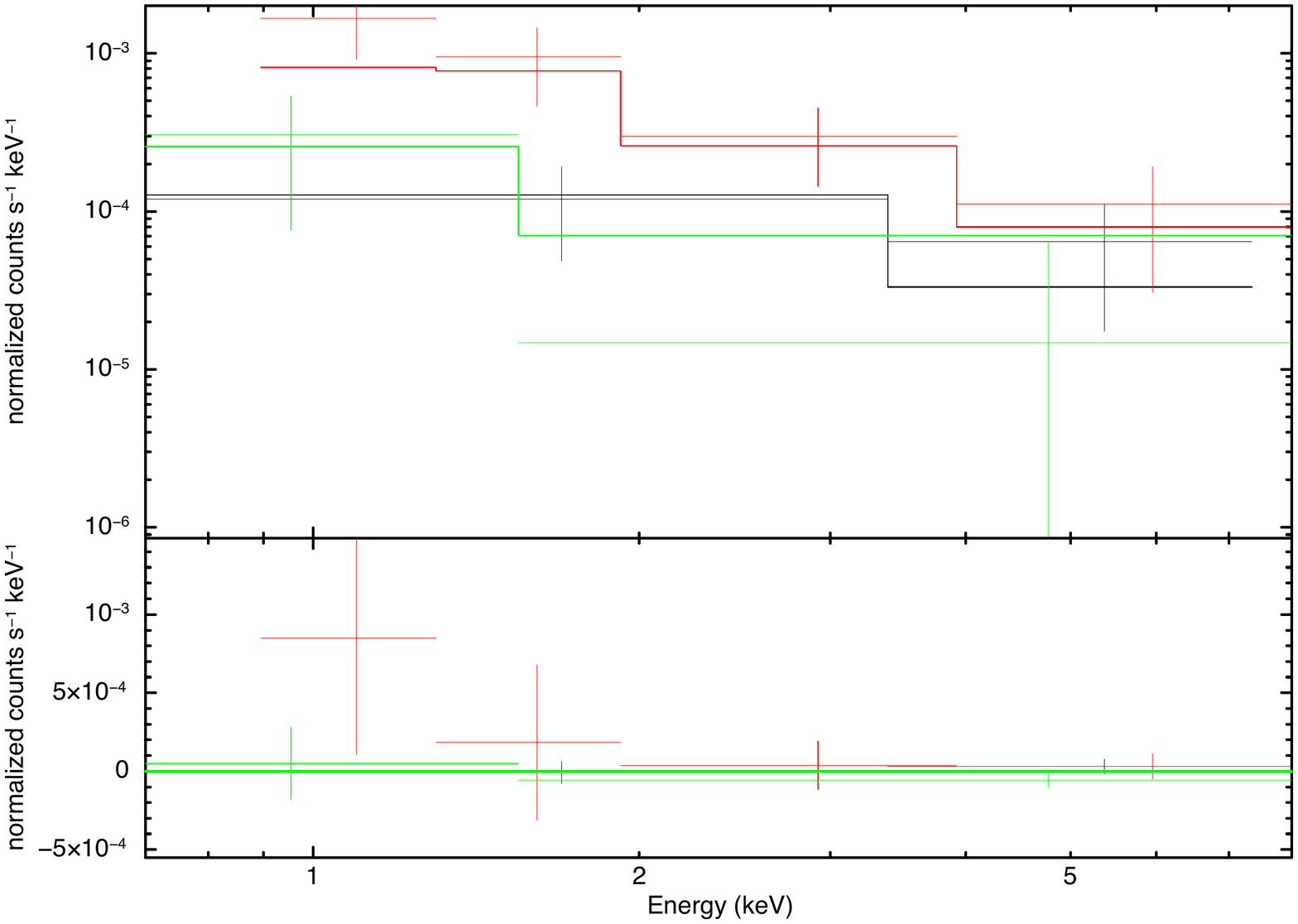}
\includegraphics[height=0.84\columnwidth]{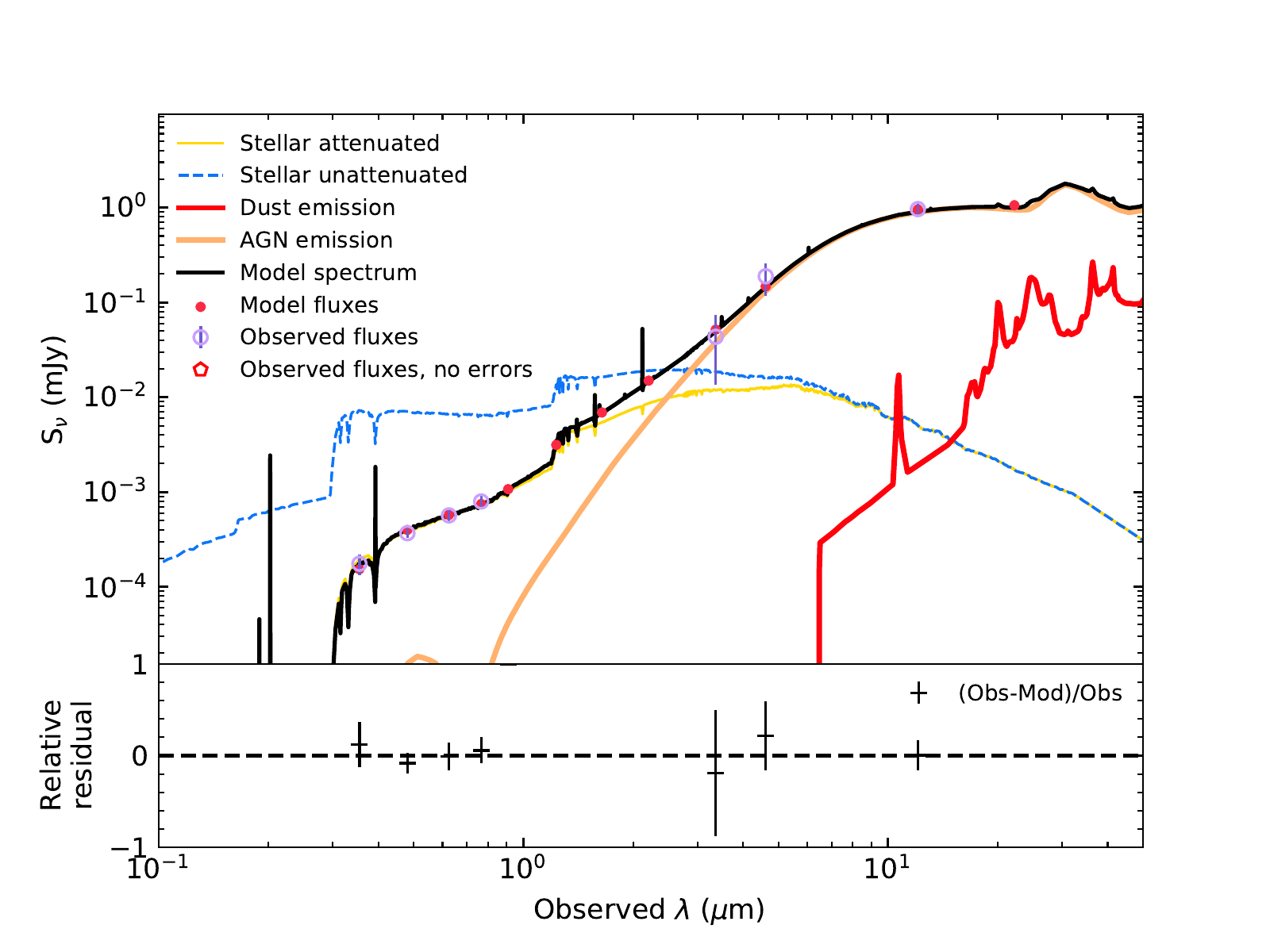}
\caption{J020823.4-040652~(1,2), z=2.23}
\label{}
\end{figure}
\begin{figure}
\includegraphics[height=0.85\columnwidth]{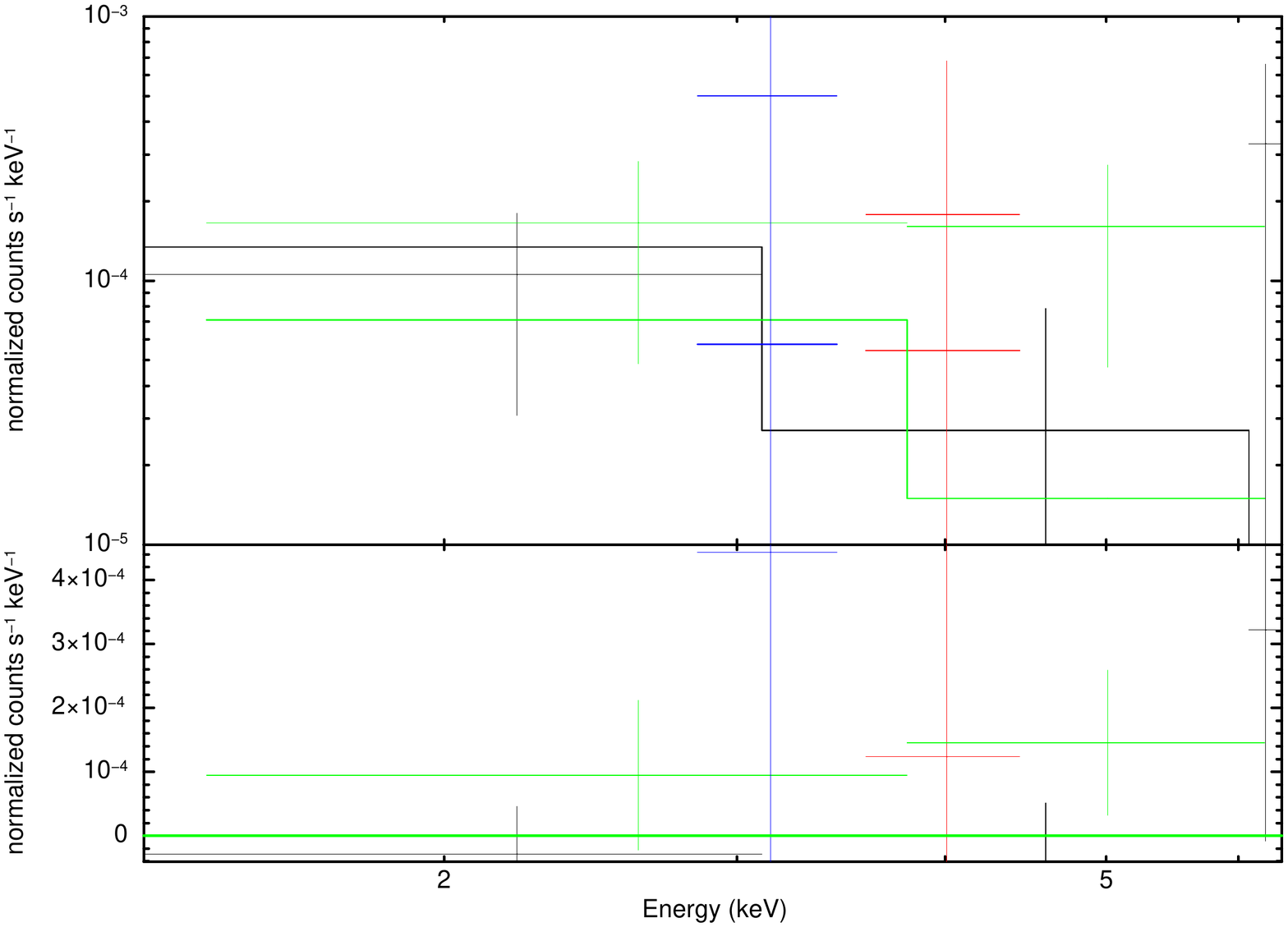}
\includegraphics[height=0.84\columnwidth]{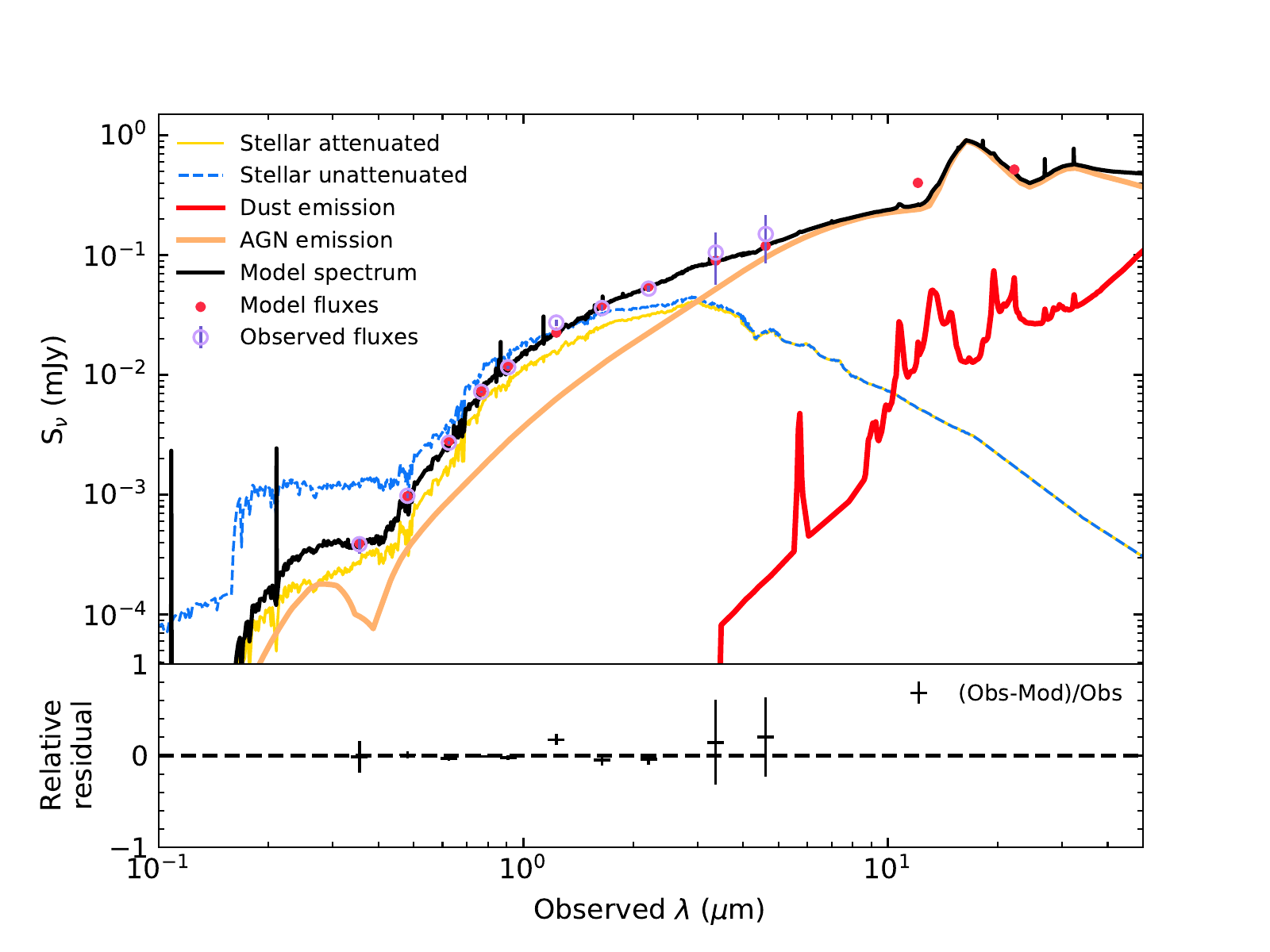}
\caption{J020135.4-050847~(1,2), z=0.73}
\label{}
\end{figure}
\begin{figure}
\includegraphics[height=0.85\columnwidth]{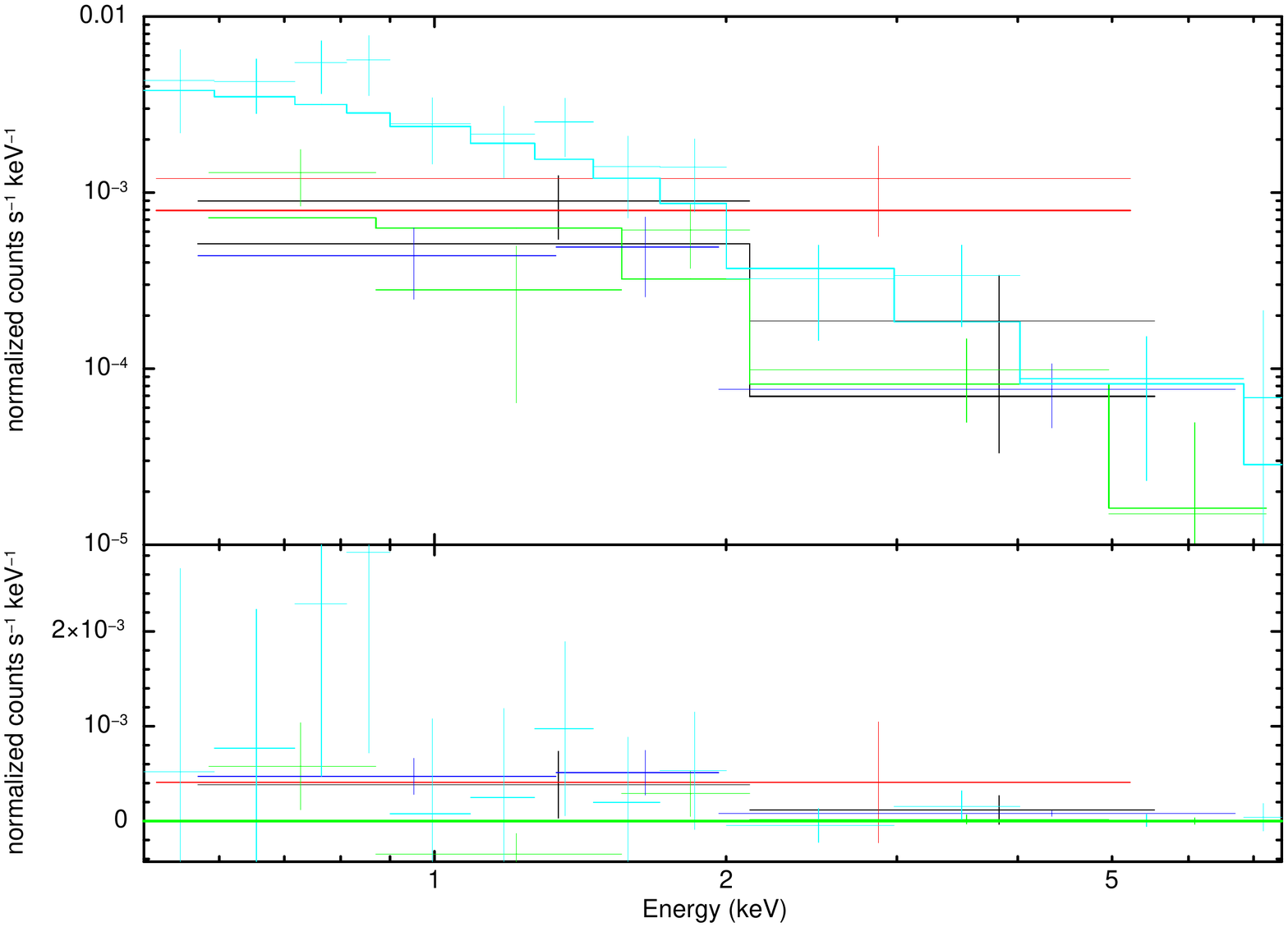}
\includegraphics[height=0.84\columnwidth]{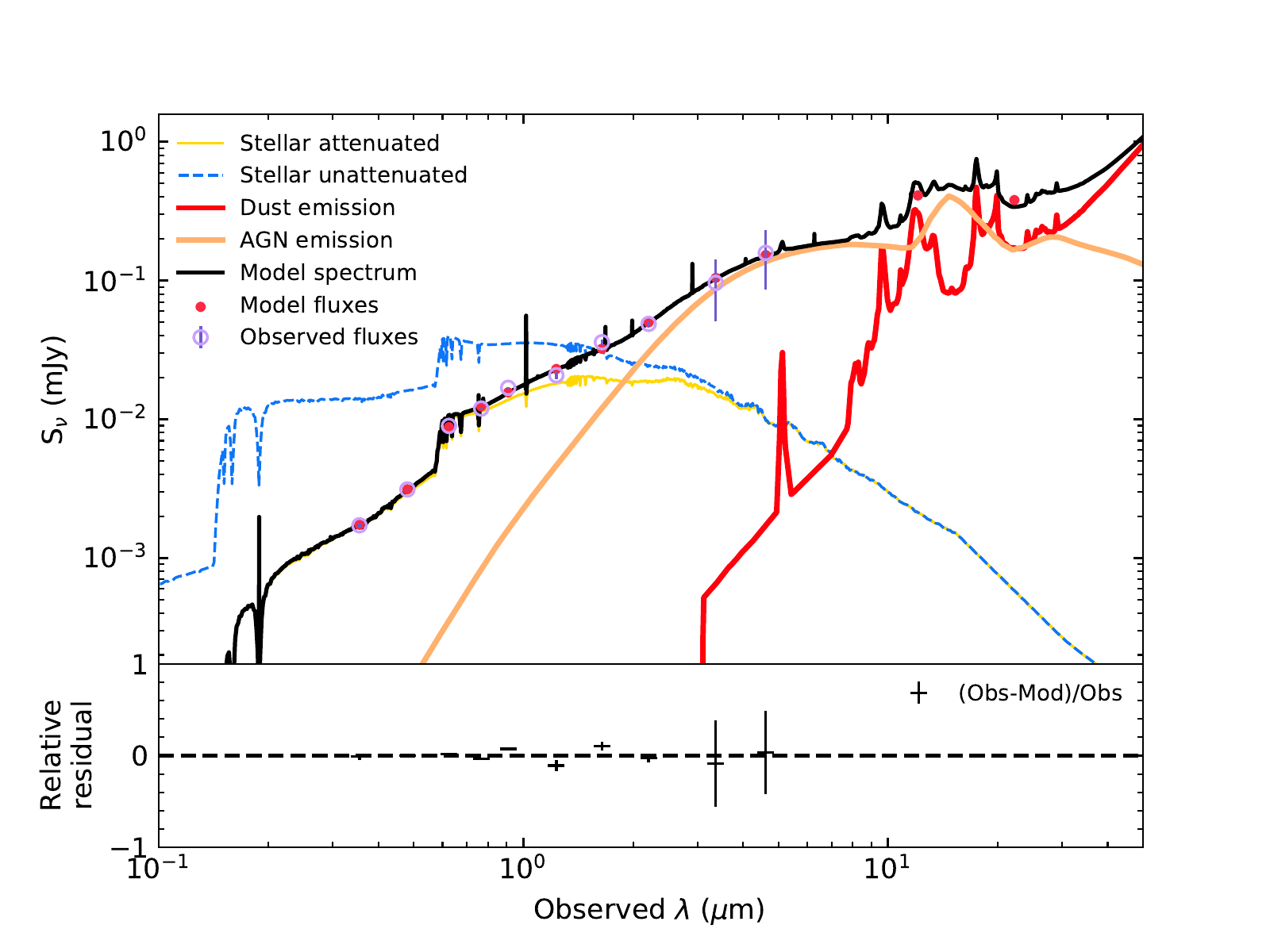}
\includegraphics[height=0.74\columnwidth]{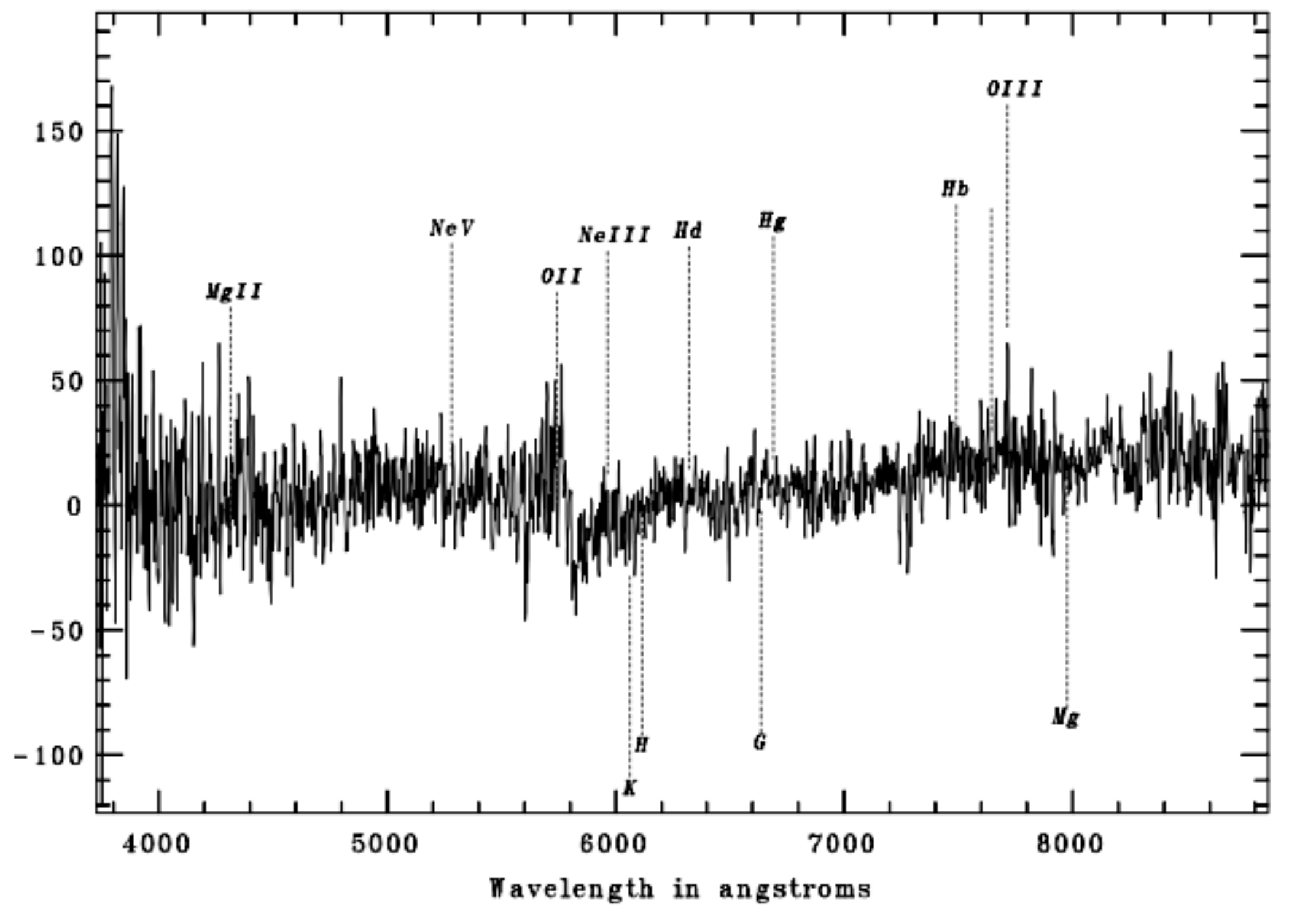}
\caption{J021837.2-060654~(2,2,0), z=0.943}
\label{}
\end{figure}
\begin{figure}
\includegraphics[height=0.85\columnwidth]{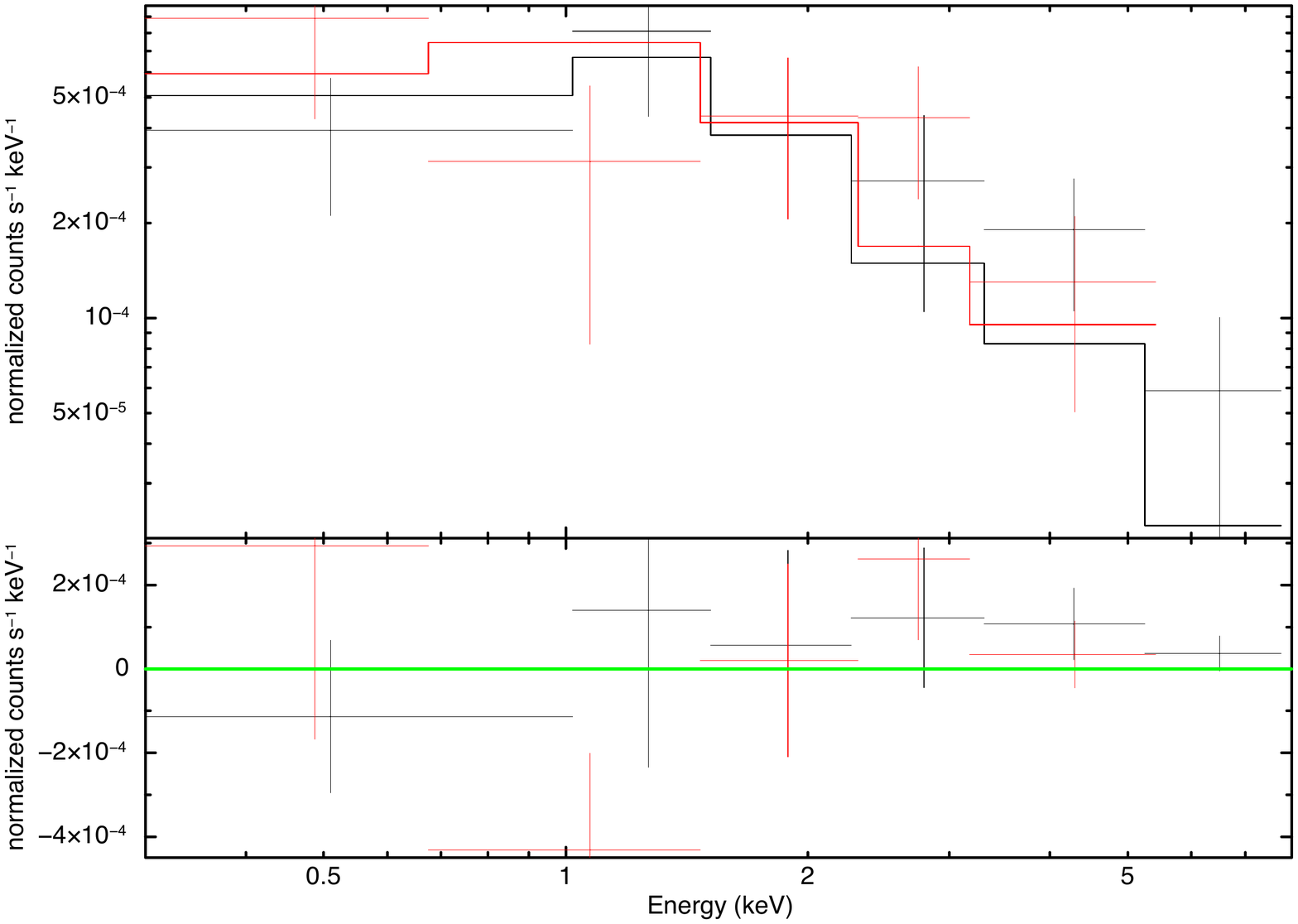}
\includegraphics[height=0.84\columnwidth]{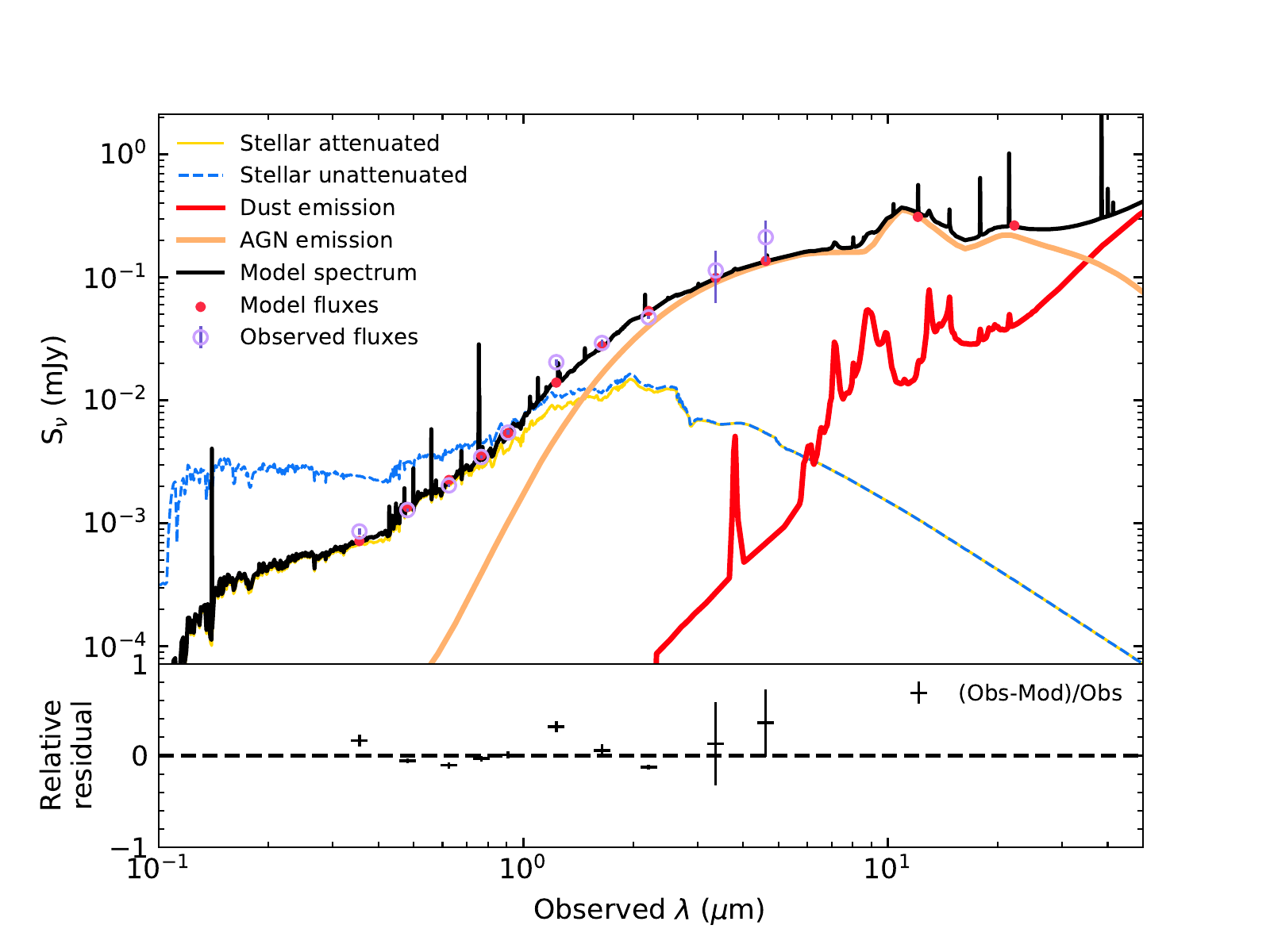}
\includegraphics[height=0.74\columnwidth]{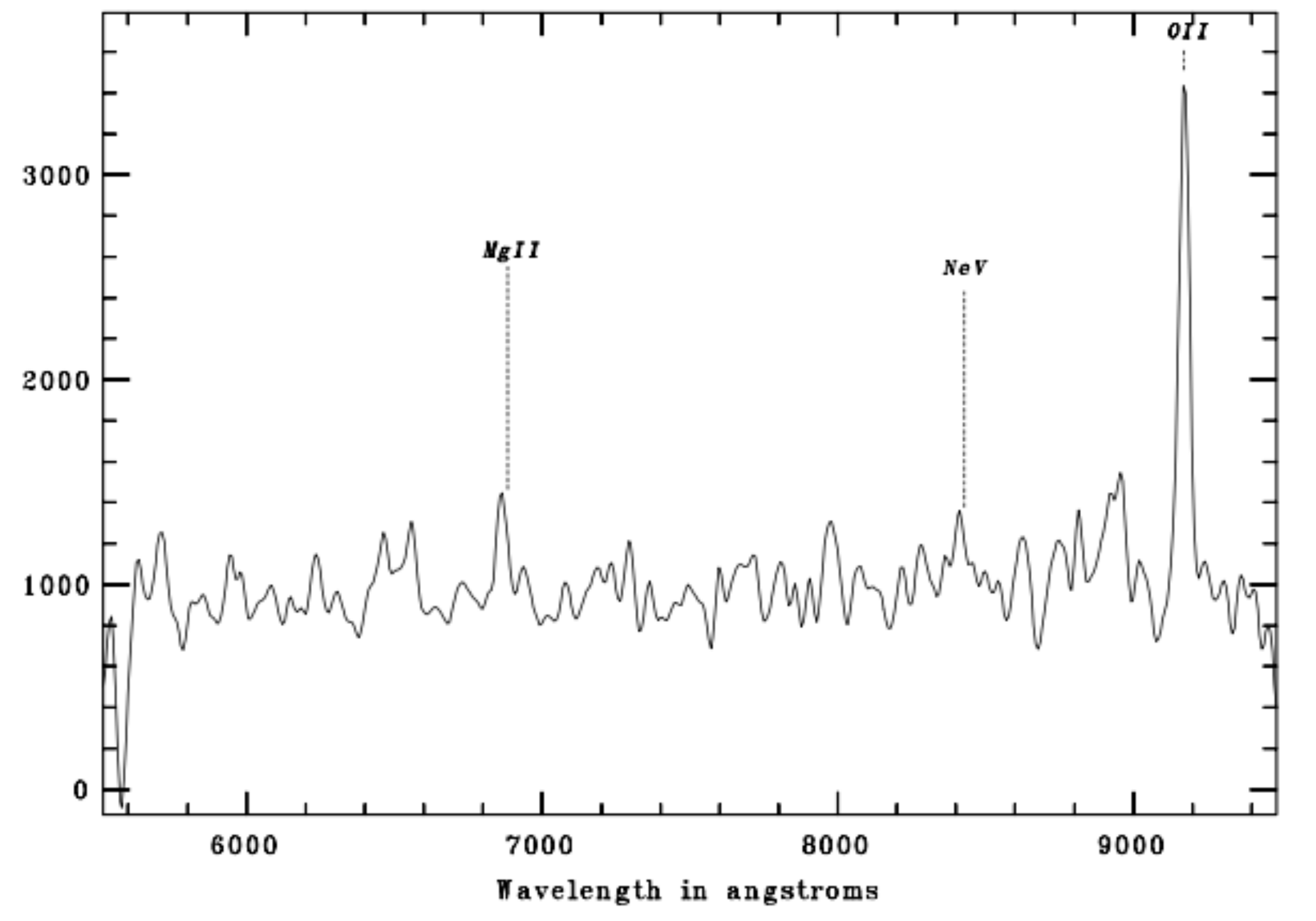}
\caption{J022149.9-045920~(2,2,2), z=1.461}
\label{}
\end{figure}
\begin{figure}
\includegraphics[height=0.85\columnwidth]{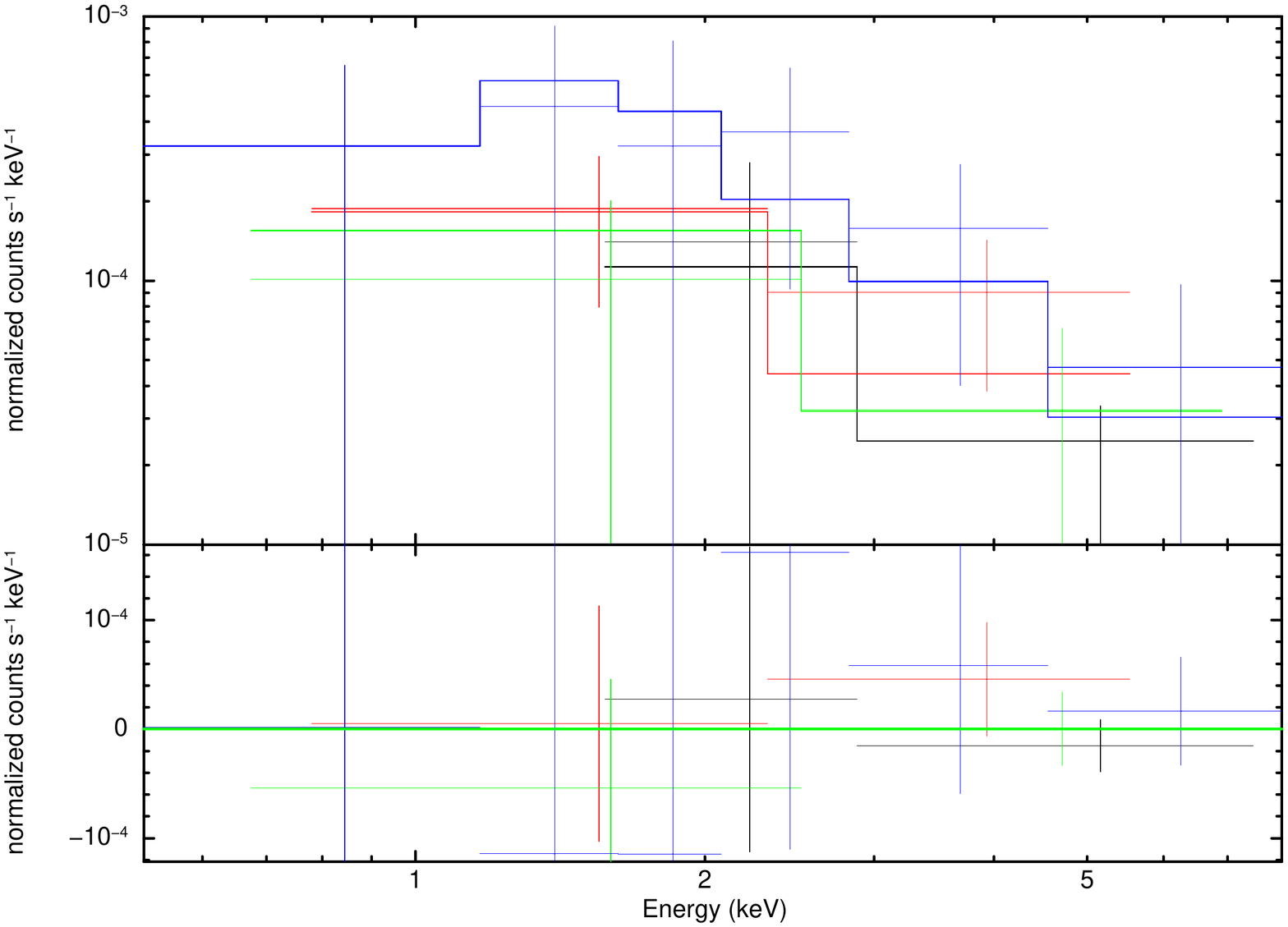}
\includegraphics[height=0.84\columnwidth]{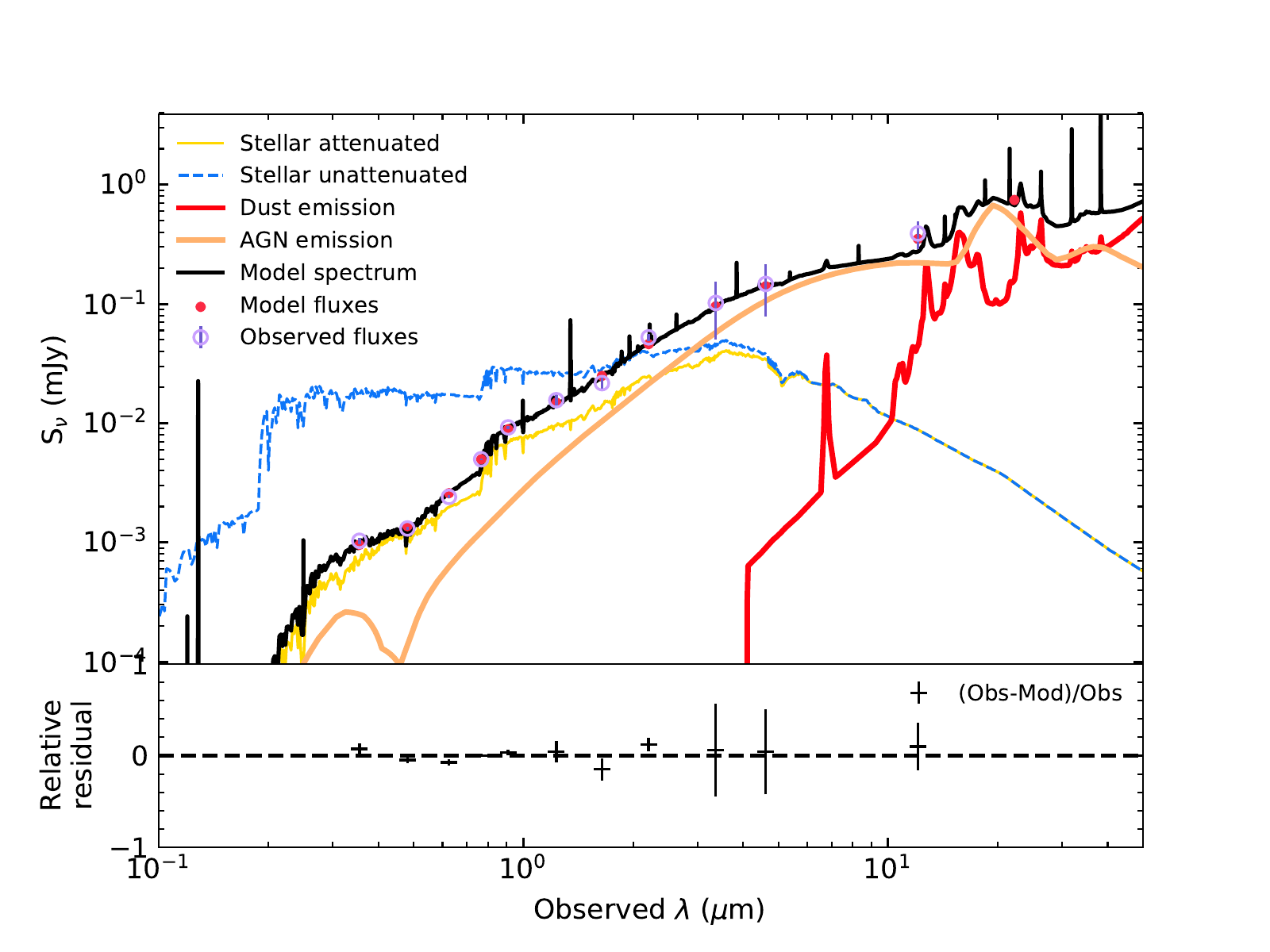}
\caption{J020311.3-063534~(2,2), z=1.05}
\label{50}
\end{figure}

\end{document}